%% file: Thesis.tex
\documentclass[a4paper, 11pt, twoside]{Thesis}  
\graphicspath{{Figures/}}  

\usepackage[sectionbib]{natbib}
\usepackage{xspace}
\usepackage{amssymb}
\usepackage{amsmath}
\usepackage{graphicx}
\usepackage{verbatim} 
\usepackage{fancyhdr}
\usepackage{floatflt}

\def\aj{AJ}%
\def\araa{ARA\&A}%
\def\apj{ApJ}%
\def\apjl{ApJL}%
\def\apss{Ap\&SS}%
\def\aap{A\&A}%
\def\icarus{Icarus}%
\def\mnras{MNRAS}%
\def\pre{Phys.~Rev.~E}%
\def\prl{Phys.~Rev.~Lett.}%
\def\nat{Nature}%
\begin{document}

\title  {Monte Carlo models of dust coagulation}
\authors  {\texorpdfstring
            {\href{zsom@mpia.de}{Andr\'as Zsom}}
            {Andr\'as Zsom}
            }

\maketitle

\newpage

\thispagestyle{empty}
\begin{table}
\begin{center}
\begin{tabular}{c}   
\sffamily\Huge{\bfseries{Dissertation}}\\\\
\sffamily\bfseries{submitted to the}\\
\sffamily\bfseries{Combined Faculties for the Natural Sciences and for Mathematics}\\
\sffamily\bfseries{of the Ruperto-Carola University of Heidelberg (Germany)}\\
\sffamily\bfseries{for the degree of}\\\\
\sffamily\Large\bfseries{Doctor of Natural Sciences}\\\\
\\\\\\\\\\\\\\\\\\\\\\\\\\\\
\sffamily\bfseries{presented by}\\\\
\sffamily\Large{\bfseries{Andr\'as Zsom}}\\\\
\sffamily\bfseries{born in P\"usp\"oklad\'any (Hungary)}\\\\
\sffamily\bfseries{Oral examination: 18, October, 2010}\\
\end{tabular}
\end{center}
\end{table}




%


\frontmatter	  

\addtotoc{Abstract}  
\abstract{
\addtocontents{toc}{\vspace{1em}}  

The thesis deals with the first stage of planet formation, namely dust coagulation from micron to millimeter sizes in circumstellar disks. For the first time, we collect and compile the recent laboratory experiments on dust aggregates into a collision model that can be implemented into dust coagulation models. We put this model into a Monte Carlo code that uses representative particles to simulate dust evolution. Simulations are performed using three different disk models in a local box (0D) located at 1 AU distance from the central star. We find that the dust evolution does not follow the previously assumed growth-fragmentation cycle, but growth is halted by bouncing before the fragmentation regime is reached. We call this the bouncing barrier which is an additional obstacle during the already complex formation process of planetesimals. The absence of the growth-fragmentation cycle and the halted growth has two important consequences for planet formation. 1) It is observed that disk atmospheres are dusty throughout their lifetime. Previous models concluded that the small, continuously produced fragments can keep the disk atmospheres dusty. We however show that small fragments are not produced because bouncing prevents fragmentation. 2) As particles do not reach the fragmentation barrier, their sizes are smaller compared to the sizes reached in previous dust models. Forming planetesimals from such tiny aggregates is a challenging task. We decided to investigate point 1) in more detail. A vertical column of a disk (1D) is modeled including the sedimentation of the particles. We find that already intermediate levels of turbulence can prevent particles settling to the midplane. We also find that, due to bouncing, the particle size distribution is narrow and homogenous as a function of height in the disk. This finding has important implications for observations. If it is reasonable to assume that the turbulence is constant as a function of height, the particles observed at the disk atmospheres have the same properties as the ones at the midplane.
}

\cleardoublepage

\addtotoc{Zusammenfassung}  
\zusammenfassung{
\addtocontents{toc}{\vspace{1em}}  

Diese Arbeit befasst sich mit der fr{\"u}hesten Phase der Planetenentstehung, n{\"a}mlich der Koagulation von mikrometer- hin zu millimetergro\ss{}en Staubpartikeln in zirkumstellaren Scheiben. Als erste Studie dieser Art simulieren wir die Staubentwicklung in `representative particle' Monte-Carlo-Simulationen unter Verwendung eines Kollisionsmodells, das die neuesten Laborexperimente ber{\"u}cksichtigt. Die Simulationen verwenden drei verschiedene Scheibenmodelle in einer lokalen Box (0D) in einem Abstand von 1 AU vom Zentralstern. Unsere Ergebnisse zeigen, dass die Staubentwicklung nicht dem bislang angenommenen Wachstums-Fragmentations-Zyklus folgt, sondern dass das Wachstum von abprallenden St{\"o}\ss{}en aufgehalten wird, bevor es das Fragmentationsregime erreicht. Wir bezeichnen dies als `bouncing barrier', ein weiteres Hindernis im ohnehin schon komplexen Entstehungsprozess von Planetesimalen. Die Abwesenheit des Wachstums-Fragmentations-Zyklus und das unterbundene Teilchenwachstum haben zwei wichtige Konsequenzen f{\"u}r die Entstehung von Planeten: 1) Beobachtungen zeigen, dass die Atmosph{\"a}ren von Scheiben w{\"a}hrend ihrer gesamten Lebenszeit staubig sind. Bisherige Modelle folgerten dass kontinuierliche Fragmentation diese kleinen Staubteilchen produziert und dadurch die Scheibenatmosph{\"a}re ``staubig" h{\"a}lt. Unsere Ergebnisse zeigen jedoch, dass kleine Fragmente gar nicht erst produziert werden, weil die Fragmentationsgrenze nicht erreicht wird. 2) Da Teilchen die Fragmentationsbarriere nicht erreichen, bleiben sie kleiner als in bisherigen Modellen. Die Entstehung von Planetesimalen aus solch kleinen Staubaggregaten ist eine herausforderungsvolle Aufgabe. Wir haben uns mit Punkt 1) n{\"a}her befasst. Hierzu modellieren wir einen vertikalen Schnitt (1D) durch die Scheibe unter Ber{\"u}cksichtigung von Staubsedimentation. Unsere Ergebnisse zeigen, dass schon eine moderat ausgeprŠgte Turbulenz die Sedimentation zur Mittelebene unterbinden kann. Des Weiteren fanden wir heraus, dass die Verteilung der Teilchengr{\"o}\ss{}e schmal und eine homogene Funktion der H{\"o}he {\"u}ber der Mittelebene ist. Dies hat wichtige Auswirkungen f{\"u}r Beobachtungen: Unter der Annahme, dass die Turbulenz h{\"o}henunabh{\"a}ngig ist, haben die in der Scheibenatmosph{\"a}re beobachteten Teilchen dieselben Eigenschaften wie diejenigen in der Mittelebene.

}

\cleardoublepage  

\setstretch{1.3}  


\pagestyle{fancy}  

\lhead{\emph{Contents}}  
\tableofcontents  

\setstretch{1.2}  

\pagestyle{empty}  

\addtocontents{toc}{\vspace{2em}}  

\mainmatter	  
\pagestyle{fancy}  


\fancyfoot[RO]{\setlength{\unitlength}{1mm}
\begin{picture}(0,0)
\put(1.9,-5){\includegraphics[angle=90,height=4.4cm]{movie2/pic\thepage.ps}}
\end{picture}}

\fancypagestyle{plain}{%
\fancyfoot[RO]{\setlength{\unitlength}{1mm}
\begin{picture}(0,0)
\put(1.9,-5){\includegraphics[angle=90,height=4.4cm]{movie2/pic\thepage.ps}}
\end{picture}}
\rhead{}
\lhead{}
\fancyfoot[RE]{}
\renewcommand{\headrulewidth}{0pt}
\renewcommand{\footrulewidth}{0pt}}

\input{./Chapters/Chapter1} 

\input{./Chapters/Chapter2} 

\input{./Chapters/Chapter3} 

\input{./Chapters/Chapter4} 

\input{./Chapters/Chapter5} 


\input{./Chapters/Chapter7} 






\pagestyle{plain}
\clearpage

\addtocontents{toc}{\vspace{2em}}  
\clearpage
\acknowledgements{
\addtocontents{toc}{\vspace{1em}}  
I am grateful to many people who helped me in many different ways during my PhD.

First of all, I'd like to say a big thanks to my supervisor, Kees Dullemond. Working with him is a lot of fun and very stimulating. It happened often during the first year of my PhD that I went to his office to talk and left hours later with several pages of project ideas and a head buzzing with questions. I hope I successfully acquired his way of thinking that will help me tremendously to do my own research later on. Most importantly perhaps, he taught me that doing research should be about fun because any research is the most successful if one does it with joy and enthusiasm. 

I'd like to thank Thomas Henning for his advices and support and for giving me an amazing opportunity to start a new journey in science.

Special thanks goes to the group (Tilman Birnstiel, Chris Ormel, S\'andor Zsolt and Frithjof Brauer) for enlightening discussions and help I get from them during my PhD. Til sometimes made me -- with a single critical question or comment -- reevaluate the direction of my work and improve it significantly. Chris is a very valuable collaborator. He has a very analytical and precise way of thinking which nicely counterbalances my (hopefully occasional) sloppiness. 

My time at the MPIA was emotionally very intense, it was not always happy. I gained and lost a very special friend, Frithjof Brauer. Apart from discussing science with him (during which he made me reconsider many aspects of my work), he continuously challenged and shaped the way I think about life, universe, and everything. His loss was, and still is, a big tragedy for me and I miss him very much. 

During the three years I visited Braunschweig countless times where I was always very warmly welcomed. These visits were incredibly productive and also a lot of fun (the first -- and probably the last -- time I had BBQ during winter, and of course the retreat with the Braunschweig group is very memorable). For these and for much more I'm grateful to J{\"u}rgen Blum and Carsten G{\"u}ttler especially, and to everyone I met in the Spacebar.

I'd like to thank Til for translating the thesis abstract to german, and to J{\"u}rgen for proofreading it. Also, I thank Ralf Klessen, Thomas and Kees for being in my IMPRS thesis committee and examination committee, special thanks to Ralf and Kees for being the referees of this thesis. I also thank Mario Trieloff for being in my examination committee. 

I thank my office mates for providing a very pleasant working atmosphere. I also thank the (IMPRS) students of the MPIA. I know I didn't participate in many events, but when I did, I had a great time.

I owe gratitude to Lilo Schleich, Ina Beckmann, Ingrid Apfel, Marco Piroth, Uli Hiller, Frank Witzel, Heide Seifert, the ladies at the canteen and to many others for making my life at the MPIA smooth. I also thank the Bergbahn conductors for providing me a very special way to the institute and for letting me practice my bad german with them.

I am grateful to many colleagues at the MPIA and around the world who were interested in my work, and helped me to understand the principles of planet formation and to see the big picture. 

I can't express my gratitude enough to Felipe Gerhard, who is the stable point in my life, to whom I can always rely on, who helped me through good times and bad ones, who is next to me when I need him most, who is very patient, kind and understanding with me even when I go through rough periods, and who is responsible for my weekend recreations.

I'd like to say thanks to Christiane, Gerd, Cita, Gilberto, Ines, Friederike, Christian and Sigrid for their kindness and for treating me as part of their family. I'll always remember the playful hours I spent with the babies Leena, Johann and Helene. I treasure every minute of it. 

I am also grateful to my parents for their love and support. I thank my sister, Brigi, for being the cheerful person she is, and taking good care of me whenever I traveled back to Hungary.

And finally, I am grateful to you, dear Reader, you who took the time and effort to go through this piece of work. Thank you very much! And I hope you liked it.

}

\backmatter


\end{document}

%% file: Chapters/Chapter1.tex

\long\def\symbolfootnote[#1]#2{\begingroup%
\def\thefootnote{\fnsymbol{footnote}}\footnote[#1]{#2}\endgroup}

\chapter[Introduction]{Introduction \protect \symbolfootnote[1]{The images at the lower right corners of this thesis show the snapshots of the sedimentation simulation presented in Chapter \ref{sec:comp_br}, Fig. \ref{fig:mass_br_1d-4_or} as a flip-cartoon.}} 
\label{chp:intro}
\lhead{Chapter 1. \emph{Introduction}}  
\rhead{}

Perhaps the most important astronomical discovery of the second half of the last century was the detection of planets orbiting around nearby sun-like stars. This discovery and the since then vastly growing number of exoplanet discoveries further trigger our interest in their formation, their atmospheres, chemical and geophysical evolution, etc. 

The theory of planet formation needs to successfully explain all the different types of exoplanets ranging from the Hot-Jupiters, which revolve around their star in a period of days, the presumably rocky Super-Earth planets, and giant planets orbiting more than 100 AU distance from the central star. This variety and the complex processes of planet formation make the field fascinating. 

The formation of stars starts with the gravitational collapse of a dense molecular cloud. Due to the initial angular momentum of the cloud, most of the infalling matter will not fall directly onto the protostar, but it forms a disk around it. The matter in the dense cloud is twofold: roughly 99\% (in mass) of the matter is present in the form of gases, the rest consists of solid, sub-micron sized dust particles. This solid matter serves as the building blocks of planets. We see that the way from the sub-micron sized particles ($10^{-12}$ g) to a planet of several 1000's of kilometers in radius ($10^{28}$ g) is a long one (see Sec. \ref{sec:formation} for more details). We do not attempt to go the whole way in this thesis, but rather concentrate on the initial stages of growth until millimeter-centimeter (1 g) in size.

To understand planet formation and dust evolution, first we need to know the typical physical conditions of the environment where it happens, i.e., the properties of protoplanetary disks. Our solar system provides only vague clues regarding its formation history, as the solar nebula has long dissipated: the distribution and properties of the planets, minor bodies, satellites and meteorites all contribute to our understanding. These results are reviewed in Sec. \ref{sec:solars}. Observations of existing protoplanetary disks in all wavelengths from UV to millimeter revealed the basic properties of planet-forming disks such as their typical masses, lifetimes, physical conditions (density and temperature structure), important processes such as accretion, photoevaporation, etc. We review these processes in Sec. \ref{sec:obs}.

Once the typical physical properties of the disk are known, we need to understand the driving force of particle growth. The solid particles in the gas disk feel the drag from the gas, and have relative velocities. As a rule of thumb, we can say that the relative velocity is increasing with mass until the particles reach a meter in size. Due to the (random and systematic) relative velocities, the particles can collide and grow, therefore it is crucial to understand what sources of relative velocity are there. One should also keep in mind that low velocity collisions produce self-similar so called fractal structures, while intermediate and high velocities lead to deformation (restructure and/or destroy) of the aggregates. Such deformation changes the aerodynamical properties of the particles and thus in turn the relative velocity (see Chapters \ref{sec:vrel} and \ref{subsec:pos}).

We have to know the outcome of each collision: will the aggregates stick, bounce or fragment? During the last fifteen years, a huge amount of laboratory data was collected about the collisions of dust aggregates. We know that the monomers (sub-micron sized particles) stick together due to surface forces (van der Waals attraction for silicates and molecular dipole interaction for ices). We can measure the strength of the surface forces, furthermore it is possible to produce fractal aggregates and perform experiments with porous, but non-fractal structures in a wide velocity range (from mm/s to several 10's of m/s, see Sec. \ref{sec:experiments}).

The laboratory experiments are crucial for our understanding of the microphysics of collisions as we cannot gain information about it in any other way. The `molecular dynamics' models can also be used to simulate collisions of aggregates, however these models also rely on measured quantities, such as surface energy, Young's modulus, etc. The difficulty with the experiments is that they produce results which are not always easily usable in theoretical modeling, as the results are rather complex. One of the main goals of this thesis is to collect all the available laboratory data and compile it into a collision model (Chapter \ref{chp:paper2}). 

Once we know what happens during individual collisions, we have to calculate how the entire population of dust particles evolves. Traditional methods integrate the Smoluchowski equation with only one particle property: the particle mass (the equation, and such solvers are introduced and described in Chapter \ref{section-test and results}). Although such methods are fast, thus the whole disk can be modeled, they have difficulty including any additional dust property. Therefore we use a Monte Carlo method (described in Chapter \ref{chp:paper1}), which is flexible and can follow the porosity of the aggregates as well (which is a dust-parameter in the experiment-based collision model), and it is straightforward to include a collision model with arbitrary complexity. The price we pay for this flexibility is the speed of the code. We can only follow dust evolution in a local box (see Chapter \ref{chp:paper3}) or in a 1D vertical column (Chapter \ref{chp:sedi}) so far. 

One should also realize how entangled dust coagulation is with the other ongoing processes in disks. A truly self-consistent model should take into account the gas and dust evolution, how the surfaces of dust grains influence the chemistry and ionization fraction of the gas, how this ionization fraction regulates the coupling of the gas with the magnetic fields, how this coupling determines the strength of turbulence, and finally closing the loop, how the strength of turbulence determines the dust evolution. The opacity of dust influences the observed properties of disks, the dust also sets the temperature of the gas at the main body of the disk (except at the low density upper layers), and the dust particles (if sufficiently concentrated) can influence the gas dynamics as well. Although such a complex model, which takes all these processes self-consistently into account, does not exist yet, the effects of the individual processes are at least approximately known (see Sec. \ref{sec:theory}).

\section{Planet formation in a nutshell}
\label{sec:formation}
Star formation mostly happens in star clusters in which roughly half of the stars are part of a binary or small multiple systems. The formation of an isolated star is easier to understand (no initial fragmentation of the parent cloud onto multiple objects, no environmental effects, such as stellar encounters and strong external radiation, have to be treated), therefore isolated star formation is better-studied and understood than star formation in an active environment. 

Star formation starts with the collapse of the parent cloud. This cloud rotates and has too much angular momentum to collapse directly onto the protostar, therefore a disk forms around the central object. It is observed that matter is accreted from the disk onto the central star \citep{Hartmann1998}. To accrete, angular momentum has to be redistributed within the disk (viscous spreading -- \cite{Shakura1973, Lynden-Bell1974}), or angular momentum has to be lost (e.g., by disk winds -- \cite{Blandford1982, Lubow1994}). These processes happen on a much longer timescale than the orbital period, therefore in many applications it is a reasonable assumption that disks are in quasi-equilibrium. How angular momentum is lost/redistributed is one of the most actively studied area of star formation and the physics of disks (see Sec. \ref{sec:theory} for more details). 

The parent cloud also contains dust particles of sub-micrometer in size \citep{Mathis:1977p88}. It is usually assumed that the dust-to-gas ratio in the interstellar matter is 1:100 by mass. This ratio is also inherited by the disk and this solid material (iron, silicates and ices) is the ingredient of terrestrial planets and the cores of some giant planets.

Currently there are two leading giant planet formation models: disk instability \citep{Boss1997} and core accretion. These models are often thought of being competitive to each other, but one can imagine that these mechanisms operate at different parts of the disk simultaneously. 

The core accretion model  \citep{Mizuno1980, Pollack1996} starts with the assembly of the protoplanetary core. This process converts the sub-micron sized dust particles into a core which has some thousands of kilometers in size. The assembly of the core happens in three steps. 1.) The dust particles grow by two-body interactions where the gravitational interaction between the bodies is negligible, in other words dust coagulation. Coagulation can easily produce aggregates of millimeter in size. 2.) How these aggregates are exactly converted into planetesimals is one of the key problems in planet formation. This step bears from many problems (radial drift barrier, fragmentation barrier, bouncing barrier), which are discussed in detail in the upcoming chapters of this thesis. Planetesimals could form via coagulation if the environment is quiet, meaning that turbulence and radial drift are low \citep{Brauer2008b}. Or millimeter-decimeter sized particles could be concentrated in turbulent eddies and/or in vortices, where these concentrations are further enhanced by self-gravity of the particles. These particles could directly coalesce into kilometer sized planetesimals via many-body interactions (see the discussion in Chapter \ref{subsec:plf}). Such a process cannot be modeled by the methods of this thesis, as we can only follow two-body interactions. 3.) Once planetesimals form, these planetesimals grow further due to gravitational agglomeration. 

Once the core is formed, and gas is still present in the disk, an atmosphere is gathered around the core. Initially the envelope around the core is in hydrostatic equilibrium. The energy gained by the impacting planetesimals and the gravitational potential energy released due to the contraction of the atmosphere is in equilibrium with the energy loss due to convection and radiative diffusion. When, however, a critical mass is exceeded, runaway gas accretion starts. This runaway accretion ends when no gas is present around the planet due to the presence of a gap or disk dissipation. After this stage, the planet goes through a Kelvin-Helmholtz contraction.

The timescales involved are a serious uncertainty of the core accretion model. A typical disk lifetime is $\sim 10^6$ yrs (e.g., \cite{Haisch2001-1} and \cite{Sicilia-Aguilar2006}) during which the core has to be assembled and sufficient amount of gas has to be accreted.

Once the planets are formed, they can migrate in the gas disk \citep{Masset:2008p554, Papaloizou2007} and interact gravitationally due to resonances \citep{Lecar2001-1}. Once the gas disk is dispersed e.g., by photo-evaporation (due to external radiation -- \cite{Adams2004}, and due to radiation from the central star -- \cite{Alexander2006, Gorti2009}), a so-called debris disk can still be present around the young star. 

\section{Observations of planet-forming systems}

The planetary system that we can examine in the greatest detail is our solar system. We know the positions and physical properties of the eight planets, we are able to observe the asteroid belt, minor bodies, and satellites in the solar system, furthermore we can directly examine meteorites in the laboratory, determine their absolute and relative ages. All this information contains hints about the formation history of the solar system.

We can also observe other planet-forming systems in various wavelengths ranging from millimeter to UV, and we can use molecular spectral lines to trace the different molecular species. These observations put constraints on the physical properties of disks. 

\subsection{The solar system}
\label{sec:solars}
We do not know the exact mass of the dust and gas from which the planets in the solar system formed. However, it is possible to derive a lower mass limit using the masses and positions of the solar system planets. This nebula model is called the minimum mass solar nebula model (MMSN) introduced by \cite{Weidenschilling1977a}. 
In the model, the masses of heavy elements of the planets are mixed with hydrogen and helium to reach solar compositions. The solar system is divided into concentric annuli each centered around the planet and extending half way until the next planet. The matter is then spread homogeneously in these annuli to obtain the gas surface density at the location of each planet. Performing these steps we get that the surface density scales as $\Sigma(r) \propto r^{-3/2}$. The most commonly used normalization is \citep{Hayashi1981}
\begin{equation}
\Sigma (r) = 1.7 \times 10^3 \left(\frac{r}{\mathrm{1\mbox{ AU}}} \right)^{-3/2} \mathrm{\mbox{g cm}}^{-2}.
\end{equation}
This is a lower limit of the solar nebula because it assumes that all the solids presented in the disk were incorporated into the planets. The model also assumes that the planets were formed in their current location, which might not be true due to the effects of migration. There are attempts to update this model by taking into account the migration of planets. Such calculations by \cite{Desch2007} found that $\Sigma(r) \propto r^{-2.2}$, although this model is debated in the literature. However, based on millimeter observations of protoplanetary disks, \cite{Andrews2007-1} found that $\Sigma(r)\propto r^{-1/2}$. Their findings are probably more representative at the outer parts of the disks ($\sim 100$ AU) and not at the inner parts of the disks. We conclude that the mass distribution in the early solar system is still very much debated.

The regular planetary satellites of Jupiter (the Galilean satellites) have similar masses, tight prograde orbits which lie close to the equatorial plane of the planet, and they are trapped in mean motion resonances. This suggests that these satellites formed in a disk which surrounded the planet similarly to the primordial solar nebula surrounding the early sun. On the other hand, Saturn has only one big regular satellite, Titan. The differences of the Jovian and Saturnian systems can be explained by the quick truncation of infalling matter in the Jovian system, which is caused by Jupiter opening a gap in the solar nebula. Due to the lower mass of Saturn, the Saturnian disk had a longer lifetime, therefore one single big moon could assemble \citep{Sasaki2010}. 

Other irregular satellites orbit in a larger distance from the host planet, sometimes in retrograde orbits. Such satellites were probably captured by the planet.

The solar system also contains many minor bodies. In the inner solar system the asteroid belt is prominent. The distribution of the semi major axes of the asteroid belt objects is not homogenous, instead, gaps can be observed (Kirkwood gaps) and the asteroid belt can be divided into three regions: inner belt (distance $<$ 2.5 AU -- 3:1 resonance with Jupiter), central belt (distance between 2.5 and 3.81 AU -- 2:1 resonance), and outer belt (beyond 3.81 AU). These asteroids were heavily perturbed by Jupiter therefore they could not assemble into a planet \citep{Petit2001-1}. Their mass is also greatly depleted compared to the primordial mass \citep{Weidenschilling1977a}, and the different asteroid types are radially mixed. It seems that in order to explain all these properties of the asteroid belt, a combination of sweeping secular resonances from the migrating Jupiter and Saturn, and embedded planetary embryos are needed that excite and scatter one another \citep{O'Brien2007}. 

Perhaps more interesting in the context of the primordial solar nebula are the Kuiper Belt Objects in the outer solar system (KBO -- for a review see \cite{Luu2002}). These objects have several classes like the classical KBOs (with low eccentricity), resonant KBOs (in 4:3, 3:2, 2:1 resonance with Neptune -- intermediate eccentricity), and the scattered KBOs (with eccentricities around 0.5). As we see, the Kuiper belt is dynamically excited by Neptune. Two other properties of the Kuiper belt are important. 1.) The Kuiper belt has a mass of 0.1 Earth mass, which is surprisingly low. Accretion models predict that a mass of 10 Earth mass must have existed to explain the growth of the objects we see now (see e.g. \cite{Kenyon1998}). 2.) The Kuiper belts ends near 50 AU. \cite{Gomes2004} examined two scenarios which could explain the low total mass of KBOs: the vast majority of KBOs crossed orbits with Neptune, therefore their orbits were scattered; fragmentation into dust removed most of the mass of the Kuiper belt. They concluded that none of these two scenarios are likely, instead they proposed that the protoplanetary disk possessed an edge about 30 AU. This edge is responsible for stopping the outward migration of Neptune, and during this migration, Neptune could have pushed the KBOs outwards \citep{Levison2008}.

We have the unique possibility to measure the ages of solar system rocks accurately with the means of radioactive dating. Although the accuracy of dating in the solar system is unmatched with any astronomical observations, one must keep in mind that this accuracy is achieved for a single (and rather special) system. One can calculate absolute and relative ages of meteorites using long-lived and short-lived radionuclides respectively (for a review see \cite{Russell2006, Wadhwa2007}). 

Primitive meteorites (chondrites) were never differentiated, they are relatively unaltered since their time of formation, therefore they preserve the early history of the solar system. Most primitive meteorites contain small inclusions which were heated to high temperatures, during which the amorphous dust became crystallized. Spectral signatures of crystalline dust in protoplanetary disks are common, such crystallization process was observed `in action' after an outburst of a Sun-like young star, EXLupi \citep{'Abrah'am2009-1}. 

Chondrules are the most abundant type of inclusions, more than 70\% of the volume of primitive meteorites are chondrules. CAIs (calcium aluminum inclusions) are rarer, but they were subject to a more extreme heating event. The formation, specifically the heating mechanism for chondrules and CAIs is the subject of active research \citep{Connolly2006}. The three main hypotheses are: 1.) heating near the young Sun (X-wind model) with strong outward transport; 2.) shock waves in the gas disk; 3.) collisions between planetesimals and/or protoplanets. 

CAIs are the oldest objects in the solar system with 4567.11 $\pm$ 0.16 Myr as determined from $^{207}$Pb-$^{206}$Pb dating \citep{Russell2006}. This age pinpoints a specific event, namely the solidification of CAIs. One has to keep in mind that other events, like the collapse of the Sun's parent cloud happened earlier, but such events cannot be dated precisely. The absolute age determination only recently became accurate enough to reliably determine the time interval between the formation of CAIs and chondrules. Such measurements suggest that some Myr passed between the formation of the oldest CAIs and the formation of chondrules. Absolute age determination methods assume that the abundance of parent and daughter isotopes is only altered via radioactive decay. 

The relative ages between chondrules and CAIs can be determined using another method using short-lived radionuclides such as $^{26}$Al. The method also assumes that the abundance of parent and daughter isotopes is altered only via radioactive decay. Furthermore, it is also assumed that $^{26}$Al was uniformly distributed in the disk (isotopic equilibration), otherwise the differences in the original $^{26}$Al/$^{27}$Al ratio can indicate different local formation environments. The $^{26}$Al method also yields to a relative age of 1-2 Myr between the CAIs and chondrules.

This age spread fits within the observed lifetime of the disk, and (as the chondrules and CAI's coexist in meteorites) it suggests that the formation of planetesimals were either delayed or ongoing for several Myr. A rapid planetesimal formation in less than 1 Myr is not supported by meteoritic data \citep{Russell2006}. 

\subsection{Observations of protoplanetary systems}
\label{sec:obs}
Most of the information about protoplanetary disks is obtained through infrared (IR) measurements. Disks have an IR excess due to the presence of dust particles. These particles can scatter or absorb the radiation of the central star, they also reemit thermal radiation in the IR. Near-IR (NIR) wavelengths map the warm dust close to the star (order of 1 AU or smaller), which originates from the upper layers of the disk. Far-IR (FIR) radiation maps colder dust further away from the star. The spectral energy distribution (SED) can be used to fit the disk parameters (such as disk geometry, dust composition, temperature and density structure -- for a 2D radiative transfer model see \cite{Dullemond2004}, a review about disk modeling can be found in \cite{Dullemond2007}). 

The IR excess of disks can be used to determine the typical lifetime of disks (e.g., \cite{Haisch2001-1} and \cite{Sicilia-Aguilar2006}). One can measure the disk fraction of young stars (e.g., how many percentage of stars have IR excess) in different young clusters and correlate this disk fraction with the estimated age of the cluster. Following this procedure it was concluded that only half of the young stars have disks in a cluster of 2 Myrs age. 

A prominent feature of the IR spectra is the 10 micron feature, which originates from small silicate particles (order of sub-micron in size) at the upper layers of disks. The shape of the 10 micron feature contains valuable information about the composition of these grains \citep{boekel2006}, although some caution is required when interpreting the data \citep{Juh'asz2009}. The most interesting aspect of the 10 micron feature for the topic of this thesis is that it can provide evidence for grain growth \citep{boekel2003}. Some sources have a flatter and broader spectral feature which suggests that in these sources the particles grew to a few micron in size at the upper layers. Theoretical models including particle settling, coagulation, and radiative transfer cannot yet convincingly reproduce this aspect of the 10 micron feature \citep{Dullemond2008}. Sedimentation driven coagulation is the topic of Chapter \ref{chp:sedi}, where these physical processes are discussed in detail.

Optical and NIR scattered light images contain lot of information about disks. The used observational technique differs for different disk inclinations: coronagraphic measurements are used for high and intermediate inclinations; observing optically thick edge-on disks require high spatial resolution, but not high contrast, therefore adaptive optics systems are advantageous; the Orion nebula provides a unique opportunity to observe silhouette disks (the disks appear dark due to the bright background of the nebula). The information that these images provide also depends on the inclination. From face-on disks we can determine the ellipticity of the disk, the dependence of the surface brightness on the radius, and the ratio of scattered and unscattered light (the relative brightness of the disk and the star). From intermediate inclinations (when the central star is still visible), one can determined the relative brightness of the disk and the star, and the outer radius of the disk (if both the upper and lower parts of the nebulae are visible). In case of the edge on disks the inclination can be very precisely determined, and the effective scale-height of dust in the outer disk. A detailed discussion of scattered light images can be found in \cite{Watson2007}.

Millimeter observations are a useful tool to estimate the mass of the solid material in the disk and these observations provide evidence for grain growth at the outer disk. At millimeter and sub-millimeter wavelengths (assuming the Rayleigh-Jeans limit and optically thin material), the observed flux is 
\begin{equation}
F_{\nu} \propto \kappa (\nu) M_d T_d,
\label{eq:Fnu}
\end{equation}
where $\kappa (\nu)$ and $T_d$ is the opacity and the temperature of the dust, $M_d$ is the mass of the dust in the disk. If $T_d$ can be obtained otherwise, and $\kappa(\nu)$ is known, the mass of the dust disk can be calculated. If the opacity has a power-law dependence ($\kappa(\nu) \propto \nu^{\beta}$), then the flux is $F_{\nu} \propto \nu^{\alpha}$ with $\alpha = 2 + \beta$. Using multi-wavelength measurements, we can obtain $\beta$. The $\beta$ parameter in the interstellar matter is $\beta_{\mathrm{ISM}} = 1.8 \pm 0.2$, but mm-observations of disks (most recently by \cite{Ricci2010}) show that $\beta$ is smaller than the ISM value, it is around 0.5--1. There can be several reasons why the $\beta$ parameter is reduced in disks \citep{Draine2006}. Some of the emission might come from optically thick regions, therefore the assumption used in deriving Eq. \ref{eq:Fnu} is not valid. The chemical composition of dust in the disks can be very different from that of the ISM dust. Grain growth can change the size distribution of particles \citep{Birnstiel2010}. The opacity model might not be correct, e.g., the opacity of fractal structures can be quite different from non-fractal, spherical particles. 

Millimeter, sub-mm observations, in a similar way as IR observations, can be used to infer disk lifetimes. Using these measurements, \cite{Andrews2005} obtained the same disk lifetime as from IR measurements. 

UV excess and magnetospheric emission lines can be used to determine accretion rates of disks \citep{Calvet2000-1}. The typical accretion rate for young stars is $10^{-6}$ M$_\odot$ yr$^{-1}$; for stars of age 1 Myr, it is $10^{-8}$ M$_\odot$ yr$^{-1}$. As the accretion rate is decreasing in time, one can infer the disk lifetime based on UV observations to be also a few $10^6$ yrs \citep{Calvet2000-1}. 

These results of disk lifetime are compelling. The IR excess measures the `survival time' of the small dust grains around 1 AU, the sub-mm measurements are based on the presence of dust at large radii, while the accretion signature means that gas can be transported onto the surface of the star directly from the gas disk. These three methods estimate the disk lifetime measuring entirely different disk material, still they obtain the same characteristic timescale. These observations imply that the disk dispersion happens across a wide range of radii in a relatively short time. 


Line emissions of molecular species provide a unique way to determine the physical conditions in disks. Using CO observations, the radial and vertical temperature profile of disks can be determined. The observation of different CO isotopologues revealed that the outer disk is smaller for $^{13}$CO and C$^{18}$O than $^{12}$CO, which suggest that photodissociation is present at the outer disk. The condensation of CO onto grain surfaces can also be observed. These results and the prospects of the Atacama Large Millimeter Array (ALMA) are summarized in \cite{Guilloteau2008}.
 
\section{Dust experiments in the laboratory}
\label{sec:experiments}
An extensive overview of the available laboratory experiments is given in Chapter \ref{sec:exp-review} and in \cite{Blum2008}. Here we shortly review the general properties of these experiments concentrating on the microphysics of aggregates.

Monomers are solid bodies which serve as building blocks of aggregates. Often these monomers are represented as spheres for simplicity, but in general they can have any shape, e.g., ellipsoids or irregular structures. If we assume that the monomers are electrically neutral and non-magnetic, these monomers stick together via surfaces forces (dielectric van der Walls forces for silicates and molecular dipole interactions for ices). The strength of the bond between the monomers can be characterized by the contact force, which can be measured in the lab \citep{Heim1999} and it is given by
\begin{equation}
F_c = 4\pi \gamma_s R,
\end{equation}
where $\gamma_s$ is the specific surface energy and $R$ is the local radius of surface curvature (for monomers of different size it is $R=a_1a_2(a_1+a_2)$, where $a_1$ and $a_2$ are the radii of the monomers). \cite{Poppe2000-1} measured the maximum velocity for sticking between monomers of different sizes impacting onto a smooth surface with different velocities and found that this threshold velocity is around 1 m/s for micron sized silicates, and that it is decreasing with increasing size. 

At intermediate relative velocities, the collision energy of the aggregates can be dissipated via restructuring and the two aggregates stick together. The most effective channel for restructuring is the rolling of two monomers \citep{Dominik:1997p89, Wada2007, Wada2008, Paszun2009}. \cite{Heim1999} also measured the rolling energy between two monomers, which is the energy needed to roll a monomer by 90 degrees, and found that it linearly depends on the local surface curvature of the monomers. 

If we want to break a contact by pulling the monomers apart, we have to pull the two monomers with a force that is higher than the contact force. The interesting feature of the contact area is that it can be stretched further apart such that the distance of the center of mass coordinates of the monomers can actually be larger than the sum of the radii. The elastic energy stored in this `neck' is then transformed into elastic waves and slowly dissipates. Two monomers can also twist and slide, but these motions seem to be less important in collisions \citep{Dominik:1997p89}.

Most of these results were obtained using silicate spheres that are smaller than 1 micron in size, thus a simple shape and a mono-disperse size distribution were used. The picture is more complicated if we assume a size distribution of monomers \citep{Dominik:1997p89} or irregular monomer shapes \citep{Blum:2004p91}. Although we have a qualitative picture how these effects change the particle properties, no exhaustive and general investigations were made. There is also uncertainty regarding the properties of different monomer-materials. It is experimentally not studied how micrometer-sized ice monomers or monomers with organic mantels behave. \cite{Kouchi2002} performed experiments with millimeter sized grains with organic mantel and found that the sticking threshold velocity was several orders of magnitude higher than for same sized silicate grains. Frost mantel also increases the sticking threshold velocity but to a lesser extent than organic mantels \citep{Hatzes1991}.

\section{The importance of dust - theoretical considerations}
\label{sec:theory}
The growth of the particles is governed by the Smoluchowski equation, which is introduced and described in Chapter \ref{section-test and results}. As discussed earlier in this chapter, the particles collide and grow because they have a relative velocity in the gas disk. The strength of the relative velocity depends on the aerodynamical properties (the stopping time) of the aggregates. These properties, as well as the different sources of relative velocities are discussed in Chapters \ref{sec:vrel} and \ref{subsec:pos}. 

It is clear from the previous section that the observations of disks are strongly influenced by the properties of the dust. In this section we review what role the dust plays in various aspects of theoretical modeling of disks and planet formation. 

Dust is the main source of opacity where the temperature is below the evaporation temperature of the dust (below 1500 K). For an individual dust particle the cross-section to radiation at a given wavelength will depend on the particle size, structure (fractal or non-fractal, spherical or more complex shape) and composition. The total opacity at a given location also depends on the particle size distribution. \cite{Semenov2003} describes how to calculate the Rosseland mean opacities for particles as a function of temperature, assuming that the particles follow a modified MRN size distribution \citep{Pollack1985} and that these particles are homogeneous and spherical. It is also possible to calculate opacities of arbitrarily complex aggregates (see e.g. \cite{Shen2008, Min2007}). However, usually the uncertainty coming from the amount and the size of the dust particles is higher than the uncertainty in opacities. Therefore, in most cases approximate opacity formulae are useful. 

The dust also effects the ionization state of the protoplanetary disk. The sources of ionization can be thermal (from the central star), and non-thermal (X-rays, cosmic rays, $^{26}$Al decay). Thermal ionization is effective at the inner disk ($<0.1$ AU from the central star). X-rays and cosmic rays can ionize a layer of thickness with 100 $g/cm^2$ column density on both sides of the disk. The decay of $^{26}$Al provides an ionization source that is present at every location in the disk. Dust affects ionization in two ways. It acts as charge carrier and it provides surface where electrons and ions can recombine \citep{Okuzumi2009}.

Although the ionization fraction is almost negligible, it has a very important affect for angular momentum redistribution in disks. To simulate viscous spreading, usually the description by \cite{Shakura1973} is used, where the turbulent viscosity is parameterized. Although this model is widely used, it does not explain what the physical source of the turbulent viscosity are. The best candidate-mechanism is currently the magneto-rotational instability (MRI), which was first discussed in the context of disks by \cite{Balbus1991-1}. The basis of the instability is that even low ionization fractions make it possible to couple the magnetic field and the gas dynamically. If there are two fluid parcels orbiting at different radii, the magnetic field acts as a spring between these two fluid parcels. This force causes the inner one to be slowed down by the outer one, and the outer one to be sped up by the inner one. As the inner one loses energy its orbit shrinks, while the outer one gains energy and moves out. During this process the distance, thus the force between these parcels increases further and the process runs away. 

There can be a region in the disk (typically between 0.2 and 4 AU -- \cite{D'Alessio1998}), where the ionization fraction is smaller than the minimum value required for the MRI. Therefore, \cite{Gammie1996} proposed a layered accretion disk model, in which a `dead zone' is present at the midplane of the disk. The dead zone has suitable properties for planet formation. A pressure bump is present at the edges of the dead zones, where dust particles can be accumulated and planetesimals can be formed (e.g. \cite{Lyra:2008p625, Dzyurkevich2010}.

Shear box simulations of MRI-driven turbulence showed that dust particles of decimeter in size can be efficiently concentrated in the turbulent eddies. This way planetesimals can be formed directly from decimeter sized bodies avoiding all size regimes in between \citep{Johansen:2007p65}. 

We restricted the discussion on MRI as a possible driving mechanism for angular momentum redistribution within disks, but there are several other candidates, like the baroclinic instability \citep{Klahr2003}, shear instability \citep{Dubrulle2005}, gravitational spiral waves \citep{Pickett2003}, or global magnetic fields threading the disk \citep{2001MNRAS.323..587S}.

Dust particles are important for chemistry in disks. From the chemical point of view, disks can be divided into three regions according to the height above the midplane. These are the photon dominated layer, warm molecular layer, and the midplane layer \citep{Bergin2009}. In these layers different type of chemistry are dominating, which are the gas phase chemistry (at the hot and highly ionized photon dominated layer), gas-grain chemistry and grain surface chemistry (in the colder regions of the disk). We also have to distinguish the inner and outer disks \citep{Semenov2010}. In the inner disk, chemical equilibrium is reached within $\sim 100$ yrs. However, the material in the outer disk can have much longer reaction timescales ($\sim 10^5$ yrs) due to the low densities and temperatures, thus kinetic chemistry must be used to obtain the abundance of different chemical species. 

\section{The outline of the thesis}
This thesis investigates the initial growth of dust particles using a Monte Carlo (MC) model which includes a collision model that is based on the currently known laboratory experiments on dust aggregates. Chapter \ref{chp:paper1} describes the basic numerical method which is used in this thesis. We test the method by comparing the results against analytical solutions and highlight the strengths and weaknesses of the MC method. 

Chapter \ref{chp:paper2} describes the laboratory based collision model. The assumptions, fitting formulas and parameters, and extrapolations that enter the model are discussed in detail.  

Chapter \ref{chp:paper3} combines the results of the previous chapters by incorporating the collision model into the MC code. In this chapter, local box simulations are performed at the midplane of three different disk models at 1 AU distance from a solar mass star. In this chapter we show the importance of bouncing collisions and how dust evolution is altered compared to previous simpler collision models. 

Chapter \ref{chp:sedi} is the extension of the previous 0D models. In this chapter we investigate the settling and growth of dust particles in a 1D vertical column of disks. 

Finally, we review the future prospects of this field in Chapter \ref{chp:concl}.


%% file: Chapters/Chapter2.tex

\chapter{A representative particle approach to the coagulation of dust particles} 
\label{chp:paper1}
\lhead{Chapter 2. \emph{A representative particle approach}} 
\rhead{}

Based on \textit{`A representative particle approach to coagulation and fragmentation of  dust aggregates and fluid droplets'} by A. Zsom \& C. P. Dullemond published in \aap, 489, 931.

\section{Introduction}
Dust particle aggregation is a very common process in various astrophysical
settings. In protoplanetary disks the aggregation of dust particles forms
the very initial step of planet formation (see e.g.\ Dominik et
al.~\citeyear{dominikblum:2007-2}). It also modifies the optical properties of
the disk, and it has influence on the chemistry and free electron abundance
in a disk (Sano et al.~\citeyear{sanomirama:2000-2}; Semenov et
al.~\citeyear{semenovwiebe:2004-2}; Ilgner \& Nelson
\citeyear{ilgnernelson:2006-2a-2}). The appearance and evolution of a
protoplanetary disk is therefore critically affected by the dust aggregation
process. In sub-stellar and planetary atmospheres the aggregation of dust
particles and the coagulation of fluid droplets can affect the
structure of cloud layers. It can therefore strongly affect the spectrum of
these objects and influence the local conditions within these
atmospheres. The process of aggregation/coagulation and the reverse process
of fragmentation or cratering are therefore important processes to
understand, but at the same time they are extremely complex.

Traditional methods solve the Smoluchowski equation for the particle mass
distribution function $f(m)$, where $f(m)$ is defined such that $f(m)dm$
denotes the number of particles per cubic centimeter with masses in the
interval $[m,m+dm]$. This kind of method has been used in many papers on
dust coagulation before (e.g.\ Nakagawa et
al.~\citeyear{nakanakahayashi:1981-2}; Weidenschilling
\citeyear{weidenschilling:1984-2,weid1997-2}; Schmitt et
al.~\citeyear{schmitthenningmucha:1997-2-2}; Suttner \& Yorke
\citeyear{suttneryorke:1999-2}; Tanaka et al.~\citeyear{tanakahimemoida:2005-2};
Dullemond \& Dominik \citeyear{duldom:2005-2}; Nomura \& Nakagawa
\citeyear{nomuranaka:2006-2}). Methods of this kind are efficient, but have
many known problems. First of all a coarse sampling of the particle mass
leads to systematic errors such as the acceleration of growth (Ohtsuki et
al.~\citeyear{ohtsuki:1990-2}). High resolution is therefore required, which
may make certain problems computationally expensive. Moreover, if one wishes
to include additional properties of a particle, such as porosity, charge,
composition etc, then each of these properties adds another dimension to the
problem. If each of these dimensions is sampled properly, this can
quickly make the problem prohibitively computationally expensive. Finally,
the traditional methods are less well suited for modeling stochastic
behavior of particles unless this stochastic behavior can be treated in an
averaged way. For instance, in protoplanetary disks if the stopping time of
a particle is roughly equal to the turbulent eddy turn-over time, then the
velocity of a particle with respect to the gas is stochastic: at the same
location there can exist two particles with identical properties but which
happen to have different velocities because they entered the eddy from
different directions (see e.g.\ the simulations by Johansen et
al.~\citeyear{johansen:2006-2}).

To circumvent problems of this kind Ormel et
al.~(\citeyear{ormelmonte:2007-2}) have presented a Monte Carlo approach to
coagulation. In this approach the particles are treated as computational
particles in a volume which is representative of a much larger volume. The simulation follows the life of $N$ particles as they collide
and stick or fragment. The collision rates among these particles are
computed, and by use of random numbers it is then determined which particle
collides with which.  The outcome of the collision is then determined
depending on the properties of the two colliding particles and their
relative impact velocity. This method, under ideal conditions, provides the true simulation of
the process, except that random numbers are used in combination with
collision rates to determine the next collision event. This method has many
advantages over the tradiational methods. It is nearly trivial to add any
number of particle properties to each particle. There is less worry of
systematic errors because it is so close to a true simulation of the system,
and it is easy to implement.  A disadvantage is that upon coagulation the
number of computational particles goes down as the particles
coagulate. Ormel et al.~solve this problem by enlarging the volume of the
simulation and hence add new particles, but this means that the method is
not very well suited for modeling coagulation within a spatially resolved
setting such as a hydrodynamic simulation or a model of a protoplanetary
disk.

It is the purpose of this chapter to present an alternative Monte Carlo
method which can quite naturally deal with extremely large numbers of
particles, which keeps the number of computational particles constant
throughout the simulation and which can be used in spatially resolved
models.

\section{The method}
\subsection{Fundamentals of the method}\label{subsec-fundamentals}
The fundamental principle underlying the method we present here is to follow
the behavior of a limited number of {\em representative particles} whose
behavior is assumed to be a good representation of all particles.  In this
approach the number of physical particles $N$ can be arbitrarily large. In
fact it {\em should} be very much larger than the number of representative
particles $n$, so that the chance that one representative particle collides
with another representative particle is negligible compared to the
collisions between a representative particle and a non-representative
particle. In other words, if $N\gg n$, we only need to consider collisions
between a representative particle and a non-representative particle. The
number of collisions among representative particles is too small to be
significant, and the collisions among non-representative particles are
not considered because we focus only on the behavior of the representative
particles.

Suppose we have a cloud of dust with $N=10^{20}$ physical particles, with a
specific size distribution, for instance, MRN (Mathis, Rumpl \& Nordsieck
\citeyear{mrn:1977-2}). Let the total mass of all these particles together be
$M_{\mathrm{tot}}$ and the volume be $V$. We randomly pick $n$ particles out
of this pool, where $n$ is a number that can be handled by a computer, for
instance, $n=1000$. Each representative particle $i$ has its own mass $m_i$
and possibly other properties such as porosity $p_i$ or charge $c_i$
assigned to it. We now follow the life of each of these $n=1000$
particles. To know if representative particle $i=20$ collides with some
other object, we need to know the distribution function of all {\em
  physical} particles with which it can collide. However, in the computer we
only have information about the $n$ representative particles. We therefore
have to make the assumption that the distribution function set up by the $n$
representative particles is representative of that of the $N$ physical
particles. We therefore assume that there exist
\begin{equation}
n_k=\frac{M_{\mathrm{tot}}}{nm_kV}
\end{equation}
physical particles per cubic cm with mass $m_k$, porosity $p_k$, charge
$c_k$ etc., and the same for each value of $k$, including $k=i$. In this
way, by assumption, we know the distribution of the $N$ physical particles
from our limited set of $n$ representative particles. One could say that
each representative particle represents a swarm of $M_{\mathrm{tot}}/nm_i$
physical particles with identical properties as the representative one. One could also say that the
true distribution of $N$ particles is, by assumption, that of the $n$
representative ones.  The rate of collisions that representative particle
$i$ has with a physical particle with mass $m_k$ etc.\ is then:
\begin{equation}
r_i(k) = n_k\sigma_{ik}\Delta v_{ik}=\frac{M_{\mathrm{tot}}}{nm_kV}\sigma_{ik}\Delta v_{ik}
\end{equation}
where $\sigma_{ik}$ is the cross-section for the collision between particles
with properties $i$ and $k$, and $\Delta v_{ik}$ is the average relative
velocity between these particles. The total rate of collisions that
representative particle $i$ has with any particle is then:
\begin{equation}
r_i = \sum_{k}r_i(k)
\end{equation}
and the total rate of collisions of any representative particle is
\begin{equation}
r = \sum_i r_i
\end{equation}

The time-evolution of the system is now done as follows. Let $t_0$ be the
current time. We now randomly choose a time step $\delta t$ according to:
\begin{equation}\label{eq-delta-t}
\delta t = -\frac{1}{r}\log(\mathrm{ran(seed)})
\end{equation}
where ran(seed) is a random number uniformly distributed between 0 and 1.
This means that a collision event happens with one of the representative
particles at time $t=t_0+\delta t$. The chance $P(i)$ that the event happens
to representative particle $i$ is:
\begin{equation}
P(i) = \frac{r_i}{r}
\end{equation}
So we can choose, using again a random number, which representative particle
$i$ has undergone the collision event. We now need to determine with what kind
of physical particle it has collided. Since the distribution of physical
particles mirrors that of the representative ones, we can write that the
chance this particle has collided with a physical particle with
properties $k$ is:
\begin{equation}
P(k|i) = \frac{r_i(k)}{r_i}
\end{equation}
With another random number we can thus determine which $k$ is involved in
the collision. Note that $k$ can be $i$ as well, i.e.\ the representative
particle can collide with a physical particle with the same properties, or
in other words: a representative particle can collide with a particle of
its own swarm of physical particles.

Now that we know what kind of collision has happened, we need to determine
the outcome of the collision. The most fundamental part of our algorithm is
the fact that only representative particle $i$ will change its properties in
this collision. Physical particle $k$ would in principle also do so (or in
fact becomes part of the new representative particle), but since we do not
follow the evolution of the physical particles, the collision will only
modify the properties of representative particle $i$. By assumption
this will then automatically also change the properties of all physical
particles associated with representative particle $i$. Statistically,
the fact that the particles $k$ are not modified is ``corrected for'' by
the fact that at some point later the {\em representative} particle $k$ will
have a collision with {\em physical} particle $i$, in which case the 
properties of the $k$ particles will be modified and not those of $i$. This
then (at least in a statistical sense) restores the ``symmetry'' of the
interactions between $i$ and $k$. If the collision
leads to sticking, then the resulting particle will have mass
$m=m_i+m_k$. This means that representative particle $i$ will from now on
have mass $m_i\leftarrow m_i+m_k$. Representative particle $k$ is left
unaffected as it is not involved in the collision. The interesting thing is
now that, because by assumption the representative particle distribution
mirrors the real particle distribution, the swarm of physical particles
belonging to the modified representative particle $i$ now contains fewer
physical particles, because the total dust mass $M=M_{\mathrm{tot}}/n$ of
the swarm remains constant.

If a collision results in particle fragmentation, then the outcome of the
collision is a distribution function of debris particles. This distribution
function can be written as a function $f_d(m)$ of debris particle mass, such
that
\begin{equation}
\int_0^{\infty}mf_d(m)dm=m_i+m_k
\end{equation}
and the function $f_d(m)$ has to be determined by laboratory experiments or
detailed computer simulations of individual particle collisions (see Dominik
et al.\ \citeyear{dominikblum:2007-2} for a review). The new value of $m_i$
for the representative particle is now randomly chosen according to this
distribution function by solving the equation
\begin{equation}\label{eq-solve-new-debris-mass}
\int_0^{\bar m}mf_d(m)dm = \mathrm{ran(seed)}(m_i+m_k)
\end{equation}
for $\bar m$ and assigning $m_i\leftarrow \bar m$. In other words: we
randomly choose a particle mass from the debris mass distribution function,
i.e.\ the choice is weighed by fragment mass, not by fragment particle
number. This can be understood by assuming that the true representative
particle before the collision is in fact just a monomer inside a larger
aggregate. When this aggregate breaks apart into for instance one big and
one small fragment it is more likely that this representative monomer
resides in the bigger chunk than in the smaller one.

After a fragmenting collision the $m_i$ will generally be smaller than
before the collision. This means that the number of physical particles
belonging to representative particle $i$ increases accordingly. Note that
although the collision has perhaps produced millions of debris particles out
of two colliding objects, our method only picks one of these debris
particles as the new representative particle and forgets all the
rest. Clearly if only one such destructive collision happens, the
representative particle is not a good representation of this entire cloud of
debris products. But if hundreds such collisions happen, and are treated in
the way described here, then the statistical nature of
Eq.~(\ref{eq-solve-new-debris-mass}) ensures that the debris products are
well represented by the representative particles.

The relative velocity $\Delta v$ can be taken to be the average relative
velocity in case of random motions, or a systematic relative velocity in
case of systematic drift. For instance, for Brownian motion there will be an
average relative velocity depending on the masses of both particles, but
differential sedimentation in a protoplanetary disk or planetary atmosphere
generates a systematic relative velocity. Also, for the Brownian motion or
turbulent relative velocity one can, instead of using an average relative
velocity, choose randomly from the full distribution of possible relative
velocities if this is known. This would allow a consistent treatment of
fluctuations of the relative velocities which could under some circumstances
become important (see e.g.\ Kostinski \& Shaw \citeyear{kostinskishaw:05-2}).

\subsection{Computer implementation of the method}
We implemented this method in the following way. For each of the
representative particles we store the mass $m_i$ and all other properties
such as porosity, charge, composition etc. Before the start of the Monte
Carlo procedure we compute the full collision rate matrix $r_i(k)$, and we
compute the $r_i$ as well as $r$. For these collision pairs $(i,k)$ we now have to determine the
cross section of particles as well as their systematic relative velocity, such as different drift speeds, and the random relative velocity, such as Brownian or turbulent motion. The random motions
can be determined with a random number from the relative velocity
probability distribution function if that is known. If that is not known in
sufficient detail, one can also take it to be the average relative velocity,
for which more often analytic formulae exist in the literature.

We determine beforehand at which times
$t_{\mathrm{sav},n}$ we want to write the resulting $m_i$ and other
parameters to a file. The simulation is now done in a subroutine with a
do-while loop. We then determine $\delta t$ using a random number (see
Eq.~\ref{eq-delta-t}), and check if $t+\delta t< t_{\mathrm{sav},n}$, where
$t_{\mathrm{sav},n}$ is the next time when the results will have to be
stored. If $t+\delta t< t_{\mathrm{sav},n}$, then a collision event occured
before $t_{\mathrm{sav},n}$. We will handle this event according to a
procedure described below, we set $t\leftarrow t+\delta t$ and then return
to the point where a new $\delta t$ is randomly determined. If, on the other
hand, $t+\delta t\ge t_{\mathrm{sav},n}$ then we stop the procedure, return
to the main program and set $t\leftarrow t_{\mathrm{sav},n}$. The main
program can then write data to file and re-call the subroutine to a time
$t_{\mathrm{sav},n+1}$ or stop the simulation altogether. Note that when the
subroutine is called again for a next time interval, it does not need to
know the time of the previously randomly determined event which exceeded
$t_{\mathrm{sav},n}$. Of course, one could memorize this time and take that
time as the time of the next event in the next time interval, but since the
events follow a Poisson distribution, we do not need to know what
happened before $t_{\mathrm{sav},n}$ to randomly determine the new time
$t+\delta t$ of the next event. 

Now let us turn to what happens if a collision event occurs, i.e.\ occurs
between time $t$ and $t+\delta t$. We then first determine which
representative particle $i$ is hit, which is done by generating a random
number and choosing from the probability distribution of collision rates, as
described in Section \ref{subsec-fundamentals}. Similarly we determine the
non-representative particle with which it collides, or in other words: we
determine the index $k$ of the ``swarm'' in which this non-representative
particle resides. Finally, we must determine the impact parameter of the collision, or assume some average
impact parameter.

Now we employ a model for the outcome of the collision. This is the
collision subroutine of our Monte Carlo method. It is here where the results of
laboratory experiments come in, and the translation of such experiments into
a coagulation kernel is a major challenge which we do not cover here.  The
collision model must be a quick formula or subroutine that roughly
represents the outcomes of the detailed laboratory collision experiments or
detailed numerical collision models. It will give a probability function
$f_d$ for the outcoming particle masses and properties. From this
distribution function we pick {\em one} particle, and from this point on our
representative particle $i$ will attain this mass and these properties. The
collision partner $k$ will not change, because it is a non-represetative
particle from that swarm that was involved in the collision, and we do not
follow the life of the non-representative particles. We therefore ignore any
changes to that particle.

We now must update $r_j(l)$ for all $l$ with fixed $j=i$ and for all $j$
with fixed $l=k$: we update a row and a column in the $r_j(l)$
matrix. Having done this, we must also update $r_j$ for all $j$. This would
be an $n^2$ process, which is slow. But in updating $r_j(l)$ we know the
difference between the previous and the new value, and we can simply add
this difference to $r_j$ for each $j$. Only for $j=i$ we must recompute the
full $r_j$ again, because there all elements of that row have been modified.
Using this procedure we assure that we limit the computational effort to
only the required updates.

\subsection{Acceleration of the algorithm for wide size distributions}
\label{subsec-acceleration}
One of the main drawbacks of the basic algorithm described above is that it
can be very slow for wide size distributions. Consider a swarm of micron
sized dust grains that are motionless and hence do not coagulate among each
other. Then a swarm of meter sized boulders moves through the dust swarm at
a given speed, sweeping up the dust. Let us assume that also the boulders
are not colliding among each other. The only mode of growth is the
meter-sized boulders sweeping up the micron sized dust. For the boulder to
grow a factor of 2 in mass it will have to sweep up $10^{18}$ micron sized
dust particles.  Each impact is important for the growth of the boulder, but
one needs $10^{18}$ such hits to grow the boulder a factor of 2 in mass. The
problem with the basic algorithm described above is that it is forced to
explicitly model each one of these $10^{18}$ impacts. This is obviously
prohibitively expensive.

The solution to this problem lies in grouping collisions into one. Each
impact of a dust grain on a boulder only increases the boulder mass by a
minuscule fraction. For the growth of the boulder it would also be fine to
lower the chance of an impact by $10^{16}$, but {\em if} it happens, then
$10^{16}$ particles impact onto the boulder at once. Statistically this
should give the same growth curve, and it accelerates the method by a huge
factor. However, it introduces a fine-tuning parameter. We must specify the
minimum increase of mass for coagulation ($dm_{max}$). If we set $dm_{max}$ to, for instance,
10\%, then we may expect that the outcome also has errors of the order
  of 10\%. This error arises because by increasing the mass of the
  bigger body in steps of 10\%, we ignore the fact that the mass at some
  time should in fact be somewhere in between, which cannot be resolved with
  this method. This is, however, not a cumulative error. While the mass of
  the bigger body may sometimes be too low compared to the real one, it
  equally probably can be too large. On average, by the Poisson nature of
  the collision events, this averages out. But it is clear that the smaller
  we take this number, the more accurate it becomes -- but also the slower
  the method becomes. It is therefore always a delicate matter to choose
  this parameter, but for problems with a large width of the size
  distribution this acceleration is of vital importance for the usability of
  the method.

\subsection{Including additional particle properties}
We mentioned briefly the possibility of adding more particle properties to
each representative particle. This is very easy to do, and it is one of the
main advantages of a Monte Carlo method over methods that directly solve the
integral equations for coagulation. One of the main properties of interest
to planet formation is porosity or fractal structure of the aggregate. Two
aggregates with the same mass can have vastly different behavior upon a
collision if they have different compactness. A fluffy aggregate may break
apart already at low impact velocities while a compact aggregate may simply
bounce. Upon collisions these properties may in fact also change. Ormel et
al.~(\citeyear{ormelmonte:2007-2}) studied the effect of porosity and how it
changes over time, and they also used a Monte Carlo approach for it.

If one wishes to include particle properties in a traditional method which
solves the integral equations of coagulation (the Smoluchowski equation),
then one increases the dimensionality of the problem by 1 for each property
one adds. With only particle mass one has a distribution function $f(m,t)$
while adding two particle properties $p_1$ and $p_2$ means we get a
distribution function $f(m,p_1,p_2,t)$, making it a 4-dimensional problem.
Methods of this kind must treat the complete phase space spanned by
$(m,p_1,p_2,t)$. This is of course possible, but computationally it is
  a very challenging task (see Ossenkopf \citeyear{ossenkopf:1993-2}). In
contrast, a Monte Carlo method only sparsely samples phase space, and it
samples it only there where a significant portion of the total dust mass
is. A Monte Carlo method focuses its computational effort automatically
there where the action is. The drawback is that if one is interested in
knowing the distribution function there where only very little mass resides,
then the method is inaccurate. For instance, in a protoplanetary disk it
could very well be that most of the dust mass is locked up in big bodies
(larger than 1 meter) which are not observable, and only a promille of the
dust is in small grains, but these small grains determine the infrared
appearance of the disk because they have most of the solid surface area and
hence most of the opacity. In such a case a Monte Carlo method, by focusing
on where most of the mass is, will have a very bad statistics for those dust
grains that determine the appearance of the disk.  For such goals it is
better to use the traditional methods. But if we are interested in following
the evolution of the dominant portion of the dust, then Monte Carlo methods
naturally focus on the interesting parts of phase space.

\begin{figure*}
  \includegraphics[width=0.5\textwidth]{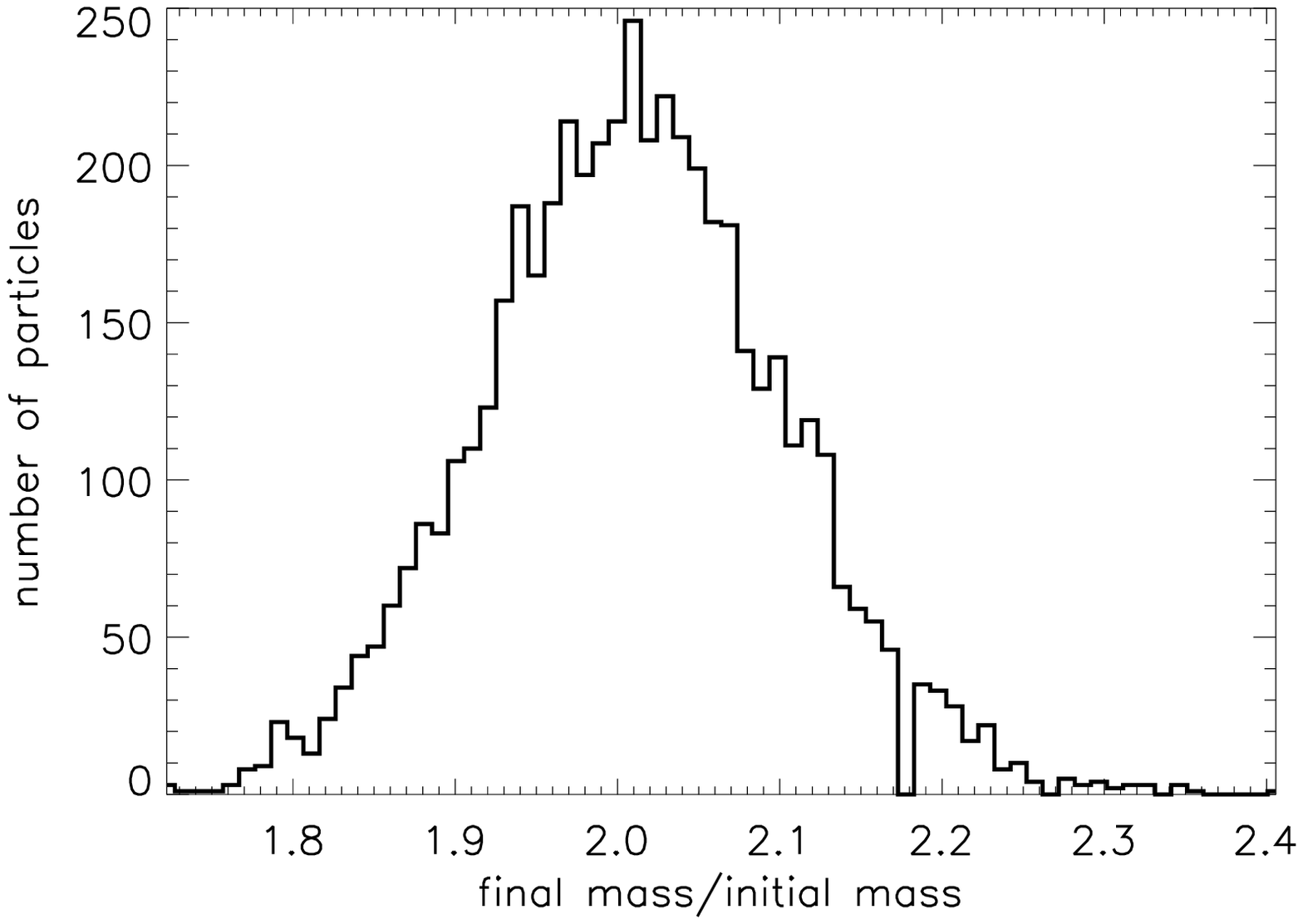}
  \includegraphics[width=0.5\textwidth]{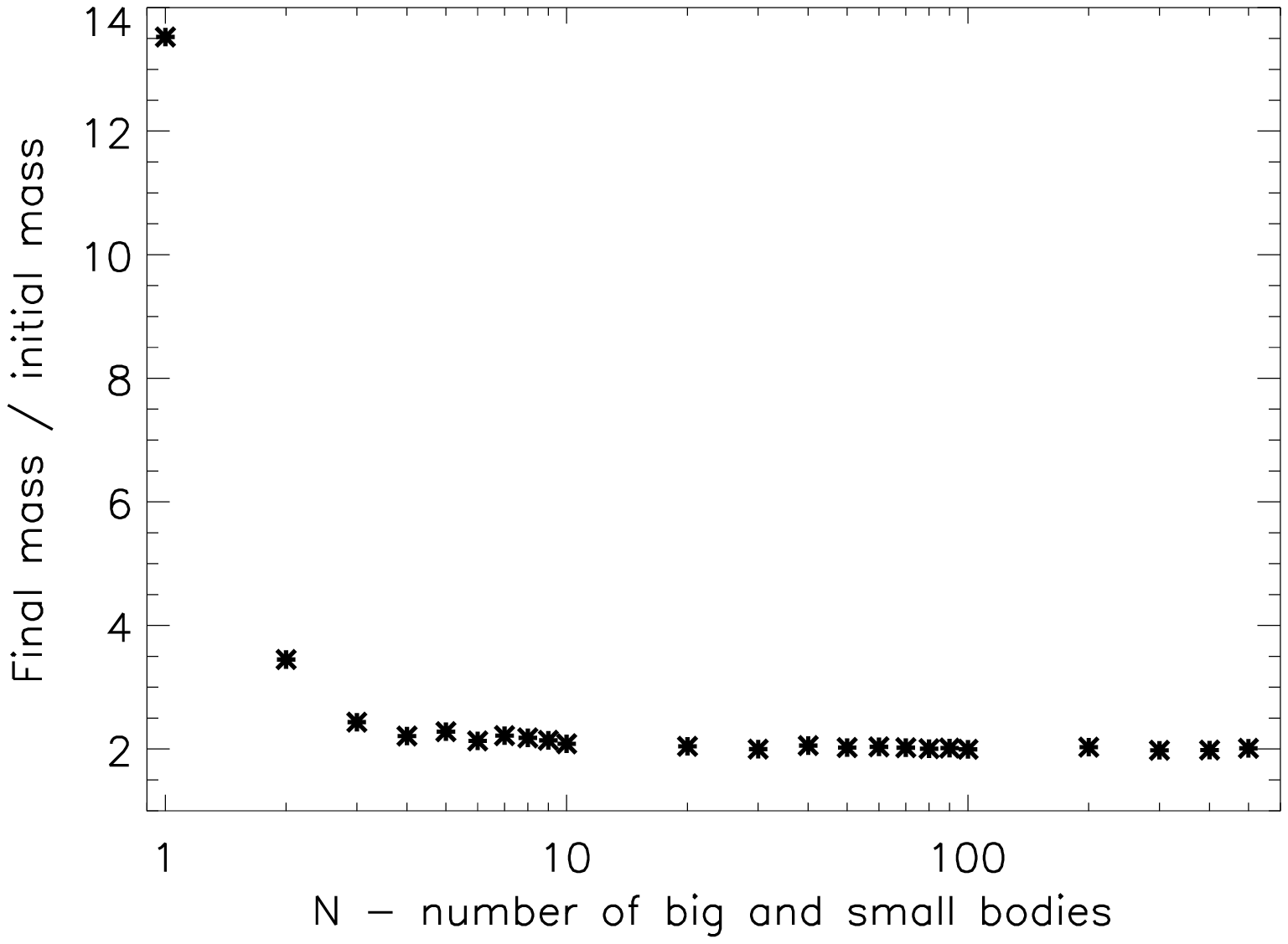}
\caption{\label{fig-exper-pairs} Results of the test problem with $N$ swarms
of small particles and $N$ swarms of big bodies, as discussed in Section \ref{subsec-cons-of-particle-number}. Left: histogram of the
final masses of the bodies relative to the initial mass of the big bodies,
for $N=500$. Right: average mass relative to the initial mass of the big bodies
as a function of $N$.}
\end{figure*}

\section{Discussion of the method}
\subsection{Conservation of particle number}
\label{subsec-cons-of-particle-number}
There are a few peculiarities of the method described here that may, at
first sight, appear inconsistent, but are statistically correct. For
instance if we return to our example of a swarm of tiny particles and a
swarm of boulders, i.e.\ $n=2$ with representative particle 1 being a micron
sized particle and representative particle 2 being a meter sized particle,
then we encounter an apparent paradox.  We again assume that collisions only
take place between 1 and 2, but not between 1 and 1 or 2 and 2. The chance
that representative particle 1 hits a meter size particle is much smaller
than the chance that representative particle 2 hits a micron size
particle. What will happen is that representative particle 2 will have very
many collision events with small micron size grains, and thereby slowly and
gradually grows bigger, while representative particle 1 will only have a
collision with big particle after a quite long time and immediately jumps to
that big size. While representative particle 2 grows in mass, the number of
big physical particles decreases in order to conserve mass. This may seem
wrong, because in reality the number of big boulders stays constant, and
these boulders simply grow by sweeping up the small dust. The solution to
this paradox is that the average time before representative particle 1 hits
a big ($k=2$) particle is of the same order as the time it takes for
representative particle 2 to grow to twice its mass by collecting small
particles. So, very roughly, by the time the big particle has doubled its
mass, and therefore the number of physical particles belonging to $k=2$ has
reduced by 50\%, the representative particle $1$ has turned into a big
particle, corresponding, statistically, to the other 50\% of big particles
that was missing. If we are a bit more precise, the statistics do not
  add up precisely in this way if we have only 1 swarm of small and 1 swarm
  of big bodies. If, however, one has N swarms of small and N swarms of big
  bodies, and again assume that only the big bodies can sweep up the smaller
  ones, then if $N\gg 1$ the statistics adds up perfectly: one finds that
  after all the growth has taken place, the average mass of the bodies is
  twice that of the original big bodies. In Figure~\ref{fig-exper-pairs} we do
  precisely this experiment, and the left panel shows that for $N=500$ the mass
  distribution of the big bodies averages to the right value, albeit with
  a spread of 10\% FWHM while in reality this spread should be 0. The
  right panel shows how the average final mass depends on $N$. For small
  $N$ the statistics clearly do not add up, but for large $N$ they do and
  produce the right value (final mass is twice initial mass of the big
  bodies). So statistically the number of big particles is restored to the
correct value, but there is then unfortunately still a large statistical
noise on it. The particle number is therefore not exactly conserved in our
method, but statistically it is.

\subsection{The number of representative particles}
\label{reppart}
It is obvious that for high number of representative particles $n$ we will
get better results than for low $n$. But there are two issues here.  First
of all, the higher $n$, the better the representative particles represent
the true physical distribution of particles. For problems that result in
wide size distributions this is all the more crucial. An inaccurate
representation of the true size distribution could lead to systematic
errors. But another reason for taking a high $n$ is simply because we want
our end-result to have as little as possible noise. If the result is too
noisy, then it is useless. Taking $n$ too big, however, makes the code slow
because more representative particles have to be followed, and for each of
these particles we must check for a larger number of possible collision
partners $k$. The problem scales therefore as $n^2$. If the expected size
distributions are not too wide, one can use an intermediately large number
of representative particles, say $\bar n$, for the simulation, but redo the
simulation $m$ times such that $n=m\bar n$, and average the results of all
$m$ simulations. This approach was also used by Ormel et
al.~(\citeyear{ormelmonte:2007-2}). This gives the same amount of noise on the
end-result, but scales as $\bar n^2m=n^2/m$, which is $m$ times faster than
the $n^2$ scaling. This works, however, only if the
coagulation/fragmentation kernel is not too sensitive to the exact
distribution of collision partners.

Interestingly, if the kernel is very insensitive to the exact distribution
of collision partners, then, in principle, one could run the model with only
a single representative particle $n=1$, because the collision partner of
representative particle $i$ could be equal to $k=i$. Of course, a single
representative particle means that we assume that all physical particles
have the same size, or in other words: that we have an infinitely narrow
size distribution. 

To decide about the sufficient number of representative particles, one has to compare the results of the MC code with the analytical solutions of the three test kernels (see Section \ref{section-test and results}). In a given time of the simulation the mean mass and the shape of the distribution function for all three test kernels must be followed accurately. It is especially important to reproduce the linear and product kernels accurately as the realistic kernels of dust particles are similar to these. 

Of course the progression from 'not sufficient' and 'sufficient' number of representative particles is smooth and in general the more representative particles we use, the more accurate the produced result will be. The sufficient number of representative particles ($\bar n$) as given in Section \ref{section-test and results}) are only suggestions, the error of the distribution functions were not quantified.

\subsection{Limitations of the method}
One of the fundamental limitations of the method described here is that we
assume $N\gg n$. We can model the growth of particles by coagulation in a
protoplanetary disk or in a cloud in a planetary atmosphere, but we can not
follow the growth to the point where individual large bodies start to
dominate their surroundings. For instance, if we wish to follow the growth
of dust in a protoplanetary disk all the way to small planets, then the
method breaks down, because $N$ is then no longer much bigger than $n$, and
interactions among representative particles become likely. Also, for the
same reason, run-away growth problems such as electrostatic gelation (Mokler
\citeyear{mokler:2007-2}) cannot be modeled with this method. 

Another limitation is encountered when modeling problems with strong growth
and fragmentation happening at the same time. This leads to very wide size
distributions, and the typical interval between events is then dominated by
the smallest particles, whereas we may be interested primarily in the growth
of the biggest particles. In such a situation a semi-steady-state can be
reached in which particles coagulate and fragment thousands of times over
the life time of a disk. The Monte Carlo method has to follow each of these
thousands of cycles of growth and destruction, which makes the problem very
``stiff''.  Methods using the integral form of the equations, i.e.\ the
Smoluchowski equation, can be programmed using implicit integration in time
so that time steps can be taken which are much larger than the typical time
scale of one growth-fragmentation cycle without loss of accuracy (Brauer et
al.~\citeyear{brauer:2008-2}). This is not possible with a Monte Carlo method.

\section{Standard tests and results}
\label{section-test and results}
In this section we test our coagulation model with kernels that have
analytical solutions. Furthermore we show the first results of applying this
model to protoplanetary disks introducing Brownian motion and turbulence
induced relative velocities as well as a new property of dust particles
namely the porosity (or enlargement factor, see Ormel et al
\citeyear{ormelmonte:2007-2}), and a simple fragmentation model.

\begin{figure}
  \centering
  \includegraphics[width=0.7\textwidth]{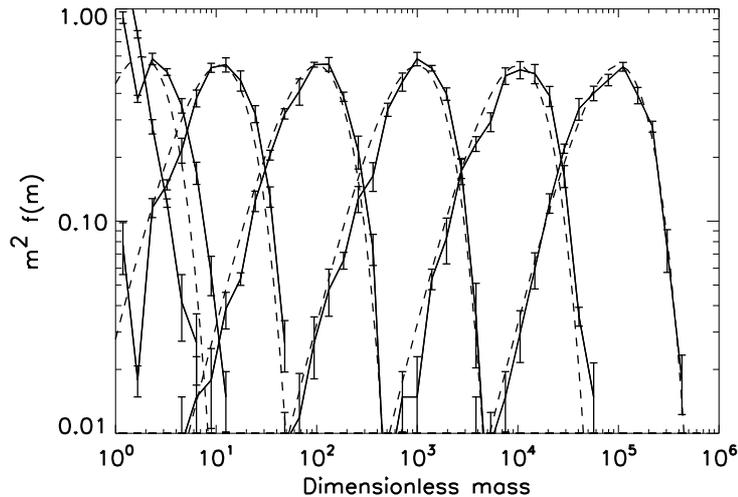}
  \caption{Test against the constant kernel ($K_{i,j}=1$). The particles were binned and the distribution function was produced at dimensionless times $t=0,10^0,10^1,10^2,10^3,10^4,10^5$. The dashed lines show the analytical solution. This run was produced by simulating 200 representative particles five times and producing the average of these. In this case $dm_{max}$ is 0.1.}
  \label{const}
\end{figure}

To follow dust coagulation and fragmentation, one has to follow the time
evolution of the particle distribution function at a given location in the
disk ($f(\bf{y},t)$), where $\bf{y}$ contains the modeled properties of the
dust grains, in our case these will be the mass ($m$) and the enlargement
factor ($\Psi$), $f(m,\Psi,t)$.

In most of the coagulation models so far the only used dust-property was the
particle mass. Then one can use the so called Smoluchowski equation
(\cite{smoluchowski:1916-2}) to describe the time-evolution of $f(m)$:
\begin{multline}
\frac{\partial f(m)}{\partial t}=-f(m)\int dm' K(m,m')f(m') + \\
 \frac{1}{2}\int dm' K(m',m-m')f(m')f(m-m').
\end{multline}
The first term on the right hand side represents the loss of dust in the
mass bin $m$ by coagulation of a particle of mass $m$ with a particle of
mass $m'$. The second term represents the gain of dust matter in the mass
bin $m$ by coagulation of two grains of mass $m'$ and $m - m'$. $K$ is the
coagulation kernel, it can be written as
\begin{equation}
K(m_1,m_2)=\sigma_c(m_1,m_2)\times \Delta v (m_1,m_2),
\end{equation} 
the product of the the cross-section of two particles and their relative
velocity. We consider all the three kernels for which there exist
  analytical solutions: The constant kernel ($K_{i,j}=1$), the linear
kernel ($K_{i,j}=m_i+m_j$) and the product kernel
  ($K_{i,j}=m_i\times m_j$). The analytical solutions are described e.g.
in \cite{ohtsuki:1990-2} and Wetherill (\citeyear{wetherill:1990-2}).

\begin{figure}
  \centering
  \includegraphics[width=0.7\textwidth]{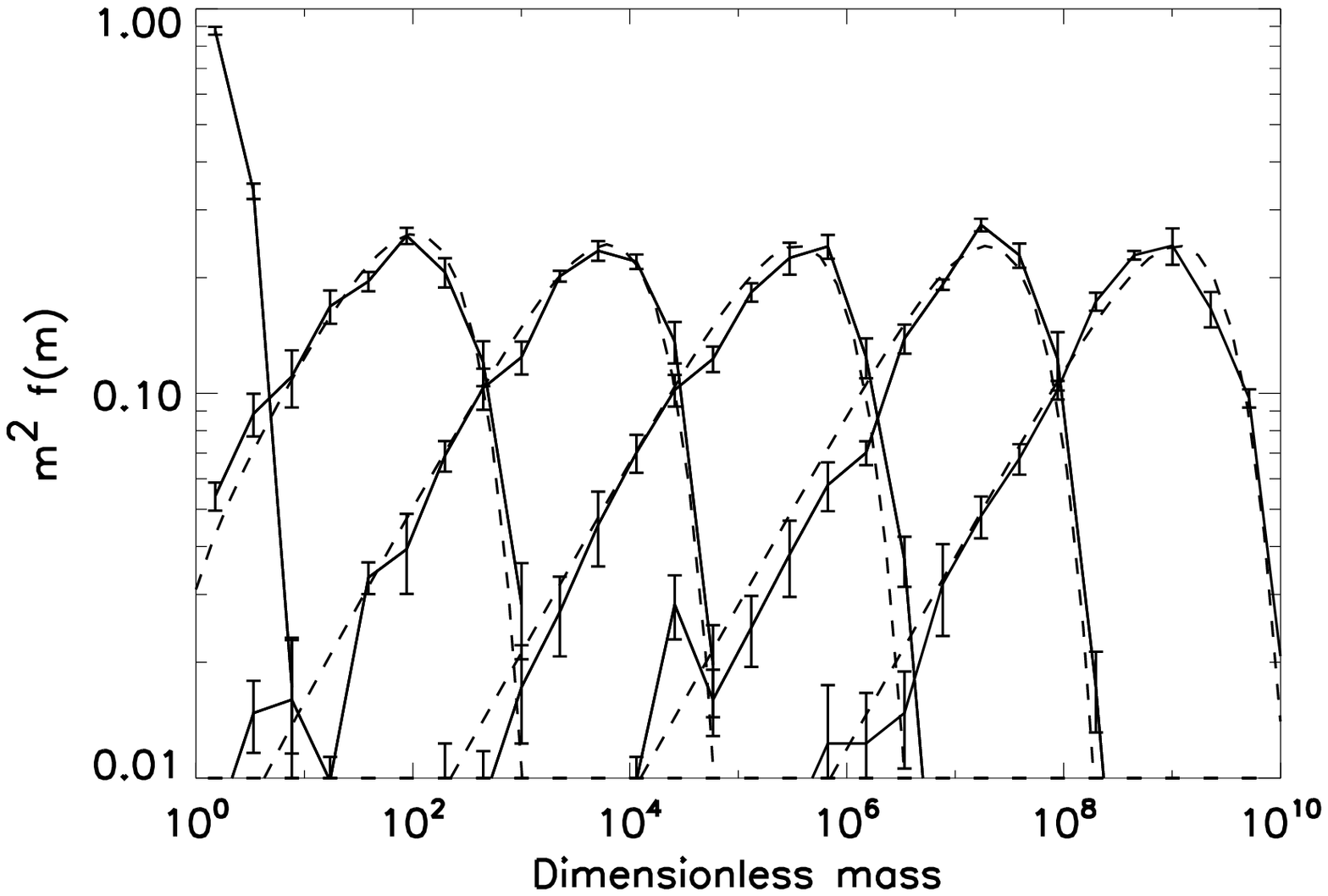}
  \caption{Test against the linear kernel ($K_{i,j}=m_i+m_j$). The particles were binned and the distribution function was produced at dimensionless times $t=0, 4, 8, 12, 16, 20$. The dashed lines show the analytical solution. This run was also produced by simulating 200 representative particles five times and producing the average of these. In this case $dm_{max}$ is 0.1.}
  \label{lin}
\end{figure}

We test our method against these three kernels, leaving the
enlargement factor unchanged, always unity. Further important properties
of the dust particles, such as material density and volume density, are also
always unity. The (dimensionless) time evolution of the swarms is followed
and at given times the particles are binned by mass so that we can produce
$f(m)$. On Figures \ref{const} and \ref{lin} the y axis shows $f(m)\times
m^2$, the mass density per bin. The analytical solutions, taken from
\cite{ohtsuki:1990-2} and Wetherill
  (\citeyear{wetherill:1990-2}), are overplotted with dashed
line. The number of particles were chosen to be $\bar n=200$, $m=5$, so
altogether 1000 representative particles were used in the model
except for the product kernel where more representative particles
  were used to achieve better results.

\begin{figure}
  \centering
  \includegraphics[width=0.7\textwidth]{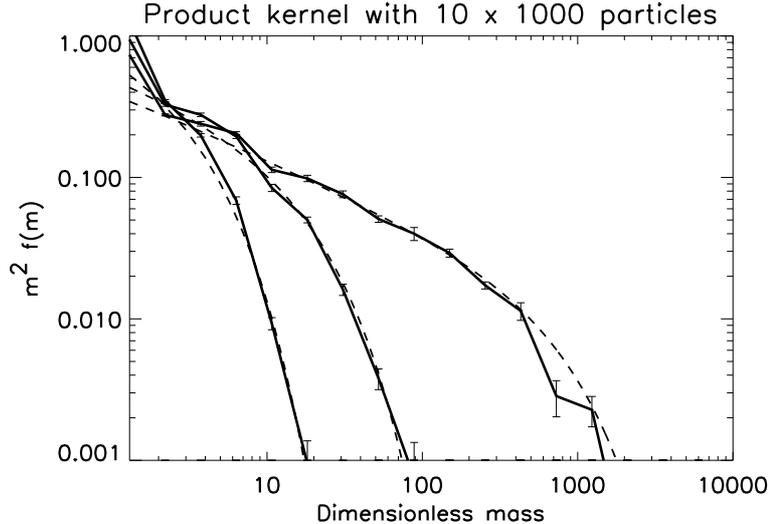}
  \caption{Test against the product kernel ($K_{i,j}=m_i \times m_j$). The particles were binned and the distribution function was produced at dimensionless times $t=0.4,0.7, 0.95$. The dashed lines show the analytical solution. This run was produced by simulating 1000 representative particles ten times and producing the average of these. In this case $dm_{max}$ is set to be  0.05.}
  \label{prod}
\end{figure}

In the case of the constant kernel (Figure~\ref{const}), we started our
simulation with MRN size distribution ($n(a)\propto a^{-3.5}$), the results
were saved at $t=0,10^0,10^1,10^2,10^3,10^4,10^5$. It is interesting to note
that this kernel is not sensitive to the initial size distribution. As the
system evolves, it forgets the initial conditions. Another interesting
property of this kernel that our model can reproduce the analytical solution
even with very limited number of representative particles (even for $\bar
n=5$!) but of course with higher noise. It is possible to use only one
representative particle, which means that the representative particle
collides with particles from its own swarm which basically results in pure
CCA growth (Cluster-Cluster Aggregation). Interestingly, the mean mass of the distribution function is
followed correctly but the shape of the function changes, additional spikes
appear on it. 

The linear kernel is known to be more problematic because the mean mass of
the particles grows exponentially with time. Our model, however reproduces
this kernel very well, too, as it can be seen in Figure \ref{lin}. The results
were saved at $t=0, 4, 8, 12, 16, 20$. We note that using low number of
representative particles with this kernel also works relatively well, the
minimum number of swarms needed to reproduce the exponential time evolution
of the mean mass is $\bar n \approx 100$.  This is larger than for the
constant kernel. It shows that for the linear kernel collisions between
particles of unequal mass are contributing significantly to the growth,
whereas for the constant kernel the growth is dominated by collisions
between roughly equal size particles. Using $\bar n \ll 100$ results in
distorted distribution function: neither the mean mass nor the actual shape
of the distribution function is correct.

The product kernel is the hardest to reproduce. The peculiarity of
  this kernel is the following: Using dimensionless units, a 'run-away'
  particle is produced around $t=1$, which collects all the other particles
  present in the simulation (Wetherill \citeyear{wetherill:1990-2}). The
  difficulty arises in our Monte Carlo code when the mass of the
  representative 'run-away' particle reaches the mass of its swarm. In other
  words, the number of physical particles belonging to the
  representative 'run-away' particle is close to unity. In this case the
  original assumption of our method (we only need to consider collisions
  between a representative particle and a physical particle) is
  not valid anymore. However, as Figure \ref{prod} shows, we can relatively
  well reproduce this kernel before $t=1$. In the case of this kernel, we
  need approximately $\bar n \approx 500$ representative particles to
  correctly reproduce it.

The required CPU time for these test cases is very low, some seconds
  only.

We conclude that our Monte Carlo method reproduces the constant and linear test kernels without any problem even with low number of representative particles. On the other hand the method has difficulties with the product kernel, but before the formation of the 'run-away' particle, we can reproduce the kernel. The relatively low number of representative particles needed to sufficiently reproduce the test kernels is very important for future applications where whole disk simulations will be done
and there will likely be regions containing low numbers of particles.

\section{Applications to protoplanetary disks}
We use the Monte Carlo code to follow the coagulation and fragmentation of
dust particles in the midplane of a protoplanetary disk at 1 AU from the
central star. Our disk model is identical with the one used by
\cite{brauer:2007-2}. We proceed step by step. First relative velocities
induced by Brownian motion and turbulence without the effects of porosity
are included (Sec.~\ref{relv}).

The next step is to include a fragmentation model (Sec.~\ref{fragmod}). 

In the final step porosity is included (Sec.~\ref{poro}). We use the
porosity model described in Ormel et al.~(\citeyear{ormelmonte:2007-2}). At this point we compare and check again our code with Ormel et al.~(\citeyear{ormelmonte:2007-2}) using their input parameters but not including the rain out of particles.

\subsection{Relative velocities}
\label{relv}
We include two processes in calculating the relative velocities: Brownian
motion and turbulence. 

Brownian motion strongly depends on the mass of the two colliding
particles. The smaller their masses are, the more they can be influenced by
the random collisions with the gas molecules/atoms. One can calculate an
average velocity given by
\begin{equation}
 \Delta v_B (m_1,m_2) = \sqrt \frac{8kT(m_1+m_2)}{\pi m_1 m_2}.
\end{equation}
For micron sized particles, relative velocity can be in the order of
magnitude of 1 cm/s, but for cm sized particles this value drops to
$10^{-7}$ cm/s. If growth is only governed by Brownian motion, it leads to
very slow coagulation, a narrow size distribution and fluffy dust particles,
so called cluster-cluster aggregates (CCA).

The gas in the circumstellar disk is turbulent, thus the dust particles
experience acceleration from eddies with different sizes and turnover
times. This process is very complex, but Ormel and Cuzzi
(\citeyear{ormelcuzzi:2007-2}) provided limiting closed-form
expressions for average relative turbulent velocities between two dust
particles. Their results are also valid for particles with high Stokes
numbers. They distinguished three regimes: a.) the stopping times of both
dust particles are smaller than the smallest eddy-turnover time ($t_1,t_2 <
t_{\eta}$, tightly coupled particles); b.) the stopping time is between the
smallest and largest turnover time ($t_{\eta} \le t_1 \le t_L$, intermediate
regime); c.) the stopping time is bigger than the largest turnover time
($t_1 > t_L$, heavy particles). For details see Ormel and Cuzzi
(\citeyear{ormelcuzzi:2007-2}). We used $ \alpha = 10^{-3}$ for the turbulence
parameter.

\begin{figure}
  \centering
  \includegraphics[width=0.7\textwidth]{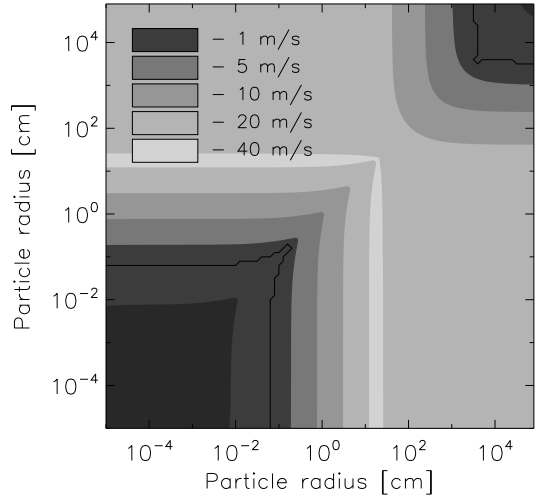}
  \caption{The relative velocity caused by Brownian motion and turbulence
    for different sized particles. The black line shows the fragmentation
    barrier. Collision events situated between these two lines result in
    fragmentation if porosity is not included. Physical parameters of the
    disk: the distance from the central star is 1 AU, temperature is 200 K,
    the density of the gas is $8.73\times 10^{12}$ particle/cm$^3$, and the
    turbulent parameter, $\alpha=10^{-3}$. Parameters of the dust: monomer
    radius is $a_0 =0.4 \mu$m, material density is $\rho = 1.6$ g/cm$^3$.}
  \label{relcont}
\end{figure}

To illustrate the relative velocity of dust particles without the effects of
porosity, we provide Figure \ref{relcont}. This contour plot includes Brownian
motion and turbulent relative velocities. The Brownian motion is negligible
for particles bigger than $10^{-2}$ cm.

\subsection{Fragmentation model} 
\label{fragmod}
The collision energy of the particles is
\begin{equation}
  E=\frac{1}{2} \frac{m_1 m_2}{m_1 + m_2} \Delta v^2 = \frac{1}{2} \mu \Delta v^2,
\end{equation}
where $\mu$ is the reduced mass. We need to define some quantities of the
dust particles. $E_{roll}$ is the rolling energy of two monomers. For
monomers of the same size it is given by (Dominik \& Tielens
\citeyear{dominiktielens:1997-2-2}; Blum \& Wurm \citeyear{blumwurm:2000-2})
\begin {equation}
  E_{roll}=\frac{1}{2}\pi a_0 F_{roll},
  \label{eq:Eroll}
\end{equation}
where $a_0$ is the monomer radius, $F_{roll}$ is the rolling force measured
by Heim et al.~(\citeyear{heim:1999-2}). Its value is $F_{roll}=(8.5 \pm 1.6)
\times 10^{-5}$ dyn for SiO$_2$ spheres.

The fragmentation energy is then defined as follows:
\begin{equation}
E_{frag} = N_c \times E_{break} \simeq 3 N \times E_{roll},
\label{eq:Efrag}
\end{equation}
where $N_c$ is the total number of contact surfaces between monomers (for
simplicity it is taken to be 3N, where N is the number of monomers in the
particle), $E_{break}$ is the energy needed to break the bond between two
monomers (its order of magnitude is similar to $E_{roll}$ for these parameters).

If the collision energy of two particles is higher than the corresponding
fragmentation energy, then the aggregate is destroyed and monomers are
produced. Note that although assuming a complete destruction of the
  collided dust particles, we are interested in the critical energy where
  the first fragmentation event happens. This is the reason why the
  fragmentation energy is assumed to be lower than the energy needed for
  catastrophic fragmentation. It is a simplification of the model to
  assume that the debris particles will be monomers. This is a very
simplified fragmentation model used previously by Dullemond \& Dominik
(\citeyear{duldom:2005-2}). A more realistic model would be the one used by
\cite{brauer:2007-2}.

We show the fragmentation barrier in Figure \ref{relcont} with black lines. If
collision happens in the regime between these two lines, that results in
fragmentation.

\subsubsection{Results}
A simulation was made including these effects in a specified location of the
disk. We choose the location to be 1 AU distance from the central solar type
star. Using the disk model of \cite{brauer:2007-2}, the temperature at this
distance is approximately 200 K, the density of the gas is $8.73\times
10^{12}$ cm$^{-3}$, the gas-to-dust ratio is 100 and we choose the
turbulent parameter to be $\alpha=10^{-3}$, the Reynolds number is
$Re=10^{8}$ (based on Ormel \& Cuzzi \citeyear{ormelcuzzi:2007-2}). The dust
monomers have the following properties: the monomer radius is $a_0 =0.4$ $\mu$m, material density is $\rho = 1.6$ g/cm$^3$. With the used parameters
the fragmentation velocity is $\Delta v_{frag} \approx 8$ m/s, though it is
somewhat larger for equal sized agglomerates. It is important to note that
this value is very sensitive to the monomer radius ($a_0$) and material
density ($\rho$), because smaller/lighter monomers mean more contact
surfaces (higher N for the same mass) and therefore higher fragmentation
energy.

\begin{figure}
\centering
\includegraphics[width=0.7\textwidth]{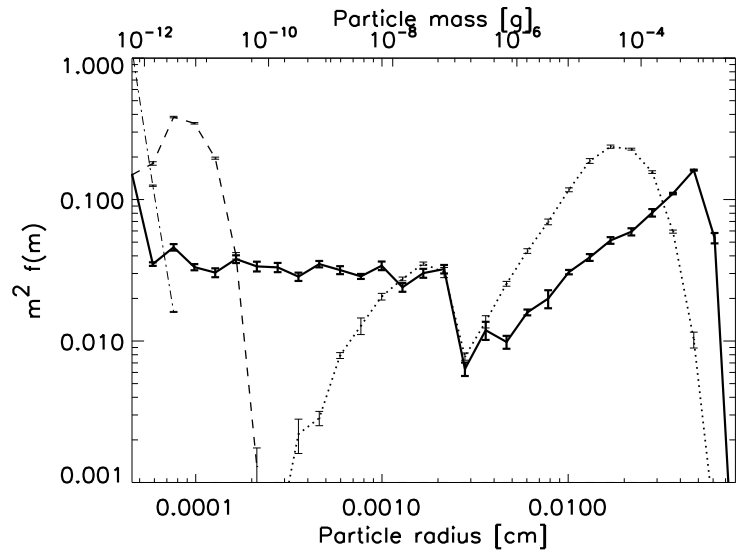}
\caption{The evolution of dust particles including the effects of Brownian
  motion and turbulence. Porosity is not included in this model. The
  particle distribution is saved after $t=3\times 10^0$ years - dash-dot
  line, $3\times 10^1$ years - dashed line, $3\times 10^2$ years - dotted
  line, and $3\times 10^3$ years - continuous line.}
\label{res_nocomp}
\end{figure}

Using these input parameters we simulated the evolution of the dust
particles for $3\times 10^3$ years so that we reach an equilibrium between
coagulation and fragmentation. Figure~\ref{res_nocomp} shows the resulting
normalized size distributions in times after $t=3\times 10^0$, $3\times
10^1$, $3\times 10^2$ and $3\times 10^3$ years. We used $\bar n=100$
particles averaging over $m=100$ times ($10^4$ particles
  altogether). The required CPU time to perform this simulation is 1.5 hours
  approximately. $dm_{max}$ is set to be 0.001 from now on in every simulation. We would like to note that giving $dm_{max}$ (Section \ref{subsec-acceleration}) a higher value would decrease the CPU time.

One can see that coagulation happens due to Brownian motion in the beginning
of the simulation (until $3\times 10^1$ years) but after that turbulence
takes over and the first fragmentation event happens after roughly $10^3$
years. After this event the "recycled" monomers start to grow again, but as
we see in Figure \ref{relcont}, particles can not reach bigger sizes than 0.07 cm.

We would like to draw attention to the sudden decrease of particles
  around 0.002 cm in Figure~\ref{res_nocomp}. This is the result of the
  turbulent relative velocity model used here (discussed in
  Sect. \ref{relv}). At this point the particles leave the 'tightly coupled
  particles' regime and enter the 'intermediate' regime. But the transition
  in relative velocity between these regimes is not smooth, there is a jump
  in relative velocity from $\sim$20 cm/s to $\sim$60 cm/s. As a result,
  particles coagulate suddenly faster and leave this part of the size
  distribution rapidly. Similar 'valleys' can be seen in the following
  figures with porosity, but the feature is less distinct as the stopping
  times can be different for particles with same mass.


 
\subsection{Porosity}
\label{poro}

 \begin{figure}
\centering
\includegraphics[width=0.7\textwidth]{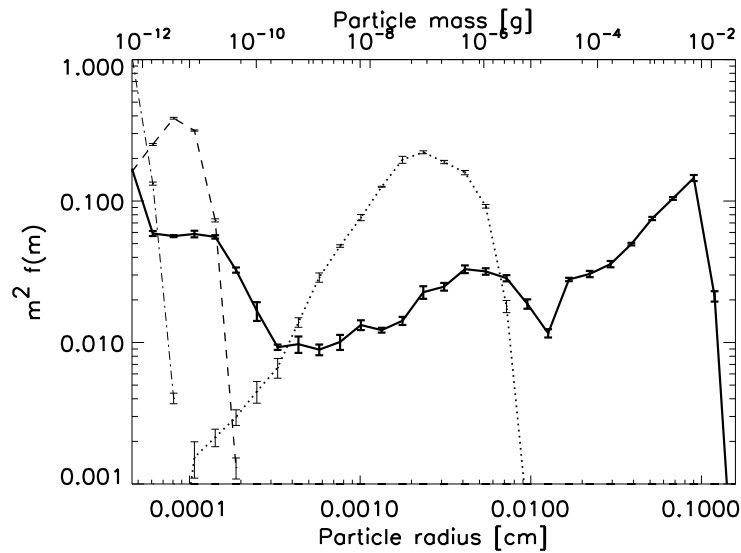}
\caption{The evolution of dust particles including the effects of Brownian
  motion and turbulence. Porosity is included in this model! The x axis
  shows the compact radius. The particle distribution is saved after
  $t=3\times 10^0$ years - dash-dot line, $3\times 10^1$ years - dashed
  line, $3\times 10^2$ years - dotted line, and $3\times 10^3$ years -
  continuous line. Note that the scaling of the x axis is different
    from Figure~\ref{res_nocomp}.}
\label{res_comp}
\end{figure} 

To be able to quantitatively discuss the effect of porosity, we have to
define the enlargement parameter following the discussion of Ormel et
al.~(\citeyear{ormelmonte:2007-2}). If $V$ is the extended volume of the grain
and $V^*$ is the compact volume, than one can define the enlargement
parameter ($\Psi$) as

\begin{equation}
\Psi = \frac{V}{V^*}.
\end{equation}  
Compact volume is the volume occupied by the monomers not taking into
account the free space between the monomer spheres. One can think of it as
melting all the monomers into a single sphere, the volume of this sphere is
the compact volume. We use compact radius later on, which is the radius of
this sphere. In the previous section the mass/volume ratio was constant for
the particles. Therefore we could automatically calculate the mass of the
particle if the radius was known or vice-versa. But from now on a particle
with given mass $m$ can have a wide range of effective radii depending on
its enlargement parameter.

It is essential to know how the enlargement parameter changes upon
collisions. We have to refine our fragmentation model and introduce two more
regimes regarding to collision energy. We use the model of Ormel et 
al.~(\citeyear{ormelmonte:2007-2}) and we only summarize their model here.

The first regime is the low collision energy regime, where the collision
energy is smaller than the restructuring energy ($E < E_{restr}$, where
$E_{restr} = 5 E_{roll}$), meaning that the particles stick where they meet,
the internal structure of the grain does not change.

The recipe for the resulting enlargement factor after the collision of two
particles assuming that $m_1 > m_2$ then is
\begin{equation}
\Psi = \langle \Psi \rangle _m \left(1+\frac{m_2 \Psi_2}{m_1 \Psi_1}  \right)^{\frac{3}{2}\delta_{CCA}-1} + \Psi _{add},
\end{equation}   
where  $\langle \Psi \rangle _m$ is the mass averaged enlargement factor of
the colliding particles:
\begin{equation}
\langle \Psi \rangle _m = \frac{m_1 \Psi_1 + m_2 \Psi_2}{m_1 + m_2}.  
\end{equation}  
Furthermore $\delta_{CCA}$ is the CCA-characteristic exponent calculated by
detailed numerical studies such as Paszun \& Dominik (\citeyear{paszun:2006-2})
($\delta_{CCA}=0.95$). $\Psi _{add}$ is a necessary additional factor for
the enlargement factor (for details see Ormel et
al.~\citeyear{ormelmonte:2007-2}): 
\begin{equation}
  \Psi _{add} = \frac{m_2}{m_1} \Psi_1 \exp \left[ \frac{-\mu}{10 m_0} \right],
\end{equation}
where $m_0$ is the monomer mass.

The second regime is the regime of compaction. The internal structure of the
monomers inside the particle changes, this causes a decreasing porous
volume. If the collision energy $E_{restr} \le E \le E_{frag}$, we talk
about compaction. In this case the porosity after the collision becomes
\begin{equation}
 \Psi = (1-f_C)(\langle \Psi \rangle _m - 1 ) +1,
 \label{eq:comp}
\end{equation}
where $f_C=E/(N E_{roll})=-\Delta V / V$ is the relative compaction. One can
see that $f_C$ has to be smaller than unity otherwise  $\Psi$ in Eq. \ref{eq:comp}
becomes less than unity. But it can theoretically happen that $E> N
E_{roll}$. In this case, as long as the total collision energy remains below
the fragmentation threshold, we assume that after compaction this excess
energy goes back into the kinetic energy of the two colliding aggregates.
The two aggregates therefore compactify and bounce, without exchanging mass
or being destroyed. Bouncing is therefore included in this model, albeit in
a crude way.

The third regime is fragmentation as it was discussed in the previous
section ($E > E_{frag}$). We use the same fragmentation model as before so
the result of a fragmenting collision are monomers.

\subsubsection{Results}
\label{subsec:res_por}
We performed a simulation with exactly the same initial conditions as in the
last section but we included the porosity as an additional dust property in
the model. The result can be seen in Figure~\ref{res_comp} (the
  required CPU time here is also 1.5 hours). One can immediately see that
including porosity increases the maximum particle mass by two orders of
magnitude (five times larger particles in radius). This was already expected
based on the work of Ormel et al.~(\citeyear{ormelmonte:2007-2}), although due
to rain out of bigger particles, they did not simulate particles bigger than
0.1 cm.

\begin{figure}
\centering
\includegraphics[width=0.7\textwidth]{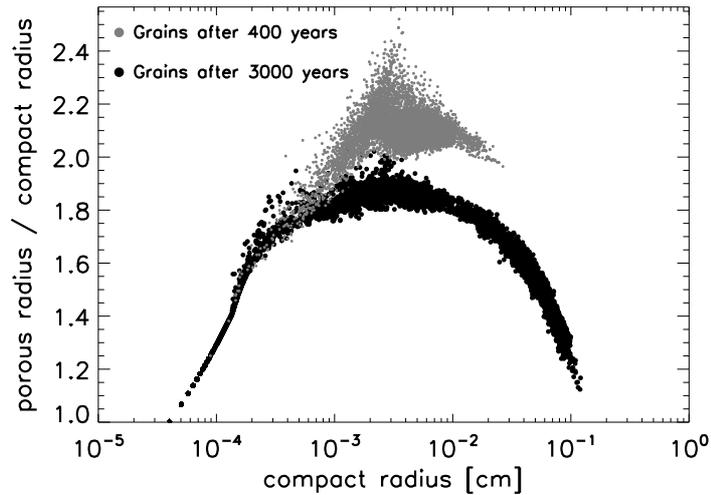}
\caption{This figure shows the radial enlargement of the dust aggregates after 400 and 3000 years. The x axis is the compact radii of the particles, the y axis is the ratio between the compact and the porous radius, this quantity is basically equal to $\Psi ^{\frac{1}{3}} $.}
\label{res_fluffy}
\end{figure}

We provide Figure~\ref{res_fluffy} to give an impression how the
porosity of the agglomerates change during the simulation. The x axis is the
compact radius of the particles, the y axis is the ratio between the compact
and the porous radii. This quantity is basically equal to $\Psi
^{\frac{1}{3}}$. Fractal growth is important for small particles creating
fluffy agglomerates (until $10^{-3}-10^{-2}$ cm approximately), after this
point the relative velocities become high enough so compactness becomes
important. Before the particles reach a fully compacted stage they fragment,
become monomers and a new cycle of growth starts. It is important to note that the porosity of the aggregates before the first fragmentation event is usually higher than the porosity values after equilibrium is reached. This can be seen in Figure~\ref{res_fluffy} (grains after 400 years and 3000 years). The reason is that before the first fragmentation event, particles involved in collisions are typically equal sized so these particles produce fluffy structures. However, when the distribution function relaxes in equilibrium, there are collisions between smaller and bigger aggregates as well which results in somewhat compacted aggregates.

\subsubsection{Model comparison with Ormel et 
al.~(\citeyear{ormelmonte:2007-2})}
We compare our Monte Carlo code with the one developed by Ormel et
al.~(\citeyear{ormelmonte:2007-2}). They use the Minimum Mass Solar Nebula
disk model (MMSN) and somewhat different dust-parameters which we changed
accordingly (distance from the central star = 1 AU, temperature = 280 K,
density of the gas is 8.5 $\times$ 10$^{-10}$ g/cm$^3$, gas to dust ratio =
240, $\alpha$ = 10$^{-4}$; monomer radius = 0.1 $\mu$m, monomer density = 3
g/cm$^3$, surface energy density of the monomers = 25 ergs/cm$^{-2}$).

They follow particle coagulation at one pressure scale height above the
midplane of the disk. Because of this if the particles reach a critical
stopping time ($\tau_{rain}=\alpha / \Omega$, where $\Omega$ is the Kepler
frequency), the particles rain out meaning that these particles leave the
volume of the simulation, the distribution function of the dust particles is
collapsing as it can be seen in their figures (Figure 10 and 11 in Ormel et
al.~(\citeyear{ormelmonte:2007-2})).

We do not include this effect in our model but we stop the simulation at the
first rain out event and compare our distribution functions until this
point. We use $10^4$ representative particles ($100\times 100$)
  during the simulation.

\begin{figure}
\centering
\includegraphics[width=0.7\textwidth]{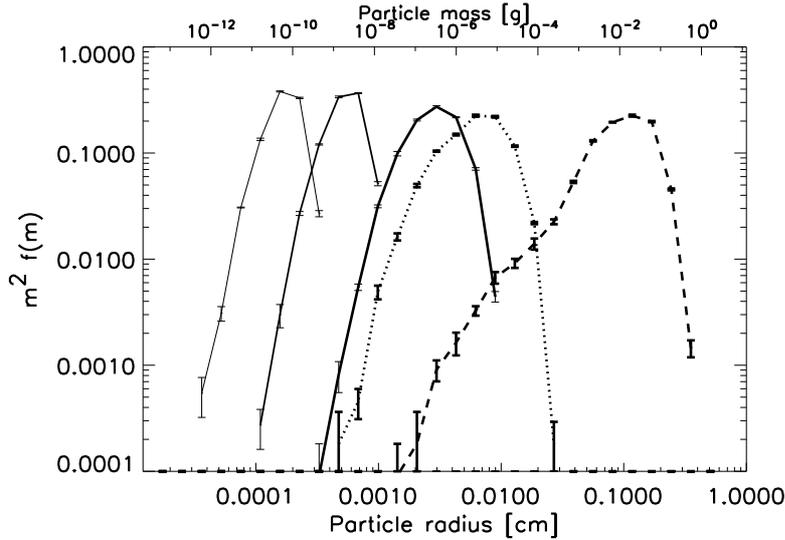}
\caption{Distribution functions obtained by using Ormel et
  al.~(\protect \citeyear{ormelmonte:2007-2}) input parameters. The continuous lines
  show the distribution functions at t = 10 years (thin line), 100 years
  (thicker line), 1000 years (thickest line). The dotted line shows the
  distribution function at the time of the first compaction event (t = 1510
  years), the dashed line shows the distribution function at the first rain
  out event (t = 2900 years).}
\label{ormel}
\end{figure}

This can be seen at Figure~\ref{ormel}. The reader is advised to examine this
figure together with Figure 10. c. from Ormel et
al.~(\citeyear{ormelmonte:2007-2}) because this is the figure we reproduced
here. Furthermore we would like to point out that the scale of the y axis is
different in the two figures. Our figure shows two orders of magnitude from
the normalized distribution functions whereas their figure covers more than
10 orders of magnitude from the real distribution function.

Keeping these in mind, one can compare the results of the two Monte Carlo
codes.

The continuous lines at Figure~\ref{ormel} in this chapter show the distribution
functions at t = 10 years (thin line), 100 years (thicker line), 1000 years
(thickest line). The dotted line shows the distribution function at the time
of the first compaction event (t = 1510 years), the dashed line shows the
distribution function at the first rain out event (t = 2900 years). The same
notation is used by Ormel et al.~(\citeyear{ormelmonte:2007-2}) at Figure 10. c.

We compared the position of the peaks of the distribution functions and the approximate shape of the curves. We can conclude that our code reproduces the results of Ormel et al.~(\citeyear{ormelmonte:2007-2}) very well.  

The required CPU time to perform this simulation is only 10
  minutes. One might ask why the CPU time is almost ten times smaller now?
  Why do the previous simulations, which used the same number of
  representative particles ($10^4$) and simulated approximately the same
  time interval (3000 years), take so long? The required CPU time does not
  scale linearly with the used number of particles. It scales linearly with
  the number of collisions simulated. The difference between this run and
  the previous two simulations is fragmentation. In the simulations of Ormel
  et al.~(\citeyear{ormelmonte:2007-2}) no fragmentation is happening because
  the growth timescales are longer. Using our initial parameters, the first
  fragmentation event happens around 1000 years, the number of small
  particles are never completely depleted after this time. As the small
  particles thereafter are always present, the number of collisions will be
  much higher than before.

Also note that the porosities of these particles would be smaller if
  the model of Ormel et al.~(\citeyear{ormelmonte:2007-2}) included
  fragmentation (for the reason see Sect. \ref{subsec:res_por}).

\subsection{Monomer size distribution}

\begin{figure}
\centering
\includegraphics[width=0.7\textwidth]{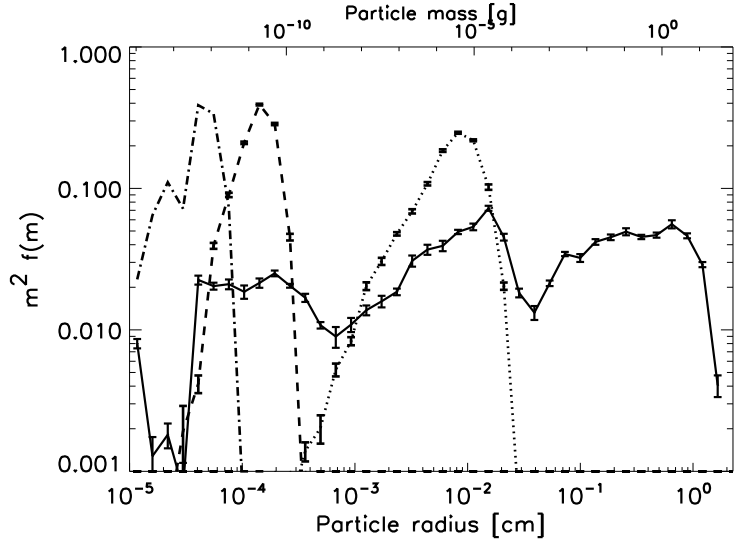}
\caption{The evolution of dust particles including the effects of Brownian
  motion and turbulence, porosity and using two different monomer sizes
  ($a_1=0.1 \mu$m and $a_2=0.4 \mu$m). The particle distribution is saved
  after $t=3\times 10^0$ years - dash-dot line, $3\times 10^1$ years -
  dashed line, $3\times 10^2$ years - dotted line, and $3\times 10^3$ years
  - continuous line.}
\label{2mon}
\end{figure}  

An interesting question which can easily be answered with our method
  is: How the mixture of different sized monomers change the maximum
  agglomerate size which can be reached? As we can see from Equations
  \ref{eq:Eroll} and \ref{eq:Efrag}, the rolling energy is lower for smaller
  monomers and of course the number of monomers in an agglomerate is much
  higher if the same agglomerate is built up of lighter monomers. This would
  mean higher fragmentation energy and one would expect that the particles
  would be harder to fragment resulting in bigger grains.

We performed a simplified simulation to be able to answer this
  question. Only two different monomer sizes are considered here, $a_1=0.1
  \mu$m and $a_2=0.4 \mu$m assuming that half of the mass (or representative
  particles) belongs to the small monomers, the other half belongs to the
  big monomers.

One problem arises here with the rolling energy. The rolling energy
  changes with monomer size, and as our method cannot follow exactly the
  number of contacts in an aggregate and what kind of monomers are
  connected, we are forced to use an averaged rolling energy. One has to
  carefully consider what the average rolling energy should be. In our case,
  the big monomer is 64 times heavier than the small monomer. Let's assume
  that 50\% of the mass of an aggregate is built up from small monomers; on
  the other hand, if we compare the number of different monomers, the small
  monomers will be 64 times more numerous than the big ones. This means that
  the contribution of small monomers in the average rolling energy ($\bar
  E_{roll}$) should be higher. This can be achieved by using the following
  weighting:
\begin{equation}
\bar E_{roll} = \frac{a_1}{a_1+a_2} E_{roll_2}+ \frac{a_2}{a_1+a_2} E_{roll_1},
\end{equation}
where $E_{roll_1}$ is the rolling energy between monomers with
  radius $a_1$, $E_{roll_2}$ is the rolling energy between monomers with
  radius $a_2$.

As we can see in Figure \ref{2mon}, the maximum aggregate sizes
  reached are approximately an order of magnitude higher than on Figure
  \ref{res_comp} as it was predicted earlier in this Section.

\section{Conclusions and outlook}
We have shown that our representative particle method for aggregation of
particles in astrophysical settings works well for standard kernels. It has
the usual advantages of Monte Carlo methods that one can add particle
properties easily and without loss of computational speed. Moreover, it
naturally conserves the number of computational elements, so there is no
need to ``add'' or ``remove'' particles. Each representative particle
represents a fixed portion of the total mass of solids. 

Our method may have various possible interesting extensions and
applications. Here we speculate on a few of these. For instance, the fact
that each representative particle corresponds to a fixed amount of solid
mass makes the method ideal for implementation into spatially resolved
models such as hydrodynamic simulations of planetary atmospheres or
protoplanetary disks. We can then follow the exact motion of each
representative particle through the possibly turbulent environment, and
thereby automatically treat the stochastic nature and deviation from a
Boltzmann distribution of the motion of particles with stopping times of the
same order as the turbulent eddy turn-over time. It is necessesary,
  however, to assure that a sufficiently large number of representative
  particles is present in each grid cell of the hydrodynamic simulation.
  For large scale hydrodynamic simulations this may lead to a very large
  computational demand for the coagulation computation, as well as for
  tracking the exact motion of these particles. If strong clumping of the
  particles happens, however, much of the ``action'' anyway happens in these
  ``clumps'', and it may then not be too critical that other grid cells are
  not sufficiently populated by representative particles. This, however, is
  something that has to be experimented.

Our representative particle
method can in principle also be used to model the sublimation and
condensation of dust grains. If a particle sublimates then the
representative particle becomes simply an atom or molecule of the vapor of
this process. It will then follow the gas motion until the temperature
becomes low enough that it can condense again. Other representative
particles which are still in the solid phase may represent physical
particles that can act as a condensation nucleus. Finally, in our method the
properties of the particle can not only change due to collisions, but we can
easily implement other environmental factors in the alteration of particle
properties.

There are two main drawbacks of the method. First, it only works for large
particle numbers, i.e.\ it cannot treat problems in which individual
particles start dominating their immediate environment. Ormel's method and
its expected extension do not have this problem. Secondly, the method
cannot be accelerated using implicit integration, while Brauer's method can.

All in all we believe that this method may have interesting applications 
in the field of dust aggregation and droplet coagulation in protoplanetary
disks and planetary atmospheres.



%% file: Chapters/Chapter3.tex

\chapter{Mapping the zoo of laboratory experiments} 
\label{chp:paper2}
\lhead{Chapter 3. \emph{Mapping the zoo of experiments}} 
\rhead{}

Based on \textit{`The outcome of protoplanetary dust growth: pebbles, boulders, or planetesimals? I. Mapping the zoo of laboratory collision experiments'} by C. G\"uttler, J. Blum, A. Zsom, C. W. Ormel \& C. P. Dullemond published in \aap, 513, 56.

\def\pressure{dyn~cm$^{-2}$}
\def\cms{cm~s$^{-1}$}
\def\heisselmann{D. Hei{\ss}elmann et al. (in prep.)}
\def\schraepler{R. Schr{\"a}pler \& J. Blum (in prep.)}
\def\heisselmannpar{D. Hei{\ss}elmann et al., in prep.}
\def\schraeplerpar{R. Schr\"apler \& J. Blum, in prep.}

\def\Sa{hit-and-stick (S1)}
\def\Sb{sticking through surface effects (S2)}
\def\Sc{sticking by deep penetration (S3)}
\def\Sd{fragmentation with mass transfer (S4)}
\def\Ba{bouncing with compaction (B1)}
\def\Bb{bouncing with mass transfer (B2)}
\def\Fa{fragmentation (F1)}
\def\Fb{erosion (F2)}
\def\Fc{fragmentation with mass transfer (F3)}
\def\pp{\textit{`pp'}}
\def\pP{\textit{`pP'}}
\def\cc{\textit{`cc'}}
\def\cC{\textit{`cC'}}
\def\pc{\textit{`pc'}}
\def\pC{\textit{`pC'}}
\def\cp{\textit{`cp'}}
\def\cP{\textit{`cP'}}

\section{Introduction\label{sec:introduction}}
The first stage of protoplanetary growth is still not fully understood. Although our empirical knowledge on the collisional properties of dust aggregates has considerably widened over the past years \citep{BlumWurm:2008}, there is no self-consistent model for the growth of macroscopic dust aggregates in protoplanetary disks (PPDs). A reason for such a lack of understanding is the complexity in the collisional physics of dust aggregates. Earlier assumptions of perfect sticking have been experimentally proven false for most of the size and velocity ranges under consideration. Recent work also showed that fragmentation and porosity play important roles in mutual collisions between protoplanetary dust aggregates. In their review paper, \citet{BlumWurm:2008} show the complex diversity that is inherent to the collisional interaction of dust aggregates consisting of micrometer-sized (silicate) particles. This complexity is the reason why the outcome of the collisional evolution in PPDs is still unclear and why no `grand' theory on the formation of planetesimals, based on firm physical principles, has so far been developed.

The theoretical understanding of the physics of dust aggregate collisions has seen major progress in recent decades. The behavior of aggregate collisions at low collisional energies -- where the aggregates show a fractal nature -- is theoretically described by the molecular dynamics simulations of \citet{DominikTielens:1997}. The predictions of this model -- concerning aggregate sticking, compaction, and catastrophic disruption -- could be quantitatively confirmed by the laboratory collision experiments of \citet{BlumWurm:2000}. Also, the collision behavior of macroscopic dust aggregates was successfully modeled by a smooth particle hydrodynamics method, calibrated by laboratory experiments \citep{GuettlerEtal:2009a, GeretshauserEtal:preprint}. These simulations were able to reproduce bouncing collisions, which were observed in many laboratory experiments \citep{BlumWurm:2008}.

As laboratory experiments have shown, collisions between dust aggregates at intermediate energies and sizes are characterized by a plethora of outcomes: ranging from (partial) sticking, bouncing, and mass transfer to catastrophic fragmentation \citep[see][]{BlumWurm:2008}. From this complexity, it is clear that the construction of a simple theoretical model which agrees with all these observational constraints is very challenging. But in order to understand the formation of planetesimals, it is imperative to describe the entire phase-space of interest, i.e., to consider a wide range of aggregate masses, aggregate porosities, and collision velocities. Likewise, the collisional outcome is a key ingredient of any model that computes the time evolution of the dust size distribution. These collisional outcomes are mainly determined by the collision velocities of the dust aggregates, and these depend on the disk model, i.e. the gas and material density in the disk and the degree of turbulence. Thus, the choice of the disk model (including its evolution) is another major ingredient for dust evolution models.

These concerns lay behind the approach we adopt in this and subsequent chapters. That is, instead of first `funneling' the experimental results through a (perhaps ill-conceived) theoretical collision model and then to calculate the collisional evolution, we will directly use the experimental results as input for the collisional evolution model. The drawback of such an approach is of course that experiments on dust aggregate collisions do not cover the whole parameter space and therefore need to be extrapolated by orders of magnitude, based on simple physical models whose accuracy might be challenged. We still feel that this drawback is more than justified by the prospects that our new approach will provide: through a direct mapping of the laboratory experiments, collisional evolution models can increase enormously in their level of realism.

In this chapter, we will classify all existing dust-aggregate collision experiments for silicate dust, including three additional original experiments not published before, according to the above parameters (Sect. \ref{sec:exp-review}). We will show that we have to distinguish between nine different kinds of collisional outcomes, which we physically describe in Sect. \ref{sec:exp_types}. For the later use in a growth model, we will sort these into a mass-velocity parameter space and find that we have to distinguish between eight regimes of porous and compact dust-aggregate projectiles and targets. We will present our collision model in Sect. \ref{sec:collision_regimes} and the consequences for the porosities of the dust aggregates in Sect. \ref{sec:porosities}. In Sect. \ref{sec:conclusion}, we conclude our work and give a critical review on our model and the involved necessary simplifications and extrapolations.

In Chapter \ref{chp:paper3} \citep{ZsomEtal:2009} we will then, based upon the results presented here, follow the dust evolution using a recently invented Monte-Carlo approach \citep{ZsomDullemond:2008} for three different disk models. This is the first fully self-consistent growth simulation for PPDs. The results presented in Chapter \ref{chp:paper3} represent the state-of-the-art modeling and will give us important insight into questions, such as if the meter-size barrier can be overcome and what the maximum dust-aggregate size in PPDs is, i.e. whether pebbles, boulders, or planetesimals can be formed.

\section{\label{sec:exp-review}Collision experiments with relevance to planetesimal formation}
\begin{table*}[t]
\center%
\caption{\label{tab:experiments}Table of the experiments which are used for the model.}
\scriptsize
\begin{tabular}{lccccl}
    \hline
           &  projectile mass    & collision velocity & micro-  & collisional outcome                            & reference \\%
           &  $m_\mathrm{p}$ [g] & $v$ [\cms]  & gravity & (see Fig. \ref{fig:pictograms}) & \\%
    \hline
    Exp 1  & $7.2\cdot 10^{-12}$ -- $7.2\cdot 10^{-9}$     & 0.1 -- 1             & yes & S1         & \citet{BlumEtal:1998, BlumEtal:2002},\\
           &                                               &                      &     &            & \citet{WurmBlum:1998}\\
    Exp 2  & $7.2\cdot 10^{-12}$ -- $2.0\cdot 10^{-10}$    & 10 -- 50             & yes & S1         & \citet{WurmBlum:1998}\\
    Exp 3  & $3.5\cdot 10^{-12}$ -- $3.5\cdot 10^{-10}$    & 0.02 -- 0.17         & yes & S1         & \citet{BlumEtal:2000},\\
           & $1.0\cdot 10^{-12}$ -- $1.0\cdot 10^{-10}$    & 0.04 -- 0.46         & yes & S1         & \citet{KrauseBlum:2004}\\
    Exp 4  & $1.2\cdot 10^{-10}$ -- $4.3\cdot 10^{-10}$    & 7 -- $1\,000$        & yes & S2         & \citet{BlumWurm:2000}\\
    Exp 5  & $2\cdot 10^{-3}$ -- $7\cdot 10^{-3}$          & 15 -- 390            & yes & B1, F1     & \citet{BlumMuench:1993}\\
           & $10^{-5}$-- $10^{-4}$                         & 15 -- 390            & yes & B1, F1     & \\
    Exp 6  & $10^{-6}$ -- $10^{-4}$                        & 10 -- 170            & yes & S2, S3     & \citet{LangkowskiEtal:2008}\\
           & $10^{-4}$ -- $3\cdot 10^{-3}$                 & 50 -- 200            & yes & B2, S2, S3 & \\
           & $2.5\cdot 10^{-5}$ -- $3\cdot 10^{-3}$        & 200 -- 300           & yes & S3         & \\
    Exp 7  & $10^{-3}$ -- $3\cdot 10^{-2}$                 & 20 -- 300            & yes & S3         & \citet{BlumWurm:2008}\\
    Exp 8  & $10^{-3}$ -- $3.2\cdot 10^{-2}$               & 16 -- 89             & no  & S3         & \citet{GuettlerEtal:2009a}\\
    Exp 9  & $10^{-3}$ -- $10^{-2}$                        & 10 -- 40             & yes & B1         & \heisselmann\\
           & $10^{-3}$ -- $10^{-2}$                        & 5 -- 20              & yes & B1         & \\
    Exp 10 & $2\cdot 10^{-3}$ -- $5\cdot 10^{-3}$          & 1 -- 30              & no  & B1         & \citet{WeidlingEtal:2009}\\
    Exp 11 & $1.6\cdot 10^{-4}$ -- $3.4 \cdot 10^{-2}$     & 320 -- 570           & yes & F1         & \citet{Lammel:2008}\\
    Exp 12 & $3.5\cdot 10^{-15}$                           & $1\,500$ -- $6\,000$ & no  & F2         & \schraepler\\
    Exp 13 & 0.2 -- 0.3                                    & $1\,650$ -- $3\,750$ & no  & F2         & \citet{WurmEtal:2005a}\\
    Exp 14 & 0.2 -- 0.3                                    & 350 -- $2\,150$      & yes & F2         & \citet{ParaskovEtal:2007}\\
    Exp 15 & 0.39                                          & 600 -- $2\,400$      & no  & S4         & \citet{WurmEtal:2005b}\\
    Exp 16 & $4\cdot 10^{-7}$ -- $5\cdot 10^{-5}$          & 700 -- 850           & no  & S4         & \citet{TeiserWurm:2009b}\\
    Exp 17 & $1.6\cdot 10^{-4}$ -- $2.0\cdot 10^{-2}$      & 100 -- $1\,000$      & no  & S4         & Sect. \ref{sec:new_exp_1}\\
    Exp 18 & $10^{-9}$ -- $10^{-4}$                        & 10 -- $1\,000$       & no  & B1, S2, S4 & Sect. \ref{sec:new_exp_2}\\
    Exp 19 & $1.5\cdot 10^{-3}$ -- $3.2\cdot 10^{-3}$      & 200 -- 700           & yes & S4, F3     & Sect. \ref{sec:new_exp_3}\\
    \hline
\end{tabular}
\end{table*}

In the past years, numerous laboratory and space experiments on the collisional evolution of protoplanetary dust have been performed \citep{BlumWurm:2008}. Here, we concentrate on the dust evolution around a distance of 1 AU from the solar-type central star where the ambient temperature is such that the dominating material class are the silicates. This choice of 1 AU reflects the kind of laboratory experiments that are included in this chapter, which were all performed with SiO$_2$ grains or other refractory materials. The solid material in the outer solar nebula is dominated by ices, which possibly have very different material properties than silicates, but only a small fraction of laboratory experiments have dealt with these colder (ices, organic materials) or also warmer regions (oxides). In Sect. \ref{sec:material_influence}, we will discuss the effect that another choice of material might potentially have, but as we are far away from even basically comprehending the collisional behavior of aggregates consisting of these materials, we concentrate in this study on the conditions relevant in the inner solar nebula around \mbox{1 AU}.

\begin{figure*}[t]
    \center
    \includegraphics[width=1\textwidth]{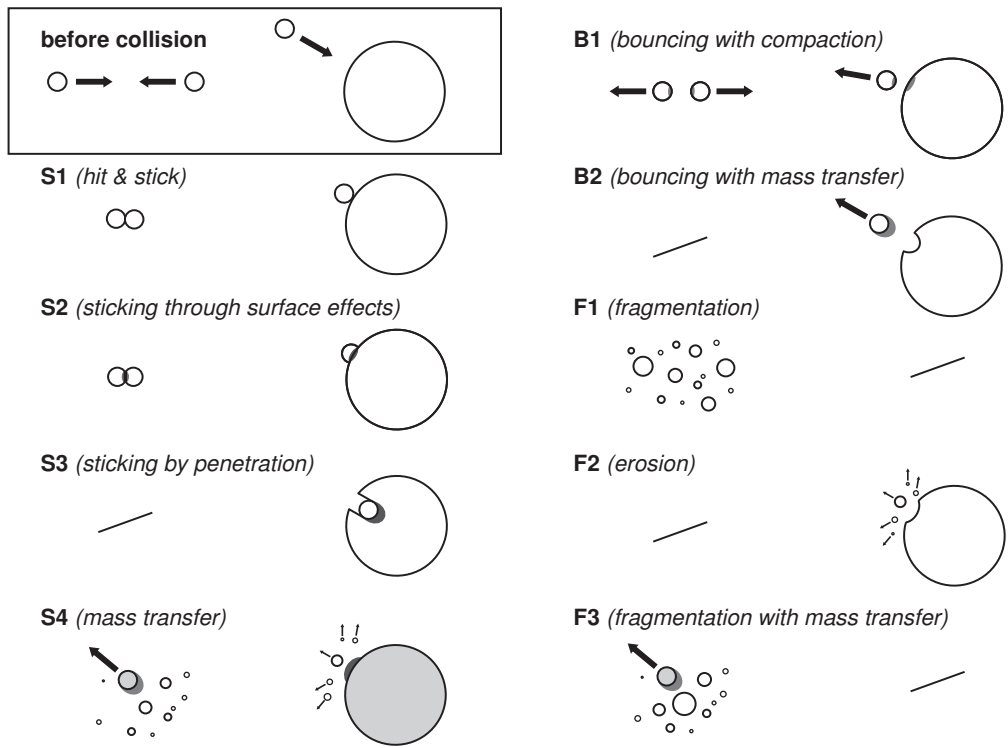}
    \caption{\label{fig:pictograms}We classify the variety of laboratory experiments into nine kinds of collisional outcomes, involving sticking (S), bouncing (B) and fragmenting (F) collisions. All these collisional outcomes have been observed in laboratory experiments, and detailed quantities on the outcomes are given in Sect. \ref{sec:exp_types}.}
\end{figure*}

Table \ref{tab:experiments} lists all relevant experiments that address collisions between dust aggregates of different masses, mass ratios, and porosities, consisting of micrometer-sized silicate dust grains, in the relevant range of collision velocities. Experiments 1 -- 16 are taken from the literature (cited in Table \ref{tab:experiments}), whereas experiments 17 -- 19 are new ones not published before. In the following two subsections we will first review the previously published experiments (Sect. \ref{sec:exp-literature}) and then introduce the experimental setup and results of new experiments that were performed to explore some regions of interest (Sect. \ref{sec:new_experiments}). All these collisions show a diversity of different outcomes for which we classify nine different collisional outcomes as displayed in Fig. \ref{fig:pictograms}. Details on these collisional outcomes are presented in Sect. \ref{sec:exp_types}.

\subsection{\label{sec:exp-literature}A short review on collision experiments}
We briefly review published results of dust-collision experiments here since these determine the collisional mapping in Sect. \ref{sec:exp_types} and \ref{sec:collision_regimes}. The interested reader is referred to the review by \citet{BlumWurm:2008} for more information. All experiments are compiled and referenced in Table \ref{tab:experiments} where we also list the collision velocities and projectile masses, as these will be used in Sect. \ref{sec:collision_regimes}. Most of the experiments in Table \ref{tab:experiments} (exception: Exp 10) were performed under low gas pressure conditions to match the situation in PPDs, and most of the experiments were carried out in the absence of gravity (i.e. free falling aggregates or micro-gravity facilities), see Col. 4 of Table \ref{tab:experiments}. For the majority of the experiments, spherical monodisperse SiO$_2$ monomers with diameters between 1.0 $\rm \mu m$ and 1.9 $\rm \mu m$ were used; some experiments used irregular SiO$_2$ grains with a wider size distribution centered around $\sim 1.0 ~\rm \mu m$, and Exp 5 used irregular $\rm ZrSiO_4$ with monomer diameters in the range $0.2 \ldots 1.0\ \mathrm{\mu m}$.

\emph{Exp 1 -- 4:} A well-known growth mechanism for small dust aggregates is the hit-and-stick growth, in which the aggregates collide with such a low kinetic energy that they stick at each other upon first contact without any restructuring. The first experiments to unambiguously show that the hit-and-stick process is relevant to protoplanetary dust aggregation were those by \citet{WurmBlum:1998}, \citet{BlumEtal:1998, BlumEtal:2000, BlumEtal:2002} and \citet{KrauseBlum:2004}. These proved that, as long as the collision velocities for small dust aggregates stay well below 100~\cms, sticking collisions lead to the formation of fractal aggregates. This agrees with the molecular-dynamics simulations by \citet{DominikTielens:1997} and \citet{WadaEtal:2007, WadaEtal:2008, WadaEtal:2009}. The various experimental approaches for Exp 1 -- 3 used all known sources for relative grain velocities in PPDs, i.e. Brownian motion (Exp 3), relative sedimentation (Exp 1), and gas turbulence (Exp 2). In these papers it was also shown that the hit-and-stick growth regime leads to a  quasi-monodisperse evolution of the mean aggregate masses, depleting small grains efficiently and rapidly. For collisions between these fractal aggregates and a solid or dusty target, \citet[Exp 4]{BlumWurm:2000} found growth at even higher velocities, in which the aggregates were restructure. This also agrees with molecular-dynamics simulations \citep{DominikTielens:1997}, and so this first stage of protoplanetary dust growth has so far been the only one that could be fully modeled.

\emph{Exp 5:} \citet{BlumMuench:1993} performed collision experiments between free falling ZrSiO$_4$ aggregates of intermediate porosity ($\phi = 0.35$, where $\phi$ is the volume fraction of the solid material) at velocities in the range of 15 -- 390~\cms. They found no sticking, but, depending on the collision velocity, the aggregates bounced ($v < 100$~\cms) or fragmented into a power-law size distribution ($v > 100$~\cms). The aggregate masses were varied over a wide range ($10^{-5}$ to $7 \times 10^{-3}$~g), and the mass ratio of the two collision partners also ranged from 1:1 to 1:66. The major difference to experiments 1 -- 4, which inhibited sticking in these collisions, were the aggregate masses and their non-fractal but still very porous nature.

\emph{Exp 6 -- 8:} A new way of producing highly porous, macroscopic dust aggregates ($\phi=0.15$ for 1.5~$\mu$m diameter SiO$_2$ monospheres) as described by \citet{BlumSchraepler:2004} allowed new experiments, using the 2.5~cm diameter aggregates as targets and fragments of these as projectiles \citep[Exp 6]{LangkowskiEtal:2008}. In their collision experiments in the Bremen drop tower, \citet{LangkowskiEtal:2008} found that the projectile may either bounce off from the target at intermediate velocities (50 -- 250~\cms) and aggregate sizes (0.5 -- 2~mm), or stick to the target for higher or lower velocities and bigger or smaller sizes, respectively. This bouncing went with a previous slight intrusion and a mass transfer from the target to the projectile. In the case of small and slow projectiles, the projectile stuck to the target, while large and fast projectiles penetrated into the target and were geometrically embedded. They also found that the surface roughness plays an important role for the sticking efficiency. If a projectile hits into a surface depression, it sticks, while it bounces off when hitting a hill with a small radius of curvature comparable to that of the projectile. A similar behavior for the sticking by deep penetration was also found by \citet[Exp 7]{BlumWurm:2008} when the projectile aggregate is solid -- a mm-sized glass bead in their case. Continuous experiments on the penetration of a solid projectile (1 to 3~mm diameter) into the highly porous target \citep[$\phi=0.15$,][]{BlumSchraepler:2004} were performed by \citet[Exp 8]{GuettlerEtal:2009a} who studied this setup for the calibration of a smoothed particle hydrodynamics (SPH) collision model. We will use their measurement of the penetration depth of the projectile.

\emph{Exp 9 -- 10:} As a follow-up experiment of the study of \citet{BlumMuench:1993}, D. Hei{\ss}elmann, H.J. Fraser and J. Blum (in prep., Exp 9) used 5~mm cubes of these highly porous ($\phi=0.15$) dust aggregates and crashed them into each other ($v=40$~\cms) or into a compact dust target with $\phi=0.24$ ($v=20$~\cms). In both cases they too found bouncing of the aggregates and were able to confirm the low coefficient of restitution ($v_\mathrm{after} / v_\mathrm{before}$) of $\varepsilon = 0.2$ for central collisions. In their experiments they could not see any deformation of the aggregates, due to the limited resolution of their camera, which could have explained the dissipation of energy. This line of experiments was taken up again by \citet[Exp 10]{WeidlingEtal:2009} who studied the compaction of the same aggregates which repeatedly collided with a solid target. They found that the aggregates decreased in size (without losing significant amounts of mass), which is a direct measurement of their porosity. After only $1\,000$ collisions the aggregates were compacted by a factor of two in volume filling factor, and the maximum filling factor for the velocity used in their experiments (1 -- 30~\cms) was found to be $\phi=0.36$. In four out of 18 experiments, the aggregate broke into several pieces, and they derived a fragmentation probability of $P_\mathrm{frag}=10^{-4}$ for the aggregate to break in a collision.

\emph{Exp 11:} The same fragments of the high porosity ($\phi=0.15$) dust aggregates of \citet{BlumSchraepler:2004} as well as intermediate porosity ($\phi = 0.35$) aggregates were used by \citet[Exp 11]{Lammel:2008} who continued the fragmentation experiments of \citet{BlumMuench:1993}. For velocities from 320 to 570~\cms\ he found fragmentation and measured the size of the largest fragment as a measure for the fragmentation strength.

\emph{Exp 12 -- 14:} Exposing the same highly porous ($\phi=0.15$) dust aggregate to a stream of single monomers with a velocity from $1\,500$ to $6\,000$~\cms, R. Schr\"apler and J. Blum (in prep., Exp 12) found a significant erosion of the aggregate. One monomer impact can easily kick out tens of monomers for the higher velocities examined. They estimated the minimum velocity for this process in an analytical model to be approx. 350~\cms. On a larger scale, \citet[Exp 13]{WurmEtal:2005a} and \citet[Exp 14]{ParaskovEtal:2007} impacted dust projectiles with masses of 0.2 to 0.3~g and solid spheres into loosely packed dust targets. \citet{ParaskovEtal:2007} were able to measure the mass loss of the target in drop-tower experiments which was -- velocity dependent -- up to 35 projectile masses. The lowest velocity in these experiments was 350~\cms.

\emph{Exp 15 -- 16:} In a collision between a projectile of intermediate porosity and a compressed dust target at a velocity above 600~\cms, \citet[Exp 15]{WurmEtal:2005b} found fragmentation of the projectile but also an accretion of mass onto the target. This accretion was up to 0.6 projectile masses in a single collision depending on the collision velocity. \citet[Exp 16]{TeiserWurm:2009b} studied this partial sticking in many collisions, where solid targets of variable sizes were exposed to 100 to 500~$\mu$m diameter dust aggregates with a mean velocity of 770~\cms. Although they cannot give an accretion efficiency in a single collision, they found a large amount of mass accretion onto the targets, which is a combination of the pure partial sticking and the effects of the Earth's gravity. \citet{TeiserWurm:2009b} argue that this acceleration is equivalent to the acceleration that micron-sized particles would experience as a result of their erosion from a much bigger body which had been (partially) decoupled from the gas motion in the solar nebula.

\subsection{New experiments\label{sec:new_experiments}}
In this section, we will present new experiments which we performed to explore some parameter regions where no published data existed so far. All experiments cover collisions between porous aggregates with a solid target and were performed with the same experimental setup, consisting of a vacuum chamber (less than 0.1~mbar pressure) with a dust accelerator for the porous projectiles and an exchangeable target. The accelerator comprises a 50~cm long plastic rod with a diameter of 3~cm in a vacuum feed through. The pressure difference between the ambient air and the pressure in the vacuum chamber drives a constant acceleration, leading to a projectile velocity of up to 900~\cms, at which point the accelerator is abruptly stopped. The porous projectile flies on and collides either with a solid glass plate (Sect. \ref{sec:new_exp_1} and \ref{sec:new_exp_2}) or with a free falling glass bead, which is dropped when the projectile is accelerated (Sect. \ref{sec:new_exp_3}). The collision is observed with a high-speed camera to determine aggregate and fragment sizes and to distinguish between the collisional outcomes (i.e. sticking, bouncing, and fragmentation). The experiments in this section are also listed in Table \ref{tab:experiments} as Exp 17 to 19.

\subsubsection{\label{sec:new_exp_1}Fragmentation with mass transfer (Exp 17)}
\begin{figure}[t]
    \center
    \includegraphics[width=0.7\textwidth]{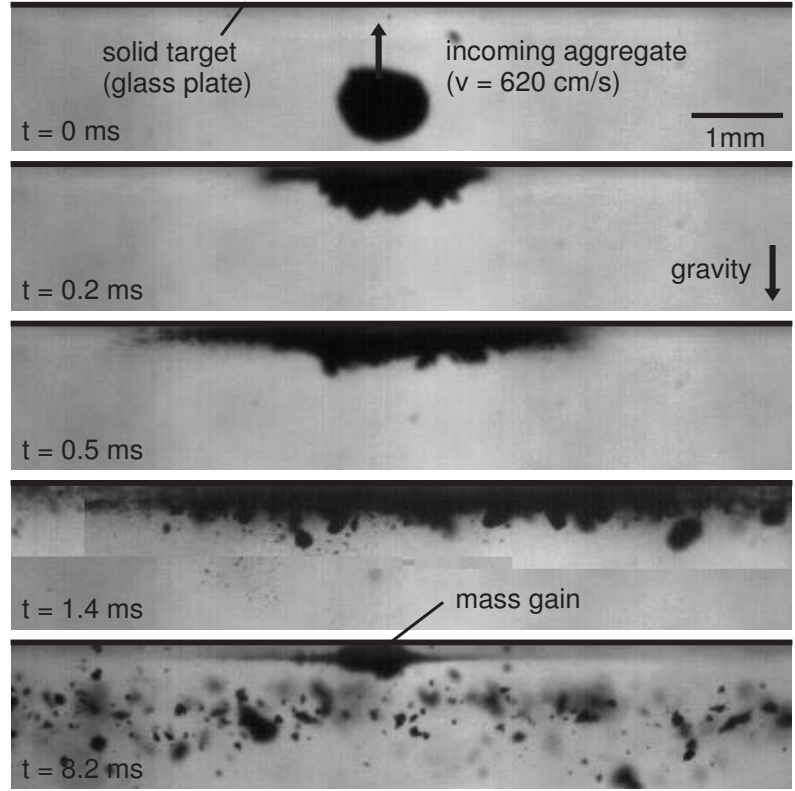}
    \caption{\label{fig:frag_img}Example for a collision of a porous ($\phi=0.35$) aggregate with a solid target at a velocity of 620~\cms. The aggregate fragments according to a power-law size distribution and some mass sticks to the target (bottom frame).}
\end{figure}
\begin{figure}[t]
    \center
    \includegraphics[width=0.5\textwidth,angle=90]{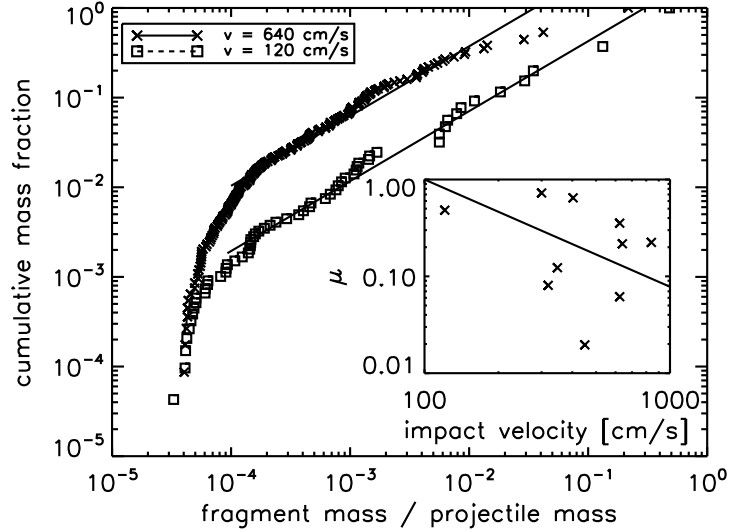}
    \caption{\label{fig:mass_dist}Mass distribution for two experiments at the velocities of 120 and 640 \cms. For the higher masses, the distribution follows a power-law, while the lower masses are depleted due to the finite camera resolution. The slopes are the same for both experiments, and there is only an offset (pre-factor) between the two. The inset describes this pre-factor $\mu$ (cf. Eq. \ref{eq:mass_dist_cum}) which is a measure for the strength of the fragmentation. The value clearly decreases with increasing velocity (Eq. \ref{eq:mu_v_exp}).}
\end{figure}
In this experiment, mm-sized aggregates of different volume filling factors ($\phi=0.15$ and $\phi=0.35$) collided with a flat and solid glass target and fragmented as the collision velocity was above the fragmentation threshold of approx. 100~\cms. The projected projectile size and its velocity were measured by a high-speed camera (see Fig. \ref{fig:frag_img}). In few experiments, the sizes of the produced fragments were measured for those fragments that were sharply resolved, which yielded a size distribution of a representative number of fragments (the number of resolved fragments varied from 100 to 400). Assuming a spherical shape of the fragments and an unchanged porosity from the original projectile, we calculated a cumulative mass distribution as shown in Fig. \ref{fig:mass_dist}, where the cumulative mass fraction $\sum_{i=0}^k (m_\mathrm{i}/M_\mathrm{F})$ is plotted over the normalized fragment mass $m_\mathrm{k}/m_\mathrm{p}$. Here, $m_\mathrm{i}$ and $M_\mathrm{F}=\sum_{i=1}^N m_\mathrm{i}$ are the mass of the $i$-th smallest fragment, and the total mass of all visible fragments and $N$ is the total number of fragments. We found that the cumulative distribution can be well described by a power law
\begin{equation}
    \int_0^m n(m')m'\;\mathrm{d}m' = \left( \frac{m}{\mu} \right)^\kappa, \label{eq:mass_dist_cum}
\end{equation}
where $m'$ and $m$ are the mass of the fragments in units of the projectile mass and $\mu$ is a parameter to measure the strength of fragmentation, defined as the mass of the largest fragment divided by the mass of the original projectile. The deviation between data and power-law for low masses (see Fig. \ref{fig:mass_dist}) is due to the finite resolution of the camera, which could not detect fragments with sizes $\ll 50~\rm \mu m$. In the ten experiments where the mass distribution was determined, the power-law index $\kappa$ was nearly constant from 0.64 to 0.93, showing no dependence on the velocity, which varied from 120 to 840~\cms. However, a clear dependence on the velocity was found for the parameter $\mu$, which decreased with increasing velocity as shown in the inset of Fig. \ref{fig:mass_dist}. This increasing strength of fragmentation can be described as
\begin{equation}
    \mu(v) = \left( \frac{v}{100\ \mathrm{cm\ s^{-1}}} \right)^{-1.1} \; \label{eq:mu_v_exp},
\end{equation}
where the exponent has an error of $\pm 0.2$. The curve was fitted to agree with the observed fragmentation threshold of 100~\cms.

It is important to know that the number density of fragments of a given mass follows from Eq. \ref{eq:mass_dist_cum} as
\begin{equation}
    n(m') = \frac{\kappa}{\mu^\kappa} m'^{\kappa-2}, \label{eq:mass_dist}
\end{equation}
and that the power law for this mass distribution can be translated into a power-law size distribution $n(a) \propto a^\lambda$ with $\lambda = 3\kappa - 4$. This yields $\lambda$ values from $-2.1$ to $-1.2$, which is much flatter than the power-law index of $-3.5$ from the MRN distribution \citep{MathisEtal:1977}, which is widely used for the description of high-speed fragmentation of {\em solid} materials. Moreover, this power-law index is consistent with measurements of \citet{BlumMuench:1993} who studied aggregate-aggregate collisions between millimeter-sized ZrSiO$_4$ aggregates (see Sect. \ref{sec:exp-review}). Their power-law index equivalent to $\lambda$ was $-1.4$, and for different velocities they also found a constant power-law index and a velocity-dependent pre-factor (their Fig. 8a).

\begin{figure}[t]
    \center
    \includegraphics[width=0.5\textwidth,angle=90]{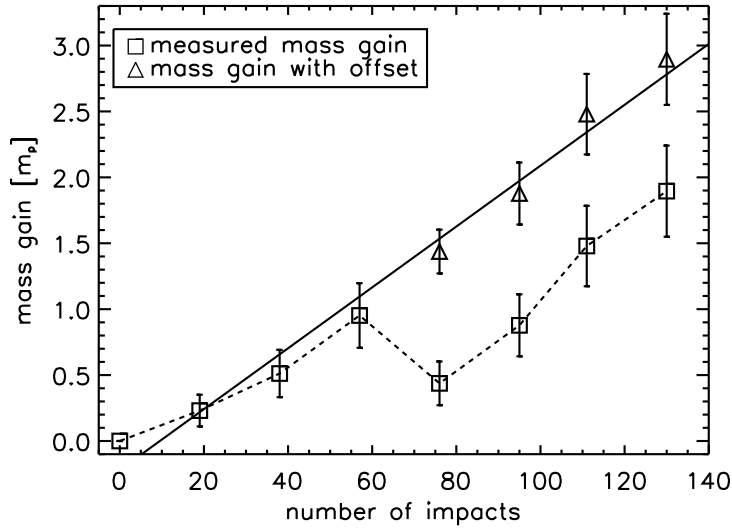}
    \caption{\label{fig:S4_mass_gain}Mass gain of a solid target in 133 collisions (S. Kothe, C. G\"uttler \& J. Blum, unpublished data). The target was weighed after every 19 collisions. After 57 collisions, one projectile mass of dust was chipped off the target, which is a clear effect of gravity. Thus, we added this mass to the following measurements (triangles) and fitted a linear mass gain, which is $0.023 \times m_\mathrm{p}$ in every collision (solid line).}
\end{figure}

While most of the projectile mass fragmented into a power-law distribution, some mass fraction stuck to the target (see bottom frame in Fig. \ref{fig:frag_img}). Therefore, the mass of the target was weighed before the collision and again after 19 shots on the same spot. The mass of each  projectile was weighed and yielded a mean value of $3.34 \pm 0.84$ mg per projectile. The increasing mass of the target in units of the projectile mass is plotted in Fig. \ref{fig:S4_mass_gain}. After 57 collisions, dust chipped off the target, which can clearly be credited to the gravitational influence. For the following measurements we therefore added one projectile mass to the target because we found good agreement with the previous values for this offset. The measurements were linearly fitted and the slope, which determines the mass gain in a single collision, was 2.3~\% (S. Kothe, C. G\"uttler \& J. Blum, unpublished data).

\subsubsection{\label{sec:new_exp_2}Impacts of small aggregates (Exp 18)}
\begin{figure}[t]
    \center
    \includegraphics[width=6cm]{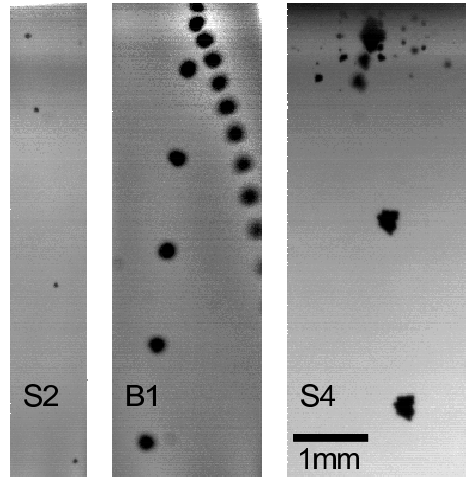}
    \caption{\label{fig:small_coll_img}Examples for the experimental outcomes in the collisions of small aggregates with a solid target. The collision can lead to sticking, bouncing, or fragmentation (from left to right). The time between two exposures is 2~ms.}
\end{figure}
Using exactly the same setup as in the previous section, we performed collision experiments with very small (20~$\mu$m to 1.4~mm diameter) but non-fractal projectiles. Those aggregates were fragments of larger dust samples as described by \citet{BlumSchraepler:2004} and had a volume filling factor of $\phi=0.15$. In this experiment we observed not only fragmentation but also bouncing and sticking of the projectiles to the solid glass target. Thus, the analysis with the high-speed camera involved the measurement of projectile size, collision velocity, and collisional outcome, where we distinguished between (1) perfect sticking, (2) perfect bouncing without mass transfer, (3) fragmentation with partial sticking, and (4) bouncing with partial sticking. The difference between the cases (3) and (4) is that in a fragmentation event at least two rebounding aggregates were produced, whereas in the bouncing collision only one aggregate bounced off.
\begin{figure}[t]
    \center
    \includegraphics[width=0.5\textwidth,angle=90]{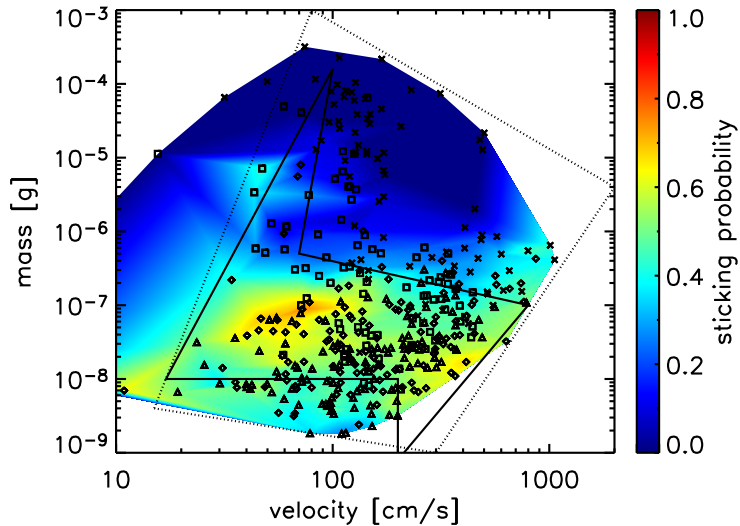}
    \caption{\label{fig:small_coll}Overview on collision experiments between 20 to 1400~$\mu$m diameter aggregates and a solid target, which leads to sticking (diamonds), bouncing (triangles), or fragmentation (crosses). The intermediate sticking-bouncing collision is indicated by the squared symbols. The color indicates the sticking probability, i.e. the fraction of sticking events in a logarithmic bin around every node. The dotted box denotes the approximated parameter range and the solid lines denote the threshold between sticking, bouncing and fragmentation as also used in Fig. \ref{fig:colored_regimes}.}
\end{figure}

For the broad parameter range in diameter (20 to 1400~$\mu$m) and velocity (10 to $1\,000$~\cms), we performed 403 individual collisions in which we were able to measure size, velocity, and collisional outcome. Examples for sticking, bouncing, and fragmentation are shown in Fig. \ref{fig:small_coll_img}. The full set of data is plotted in Fig. \ref{fig:small_coll}, where different symbols were used for different collisional outcomes. Clearly, collisions of large aggregates and high velocities lead to fragmentation, while small aggregates tend to bounce off the target. For intermediate aggregate mass (i.e. $m_\mathrm{p}=10^{-7}$~g), all kinds of collisions can occur. The background color shows a sticking probability, which was calculated as a boxcar average (logarithmic box) at every node where an experiment was performed. Blue color denotes a poor sticking probability, while a green to yellow color shows a sticking probability of approx. 50~\%. We draw the solid lines in a polygon [$(100,70,800,200,200,17)$~\cms, $(1.6\cdot 10^{-4},5\cdot 10^{-7},1\cdot 10^{-7},8\cdot 10^{-10},1\cdot 10^{-8},1\cdot 10^{-8})$~g] to mark the border between sticking and non-sticking as we will use it in Sect. \ref{sec:collision_regimes}. For the higher masses, this accounts for a bouncing-fragmentation threshold of 100~\cms\ at $1.6\cdot 10^{-4}$~g (Exp 18), and for the lower masses, we assume a constant fragmentation threshold of 200~\cms, which roughly agrees with the restructuring-fragmentation threshold of \citet[Exp 4]{BlumWurm:2000}. For lower velocities outside the solid-line polygon, bouncing collisions are expected, whereas for higher velocities outside the polygon, we expect fragmentation. Thus, an island of enhanced sticking probability for $10^{-7}$ -- $10^{-7}$~g aggregates at a broad velocity range from 30 to 500~\cms\ was rather unexpected before. The dotted box is just a rough borderline showing the parameters for which the experiments were performed as it will also be used in Sect. \ref{sec:collision_regimes}.

\subsubsection{\label{sec:new_exp_3}Collisions between similar sized solid and porous aggregates (Exp 19)}
\begin{figure}[t]
    \center
    \includegraphics[width=0.5\textwidth,angle=90]{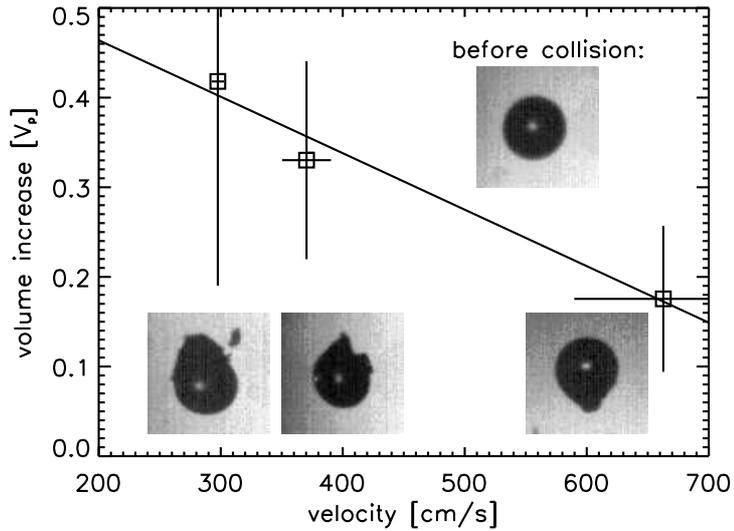}
    \caption{\label{fig:F3_plot}The volume gain of a solid particle colliding with a porous aggregate depends on the collision velocity. The data points are mean values of 11, 8, and 7 individual experiments (left to right), thus, the error bars show the $1\sigma$ standard deviation of velocities and volume gain in these. The images with a width of 1.9~mm show the original 1~mm glass bead and examples for the mass gain in the three corresponding collision velocities (S. Olliges \& J. Blum, unpublished data).}
\end{figure}
In a collision between a free falling glass bead of 1~mm diameter and a porous ($\phi=0.15$) dust aggregate of 1.5 to 8.5~mg mass, we observed fragmentation of the porous aggregate while some mass was growing on the solid and indestructible glass bead (S. Olliges \& J. Blum, unpublished data). In this case, the high-speed camera was used with a 3D optics that allowed the imaging of the collision from two angles, separated by 90$^\circ$. On the one hand this made it possible to exactly measure the impact parameter $b$ also if the offset of the two collision partners is in the line of sight of one viewing angle. Moreover, observing the mass growth of the solid projectile is not only a projection in one direction but can be reconstructed to get a 3D measurement. The relative velocity and aggregate size were accordingly measured from the images before the collision while the mass gain of the solid glass bead was measured after the collision. Figure \ref{fig:F3_plot} shows a diagram of the volume gain in units of projectile volume (projectile: porous aggregate) over the collision velocity. The three data points are averaged over a number of experiments at the same velocity. The error bars denote the $1 \sigma$ standard deviation of collision velocities and projectile volume, respectively. A clear trend shows that the volume gain of the solid particle decreases with velocity, and we fitted the data points with
\begin{equation}
    \Delta V = V_\mathrm{p} \left( 0.59 - 6.3 \times 10^{-4} \frac{v}{\mathrm{cm\ s^{-1}}} \right) \ , \label{eq:F3_vol_trans}
\end{equation}
where $V_\mathrm{p}$ is the volume of the glass bead. In this experiment we were not able to measure the size distribution of the fragments because the absolute velocity is determined by the projectile velocity (up to 600~\cms), and the faster fragments were out of the frame before they were clearly separated from each other.

\section{Classification of the laboratory experiments\label{sec:exp_types}}
In this section, the experiments outlined above will be categorized according to their physical outcomes in the respective collisions. In Sect. \ref{sec:exp-review}, we saw that various kinds of sticking, bouncing, and fragmentation can occur. Here, we will keep all these experiments in mind and classify them according to nine kinds of possible collisional outcomes that were observed in laboratory experiments. These collisional outcomes are displayed in Fig. \ref{fig:pictograms}. The denomination of the classification follows S for sticking, B for bouncing, and F for fragmentation. S and F are meant with respect to the target, i.e. the more massive of the two collision partners. We will discuss each of the pictograms in Fig. \ref{fig:pictograms}, describe the motivation for the respective collisional outcomes and physically quantify the outcome of these collisions.

(1) {\em Sticking collisions:} A well-known growth mechanism is due to \Sa\ collisions. Hit-and-stick growth was observed in the laboratory \citep{BlumWurm:2000, BlumEtal:2000} and numerically described \citep{DominikTielens:1997}. Experiments show that the mass distribution during the initial growth phase is always quasi-monodisperse. The evolution of the mean mass within an ensemble of dust aggregates due to \Sa collisions was calculated to follow a power-law in time, in good agreement with the experiments \citep{WurmBlum:1998,KrauseBlum:2004}. \citet{DominikTielens:1997} showed theoretically and \citet{BlumWurm:2000} confirmed experimentally that small fractal aggregates stick at first contact if their collision energy is smaller than a threshold energy. For higher energies, experiments showed that an aggregate is elastically and plastically deformed at the contact zone \citep{BlumMuench:1993,WeidlingEtal:2009}. This increases the number of contacts, which can then lead to sticking at higher velocities, an effect we call \Sb. \citet{LangkowskiEtal:2008} found that sticking can occur for even larger velocities if the target aggregate is porous and significantly larger than the projectile. In this case, the projectile sticks by deep penetration (S3) into the target and cannot rebound simply because of geometrical considerations. This effect holds also true if the projectile aggregate is compact, which has been shown by \citet{BlumWurm:2008} and further studied by \citet{GuettlerEtal:2009a}. In Sect. \ref{sec:new_exp_1}, we saw that the growth of a solid target can occur if a porous projectile fragments and partially sticks to the target surface (S4). This growth mechanism was already described by \citet{WurmEtal:2005b}. \citet{TeiserWurm:2009b} found it to be an efficient growth mechanism in multiple collisions.

(2) {\em Bouncing collisions:} If the collision velocity of two dust aggregates is too low for fragmentation and too high for sticking to occur, the dust aggregates will bounce (B1). \heisselmann\ found highly inelastic bouncing between similar-sized porous dust aggregates and between a dust aggregate and a dusty but rather compact target, where 95~\% of the kinetic energy were dissipated. \citet{WeidlingEtal:2009} showed that the energy can effectively be dissipated by a significant (and for a single collision undetectable) compaction of the porous aggregates after multiple collisions (collisional outcome \Ba). Another kind of bouncing occurred in the experiments of \citet{LangkowskiEtal:2008} in which a porous projectile collided with a significantly bigger and also highly porous target aggregate. If the penetration of the aggregate was too shallow for the S3 sticking to occur, the projectile bounced off and took away mass from the target aggregate. This \Bb\ was also observed in the case of compact projectiles \citep{BlumWurm:2008}.

(3) {\em Fragmenting collisions:} Fragmentation (F1), i.e. the breakup of the dust aggregates, occurs in collisions between similar-sized dust aggregates at a velocity above the fragmentation threshold. \citet{BlumMuench:1993} showed that both aggregates are then disrupted into a power-law size distribution. If a target aggregate is exposed to impacts of single monomer grains or very small dust aggregates, \schraepler\ found that the target aggregate is efficiently eroded (F2) if the impact velocities exceed $1\,500$~\cms. This mass loss of the target was also observed in the case of larger projectiles into porous targets \citep{WurmEtal:2005a, ParaskovEtal:2007}. Similar to the F1 fragmentation, it may occur that one aggregate is porous while the other one is compact. In that case, the porous aggregate fragments but cannot destroy the compact aggregate. The compact aggregate accretes mass from the porous aggregate (Sect. \ref{sec:new_exp_3}). We call this \Fc.

These nine fundamental kinds of collisions are all based on firm laboratory results. Future experiments will almost certainly modify this picture and potentially add so far unknown collisional outcomes to this list. But at the present time this is the complete picture of possible collisional outcomes. Below we will quantify the thresholds and boundaries between the different collision regimes as well as characterize physically the collisional outcomes therein.

\subsection*{S1: Hit-and-stick growth\label{sec:S1}}
Hit-and-stick growth occurs when the collisional energy involved is less than $5 \cdot E_\mathrm{roll}$ \citep{DominikTielens:1997, BlumWurm:2000}, where $E_\mathrm{roll}$ is the energy which is dissipated when one dust grain rolls over another by an angle of $90^\circ$. We can calculate the upper threshold velocity for the hit-and-stick mechanism of two dust grains by using the definition relation between rolling energy and rolling force, i.e.
\begin{equation}
    E_\mathrm{roll} = \frac{\pi}{2} a_0 F_\mathrm{roll} ~.
\end{equation}
Here, $a_0$ is the radius of a dust grain and $F_\mathrm{roll}$ is the rolling force. Thus, we are inside the hit-and-stick regime if
\begin{equation}
    \frac{1}{2} m_\mu v^2 \le 5 E_\mathrm{roll},
\end{equation}
where $m_\mu$ is the reduced mass of the aggregates. The hit-and-stick velocity range is then given by
\begin{equation}
    v \le \sqrt{5 \frac{\pi a_0 F_\mathrm{roll}}{m_\mu}}\;.\label{eq:S1_threshold}
\end{equation}

\subsection*{S2: Sticking by surface effects}
For velocities exceeding the hit-and-stick threshold velocity (Eq. \ref{eq:S1_threshold}), we assume sticking because of an increased contact area due to surface flattening and, therefore, an increased number of sticking grain-grain contacts. For the calculation of the contact area, we take an elastic deformation of the aggregate \citep{Hertz:1881} and get a radius for the contact area of
\begin{equation}
    s_0=\left[\left(\frac{15}{32}\right)\frac{m_\mu a_\mu^2v^2}{G}\right]^\frac{1}{5}\;.
\end{equation}
Here, $v$ is the collision velocity, $G$ is the shear modulus, and $a_\mu$ is the reduced radius. We estimate the shear modulus with the shear strength, which follows after \citet{Sirono:2004} as the geometric mean of the compressive strength and the tensile strength. These parameters were measured by \citet{BlumSchraepler:2004} to be $4\, 000$~\pressure\ (compressive strength) and $10\; 000$~\pressure\ (tensile strength), so we take $6\, 320$~\pressure\ for the shear modulus, which is consistent with estimates of \citet{WeidlingEtal:2009}.

The energy of a pair of bouncing aggregates after the collision is
\begin{equation}
    E_\mathrm{rest.}=\varepsilon^2\frac{1}{2}m_\mu v^2
\end{equation}
with the coefficient of restitution $\varepsilon$. The contact energy of the flattened surface in contact is
\begin{equation}
    E_\mathrm{cont.}=s_0^2\frac{\phi^\frac{2}{3}E_0}{a_0^2},
\end{equation}
where $E_0$ is the sticking energy of a monomer grain with the radius $a_0$. We expect sticking for $E_\mathrm{cont.} \geq E_\mathrm{rest.}$, thus,
\begin{eqnarray}
    \left[\left(\frac{15}{32}\right)\frac{m_\mu a_\mu^2 v^2}{G}\right]^\frac{2}{5}\frac{\phi^\frac{2}{3}E_0}{a_0^2} \geq \varepsilon^2\frac{1}{2}m_\mu v^2 \quad \mathrm{or}\\
    v \leq \left[\left(\frac{15}{32}\right) \frac{m_\mu a_\mu^2}{G}\right]^\frac{1}{3} \left[\frac{2\phi^\frac{2}{3}E_0}{a_0^2m_\mu\varepsilon^2}\right]^\frac{5}{6}\ . \label{eq:S2_threshold}
\end{eqnarray}
This is the sticking threshold velocity for \Sb, which is based on the Hertzian deformation, which is of course a simplified model, but has proven as a good concept in many attempts to describe slight deformation of porous dust aggregates \citep{LangkowskiEtal:2008, WeidlingEtal:2009}.

We have to ensure that the centrifugal force of two rotating aggregates, sticking like above, does not tear them apart, which is the case if
\begin{equation}
    F_\mathrm{cent} > T\pi s_0^2,
\end{equation}
where $T$ is the tensile strength of the aggregate material. The centrifugal force in the worst case of a perfectly grazing collision is
\begin{equation}
    F_\mathrm{ cent} = \frac{m_\mu\varepsilon^2v^2}{2a_\mu}\;,
\end{equation}
where $2a_\mu$ is a conservative estimation for the radial distance of the masses with the tangential velocity $\varepsilon v$. Thus, only collisions with velocities
\begin{equation}
    v < \left[\left(\frac{15}{32}\right)\frac{m_\mu a_\mu^2}{G}\right]^\frac{1}{3}\left[\frac{2\pi T a_\mu}{m_\mu\varepsilon^2}\right]^\frac{5}{6} \label{eq:centrifugal}
\end{equation}
can lead to sticking. For the relevant parameter range (see Table \ref{tab:parameters} below), the threshold velocity in Eq. \ref{eq:centrifugal} is always significantly greater than the sticking velocity in Eq. \ref{eq:S2_threshold}, thus, we can take Eq. \ref{eq:S2_threshold} as the relevant velocity for the process S2.

We will use this kind of sticking not only within the mass and velocity threshold as defined by Eq. \ref{eq:S2_threshold}, but also for collisions where we see sticking which cannot so far be explained by any model, like in experiment 6 or 18. For all these cases, we assume the porosity of target and projectile to be unchanged, disregarding any slight compaction as needed for the deformation. One exception is the sticking of small, fractal aggregates, which clearly goes together with a compaction of the projectile \citep{DominikTielens:1997, BlumWurm:2000}. In these cases we assume a projectile compaction by a factor of 1.5 in volume filling factor as there is no precise measurement on this compaction.

\subsection*{S3: Sticking by deep penetration}
If the target aggregate is much larger than the projectile, porous and flat, an impact of a (porous or compact) projectile results in its penetration into the target. Sticking is inevitable if the penetration of the projectile is deep enough, i.e. deeper than one projectile radius. In that case, the projectile cannot bounce off the target from geometric considerations. This was found in experiments of \citet{LangkowskiEtal:2008} in the case of porous projectiles and by \citet{BlumWurm:2008} in the case of solid projectiles. The result of the collision for penetration depths $D_\mathrm{p} \geq a_\mathrm{p}$ is that the mass of the target is augmented by the mass of the projectile, and the volume of the new aggregate reads
\begin{eqnarray}
    V &=& V_\mathrm{t} - \pi a_\mathrm{p}^2\left(D_\mathrm{p}-a_\mathrm{p}\right) + \frac{1}{2} V_\mathrm{p}\\
      &=& V_\mathrm{t} + \frac{5}{4}V_\mathrm{p}-\pi a_\mathrm{p}^2D_\mathrm{p}\;, \label{eq:S3_new_vol_porous}
\end{eqnarray}
with $V_\mathrm{p}$ and $V_\mathrm{t}$ being the volume of the projectile and target, respectively. We distinguish between compact and porous projectiles and take the experiments of \citet{GuettlerEtal:2009a} and \citet{LangkowskiEtal:2008} for impacts into $\phi=0.15$ dust aggregates and calculate the sticking threshold velocities.

For \textit{compact} projectiles, we use the linear relation for the penetration depth of \citet{GuettlerEtal:2009a}
\begin{equation}
    D_\mathrm{p} = \gamma \frac{m_\mathrm{p}v}{A_\mathrm{p}}\;, \label{eq:penetration_depth}
\end{equation}
where $m_\mathrm{p}=\frac{4}{3}\pi \rho_0 \phi_\mathrm{p}a_\mathrm{p}^3$ and $A_\mathrm{p}=\pi a_\mathrm{p}^2$ are the projectile mass and cross section, respectively. Although \citet{GuettlerEtal:2009a} suggest a power-law relation for the penetration depth, i.e. $D_\mathrm{p} = \gamma m_\mathrm{p}^{0.23\pm0.13} v^{0.89\pm0.34}$, we choose the linear relation in Eq. \ref{eq:penetration_depth} for simplicity, which also agrees with the data within the error bars. For such a linear fit, the slope to the data in \citet{GuettlerEtal:2009a} is $\gamma = 8.3 \cdot 10^{-3}\;\mathrm{cm^2\;s\;g^{-1}}$. We assume sticking for $D_\mathrm{p} \geq a_\mathrm{p}$ and get sticking due to process S3 in the velocity range
\begin{equation}
    v \geq \left(\frac{4}{3}\gamma\rho_\mathrm{0}\phi_\mathrm{p}\right)^{-1}\;,\label{eq:S3_threshold_compact}
\end{equation}
which only depends on the projectile bulk density $\rho_\mathrm{0}$ and filling factor $\phi_\mathrm{p}$ and not on the projectile radius.

A \textit{porous} projectile, colliding with a porous target, makes a visible indentation into the target aggregate if the kinetic energy is $E > E_\mathrm{min}$, with a material-dependent minimum energy $E_\mathrm{min}$. The crater volume is then given by
\begin{equation}
    V_\mathrm{cr.}=\left(\frac{E}{E_\mathrm{t}}\right)^\frac{3}{4}\;\mathrm{cm^3}\;, \label{eq:S3_crater_volume}
\end{equation}
\citep[see Fig. 15 in][]{LangkowskiEtal:2008}. Again, from geometrical considerations, we assume that sticking occurs if the projectile penetrates at least one radius deep, thus, $V_\mathrm{cr.} \geq 0.5 V_\mathrm{p}$, where $V_\mathrm{p}=\frac{4}{3}\pi a_\mathrm{p}^3$ is the
volume of the projectile. Thus,
\begin{eqnarray}
    \left(\frac{E}{E_\mathrm{t}}\right)^\frac{3}{4} \geq \frac{1}{2}V_\mathrm{p}\\
    \frac{1}{2}mv^2 \geq E_\mathrm{t} \left(\frac{1}{2}\frac{m}{\rho}\right)^\frac{4}{3}\\
    v \geq \left(\frac{mE_\mathrm{t}^3}{2\rho_0^4\phi_\mathrm{p}^4}\right)^\frac{1}{6} \;. \label{eq:S3_threshold_porous_1}
\end{eqnarray}
For these velocities, the projectile is inevitably embedded into the target aggregate. However, if the impact energy is less than $E_\mathrm{min}$, the collision will not lead to a penetration so that the final condition for sticking of a porous projectile according to
process S3 is
\begin{equation}
    v \geq \max\left({\sqrt{\frac{2 E_\mathrm{min}}{m}} ,\left(\frac{mE_\mathrm{t}^3}{2\rho_0^4\phi_\mathrm{p}^4}\right)^\frac{1}{6}}\right) \;. \label{eq:S3_threshold_porous_2}
\end{equation}

\subsection*{S4: Partial sticking in fragmentation events}
As introduced in Sect. \ref{sec:new_exp_1}, a fragmenting collision between a porous aggregate and a solid target can lead to a partial growth of the target. The mass transfer from the projectile to the target is typically 2.3~\% of the projectile mass (Fig. \ref{fig:S4_mass_gain}), and without better knowledge we assume that the transferred mass has a volume filling factor of $1.5 \phi_\mathrm{p}$. The remaining mass of the projectile fragments according to the power-law mass distribution given in Eq. \ref{eq:mass_dist}, with the fragmentation strength from Eq. \ref{eq:mu_v_exp}.

For a compact projectile aggregate impacting a compact target, the threshold velocity for the S4 process is $v=100$~\cms\ and thus identical to that of the F1 process. The fragmentation strength is given by Eq. \ref{eq:mu_v_ag-ag}.

\subsection*{B1: Bouncing with compaction\label{sec:B1}}
In a bouncing collision we find compaction of the two collision partners. For similar-sized aggregates, the increase of the volume filling factor was formulated by \citet[their Eq. 25]{WeidlingEtal:2009} to be
\begin{equation}
    \phi^+(\phi)=\frac{\phi_\mathrm{max}(v)-\phi}{\nu(v)}\;;\;\;\phi^+(\phi, v)>0\label{eq:weidling_ff_increase}
\end{equation}
with $\nu(v)=\nu_0\cdot\left(v/20\;\mathrm{cm~s^{-1}}\right)^{-4/5}$, $\phi_\mathrm{max}(v) = \phi_0 + \Delta \phi \cdot\left(v/20\;\mathrm{cm~s^{-1}}\right)^{4/5}$ and $\nu_0=850$, $\phi_0=0.15$, $\Delta \phi =0.215$ for $v \leq 50$ \cms. Here, $\phi_\mathrm{max}$ is the saturation of the filling factor after many collisions, which follows an exponential function with the e-folding width $\nu$ \citep{WeidlingEtal:2009}. In their experiments, $v$ was the velocity of a porous projectile colliding with a solid target (infinite mass). In the case of similar-sized colliding aggregates, the velocity would be $0.5 \cdot v$ for each aggregate in a center-of-mass system. Therefore, we scale the velocity as
\begin{eqnarray}
    v_\mathrm{p}&=&\frac{v}{1+\frac{m_\mathrm{p}}{m_\mathrm{t}}} \label{eq:red_mass_1} \\
    v_\mathrm{t}&=&\frac{v}{1+\frac{m_\mathrm{t}}{m_\mathrm{p}}} \;, \label{eq:red_mass_2}
\end{eqnarray}
where $v_\mathrm{p}$ ($v_\mathrm{t}$) is the center-of-mass velocity of the projectile (target). In the case of $m_\mathrm{p} \ll m_\mathrm{t}$ we have the situation of \citet{WeidlingEtal:2009} with $v_\mathrm{p} = v$, thus, these velocities are chosen to calculate the scaling of $\nu(v)$ and $\phi_\mathrm{max}(v)$ for projectile and target compaction, respectively. This means that a projectile with a negligible mass with respect to the target cannot compact the target but is only compacted by itself, while two aggregates of the same mass are equally compacted.

\begin{figure}[t]
    \center
    \includegraphics[width=0.5\textwidth,angle=90]{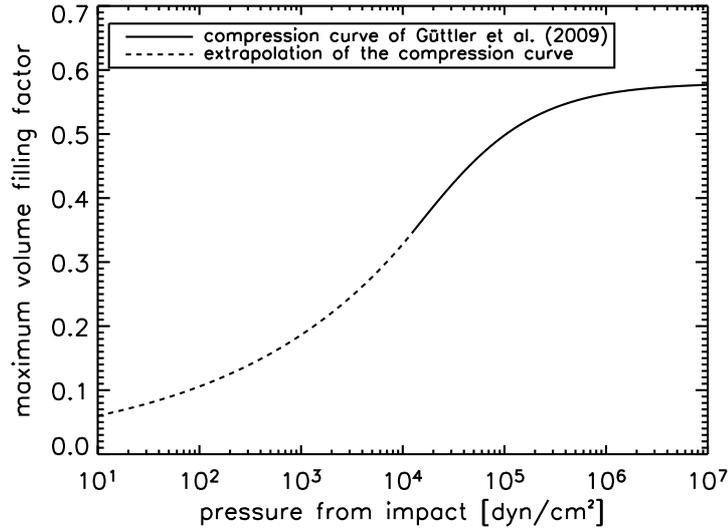}
    \caption{\label{fig:phi_max-pressure}The original compressive strength curve measured by \protect \citet{GuettlerEtal:2009a} (Eq. \ref{eq:guettler_compr_curve}, solid line) is biased by the dust samples used in the experiments. To describe also the compression of dust aggregates with a volume filling factor lower than those used by \protect \citet{GuettlerEtal:2009a}, we extrapolate the curve with a power-law (Eq. \ref{eq:compr_curve_power_law}, dashed line) for $p<p_\mathrm{m}$.}
\end{figure}

For $\phi_\mathrm{max}(v)$, \citet{WeidlingEtal:2009} gave the above relation which is biased by the experimentally used dust samples and overestimates the compression for very low velocities. Therefore, we propose an alternative scaling relation for $\phi_\mathrm{max}(v)$. In a collision with a velocity $v$ we can calculate a dynamic pressure
\begin{equation}
    p_\mathrm{dyn}=\nu(v)\cdot\frac{1}{2}\rho v^2\;.\label{eq:enh_dyn_pressure}
\end{equation}
This pressure is increased by a factor $\nu(v)$, as we know from the experiments of \citet{WeidlingEtal:2009} that the contact area is very small (factor $1/\nu$ of the aggregate surface) and that only a very confined volume is compressed. For $v=20$~\cms\ the pressure calculated from Eq. \ref{eq:enh_dyn_pressure} is very close to the value given by \citet{WeidlingEtal:2009}. From this pressure we calculate the compression from the compressive strength curve, which \citet{GuettlerEtal:2009a} derived for collisions:
\begin{equation}
    \phi_\mathrm{comp}(p)=\phi_2-\frac{\phi_2-\phi_1}{\exp\left(\frac{\lg p-\lg p_\mathrm{m}}{\Delta}\right)+1}\label{eq:guettler_compr_curve}
\end{equation}
with $\phi_1=0.12$, $\phi_2=0.58$, $\Delta=0.58$, and $p_\mathrm{m}=1.3\times 10^4$~\pressure. This compressive strength curve is also biased from the experiments, as its lowest value is $\phi_1=0.12$. Assuming the saturation part of the compressive strength curve to be general, we propose a power law for $p<p_\mathrm{m}$ with the same slope as in Eq. \ref{eq:guettler_compr_curve} for $\phi_\mathrm{comp}(p_\mathrm{m})$, which is then given by
\begin{equation}
    \phi_\mathrm{comp}(p)=\frac{\phi_1+\phi_2}{2}\cdot\left(\frac{p}{p_\mathrm{m}}\right)^{ \frac{\phi_2-\phi_1}{\phi_2+\phi_1}\cdot\frac{1}{2\Delta \ln 10}} \label{eq:compr_curve_power_law}
\end{equation}
and is able to treat the lowest filling factors and pressures. Equations \ref{eq:guettler_compr_curve} and \ref{eq:compr_curve_power_law} determine the compression in a confined volume. Taking into account that after many collisions only an outer rim of the aggregate is compressed, we reduce the compression by a factor $f_\mathrm{c}=0.79$ to fit the $\phi_\mathrm{max}(v=20\;\mathrm{cm~s^{-1}})=0.365$ experimentally measured by \citet{WeidlingEtal:2009}.

Conclusively, we calculate the increase of the volume filling factor from Eq. \ref{eq:weidling_ff_increase}, where $\phi_\mathrm{max}$ is now provided by the dynamical pressure curve as
\begin{equation}
    \phi_\mathrm{max}(v) = f_\mathrm{c} \cdot \phi_\mathrm{comp} (p_\mathrm{dyn})\; , \label{eq:B1_phi_max_scale}
\end{equation}
where $\phi_\mathrm{comp}$ is given by Eqs. \ref{eq:guettler_compr_curve} and \ref{eq:compr_curve_power_law}. For the pressure we use Eq. \ref{eq:enh_dyn_pressure} and for the corresponding velocities we use Eqs. \ref{eq:red_mass_1} and \ref{eq:red_mass_2} to calculate the projectile and target compression, respectively. The maximum compression $\phi_\mathrm{max}(v)$, which an aggregate can achieve in many collisions at a given velocity, is shown in Fig. \ref{fig:phi_max-pressure}.

\citet{WeidlingEtal:2009} found that in this bouncing regime, the aggregates can also fragment with a low probability. We adopt this fragmentation probability of
\begin{equation}
    P_\mathrm{frag} = 10^{-4}
\end{equation}
and assume that an aggregate breaks into two similar-sized fragments as suggested by their Fig. 5.

\subsection*{B2: Bouncing with mass transfer}
\citet{LangkowskiEtal:2008} and \citet{BlumWurm:2008} found that the collision between a projectile (porous or solid) and a porous target aggregate can lead to a slight penetration of the projectile into the target followed by the bouncing of the projectile. This leads to a mass transfer from the target to the projectile \citep[see Fig. 7 in][]{LangkowskiEtal:2008}. We assume that the transferred mass is one projectile mass \citep[Fig. 8 in][]{LangkowskiEtal:2008}, thus,
\begin{equation}
    \Delta m_{\mathrm{t}\rightarrow\mathrm{p}} = m_\mathrm{p}\ ,
\end{equation}
and that the filling factor of the transferred (compacted) material is 1.5 times that of the original target material, i.e.
\begin{equation}
    \phi_{\mathrm{t}\rightarrow\mathrm{p}} = 1.5 \times \phi_\mathrm{t} \;.
\end{equation}
Although the filling factor of the transferred material was not measured, we know that the material is significantly compacted in the collision \citep[see x-ray micro tomography (XRT) analysis of][]{GuettlerEtal:2009a}, so that the above assumption seems justified.

\subsection*{F1: Fragmentation}
\begin{figure}[t]
    \center
    \includegraphics[height=9cm,angle=90]{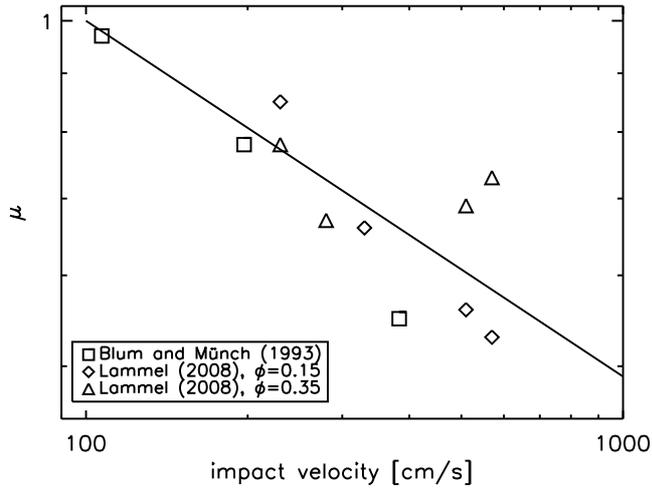}
    \caption{\label{fig:mu_f1}The impact strength for aggregate-aggregate collision also increases for higher velocities (decreasing $\mu$, cp. inset in Fig. \ref{fig:mass_dist}). The fitted power-law is given by Eq. \ref{eq:mu_v_ag-ag}.}
\end{figure}
When two similar-sized dust aggregates collide at a velocity which is greater than the fragmentation velocity of
\begin{equation}
    v_\mathrm{frag} = 100\ \mathrm{cm\ s^{-1},} \label{eq:canonic_frag_threshold}
\end{equation}
they will both be disrupted. \citet{BlumMuench:1993} found fragmentation for mm-sized ZrSiO$_4$ dust aggregates with a porosity of $\phi=0.35$ at a velocity greater than 100~\cms. In their experiments, the aggregates fragmented according to a power-law size distribution with an exponent of $\lambda = -1.4$ (see Sect. \ref{sec:new_exp_1}), which we will use hereafter. The two largest fragments together have a mass of $\mu(v) (m_\mathrm{p}+m_\mathrm{t})$, where we can determine $\mu(v)$ from the experiments of \citet[ZrSiO$_4$ aggregate collisions with $\phi=0.35$]{BlumMuench:1993} and \citet[SiO$_2$ aggregates of different porosities]{Lammel:2008}. These values are plotted in Fig. \ref{fig:mu_f1} and a power-law fit for velocities $v \geq 100$ \cms
\begin{equation}
    \mu(v) = \left(\frac{v}{100\ \mathrm{cm\ s^{-1}}}\right)^{-0.31} \label{eq:mu_v_ag-ag}
\end{equation}
is shown by the solid line, which is again fitted to match the fragmentation threshold of 100~\cms\ (cp. Eq. \ref{eq:mu_v_exp}). Here, the error in the exponent is $\pm 0.02$.

\subsection*{F2: Erosion}
If a projectile collides with a significantly larger {\em porous} target aggregate at a sufficiently high impact velocity, the target may be eroded. \schraepler\ found erosion of porous ($\phi=0.15$) aggregates which were exposed to 1.5~$\mu$m diameter SiO$_2$ monomers (mass $m_0$) at velocities from $1\,500$ to $6\,000$~\cms. Their numerical model, which fits the experimental data very well, predicts an onset of erosion for a velocity of 350~\cms. The eroded mass grows roughly linear with impact velocity, i.e.
\begin{equation}
    \frac{\Delta m}{m_\mathrm{p}} = \frac{8}{60} \left(\frac{v}{100\ \mathrm{cm\ s^{-1}}}\right)\ ,
\end{equation}
where $\Delta m$ is the amount of eroded mass and $m_\mathrm{p} = m_0$ is the projectile mass. \citet{ParaskovEtal:2007} also found mass loss of a porous target aggregate for velocities from 350 to $2\,150$~\cms, although the process involved is widely different. They used porous and solid projectiles, and their results \citep[Fig. 4 in][]{ParaskovEtal:2007} are consistent with
\begin{equation}
    \frac{\Delta m}{m_\mathrm{p}} = \frac{15}{20} \left(\frac{v}{100\ \mathrm{cm\ s^{-1}}}\right)\ ,
\end{equation}
which agrees with non-zero-gravity experiments of \citet{WurmEtal:2005a}, who estimated a mass loss of 10 projectile masses for velocities of more than $1650$~\cms. Due to the small variation in projectile mass within each of the two experiments, we apply a power-law in mass and merge both experiments to
\begin{equation}
    \frac{\Delta m}{m_\mathrm{p}} = \frac{6}{80} \left(\frac{v}{100\ \mathrm{cm\ s^{-1}}}\right) \left( \frac{m_\mathrm{p}}{m_0} \right)^{0.092}\ .\label{eq:mass_loss_erosion_porous}
\end{equation}
The velocity range for erosion is therefore
\begin{equation}
    v_\mathrm{er} \geq 350\ \mathrm{cm\ s^{-1}}
\end{equation}
and is consistent in both experiments.

For {\em compact} targets, \schraepler\ were able to measure the velocity range for erosion at
\begin{equation}
    v_\mathrm{er} \geq 2\,500\ \mathrm{cm\ s^{-1}} \label{eq:25ms}.
\end{equation}
Due to the nature of the compact target, far less material was eroded, i.e.
\begin{equation}
    \frac{\Delta m}{m_\mathrm{p}} = \frac{8}{550} \left(\frac{v}{100\ \mathrm{cm\ s^{-1}}}\right) \left( \frac{m_\mathrm{p}}{m_0} \right)^{0.092}\ .\label{eq:mass_loss_erosion_compact}
\end{equation}
Here, we applied the same power-law index as in Eq. \ref{eq:mass_loss_erosion_porous} due to the absence of large-scale experiments in this case. We assume a mass distribution of the eroded material according to Eq. \ref{eq:mu_v_exp}.

\subsection*{F3: Fragmentation with mass transfer\label{sec:F3}}
In Sect. \ref{sec:new_exp_3} we described the volume transfer from a porous aggregate to a solid sphere (assumed to be representative for a compact aggregate) above the fragmentation threshold velocity (see Eq. \ref{eq:F3_vol_trans}). Without better knowledge, we assume that the transferred mass has a volume filling factor of $1.5$ times that of the porous collision partner ($\phi_\mathrm{p}$) and cannot exceed the mass of the porous aggregate, thus
\begin{equation}
    \Delta m = m_\mathrm{p(t)} 1.5 \phi_\mathrm{p} \left( 0.59 - 6.3 \times 10^{-4} \frac{v}{\mathrm{cm\ s^{-1}}} \right)
    \;, \label{eq:F3_mass_trans}
\end{equation}
where $m_\mathrm{p(t)}$ is the mass of the porous aggregate, which can either be projectile or target in our definition, depending on its actual mass. For the fragmentation of the porous aggregate we assume a power-law distribution following the F1 case. If the collision velocity is higher than 940~\cms, Eq. \ref{eq:F3_mass_trans} yields no mass gain for the compact aggregate, thus, the mass of the compact aggregate is conserved and only the porous aggregate fragments.

\section{Collision regimes\label{sec:collision_regimes}}
In this section we intend to build on the physical descriptions, which we have derived in the previous section, and develop a complete collision model for the determination of the collisional outcome in protoplanetary dust interactions (Fig. \ref{fig:pictograms}). This means that for each collision which may occur, a set of collision parameters will be provided as input for a numerical model of the evolution of protoplanetary dust (see Chapter \ref{chp:paper3}). The most crucial parameters that mainly determine the fate of the colliding dust aggregates in each collision are the respective dust-aggregate masses and their relative velocity.

\begin{figure}[t]
    \center
    \includegraphics[width=0.7\textwidth]{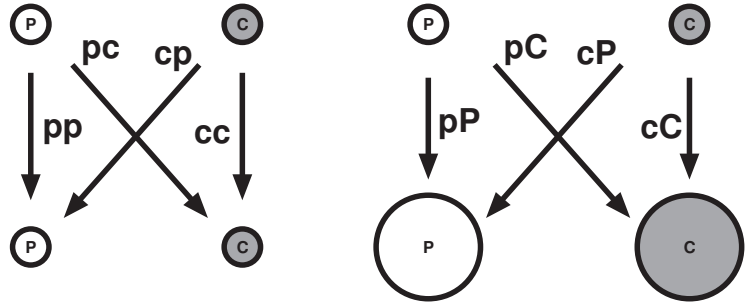}
    \caption{\label{fig:categorization}Our model distinguishes between porous and compact aggregates, which leads to the displayed four types of collisions (\pp, \pc, \cp, \cc) if the collision partners are not too different in size (left). The size ratio of projectile and target aggregate was identified as another important parameter and we distinguish between similar-sized and different-sized collision partners. Thus, in addition to the four collision types on the left, impacts of projectiles into much larger targets (\pP, \pC, \cP, \cC; the target characterized by a capital letter) can also occur (right). The boundary between similar-sized and different-sized aggregates is given by the critical mass-ratio parameter $r_\mathrm{m}$. Collisions on the left are restricted to $m_{\mathrm{p}} \leq m_{\mathrm{t}} \leq r_\mathrm{m} m_{\mathrm{p}}$, collisions on the right happen for $m_{\mathrm{t}} > r_\mathrm{m} m_{\mathrm{p}}$.}
\end{figure}

Moreover, in Sect. \ref{sec:exp-review} and \ref{sec:exp_types}, we saw that the porosity difference between the two collision partners also has a big impact on the collisional outcome. The only difference between the outcomes F1 and F3 (and between S3 and S4) is that the target aggregate is either porous or compact. Thus, we define a critical porosity $\phi_\mathrm{c}$ to distinguish between porous or compact aggregates. This value can only roughly be confined between $\phi=0.15$ \citep[S3 sticking, clearly an effect of porosity,][]{LangkowskiEtal:2008} and $\phi=0.64$ \citep[random close packing, clearly compact][]{TorquatoEtal:2000}, and without better knowledge we will choose $\phi_\mathrm{c}=0.4$.

Another important parameter is the mass ratio of the collision partners. Again, the \Sc\ occurs for the same set of parameters as the \Fa, and only the critical mass ratio $r_\mathrm{m} = m_\mathrm{t} / m_\mathrm{p}$ is different. From the work of \citet{BlumMuench:1993} and \citet{LangkowskiEtal:2008}, we can confine this parameter to the range $10 \leq r_\mathrm{m} \leq 1\,000$ and will also treat it in Chapter \ref{chp:paper3} as a free parameter (with fixed values $r_\mathrm{m} = 10, 100, 1\,000$).

A further parameter, which has an impact on the collisional outcome, is the impact angle, but at this stage we will treat all collisions as central collisions due to a lack of information of the actual influence of the impact angle on the collisional result. Experiments by \citet{BlumMuench:1993}, \citet{LangkowskiEtal:2008}, or \citet{Lammel:2008} indicate rather small differences between central and grazing collisions, so that we feel confident that the error due to this simplification is small. Another parameter, which we also neglect at this point due to a lack of experimental data, is the surface roughness of the aggregates. \citet{LangkowskiEtal:2008} showed its relative importance, but a quantitative treatment of the surface roughness is currently not possible.

The binary treatment of the parameters $\phi_\mathrm{c}$ and $r_\mathrm{m}$ leads to Fig. \ref{fig:categorization}, whereafter we have four different porous-compact combinations and, if we take into account that the collision partners can either be similar-sized or different-sized, we have a total of eight collision combinations. We will call these \pp, \pP, \cc, \cC, \cp, \cP, \pc, and \pC. Here, the first small letter denotes the porosity of the projectile (\textit{'p'} for porous and \textit{'c'} for compact) and the second letter denotes the target porosity, which can be either similar-sized (small letter) or different-sized (capital letter). Aggregates with porosities $\phi < \phi_\mathrm{c}$ are \textit{'porous'}, those with $\phi \geq \phi_\mathrm{c}$ are \textit{'compact'}. If the mass of the target aggregate $m_{\mathrm{t}} \leq r_\mathrm{m} m_{\mathrm{p}}$, we treat the collisions as equal-sized, for $m_{\mathrm{t}} > r_\mathrm{m} m_{\mathrm{p}}$, the collisions are treated as different-sized.

\begin{figure*}[!t]
    \center
    \includegraphics[width=1\textwidth]{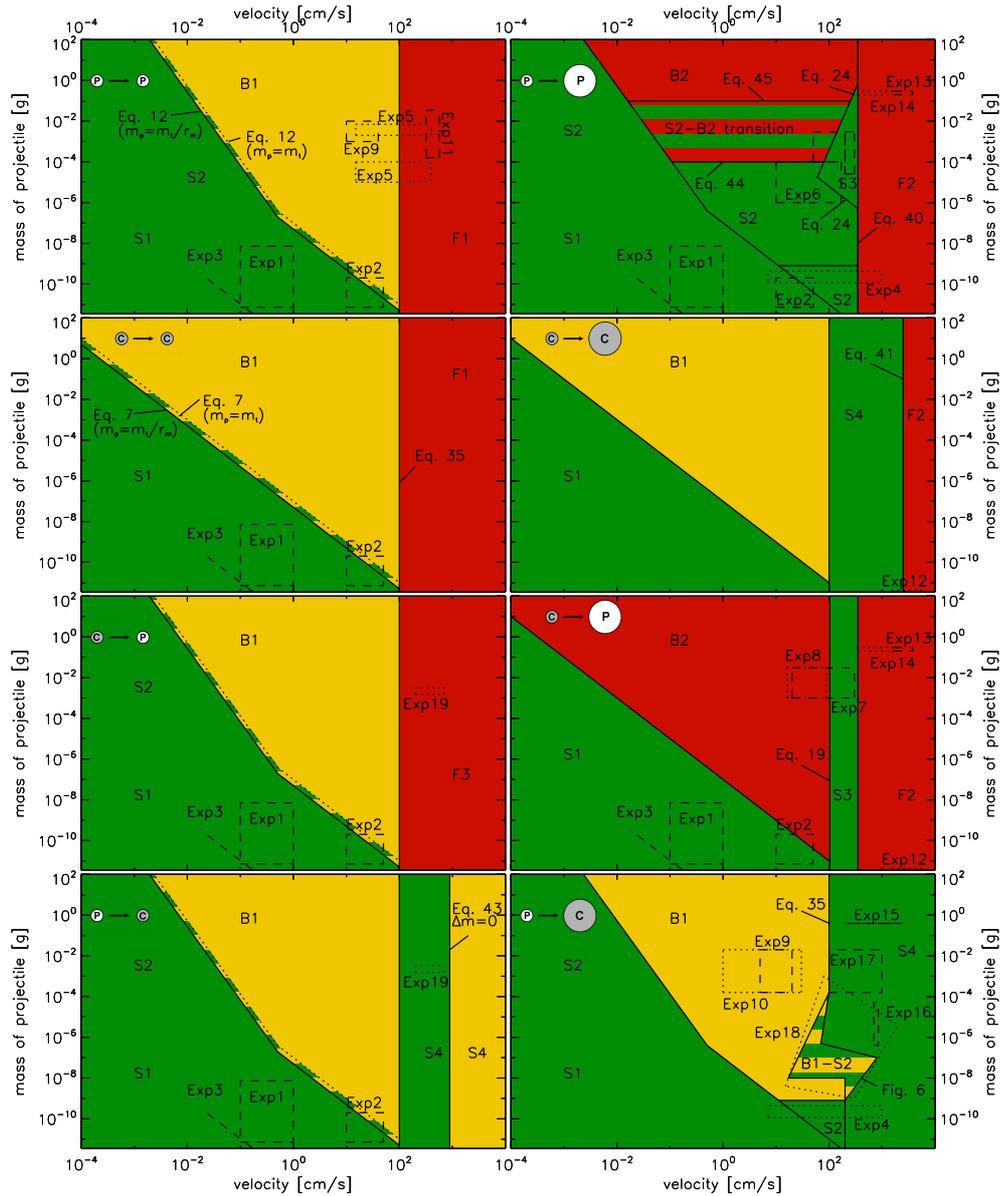}
    \caption{\label{fig:colored_regimes}The resulting collision model as described in this chapter. We distinguish between similar-sized (left column) and different-sized (right column) collision partners, which are either porous or compact (also see Fig. \ref{fig:categorization}). For each case, the important parameters to determine the collisional outcome are the projectile mass and the collision velocity. Collisions within green regions can lead to the formation of larger bodies, while red regions denote mass loss. Yellow regions are neutral in terms of growth. The dashed and dotted boxes show where experiments directly support this model.}
\end{figure*}

\begin{table}[t]
\center%
\caption{\label{tab:parameters}Particle and aggregate material properties used for generating Fig. \ref{fig:colored_regimes}.}
\begin{tabular}{llp{3.8cm}}
    \hline
    symbol            & value                                  & reference \\
    \hline
\multicolumn{3}{l}{\underline{monomer-grain properties:}}\\
    $a_0$             & 0.75~$\mu$m                            & \\
    $m_0$             & $3.18 \times 10^{-12}$~g               & \\
    $\rho_0$          & 2~g~cm$^{-3}$                          & \\
    $E_0$             & $2.2 \times 10^{-8}$~erg               & \citet{BlumWurm:2000}, \citet{PoppeEtal:2000a} \\
    $F_\mathrm{roll}$ & $10^{-4}$~dyn                          & \citet{HeimEtal:1999} \\
    \hline
\multicolumn{3}{l}{\underline{aggregate properties:}}\\
    $\varepsilon ^2$     & 0.05                                   & \citet{BlumMuench:1993}, \heisselmann\\
    $G$               & 6320~\pressure                         & this work \\
    $T$               & $10^4$~\pressure                       & \citet{BlumSchraepler:2004} \\
    $\phi_\mathrm{c}$ & 0.40                                   & this work \\
    $r_\mathrm{m}$    & 10 -- $1\,000$                         & this work \\
    $\gamma$          & $8.3 \times 10^{-3}$~s~cm$^2$~g$^{-1}$ & \citet{GuettlerEtal:2009a} \\
    $E_\mathrm{t}$    & $3.5 \times 10^4$~erg                  & \citet{LangkowskiEtal:2008} \\
    $E_\mathrm{min}$  & $3.1 \times 10^{-2}$~erg               & \citet{LangkowskiEtal:2008} \\
    $\phi_1$          & 0.12                                   & \citet{GuettlerEtal:2009a} \\
    $\phi_2$          & 0.58                                   & \citet{GuettlerEtal:2009a} \\
    $\Delta$          & 0.58                                   & \citet{GuettlerEtal:2009a} \\
    $p_\mathrm{m}$    & $1.3 \times 10^{4}$~\pressure          & \citet{GuettlerEtal:2009a} \\
    $f_\mathrm{c}$    & 0.79                                   & this work \\
    $\nu_0$           & 850                                    & \citet{WeidlingEtal:2009}\\
    $\lambda$         & -1.4                                   & this work \\
    \hline
\end{tabular}
\end{table}

For each combination depicted in Fig. \ref{fig:categorization}, we have the most important parameters (1) projectile mass $m_\mathrm{p}$ and (2) collision velocity $v$, which then determine the collisional outcome. As shown in Fig. \ref{fig:colored_regimes}, we treat each combination from Fig. \ref{fig:categorization} separately and define the collisional outcome as a function of projectile mass and collision velocity. For the threshold lines and the quantitative collisional outcomes we use a set of equations, which were given in Sect. \ref{sec:exp_types}. For a quantitative analysis and application to PPDs (see Chapter \ref{chp:paper3}), knowledge of the material parameters of the monomer dust grains and dust aggregates is required. In Table \ref{tab:parameters} we list all relevant parameters for 1.5 $\rm \mu m$ $\rm SiO_2$ spheres, for which most experimental data are available. However, we believe that the data in Table \ref{tab:parameters} are also relevant for most types of micrometer-sized silicate particles.

The only collisional outcome, which is the same in all regimes, is the \Sa\ process, which, due to its nature, does not depend on porosity or mass ratio but only on mass and collision velocity. Thus, all collision combinations in Fig. \ref{fig:colored_regimes} have the same region of sticking behavior for a mass-velocity combination smaller than defined by Eq. \ref{eq:S1_threshold}. This parameter region is marked in green because \Sa\ can in principle lead to the formation of arbitrary large aggregates. Marked in yellow are collisional outcomes, which do not lead to further growth of the \textit{target} aggregate, but conserve the mass of the target aggregate, which is only the case for \Ba. For simplicity, the weak fragmentation probability of $P_\mathrm{frag}=10^{-4}$ (see Sect. \ref{sec:B1}) has been neglected in the coloring. The red-marked regions are parameter sets for which the \textit{target} aggregate loses mass.

The dashed and dotted boxes in Fig. \ref{fig:colored_regimes} mark the mass and velocity ranges of the experiments from Table \ref{tab:experiments}. In Chapter \ref{chp:paper3}, this plot will help us to see in which parameter regions collisions occur and how well they are supported by experiments. We will now go through all of the eight plots in Fig. \ref{fig:colored_regimes} and explain the choice for the thresholds between the collisional outcomes.

\pp: In addition to the omnipresent \Sa\ regime, which is backed by experiments 1 -- 3 in Table \ref{tab:experiments}, collisions of porous projectiles can also lead to \Sb, whose threshold is determined by Eq. \ref{eq:S2_threshold}. For higher velocities ($v>100$~\cms, Eq. \ref{eq:canonic_frag_threshold}), fragmentation sets in. Bouncing (B1) and fragmentation (F1) in this regime are well-tested by experiments 5, 9, and 11 in Table \ref{tab:experiments}.

\pP: As the projectiles are also porous here, we have the same \Sb\ threshold as in \pp. The same collisional outcome (but with compaction of the projectile) was found for collisions of small aggregates \citep[experiment 4 in Table \ref{tab:experiments}]{BlumWurm:2000}. \citet{LangkowskiEtal:2008} (experiment 6) found the S2 collisional outcome for projectile masses
\begin{equation}
    m_\mathrm{p} < 10^{-4}\ \mathrm{g} \ ,
\end{equation}
thus we have a horizontal upper limit for S2 in the \pp\ plot of Fig. \ref{fig:colored_regimes}. Extrapolation of Exp. 6 to large aggregate masses
\begin{equation}
    m_\mathrm{p} > 0.1\ \mathrm{g}
\end{equation}
results in \Bb. A linear interpolation between perfect sticking for $m_\mathrm{p} < 10^{-4}\ \mathrm{g}$ and perfect bouncing for $m_\mathrm{p} > 0.1\ \mathrm{g}$, justified by the sticking probabilities shown in Fig. 5 of \citet{LangkowskiEtal:2008}, gives a sticking probability for the mass range $10^{-4}\ \mathrm{g} \leq m_\mathrm{p} \leq 0.1\ \mathrm{g}$ (striped region in the \pP\ of Fig. \ref{fig:colored_regimes}) of
\begin{equation}
    P_\mathrm{stick} = - \frac{1}{3} \log_{10} \left( \frac{m_\mathrm{p}}{0.1\ \mathrm{g}} \right) \ .
\end{equation}
In Sect. \ref{sec:exp_types} we defined the threshold for \Sc\ by Eqs. \ref{eq:S3_threshold_porous_1} and \ref{eq:S3_threshold_porous_2}, which are prominent in the \pP\ plot for high velocities. For even higher velocities, we have erosion of the porous aggregate (F2), defined by the threshold velocity in Eq. \ref{eq:mass_loss_erosion_porous} and based on experiments 12 -- 14 in Table \ref{tab:experiments}.

\cc: Our knowledge about collisions between similar-sized, compact dust aggregates is rather limited. \citet{BlumMuench:1993} performed collisions between similar-sized aggregates with $\phi=0.35$. Although this is lower than the critical volume filling factor $\phi_\mathrm{c}$ as defined in Table \ref{tab:parameters}, we assume a similar behavior also for aggregates with higher porosity. Therefore, without better knowledge, we define a fragmentation threshold as in the \pp\ regime, and take the \Sa\ threshold for low energies. We omit the \Sb\ in this regime because of the significantly lower compressibility of the compact aggregates.

\cC: In this collision regime the experimental background is also very limited. For low collision energies we assume a \Sa\ growth, for higher velocities \Ba\ and, if the fragmentation threshold ($v>100$~\cms, Eq. \ref{eq:canonic_frag_threshold}) is exceeded, \Sd. Based on experiment 12, we have an \Fb\ limit for velocities higher than $2\,500$~\cms\ (Eq. \ref{eq:25ms}).

\cp and \pc: These two cases are almost identical, with the only difference that the compact aggregate can either be the projectile or the target (i.e. slightly lower or higher in mass than the target aggregate). The mass ratio of both aggregates is however within the critical mass ratio $r_\mathrm{m}$. Besides the already-discussed cases S1, S2, and B1, we assume fragmentation 100~\cms\ (Eq. \ref{eq:canonic_frag_threshold}). Due to the nature of the collision between a compact and a porous aggregate, only the porous aggregate is able to fragment, whereas the compact aggregate stays intact. If the compact aggregate is the projectile, the target mass is always reduced, thus we have \Fc\ from the target to the projectile. If the target is compact, it grows by \Sd\ if the velocity is less than 940~\cms\ (see Eq. \ref{eq:F3_mass_trans}). For higher velocities, Eq. \ref{eq:F3_mass_trans} yields no mass gain and so this region is neutral in terms of growth. Collisions at high velocities are confirmed by Exp. 19 in this regime.

\cP: While small collision energies lead to \Sa, higher energies result in \Bb\ \citep[Exp. 8,][]{BlumWurm:2008}. This region is confined by the \Sc\ threshold velocity as defined in Eq. \ref{eq:S3_threshold_compact}, based on Exp. 7 \citep{GuettlerEtal:2009a}. At even higher velocities of above 350~\cms\ (Eq. \ref{eq:mass_loss_erosion_porous}), we get erosion of the target aggregate as seen in Exp. 12 -- 14.

\pC: This plot in Fig. \ref{fig:colored_regimes} looks the most complicated but it is supported by a large number of experiments. For low collision velocities, we again have \Sa\ and \Sb as well as a transition to \Ba\ for larger collision energies. The existence of the B1 bouncing region has been shown in Exp. 9 and 10 \citep[\heisselmannpar;][]{WeidlingEtal:2009}. For higher velocities and masses above $1.6 \cdot 10^{-4}$~g we assume a fragmentation threshold of 100~\cms\ with a mass transfer to the target (S4), as seen in Exp. 16 (Sect. \ref{sec:new_exp_1}). For lower masses, the odd-shaped box of Exp. 18 is a direct input from Sect. \ref{sec:new_exp_2} (see Fig. \ref{fig:small_coll}). In the striped region between B1 and S4, we found a sticking probability in Exp. 18 of $P_\mathrm{stick} = 0.5$. For lower masses, Exp. 4 showed \Sb\ with a restructuring (compaction) of the projectile. As in the \pP\ regime, we set the threshold for a maximum mass to $8 \cdot 10^{-10}$~g, while the upper velocity threshold -- which must be a transition to a fragmentation regime \citep{BlumWurm:2000} -- is 200~\cms\ from Exp. 4 and 18.

\section{Porosity evolution of the aggregates\label{sec:porosities}}
Since the porosity of dust aggregates is a key factor  for the outcome of dust aggregate collisions \citep{BlumWurm:2008}, it is paramount that collisional evolution models follow its evolution \citep[Paper II]{OrmelEtal:2007}. Therefore we want to concentrate on the evolution of the dust aggregates' porosities and recapitulate the porosity recipe as used in Sect. \ref{sec:exp_types}. In this chapter we have used the volume filling factor $\phi$ as a quantitative value, defined as the volume fraction of material (one minus porosity). In Chapter \ref{chp:paper3}, we will also use the enlargement parameter $\Psi$ as introduced by \citet{OrmelEtal:2007}, which is the reciprocal quantity
$\Psi=\phi^{-1}$.

Starting the growth with solid dust grains, we have a volume filling factor of 1, which will however rapidly decrease due to the \Sa\ growth, producing highly porous, fractal aggregates. Here, we use the porosity recipe of \citet{OrmelEtal:2007}, who describe this fractal growth by their enlargement parameter as
\begin{equation}
    \Psi_\mathrm{new} = \frac{m_\mathrm{p}\Psi_\mathrm{p} + m_\mathrm{t}\Psi_\mathrm{t}}{m_\mathrm{p}+m_\mathrm{t}} \times
    \left(1+\frac{m_\mathrm{t}\Psi_\mathrm{t}}{m_\mathrm{p}\Psi_\mathrm{p}}\right)^{0.425} + \Psi_\mathrm{add}\ , \label{eq:porosity_S1}
\end{equation}
where $\Psi_\mathrm{add}$ is a correction factor in case of $m_\mathrm{p} \approx m_0$ and otherwise zero (for details see their Sect. 2.4). This equation predicts an increasing porosity in every \Sa\ collision. In collisions that lead to \Sb, we assume that the compaction of the aggregates is so small, that their porosity is unaffected. So the aggregates are merged and only the mass and volume of both are being added, thus,
\begin{equation}
    \phi_\mathrm{new} = \frac{V_\mathrm{t}\phi_\mathrm{t} + V_\mathrm{p}\phi_\mathrm{p}}{V_\mathrm{t} + V_\mathrm{p}}\ .\label{eq:porosity_S2a}
\end{equation}
One exception for the \Sb\ occurs in a small parameter space which is determined by the experiments of \citet{BlumWurm:2000}. For the smallest masses and a velocity around 100~\cms, \citet{BlumWurm:2000} found sticking of fractal aggregates in the \pP\ and \pC\ regimes, which goes with a restructuring and, thus, compaction of the projectiles. In this case, we assume a compaction of the projectile by a factor of 1.5 in volume filling factor, thus
\begin{equation}
    \phi_\mathrm{new} = \frac{V_\mathrm{t}\phi_\mathrm{t} + \mathrm{min}\left(1.5V_\mathrm{p}\phi_\mathrm{p},\ \phi_\mathrm{c}\right)}{V_\mathrm{t}+V_\mathrm{p}}\ . \label{eq:porosity_S2b}
\end{equation}
An increasing filling factor is also applied for \Sc. Here, the mass of the projectile is added to the target while the new volume must be less than $V_\mathrm{t} + V_\mathrm{p}$. The new volume filling factor will be
\begin{equation}
    \phi_\mathrm{new} = \frac{V_\mathrm{t}\phi_\mathrm{t} + V_\mathrm{p}\phi_\mathrm{p}}{V_\mathrm{new}}\ ,\label{eq:porosity_S3}
\end{equation}
where $V_\mathrm{new}$ is taken from Eq. \ref{eq:S3_new_vol_porous} (compact projectile) or as $V_\mathrm{new} = V_\mathrm{t}-V_\mathrm{cr.}$ with $V_\mathrm{cr.}$ from Eq. \ref{eq:S3_crater_volume} (porous projectile). In the cases where we transfer mass from one aggregate to the other, we always assume that this mass is previously compacted by a factor of 1.5 in volume filling factor, but cannot be compacted to more than the critical filling factor $\phi_\mathrm{c}$. For the \Bb\ we have good arguments for this assumption as this compaction is consistent with XRT measurements of \citet{GuettlerEtal:2009a}, who also showed that it is likely that this compacted material is transferred to the projectile (see their Figs. 7 and 9). Without better knowledge, we assume the same compaction of transferred material for fragmentation with mass transfer (F3 and S4), and for these three cases we again use Eq. \ref{eq:porosity_S2b}. It is necessary to swap the indices of target and projectile in the case of \Bb\ and \Fc, as the projectile is accreting mass in this collisional outcome. For the fragments in S4 and F3 as well as for those in the case of F1 and F2, we assume an unchanged porosity with respect to the destroyed aggregate. The most sophisticated compaction model is used for collisions that lead to \Ba. Although \citet{WeidlingEtal:2009} measured the compaction only for a small range of aggregate sizes and collision velocities, they derived an analytic model to scale this compaction in collision velocity and showed that it is independent in aggregate mass. We follow this model but release it from the experimental bias due to the $\phi=0.15$ samples they used. As outlined in detail in Sect. \ref{sec:exp_types}, we basically use Eq. \ref{eq:weidling_ff_increase} and scale the $\phi_\mathrm{max} (v)$ according to Eq. \ref{eq:B1_phi_max_scale} (furthermore using Eqs. \ref{eq:red_mass_1} -- \ref{eq:compr_curve_power_law}).

\begin{table}[t]
\center%
\caption{\label{tab:porosity_evolution}Overview of the porosity evolution in the different collisional outcomes.}
\begin{tabular}{lll}
    \hline
    collisional outcomes & porosity evolution & equation \\
    \hline
    S1              & fluffier   & \ref{eq:porosity_S1} \\
    S2              & neutral or compaction   & \ref{eq:porosity_S2a} or \ref{eq:porosity_S2b}\\
    S3              & compaction & \ref{eq:S3_new_vol_porous}, \ref{eq:S3_crater_volume}, \ref{eq:porosity_S3}\\
    S4 (target)     & fluffier   & \ref{eq:porosity_S2b} \\
    S4 (projectile) & neutral    & -- \\
    B1              & compaction & \ref{eq:weidling_ff_increase} -- \ref{eq:B1_phi_max_scale}\\
    B2 (target)     & neutral    & -- \\
    B2 (projectile) & both       & \ref{eq:porosity_S2b}$^\mathrm{a}$ \\
    F1              & neutral    & -- \\
    F2              & neutral    & -- \\
    F3 (target)     & fluffier   & \ref{eq:porosity_S2b}$^\mathrm{a}$ \\
    F3 (projectile) & neutral    & -- \\
    \hline
\end{tabular}
\\
$^\mathrm{a}$The indices of target and projectile must be swapped here.
\end{table}

In summary, one can say that the aggregates' porosities can only be increased by the collisional outcomes S1, S4, and F3 (see Table \ref{tab:porosity_evolution}), where the \Sa\ collisions will have the most effect. While some collisional outcomes are neutral in terms of porosity evolution (F1 and F2), the main processes which lead to more compact aggregates are S3 and B1.

\section{Discussion\label{sec:conclusion}}
In the previous sections we have developed a comprehensive model for the collisional interaction between protoplanetary dust aggregates. The culmination of this effort is Fig. \ref{fig:colored_regimes}, which presents a general collision model based on 19 different dust-collision experiments, which will be the basis for Chapter \ref{chp:paper3}. Since it plays a vital role, it is worth a critical appraisal. We want to discuss the main simplifications and shortcomings of our current model in a few examples.

(1) The categorization into collisions between similar-sized and different-sized dust aggregates (see Figs. \ref{fig:categorization} and \ref{fig:colored_regimes}) is well-motivated as we pointed out in Sect. \ref{sec:collision_regimes}. Still we may ask ourselves whether this binarization is fundamentally correct if we need more than two categories, or `soft' transitions between the regimes. At this stage, a more complex treatment would be impractical due to the lack of experiments treating this problem.

(2) The binary treatment of porosity (i.e. $\phi < \phi_\mathrm{c}$ for `porous' and $\phi \geq \phi_\mathrm{c}$ for `compact' dust aggregates) is also a questionable assumption. Although we see fundamental differences in the collision behavior when we use either porous or compact targets, there might be a smooth transition from the more `porous' to the more `compact' collisions. In addition to that, the assumed value $\phi_\mathrm{c} = 0.4$ is reasonable but not empirically affirmed. On top of that, the maximum compaction that a dust aggregate can achieve in a collision depends on many parameters, such as, e.g., the size distribution of the monomer grains \citep{BlumEtal:2006} and the ability of the granular material to creep sideways inside a dust aggregate \citep{GuettlerEtal:2009a}.

(3) Although the total number of experiments upon which our model is based is unsurpassedly large, the total coverage of parameter space is still small (see the experiment boxes in Fig. \ref{fig:colored_regimes}). Thus, we sometimes apply extrapolations into extremely remote parameter-space regions. Although not quantifiable, it must be clear that the error of each extrapolation grows with the distance to the experimentally confirmed domains (i.e. the boxes in Fig. \ref{fig:colored_regimes}). Clearly, more experiments are required to fill the parameter space, and the identification of the key regions in the mass-velocity plane is exactly one of the goals of Chapter \ref{chp:paper3}.

(4) With such new experiments, performed at the `hot spots' predicted in Chapter \ref{chp:paper3}, we will not only close gaps in our knowledge of the collision physics of dust aggregates but will most certainly reveal completely new effects. That the \cc\ panel in Fig. \ref{fig:colored_regimes} is rather simple compared to the more complex \pC\ is due to the fact that there are hardly any experiments that back-up the \cc\ regime, whereas in the \pC\ case we have a pretty good experimental coverage of the parameter space.

In summary, the sophisticated nature of our collision model is both its strength and its weakness. The drawbacks of identifying four parameters that shape the collision outcome are that rather crude approximations and extrapolations have to be made. But it is still preferable to acknowledge the role of, e.g., porosity through a binary treatment than not to treat this parameter at all. Our new collision model represents the first attempt to include all existing laboratory experiments (for the material properties of interest); collisional evolution models can enormously profit from this effort.

\subsection{The bottleneck for protoplanetary dust growth\label{sec:outlook}}
We have presented the framework and physical background for an extended growth simulation. What is to be expected from this? This is the place to speculate under which conditions growth in PPDs is most favorable. A look at Fig. \ref{fig:colored_regimes} immediately shows that large dust aggregates can preferentially grow for realistic collision velocities in the \cC\ and \pC\ collision regimes (and to a lesser extent in the \pc\ case), due to \Sd. A broad mass distribution of protoplanetary dust must be present to make this possible. This prerequisite for efficient growth towards planetesimal sizes has also been suggested by \citet[][see their Fig. 11]{TeiserWurm:2009a}. Agglomeration experiments with micrometer-sized dust grains and a sticking probability of unity (experiments 1 -- 3 in Table \ref{tab:experiments}) have shown that nature chooses a rather narrow size distribution for the initial fractal growth phase. To see if this changes when the physical conditions leave no room for growth under quasi-monodisperse conditions, i.e. whether nature is so `adaptive' and `target-oriented' to find out that growth can only proceed with a wide size distribution, will be the subject of Chapter \ref{chp:paper3}, in which we apply the findings of this chapter to a collisional evolution model.

\subsection{Influence of the adopted material properties\label{sec:material_influence}}
The choice of material in our model is 1.5~$\mu$m diameter silica dust, as most of the underlying experiments were performed with this material. Many experiments \citep{BlumWurm:2000, LangkowskiEtal:2008, BlumWurm:2008} showed that this material is at least in a qualitative sense representative for other silicatic materials -- also for irregular grains with a broader size distribution. Still, the grain size of the dust material may have a quantitative influence on the collisional outcomes. For example, dust aggregates consisting of 0.1~$\mu$m are assumed to be stickier and more rigid \citep{WadaEtal:2007, WadaEtal:2008, WadaEtal:2009}, because the grain size may scale the rolling force or breaking energy entering into Eqs. \ref{eq:S1_threshold} and \ref{eq:S2_threshold}. However, due to a lack of experiments with smaller monomer sizes, we cannot give a scaling for our model for smaller monomer sizes at this point. Moreover, organic or icy material in the outer regions of PPDs or oxides and sintered material in the inner regions may have a big impact on the collisional outcome, i.e. in enhancing the stickiness of the material and thereby potentially opening new growth channels.

As for organic materials, \citet{KouchiEtal:2002} found an enhanced sticking of cm-sized bodies covered with a 1~mm thick layer of organic material at velocities as high as 500~\cms\ and at a temperature of $\sim250$~K. Icy materials are also believed to have an enhanced sticking efficiency compared to silicatic materials. \citet{HatzesEtal:1991} collided 5~cm diameter solid ice spheres, which were covered with a 10 -- 100~$\mu$m thick layer of frost. They found sticking for a velocity of 0.03~\cms, which is in a regime where our model for refractory silicatic material predicts bouncing (see \pp\ or \cc\ in Fig. \ref{fig:colored_regimes}). Sintering of porous dust aggregate may occur in the inner regions near the central star or -- triggered by transient heating events \citep[e.g. lightning,][]{GuettlerEtal:2008} -- even further out. Ongoing studies with sintered dust aggregates \citep{Poppe:2003} show an increased material strength (e.g. tensile strength) by an order of magnitude (C. G\"uttler \& J. Blum, unpublished data). This would at least make the material robust against fragmentation processes and qualitatively shift them from the porous to the compact regime in our model -- without necessarily being compact. Due to a severe lack of experimental data for all these materials, it is necessary and justified to restrict our model to silicates at around 1~AU, while it is to be kept in mind that these examples of rather unknown materials might potentially favor growth in other regions in PPDs.



%% file: Chapters/Chapter4.tex

\chapter{Introducing the bouncing barrier} 
\label{chp:paper3}
\lhead{Chapter 4. \emph{Bouncing barrier}} 
\rhead{}

Based on \textit{`The outcome of protoplanetary dust growth: pebbles, boulders, or planetesimals? II. Introducing the bouncing barrier'} by A. Zsom, C. W. Ormel, C. G\"uttler, J. Blum \& C. P. Dullemond published in \aap, 513, 57.

\def\remark#1{{{\bf remark:} \bf #1}}
\def\action#1{{\bf #1}}
\def\putin#1{{\it #1}}
\def\revised#1{{#1}}

\def\Sa{hit \& stick (S1)}
\def\Sb{sticking through surface effects (S2)}
\def\Sc{penetration (S3)}
\def\Sd{mass transfer (S4)}
\def\Ba{bouncing with compaction (B1)}
\def\Bb{bouncing with mass transfer (B2)}
\def\Fa{fragmentation (F1)}
\def\Fb{erosion (F2)}
\def\Fc{fragmentation with mass transfer (F3)}

\def\Sac{Hit \& stick (S1)}
\def\Sbc{Sticking through surface effects (S2)}
\def\Scc{Penetration (S3)}
\def\Sdc{Mass transfer (S4)}
\def\Bac{Bouncing with compaction (B1)}
\def\Bbc{Bouncing with mass transfer (B2)}
\def\Fac{Fragmentation (F1)}
\def\Fbc{Erosion (F2)}
\def\Fcc{Fragmentation with mass transfer (F3)}

\def\pp{\textit{`pp'}}
\def\pP{\textit{`pP'}}
\def\cc{\textit{`cc'}}
\def\cC{\textit{`cC'}}
\def\pc{\textit{`pc'}}
\def\pC{\textit{`pC'}}
\def\cp{\textit{`cp'}}
\def\cP{\textit{`cP'}}

\section{Introduction}
In the core accretion paradigm of planet formation \citep{Mizuno1980-4, Pollack1996-4-4}, planets are the outcome of an accretion process that starts with micron-size dust grains and covers 40 magnitudes in mass. The paradigm can be divided into three stages. The first stage involves the formation of rocky planets and the rocky cores of gas giant planets and begins with the coagulation of dust in the protoplanetary disks surrounding many pre-main-sequence stars \citep{Safronov1969, Weidenschilling1993-4, Blum2008-4}. The next stage of planet formation is the formation of protoplanetary cores from the planetesimals. The idea is that the kilometer-size planetesimals are so large that gravity begins to dominate leading to the gravitational agglomeration of these bodies to rocky planets. This scenario was studied by \cite{Safronov1969} and modeled using numerical methods by \cite{Weidenschilling1980}, \cite{Nakagawa1983}, \cite{Mizuno1988}, \cite{Schmitt1997}, \cite{Wetherill:1990p85}, \cite{Nomura2006-4}, \cite{Garaud2004-4}, \cite{Tanaka2005-4}, and several additional authors. These models solve for the size distribution of dust aggregates in the disk as a function of time, and investigate if, where, and how larger dusty bodies form, and how long this takes. Finally, in the third stage, gas accretes onto these protoplanets forming giant planets or -- in the absence of gas -- gravitational encounters occur between these protoplanets result in a chaotic, giant impact phase, until orbital stability is achieved \citep{Chambers2001-4, Kokubo2006-4, Thommes2008-4}.\\

In this study, we focus on the first phase and study the effectiveness of the dust growth by surface force, that is, how large particles become due to simple sticking processes only.  It is known that initially, for micron-size grains, the growth is driven by Brownian motion. This typically leads to slow collisions and forms aggregates of fractal structure \citep{Kempf1999, Blum1996-4}. In the current picture of dust growth, as these aggregates grow, at some point the growth will leave the fractal regime, and collisions will start to lead to the compaction and breaking of the aggregates \citep{Blum2000-4}, embedding of small bodies into larger aggregates (leading to `filling up' of these larger aggregates and compaction caused by the force of the collision \citep{Ormel:2007p93}. As the size of the dust aggregates increases, differential vertical settling \citep{Safronov1969}, radial drift \citep{Whipple1972}, and turbulence \citep{Voelk1980, Mizuno1988, Ormel:2007p92} will become important new mechanisms driving relative velocities between aggregates. The increasing relative velocities caused by these mechanisms will at least partly compensate the lower collision probability due to lower surface-over-mass ratio of large aggregates. When the aggregates grow to sizes of between millimeter and meter, however, the sticking efficiency drops strongly (e.g., \cite{Blum1993-4}) and the relative velocities become so large that aggregates can fragment (\cite{Blum2008-4}, so-called `fragmentation barrier'). Another hurdle that the particles have to circumvent is the `drift barrier' \citep{Weidenschilling1977}, namely that millimeter or centimeter-sized particles are lost to the star because of radial drift that occurs on a short timescale. \cite{Okuzumi2009-4} pointed out the existence of a `charge barrier', which possibly halts the particle
 growth at an early stage of fractal aggregates. Despite many years of efforts, it is unknown whether the coagulation process can overcome these barriers. These barriers have been and remain the main open question about the initial stages of planet formation, i.e., the growth from dust to planetesimals. \\

Several mechanisms have been proposed to overcome this problem, among which are the trapping of dust in vortices \citep{Barge1995-4, Klahr1997, Lyra2009a}, the trapping of decimeter-sized boulders in turbulent eddies and the subsequent gravitational collapse of swarms of these trapped boulders \citep{Johansen:2007p65-4}, the trapping of particles in a pressure bump caused by the evaporation front of water \citep{Kretke2007-4, Brauer2008b-4} and many other scenarios. However, the correct modeling of any of these requires detailed knowledge of the collisional physics, and these models have so far relied on either simplified input phyisics or simplified initial conditions.

Because of their complexity, collisional evolution models have to make simplifying assumptions about the outcome of dust aggregate collisions, for example that collisions always result in sticking, or otherwise use simple recipes for the collisional outcome.  Ideally, one requires detailed knowledge of the outcome of every collision.  But modeling this microphysics within an evolution model is simply impractical. There are computer programs that model these individual collisions in detail (e.g., \cite{Dominik:1997p89-4}, \cite{Suyama2008-4}; Geretshauser et al, in prep.), but each collision model takes anywhere between hours and weeks to complete on a computer. They are therefore impractical to use at run-time in a model that computes the overall time-dependent evolution of the dust size distribution inside protoplanetary disks. These collision models themselves often depend on poorly known input physics. 

Another approach to obtaining the collisional outcome of dust aggregates is to model these collisions in the laboratory. From the many experiments that have been performed, a picture emerges of the outcome of dust aggregate collision under a variety of conditions in the protoplanetary disk (PPD). In the previous chapter, we collected data from over 19 experiments, and compiled a set of formulae to describe reasonably well the outcomes of these collisions in such a way that they can be used as input to models that address the temporal evolution in the dust size distribution.

In this chapter, we directly rely on the outcome of these laboratory experiments to model the dust aggregate size distribution. As described in Chapter \ref{chp:paper2}, we have produced a mapping of all available collision experiments for silicate-like particles. The velocity range of these experiments is also sufficiently wide to cover various disk models that roughly correspond to the conditions at 1 AU in the PPD. For details of the collisional mapping, we refer to Chapter \ref{chp:paper2} but summarize the elements of our new collision model in Sect. 3.1.

We build this collision kernel into a Monte Carlo code for modeling the size and porosity distribution of dust in a protoplanetary disk (\cite{Zsom2008-4}, hereafter ZsD08). The outcome of our laboratory-driven dust coagulation model is difficult to predict a priori since the key variables involved depend on a non-trivial interplay between the collision kernel (Chapter \ref{chp:paper2}) and the velocity field.  We can, however, propose two scenarios. In the first, particle growth proceeds beyond the meter-size barrier, all the way to forming planetesimals. In the second scenario, growth terminates at an intermediate size. In this case, additional growth to planetesimal sizes may proceed by the means of the concentration and subsequent gravitational collapse of these particles \citep{Johansen:2007p65-4, Cuzzi2008-4}. Thus, our model providea the starting conditions for these concentration models. We emphasize, however, that in this work we do not in any way `optimize' the outcome by laboriously scanning all the parameter space or treating environments that may be more conducive to growth, such as nebula pressure bumps or the trapping of dust in vortices \citep{Kretke2007-4, Lyra2009a}. These are obvious extensions of this present work. But by considering the sensitivity of a few key parameters (e.g., gas density, and turbulence strength) to the outcome of the growth process, we obtain a picture of where the arrow of coagulation typically points to in protoplanetary environments: pebbles, boulders, or planetesimals.\\

In this chapter, we describe the three nebulae models used in this work and the sources of relative velocity between the aggregates (Sect. \ref{sec:2}), how we developed our coagulation/fragmentation model of Chapter \ref{chp:paper2} into a Monte Carlo code (Sect. \ref{sec:impl}), and our first results (Sect. \ref{sec:res4}). We also test the sensitivity of these results to variations in gas density, the velocity field, and other key model parameters. Section  \ref{sect:disc} reflects the importance of our result to planetesimal formation and provides suggestions for future experiments. Finally, Sect. \ref{sec:sum} lists our main conclusions. 

\section{The nebulae model}
\label{sec:2}
\subsection{Disk models}
\label{sec:disks}
We briefly describe the disk models considered in this chapter.

\paragraph{The low density model:}
Resolved millimeter emission maps of protoplanetary disks seem to indicate a shallow surface density profile (\cite{Andrews2007-4}) given by $\Sigma_g(r) \propto r^{-0.5}$. The systematic effects of some of their assumptions, such as the disk inclinations or the simplified treatment of the temperature distribution, may produce steeper profiles. Therefore, \cite{Brauer2008a-4} adopted the following profile:
\begin{equation}
\Sigma_g(r)= 45 \frac{\mbox{ g}}{\mathrm{cm}^2} \left( \frac{r}{\mathrm{AU}} \right)^{-0.8}.
\end{equation}
Here we assumed that the central star is of solar mass, the disk extends from 0.03 AU to 150 AU, and the total mass of the disk is 0.01 M$_\odot$. Assuming that the pressure scale-height is $H_p=0.05\times r$ and the vertical structure is Gaussian, we obtain:
\begin{equation}
\rho_g (z,r) = \frac{\Sigma_g(r)}{\sqrt{2 \pi} H_p}\exp(-z^2/2H_p^2),
\end{equation}
the density at 1 AU in the midplane ($z=0$) being $2.4 \times 10^{-11}$ g cm$^{-3}$, approximately two orders of magnitude lower than the minimum mass solar nebulae (MMSN) value.

\paragraph{MMSN model: }
The minimum mass solar nebulae model (MMSN) was introduced by \cite{Weidenschilling1977a-4} and \cite{Hayashi1985}. From the present state of the Solar System today, it is possible to infer a lower limit to the mass in the solar nebulae from which the planets were formed. The model assumes that the planets were formed where they are currently located (no migration included). It also assumes that all the solid material presented in the solar nebula had been incorporated into the planets. The loss of solid material  because of radial drift is not taken into account. Despite these uncertainties, the MMSN model is frequently used as a benchmark. The surface density of the MMSN disk is given by
\begin{equation}
\Sigma_g(r)= 1700 \frac{\mbox{ g}}{\mathrm{cm}^2} \left( \frac{r}{\mathrm{AU}} \right)^{-1.5},
\end{equation}
which corresponds to a total disk mass of 0.01 M$_\odot$ between 0.4 and 30 AU (between the orbits of Mercury and Neptune). Assuming that the vertical structure of the gas follows a Gaussian distribution, we infer a midplane density at 1 AU of $1.4\times 10^{-9}$ g cm$^{-3}$. 

\paragraph{The high density model:}
\cite{Desch2007-4} introduced a `revised MMSN model' by adopting the starting positions of the planets in the Nice model of planetary dynamics \citep{Tsiganis2005-4} thus taking into account planetary migration. The revised MMSN model predicts that the Solar System was initially in a far more compact configuration and its surface density profile is given by
\begin{equation}
\Sigma_g(r)= 5.1\times 10^4 \frac{\mbox{ g}}{\mathrm{cm}^2} \left( \frac{r}{\mathrm{AU}} \right)^{-2.2}.
\end{equation}
This model is consistent with that of a decretion disk that is being photoevaporated by the central star. Although the model of \cite{Desch2007-4} was defined for the outer Solar System, we extrapolate the profile to 1 AU to cover a broad range of surface density values in our calculations. Assuming, as in the MMSN model, a Gaussian vertical distribution, the density at 1 AU in the midplane is $2.7 \times 10^{-8}$ g cm$^{-3}$.\\

For simplicity, we adopt a midplane temperature of 200 K (isothermal sound speed of $c_s = 8.5\times 10^{4}$ cm s$^{-1}$) in all three models.

\subsection{Relative velocities}
\label{sec:vrel}
We consider three contributors to the relative velocities between dust aggregates: Brownian motion, radial drift and turbulence. In the following, we discuss each of these sources.

\begin{figure*}
  \includegraphics[width=0.5\textwidth]{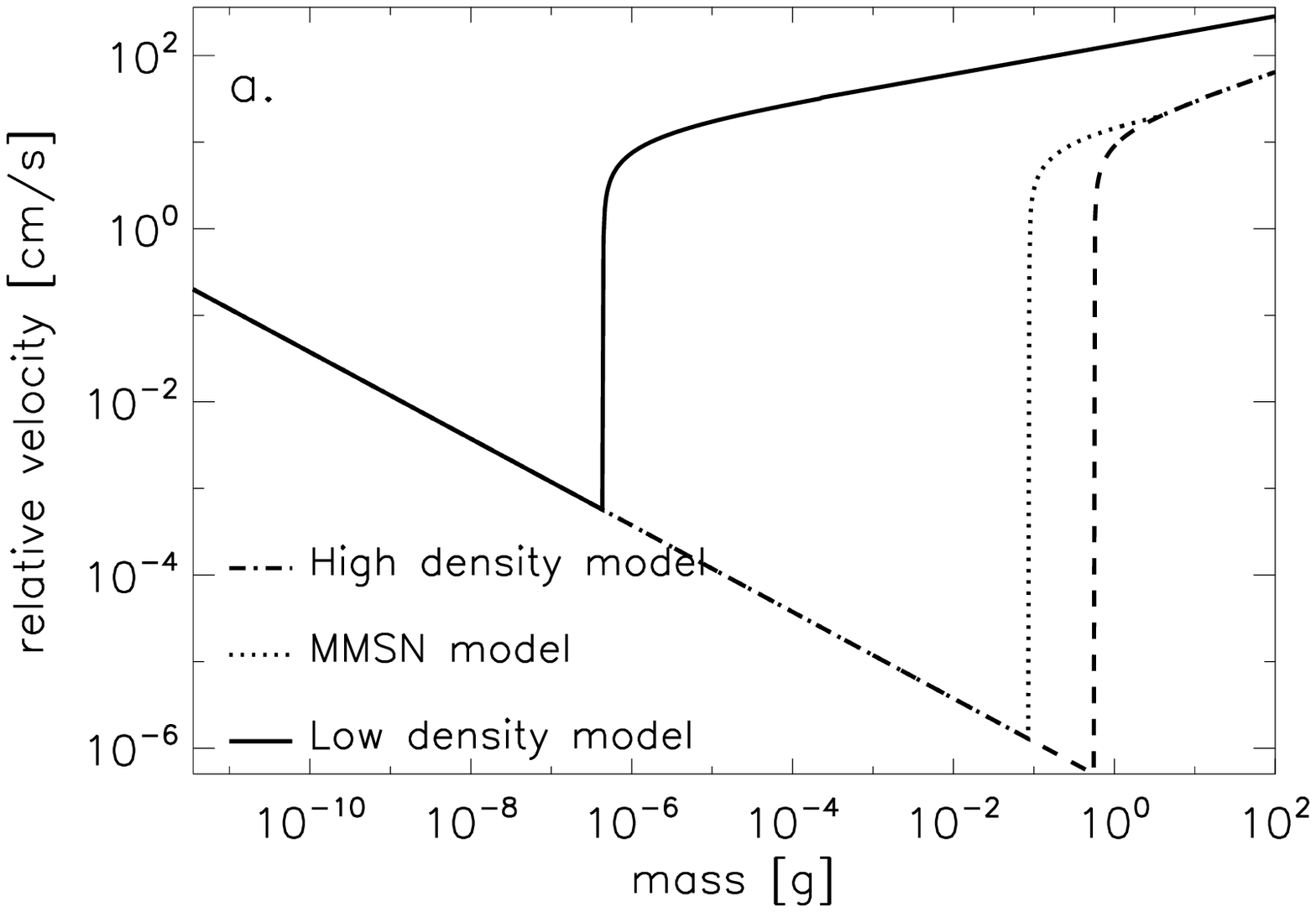}
  \includegraphics[width=0.5\textwidth]{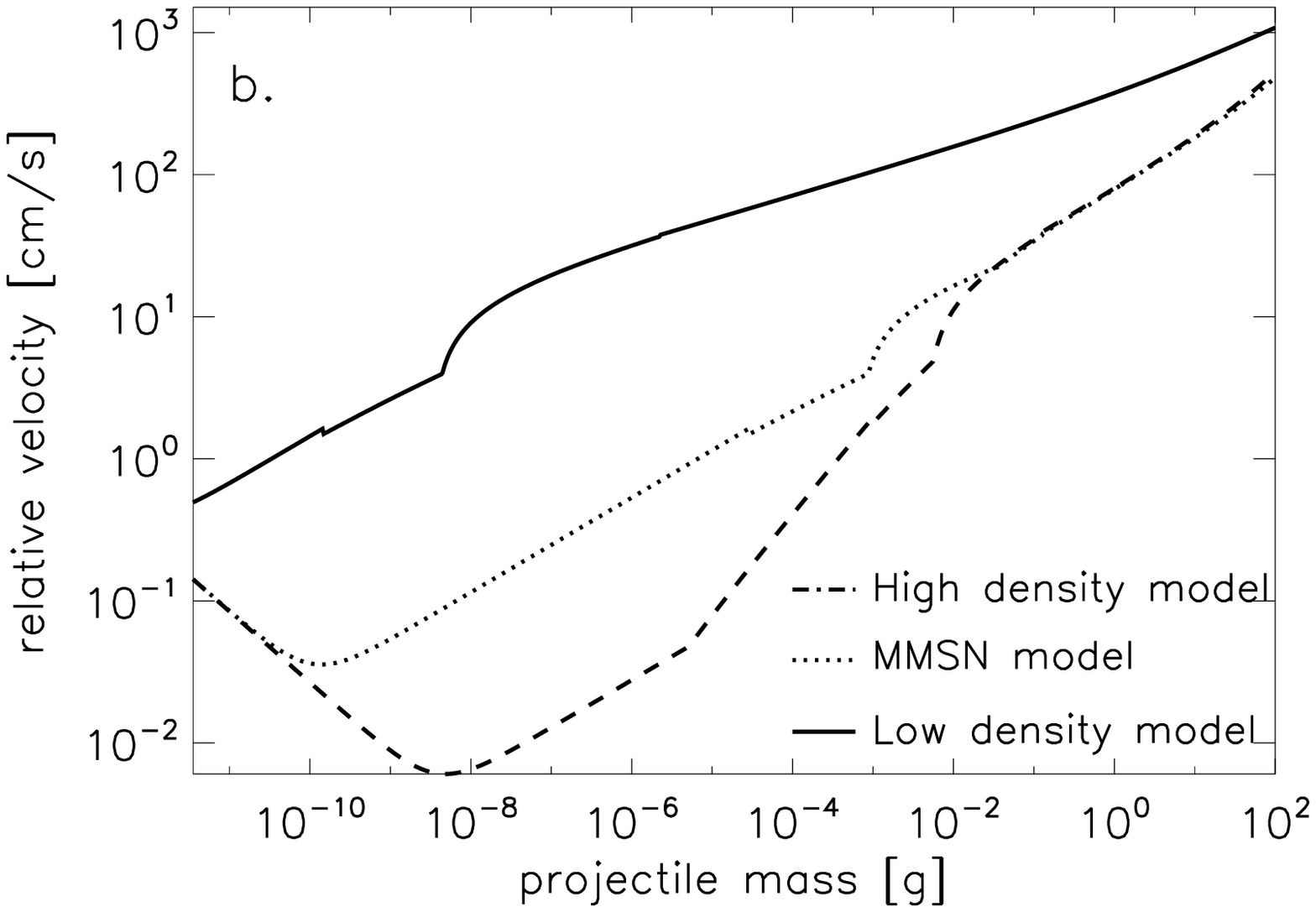}
  \caption{The combined relative velocities caused by Brownian motion, radial drift, and turbulence for fluffy particles ($\Psi = 20$) in the three disk models for equal-sized particles (a) and for different-sized particles with a mass ratio of 100 (b). The solid line indicates the low density model of \protect \cite{Brauer2008a-4}. Physical parameters of the disk: the distance from the central star is 1 AU, temperature is 200 K, the density of the gas is $2.4\times 10^{-11}$ g cm$^{-3}$, and the turbulence parameter, $\alpha=10^{-4}$. The dotted line represents the MMSN model. The density is $1.4 \times 10^{-9}$ g cm$^{-3}$, and the other parameters are the same. The dashed line corresponds to the high density disk. The gas density is $2.7\times 10^{-8}$ g cm$^{-3}$.}
  \label{relcont4}
\end{figure*}

The average relative velocity of two particles with mass $m_1$ and $m_2$ in a region of a disk with temperature $T$ due to Brownian motion is
\begin{equation}
 \Delta v_B (m_1,m_2) = \sqrt \frac{8kT(m_1+m_2)}{\pi m_1 m_2}.
\end{equation}
For micron-sized particles, the relative velocity is of the order of 0.1 cm s$^{-1}$, but for cm-sized particles this value drops several orders of magnitude. Therefore, Brownian motion is only effective for collisions between small particles during the initial stages of growth. Coagulation caused by Brownian motion results in fluffy aggregates of fractal dimensions of around 2 and 3 \citep{Blum1996-4, Kempf1999, Blum2000a, Krause2004-4}. In practice, no growth is caused by Brownian motion for aggregates larger than 100 micron. 

The second contributor to the relative velocity is turbulence. The relative velocity of aggregates produced by the random motion of turbulent eddies was calculated numerically by \cite{Voelk1980}, \cite{Mizuno1988}, and \cite{Markiewicz1991}. We use the closed form expressions presented by \cite{Ormel:2007p92}. We assume that turbulence is parameterized by the \cite{Shakura1973-4} $\alpha$ parameter
\begin{equation}
\nu_T=\alpha c_s H_g,
\label{eq:nuT4}
\end{equation}
where $\nu_T$ is the turbulent viscosity, $c_s$ is the isothermal sound speed, and $H_g$ is the pressure scale height of the disk. The value of the $\alpha$ parameter reflects the strength of the turbulence in the disk. Typical values of $\alpha$ in this chapter range between 10$^{-3}$ and 10$^{-5}$. The turbulent relative velocity is a function of the stopping times of the two colliding particles. The stopping time (or friction time) is the time the particle needs to react to the changes in the motion of the surrounding gas. As long as the radius of the particle is smaller than the mean free path of the gas ($a < \frac{9}{4}\lambda_{\mathrm{mfp}}$), the particle is in the Epstein regime, where the stopping time is (\cite{Epstein1924-4}):
\begin{equation}
t_{s} = t_{\mathrm{Ep}} = \frac{3 m}{4 v_{\mathrm{th}} \rho_g A},
\label{eq:ts1-4}
\end{equation}
where $m$ and $A$ are the mass and the cross-section of the particle, and $\rho_g$ and $v_{\mathrm{th}}$ are the gas density and the thermal velocity. At high gas densities where the mean free path is low or in the case of larger particles, the first Stokes regime applies and the stopping time is
\begin{equation}
t_s = t_{\mathrm{St}} = \frac{3 m}{4 v_{\mathrm{th}} \rho_g A} \times \frac{4}{9} \frac{a}{\lambda_{\mathrm{mfp}}}.
\label{eq:ts2-4}
\end{equation} 
In the first Stokes regime, the stopping time is independent of the gas particle relative velocity as well as the gas density. This regime can be used as long as the particle Reynolds number is smaller than unity. The particle Reynolds number is calculated to be (\cite{Weidenschilling1977})
\begin{equation}
Re_p = \frac{2 a \Delta v_{\mathrm{pg}}}{\eta},
\end{equation}
where $\Delta v_{\mathrm{pg}}$ is the relative velocity between the particle and the gas, and $\eta$ is the gas viscosity. For particles outside the Epstein regime, we can assume that the systematic velocity (radial drift) dominates over the random velocities (turbulence); therefore, $\Delta v_{\mathrm{pg}} \approx v_D$, where $v_D$ is the drift velocity of the particle, defined in the next paragraph. The particle Reynolds number never exceeds unity in our simulations. Therefore, we do not include additional Stokes regimes.

\begin{figure}
\centering
  \includegraphics[width=0.7\textwidth]{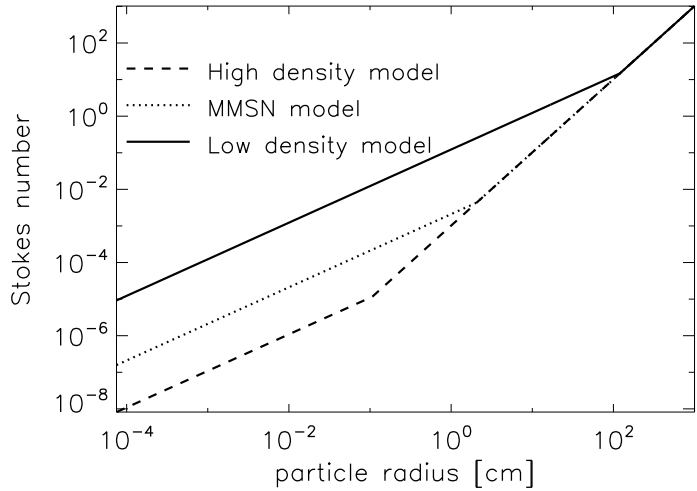}
   \caption{The Stokes number as a function of the particle radius in the three models. The parameters of the dust for all of the models are the following: monomer radius is $a_0 =0.75$ $\mu$m, material density is $\rho_0 = 2$ g cm$^{-3}$, and $\Psi = 1$.}
  \label{fig:ep-st}
\end{figure}

Radial drift also contributes to relative velocities between aggregates. Radial drift ($v_D$) has two sources: the drift of individual particles ($v_{d}$) and drift caused by accretion processes of the gas ($v_{da}$), thus the total radial drift velocity is $v_D=v_{d}+v_{da}$.\\
The radial drift of individual dust aggregates with mass $m$ is (\cite{Weidenschilling1977})
\begin{equation}
v_{d} = -\frac{2 v_N}{St+1/St} ,
\end{equation}
where $St$ is the Stokes number of the aggregate ($St=t_s\Omega$, where $\Omega$ is the orbital frequency) and $v_N$ is the maximum radial drift velocity (\cite{Whipple1972}). 

The second contribution to the radial velocity is produced by the accretion of the gas. This part of the radial velocity is calculated as follows (\cite{Kornet2001-4})
\begin{equation}
v_{da} = \frac{v_{\mathrm{gas}}}{1+St^2},
\label{eq:drift}
\end{equation}
where $v_{\mathrm{gas}}$ is the accretion velocity of the gas (\cite{Takeuchi2002-4}). 

The relative velocity due to radial drift is then simply the difference between the radial velocity of particles 1 and 2. However, as the Stokes number of the aggregates is always smaller than $10^{-3}$ (see Sect. \ref{sec:res4}), the second term of the radial velocity ($v_{da}$) can be safely neglected
\begin{equation}
\Delta v_D = | v_{D1} - v_{D2}| \approx | v_{d1} -v_{d2}|.
\end{equation}
\\

This study uses two quantities to describe the porosity of the aggregates. The volume filling factor is
\begin{equation}
\phi = V^*/V_{\mathrm{tot}}=(A^*/A)^{3/2},
\end{equation}
where $V^*$ is the volume occupied by the monomers and $V_{\mathrm{tot}}$ is the total volume of the aggregate, including pores, and $A$ and $A^*$ are the surface area equivalents of these quantities. In this way, the filling factors also enters the definition of the friction time (Eqs. \ref{eq:ts1-4} and \ref{eq:ts2-4}). The density of aggregates then follows as $\rho=\rho_0 \phi$, where $\rho_0 = 2$ g cm$^{-3}$ is the material density of the silicate. In this study, we also use the reciprocal parameter of the filling factor, which is denoted by the enlargement parameter, $\Psi = \phi^{-1}$.\\

We illustrate the relative velocity between equal-sized and different-sized aggregates with $\Psi = 20$ ($\phi = 0.05$) in Fig. \ref{relcont4} for the disk models considered in this work. Adopting a threshold (fragmentation) velocity of 1 m s$^{-1}$, the maximum particle size that can be reached in the models are 0.025 cm in the low density model, 1.4 cm in the MMSN model, and 1.7 cm in the Desch model. The Stokes numbers of these particles are identical in all three models, $4.7 \times 10^{-3}$. The constant fragmentation velocity of 1 m s$^{-1}$ is the typical velocity at which silicate particles fragment. In our collision model this is not the case for all combinations of mass ratio and porosity (Chapter \ref{chp:paper2}), but the m s$^{-1}$ threshold remains a useful proxy for the point where fragmentation processes become important.

Figure \ref{fig:ep-st} shows the Stokes number as a function of particle radii in the three models. Initially, particles are in the Epstein regime, where the stopping time, thus the Stokes number, depends on the gas density. When the particles enter the Stokes regime, the stopping time becomes independent of the gas density (see Eq. \ref{eq:ts2-4}). One can see that particles in the Desch model are in the Stokes regime at a Stokes number of $4.7 \times 10^{-3}$ (when the particles have relative velocities of 1 m s$^{-1}$), while the aggregates in the MMSN model are close to it, which explains why the maximum particle size is almost the same in these two models.

As discussed in \cite{Ormel:2007p92}, particles are initially in the `tightly coupled particle' regime, where the eddies are all of class I type, meaning that the turnover time of all eddies is longer than the friction time of the particles (\cite{Voelk1980}). Upon entering a class I eddy, a particle therefore forgets its initial motion and aligns itself with the gas motions of the eddy before the eddy decays or the particle leaves it. This regime is presented in Figs. \ref{relcont4}a and b. Different-sized particles are found in this relative velocity regime as long as their masses are less than $10^{-8}$ g in the low density model, $10^{-3}$ g in the MMSN model, and $10^{-2}$ g in the high density model assuming fluffy particles ($\Psi = 20$). If the particles leave this regime and enter the `intermediate particle' regime, their relative velocity increases. This transition affects the particle evolution, as discussed in e.g., Sect \ref{sec:mmsn}.

\section{Collision model and implementation}
\label{sec:impl}
We use a statistical or `particle in a box' method to compute the collisional evolution, that is, we assume that all particles are homogeneously distributed within a certain volume (the simulation volume). In reality, the particles could however leave the simulated volume or new particles could enter from outside because of radial drift or random motions (turbulence and Brownian motion). Since we do not resolve the spatial dependence of the aggregates, we simply assume that local conditions hold during the run. The gas and dust densities are kept constant and particles cannot leave or enter the simulation volume (hereafter `local approach').

\subsection{Short overview of the collision model}

Many laboratory experiments on dust aggregate collisions have been performed (see \cite{Blum2008-4}). The growth begins as fractal growth and we use the recipe of \cite{Ormel:2007p93} to describe this initial stage. However, once aggregates have restructured into non-fractal, macroscopic aggregates (e.g., $\gtrsim100\;\mu$m in size), laboratory experiments show that the collisional outcomes become very diverse. In this regime, many new experiments were performed with dust aggregates consisting of 1.5 $\mu$m diameter SiO$_2$ monomers of either high porosity $\phi=0.15$ (\cite{Blum2004-4}), or intermediate porosity ($\phi=0.35$). Chapter \ref{chp:paper2} compiled 19 experiments with different aggregate masses, collision velocities, and aggregate porosities.\\

\begin{figure*}
\centering
  \includegraphics[width=0.7\textwidth]{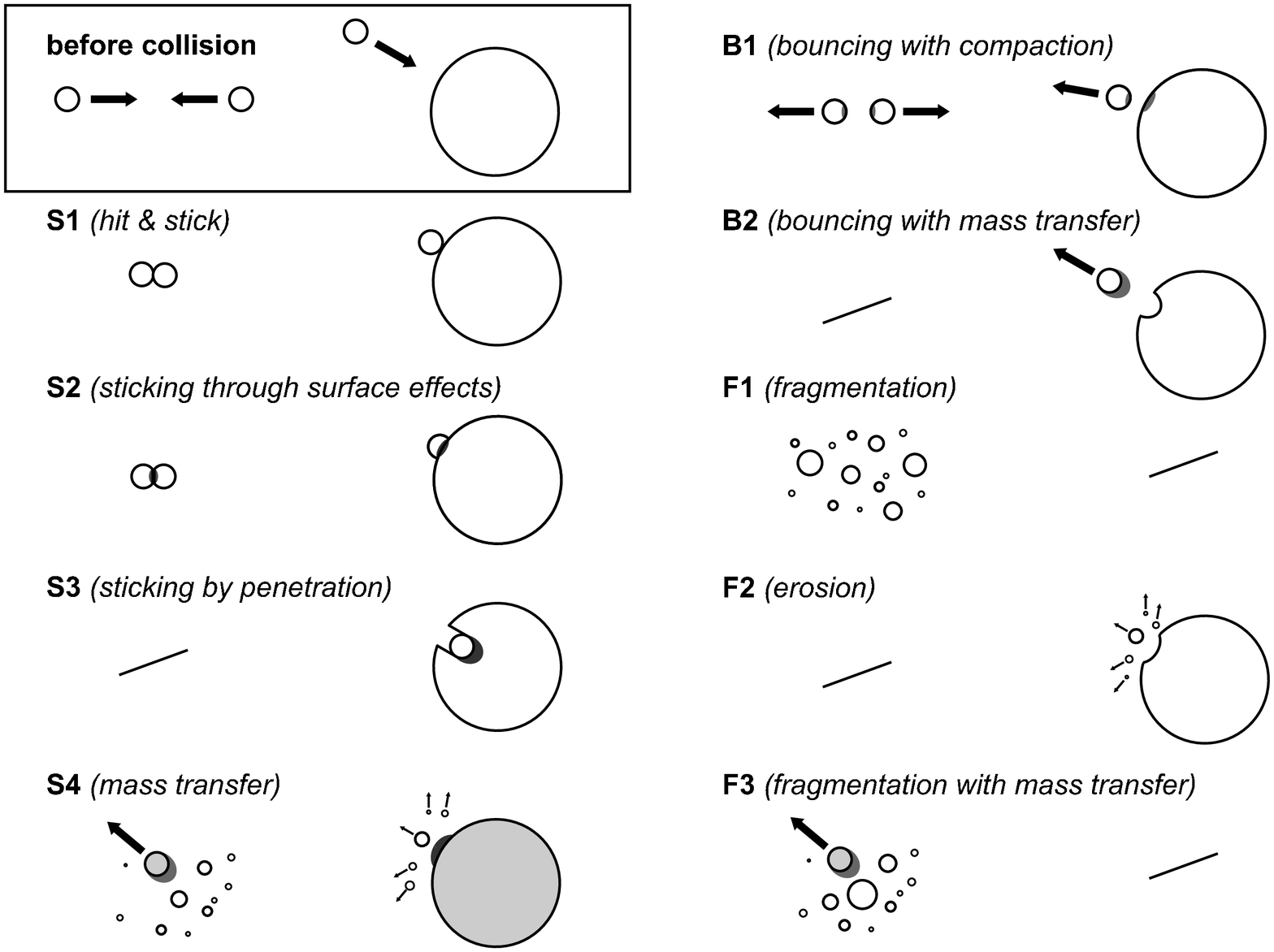}
  \caption{The collision types considered in this chapter. We distinguish between similar-sized and different-sized particles. Some of the collision types only happens for one of the mass ratios. Grey color indicates that during the given collision type the particle is compact or part of the mass is compacted.}
  \label{coltypes}
\end{figure*}

From these experiments, we identified nine different collisional outcomes involving sticking, bouncing, or fragmentation (see Fig. \ref{coltypes}). The occurrence of these regimes depends mainly on aggregate masses and collision velocities. However, it also depends on both the porosity of the particles and the critical mass ratio. For example, Chapter \ref{chp:paper2} finds fragmentation in collisions between a porous aggregate and a solid wall, whereas Langkowski et al. (2008) find sticking of a porous projectile by penetrating a target that is also porous.  Likewise, Hei\ss elmann et al., in prep. detect the bouncing of two similar-sized, porous dust aggregates, while Langkowski et al. (2008) uncover sticking of the same velocity where one collision partner (target) is significantly bigger. To address the importance of the mass ratio and porosity, we identified eight different collision regimes (look-up tables) based on a binary treatment of porosity and mass ratio i.e., (i) similar-sized or different-sized collision partners and (ii) porous or compact collision partners. The additional distinction between target, which we always define as the heavier collision partner, and projectile then defines eight different collision regimes. We denote these regimes as {\pP} (porous projectile, porous target; target significantly bigger than the projectile) and {\pc} (porous projectile, compact target; target of similar size than the projectile).  

In Chapter \ref{chp:paper2}, we classified each of these 19 experiments into one or more of these eight regimes (see Fig. 10 of Chapter \ref{chp:paper2}). Based on extrapolation of experimental findings, we decide in which mass and velocity range collisions result in sticking, bouncing, or fragmentation. These results are presented in Fig. 11 in Chapter \ref{chp:paper2}.

It should be noted that the critical mass ratio of the equal-size (e.g., {\pp}, {\cc}) to the different-size regimes (e.g., {\pP}, {\cC}) is ill-constrained by experiments. Therefore, we use critical mass ratios of $r_m$ = 10, 100, and 1000 to explore the effect of this parameter.

\subsection{Porosity}

Chapter \ref{chp:paper2} defined a binary representation of the porosity, particles are either porous or compact. Following the simple model of \cite{Weidling2009-4}, we include a continuous transition between these two `phases'. These authors showed that the compaction of particles caused by bouncing can be described by porous and compacted sites on the surface of the aggregate. A site of the aggregate is porous if it has not yet experienced any collisions (e.g., bouncing); a compacted site has encountered at least one collision but any additional collision at that part of the surface cannot change the porosity of the site. We define the probability of hitting a passive site of the aggregate with the experssion
\begin{equation}
P_p = \frac{\phi_c - \phi}{\phi_c - \phi_p}, 
\end{equation}
where $\phi$ is the volume-filling factor of the aggregate, $\phi_c$ is the critical porosity ($\phi_c=0.4$, see Chapter \ref{chp:paper2}), and $\phi_p$ is the volume filling factor of the porous site, which is chosen to be 0.15. If $\phi$ is between 0.15 and 0.4, a random number decides whether the particle collided with a porous or a compact site. This treatment of the porosity ensures a continuous transition from porous to compact aggregates. \\

During the initial {\Sa} phase, particles are in the fractal growth regime (\cite{Ossenkopf:1993p80}, \cite{Blum2000a}, \cite{Krause2004-4}). Particles grow initially because of Brownian motion and later due to turbulence. The structure of the aggregate depends on whether the collision happened between a cluster and a monomer (PCA) or between two clusters (CCA). The latter type of collision produces fluffy aggregates with fractal dimensions of 2, while the former leads to more compact structures of fractal dimension 3. The hit and stick recipe of \cite{Ormel:2007p93} attempts to interpolate between these two fractal models. At one point, collisional energies become high enough to invalidate this assumption. This occurs when the collisional energy is five times higher than the rolling energy of monomers (\cite{Dominik:1997p89-4}, \cite{Blum2000-4}). The internal structure then becomes homogenous. In our model we assume that once the fractal growth caused by S1 is over, {\Ba} restructures the aggregates producing compact structures of fractal dimension 3. Since our model cannot follow the exact shape of the particles, we assume that the aggregates are spheres and can be described with a single density ($\rho$) thus neglecting the effects of e.g., the `toothing radius' of \cite{Ossenkopf:1993p80} or the craters forming during {\Sc}.

\subsection{The Monte Carlo method}

Using the expressions for the relative velocity, the collisional cross-section between the dust particles, and the collisional outcome, we solve for the temporal evolution of the dust size distribution. Traditionally, the Smoluchowski equation is used to determine the evolution of the mass distribution function (e.g., \cite{Dullemond:2004p325}, \cite{Dullemond:2005p78}, \cite{Tanaka2005-4}, \cite{Brauer2008a-4}). The continuous form of the Smoluchowski equation used in these works lacks the stochasticity of the coagulation problem (\cite{Safronov1969}). All bodies of mass $m$ will grow in the same way thus the spatial and temporal fluctuations of the particle ensemble are averaged out. In reality, however, particles with similar masses might follow a different evolutionary path depending on which other particles they collide with. The collision model typically used in these works is, by necessity, rather simple because in the Smoluchowski formulation the collision and time evolution steps are linked together. These collision models consist of sticking and fragmentation and only the mass of the particles is followed. The advantage of such a model is that it is not computationally too expensive: the entire disk can be practically modeled. 

\cite{Ormel:2007p93} introduced a new Monte Carlo method to solve for the mass and the porosity distribution function simultaneously. Their collision model consists of sticking and compaction; ZsD08 also added a simple fragmentation model. Although these models are more detailed, one can see that they still lack the full complexity observed in ``the zoo'' of laboratory collision experiments.

The MC-approach used in this study was previously presented by ZsD08. It can be characterized by two key properties: (1) the number of MC-particles (also referred to as representative particles) is kept constant; (2) the method follows the mass of the particle distribution.

Property (1) is required to ensure statistics. Because of the $\sqrt{N}$ noise of MC-methods, a large fluctuation in $N$ would severely affect the accuracy of the method (\cite{Ormel:2008p95}). The second property states that our primary interest lies in the particles that contain most of the mass of the system.  Moreover, it has been shown that following the particle's mass distribution -- rather than its number distribution -- is also a prerequisite to preserving a close correspondence with systems that experience strong growth (\cite{Ormel:2008p95}).

Property (2) ensures that the MC method samples only the parameter space where a signiÞcant portion of the total dust mass is. However, this is not always desirable. For instance, radiative transfer calculations require as input the surface area distribution of the aggregates, which determines the opacity. If most of the particle mass is contained in large particles (which are not observable), the number of small particles (which could contain most of the surface area and determines the IR appearance of the disk) might be resolved with poor statistics. But if we are interested in following the evolution of the dominant portion of the dust, then MC methods naturally focus on these parts of phase space. 

A necessary condition for the ZsD08 method to work is that the number of the representative particles $N$ is much less than the number of actual aggregates present in the system being considered -- a condition that is safely met in any of our simulation runs. A representative particle will then collide only with the non-representative particles, whose distribution is assumed to be identical to that of the representative particles. We refer to ZsD08 for details of the precise implementation and accuracy of the method; here we concentrate further on how the method operates in the new collisional setup.\\

The collision kernel is defined as the product of the cross-section of the colliding particles and their relative velocity:
\begin{equation}
K_{i,k}=\sigma_{i,k} \Delta v_{i,k},
\label{eq:kernel}
\end{equation}
where the index $i$ corresponds to the representative particle and $k$ is the index of the non-representative particle. The kernel is proportional to the probability of a collision. The value of $K_{i,k}$ is calculated for every possible particle pair, and random numbers determine which of the collisions occur first and at which time interval.\\

The above properties and conditions specify the essence of the ZsD08 method: one of the two collision particles is a representative particle and, because of property (1), only one of the collisional products becomes the new representative particle. Because of property (2), the choice for the new representative particle is weighed by the mass of the collision products. A very helpful analogy here is that of the representative `atom', which is contained within the representative particle. The choice of new representative particle after the collision is then proportional to the probability of the representative `atom' ending up in the collision products. If, for instance, a collision leads to the production of an entire distribution of debris particles, the probability that a particular debris fragment becomes the new representative particle is proportional to the likelihood of this fragment containing the representative `atom'.

\subsection{Implementation of the collision types}
\label{sect:implement}

We describe the implementation of the collision model using the representative `atom' concept. We refer to Chapter \ref{chp:paper2} for details of the various collision types mentioned below. 

\paragraph{{\Sac}, {\Sb}, {\Sc}: }All three of these collision types cause sticking and increase the mass of the aggregate by that of the projectile, but the porosity changes in a different manner (see Chapter \ref{chp:paper2}). The new mass of the representative particle $i$ is then the sum of the original particle masses, $m_{i, \mathrm{new}}=m_i+m_k$, where $m_i$ is the mass of the representative particle and $m_k$ is the mass of the non-representative particle. 

\paragraph{\Sdc : } In the case of {\Sd}, a certain percentage of the mass of the projectile sticks to the target, while the leftover mass of the projectile fragments according to a power-law distribution (see Chapter \ref{chp:paper2}). 

There are two situations to consider:
\begin{enumerate} 
\item The representative `atom' is part of the target. The mass of the new aggregate will be the mass of the original aggregate plus the transferred mass from the non-representative particle ($m_{i, \mathrm{new}}=m_i+m_{\mathrm{trans}}$, where $m_{\mathrm{trans}}$ is the transferred mass calculated according to Chapter \ref{chp:paper2}).
\item The representative `atom' is part of the projectile. We again have two situations.
\begin{enumerate} 
\item The representative `atom' will be transferred to the non-representative particle. The mass of the new representative particle will be the mass of the non-representative particle plus the transferred material ($m_{i, \mathrm{new}}=m_k+m_{\mathrm{trans}}$). The probability of transferring (removing) the representative atom from the projectile is simply $P=m_{\mathrm{trans}}/m_i$, the ratio of the transferred mass to the mass of the projectile.
\item The representative `atom' remains in one of the fragments. The probability of this event is $P=(m_i-m_{\mathrm{trans}})/m_i$, the ratio of the fragmented mass to the original mass of the representative particle. As discussed in Chapter \ref{chp:paper2}, the fragments follow a power-law mass distribution. The distribution is defined by the maximum mass of the fragments, which is a function of the relative velocity and the total mass of the fragments. The total mass of the fragments is $m_i-m_{\mathrm{trans}}$. We randomly choose from the fragment distribution to determine the new mass of the representative particle (to find which of the fragments will contain the representative `atom'). 
\end{enumerate}
\end{enumerate}

\paragraph{{\Bac}: } In this process, particles collide and bounce. Bouncing itself does not change the mass of the particles, but it compactifies them according to Chapter \ref{chp:paper2}. As observed in laboratory experiments (\cite{Weidling2009-4}), there is a small probability ($P_{\mathrm{frag}}=10^{-4}$) that the bouncing particle will break apart. If this happens, we break the particle into two equal-mass pieces.

\paragraph{{\Bbc}: } From the implementation point of view, this is similar to {\Sd}. The recipe to define the new representative particle is as in {\Sd}. The difference is that the projectile does not fragment during the collision, and that the porosity changes differently (see Chapter \ref{chp:paper2}).  

\paragraph{{\Fac}: } Fragmentation only happens between similar-sized aggregates in the {\pp} and {\cc} regimes. The fragments follow a power-law mass distribution, where the maximum mass of the fragments is determined by the relative velocity of the particles and the total mass that goes into the fragments (Chapter \ref{chp:paper2}). We randomly choose from this distribution to determine the new mass of the representative particle.

\paragraph{{\Fbc}: } {\Fbc} happens between only different-sized particles. During the collision, the projectile ``kicks out" pieces from the target aggregate. These pieces follow a power-law distribution (see Chapter \ref{chp:paper2}). We have to consider two cases:
\begin{enumerate} 
\item The representative `atom' is in the target. Again, we have two possibilities.
\begin{enumerate} 
\item The representative `atom' remains in the target after the collision. The mass of the new particle will be $m_{i, \mathrm{new}}=m_i-m_{\mathrm{er}}$, where $m_{\mathrm{er}}$ is the eroded mass. The probability of this event is $P=(m_i-m_{\mathrm{er}})/m_i$, which is the ratio of the left-over mass (which does not erode) to the mass of the original particle.
\item The representative `atom' originates in the eroded particles. Since the eroded particles follow a power-law distribution, we randomly draw from this distribution to determine the new mass of the representative particle. The likelihood of this event is the ratio of the eroded mass to the original mass of the particle ($P=m_{\mathrm{er}}/m_i$). 
\end{enumerate} 
\item The representative `atom' is part of the small particle that caused the erosion. As the particles do not stick and the small particle does not fragment, the representative particle remains unaffected. 
\end{enumerate}

\paragraph{{\Fcc}: } The porous particle becomes destroyed by the compact one and transfers a certain amount of mass to the compact particle. {\Fc} only happens in the {\cp} regime. We again have two possibilities. 
\begin{enumerate} 
\item The representative `atom' is part of the compact particle. In this case, the representative `atom' cannot leave the particle. The new mass of the representative particle will be $m_{i, \mathrm{new}}=m_i+m_{\mathrm{trans}}$, the sum of the original mass plus the transferred mass. 
\item The representative `atom' was part of the porous aggregate. 
\begin{enumerate} 
\item The representative `atom' is part of the material that is transferred to the compact particle. In this case, the new mass of the particle will be that of the compact (non-representative) particle plus the transferred material ($m_{i, \mathrm{new}}=m_k+m_{\mathrm{trans}}$). The probability of this event is $P=m_{\mathrm{trans}}/m_i$. 
\item The representative `atom' is among the fragments. As before, the mass distribution follows a power-law and we draw randomly from this distribution to determine the new mass of the representative particle. The probability of this event is $P=(m_i-m_{\mathrm{trans}})/m_i$. 
\end{enumerate} 
\end{enumerate}

\subsection{Evolving the particle properties in time}
We summarize how the particle properties evolve in time using the aforementioned kernel. We begin with the size and porosity distribution of the particles at a given time, $t$. At $t=0$, we provide the initial size and porosity distribution (see Sect. \ref{sec:inicond}). Knowing these:
\begin{itemize}
\item We calculate both the cross-sections of all possible collision partners and their relative velocities using the equations described in Sect. \ref{sec:vrel}. Both sets of quantities are used to determine the collision rates between the particle pairs. 
\item By using random numbers, we identify from the collision rates the representative particle involved in the collision, the non-representative particle it collides with, and the time at which the collision takes place ($t+\Delta t$).
\item Knowing the masses (mass ratio) and porosities of the collision partners, we identify in which of the eight regimes the collision takes place (e.g., {\pP}, or {\pC}, etc.). 
\item Next, we identify which of the nine collision types materializes (Fig. \ref{coltypes}) using the relative velocity of the particles and the mass of the projectile (see Chapter \ref{chp:paper2}).
\item Based on the collision recipe described in Chapter \ref{chp:paper2} and Sect. \ref{sect:implement}, the new mass and new porosity of the representative particle is calculated and the new size and porosity distribution of the particles at time $t+\Delta t$ is obtained.
\item In the final step, we update the collision rates.
\end{itemize}

\subsection{Numerical issues}

As mentioned in ZsD08, a sufficiently high number of representative particles is needed to properly reproduce the physics of the collision kernel. We performed simulations with an increasing number of representative particles and found that for more than 200-300 particles the results of the simulations do not change significantly. For all simulations described in the following sections, 500 representative particles are used and we average the results of 20 simulations to decrease the numerical uncertainty of the code.\\

The required computational time strongly depends on the collision rate of the particles thus determined by the dust density and the relative velocity (the $\alpha$ turbulence parameter mostly) and the length of the simulation. On a 2.83 GHz CPU, performing the simulations on a single core, the CPU time varies between nine hours (the low density model with $\alpha = 10^{-5}$) and three days (the high density model with $\alpha=10^{-3}$). Both simulations modeled $10^6$ years of particle evolution. The high density simulation with $\alpha = 10^{-4}$, critical mass ratio of 100, and $t=10^7$ years of evolution takes twelve days to complete.

\section{Results}
\label{sec:res4}

\subsection{Initial conditions and setup of simulations}
\label{sec:inicond}
All simulations begin with silicate monomers of 1.5 $\mu$m diameter and 2 g cm$^{-3}$ material density (monodisperse size distribution). We simulate the dust evolution at the midplane of our disk models at a distance of 1 AU from the central star. The gas density is obtained from the disk models described in Sect. \ref{sec:disks}. We assume a typical 1:100 dust-to-gas ratio. We follow the history of each collision: the mass and porosity of the colliding particles, their relative velocity, the occurred collision type, and the new mass and porosity of the particles. In this way we are able to reconstruct the history of the dust evolution.

The parameters that we vary in this study are the gas density $\rho_g$ and the turbulence parameter $\alpha$. We also treat the critical mass ratio $r_m$ as a free parameter to explore its effect on the dust evolution.

We provide a detailed description of the low density model with $\alpha = 10^{-4}$ and critical mass ratio of 100 in Sect. \ref{sec:lowd}. We then compare this with the MMSN model and the high density model using the same turbulence parameter and the critical mass ratio (Sects. \ref{sec:mmsn} and \ref{sec:desch}). In Sects. \ref{sec:turb4} and \ref{sec:critmass}, we discuss the effects of changing the turbulence parameter and critical mass ratio by comparing those results with the two example runs.

\begin{figure*}
\centering
  \includegraphics[width=0.9\textwidth]{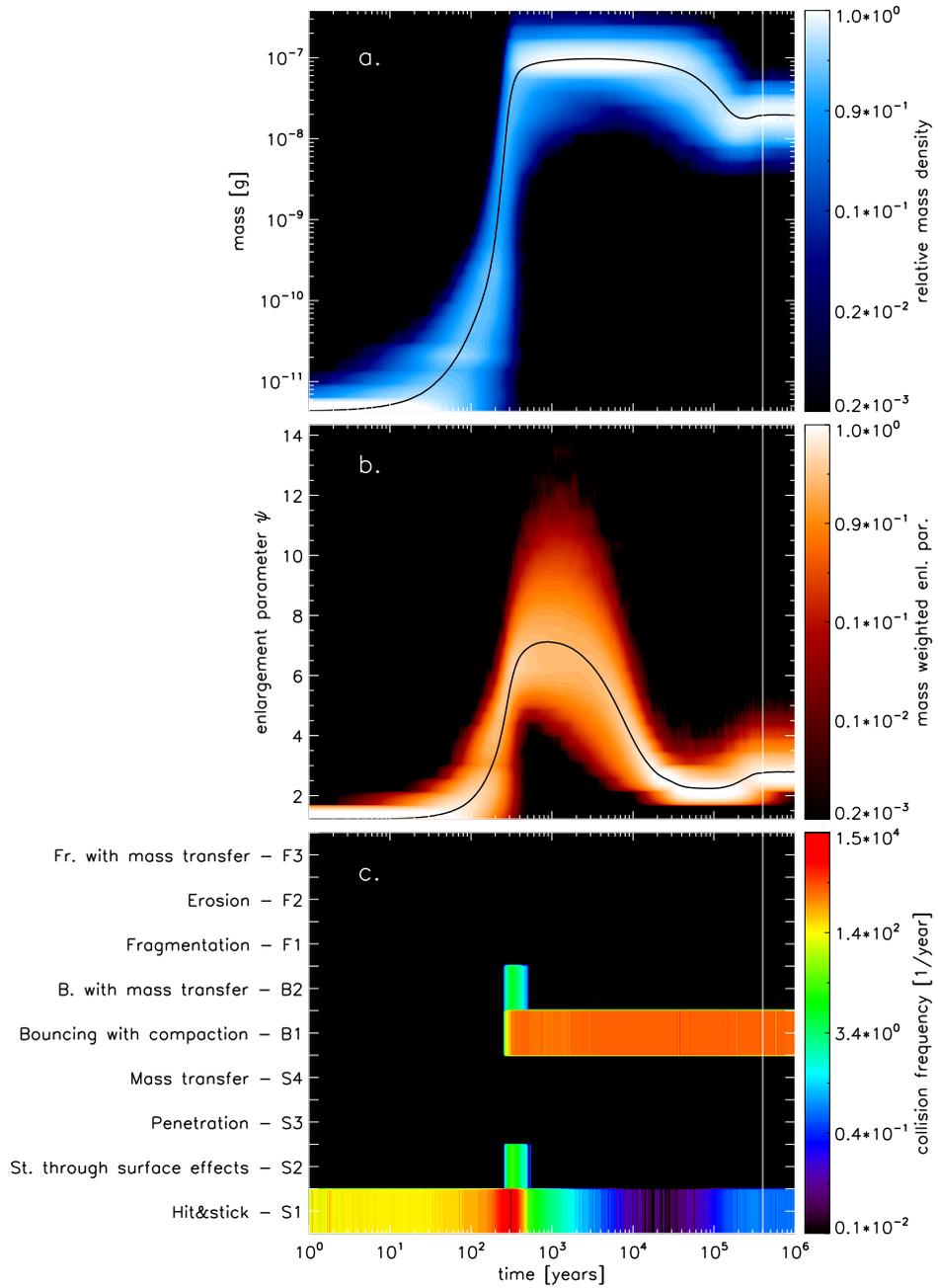}
  \caption{(a) The evolution of the mass distribution, (b) enlargement parameter distribution, and (c) the collision frequency of the nine different collision types in the low density model with $\alpha=10^{-4}$ and critical-mass ratio of 100. The x-axis shows the time. The y-axis of the (a) and (b) figures show the logarithmic mass and the linear enlargement parameter, respectively. The contours represent the normalized mass density and the mass weighted enlargement parameter. The black lines represents the average of the mass and enlargement parameter at a given time. The y-axis on the (c) figure represents the nine collision types. Each stripe shows the total collision rate of the collision types. Two distinct phases can be distinguished. During the initial 300 yr, particles grow by {\Sa}, after that the evolution is governed by {\Ba}. The white lines indicate how long our `local approach' assumption remains valid (discussed in Sect. \ref{sec:impl}).}
  \label{lowd_pics}
\end{figure*}

\begin{figure*}
\centering
  \includegraphics[width=0.88\textwidth]{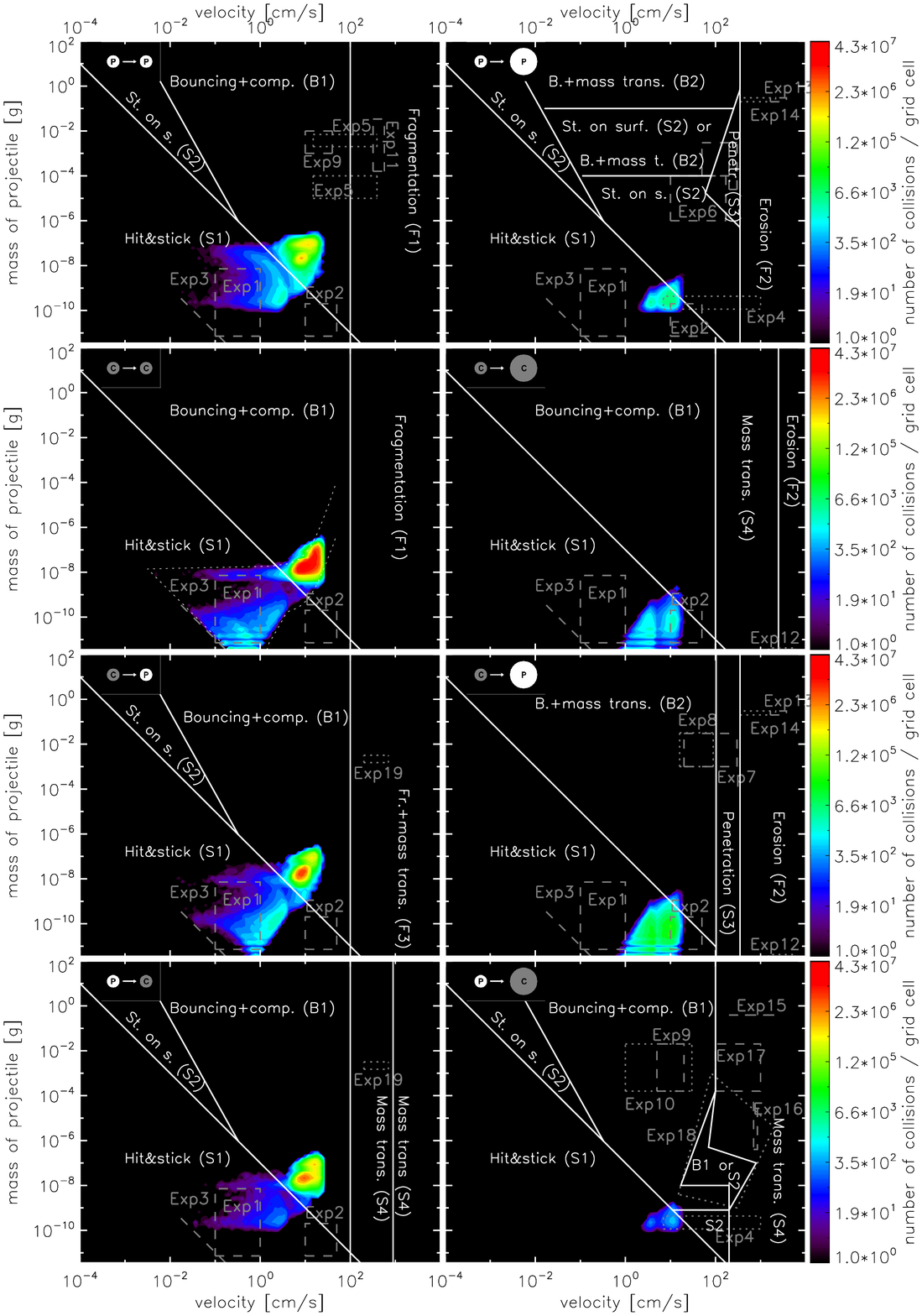}
  \caption{The collision history of the eight regimes in the low density model for $\alpha=10^{-4}$. The x-axis is the relative velocity, the y-axis shows the projectile mass. The different collision types, their border lines, as well as the areas covered with laboratory experiments (grey) are plotted. A relative velocity - mass grid is created and in these grid cells we calculate how many collisions happened until the `local approach' assumption is valid ($4\times 10^5$ yr). This is represented by the colors: yellow and red indicate a high collision frequency. The two dotted lines in the {\cc} regime are evolutionary tracks. Assuming a constant (40\%) volume-filling factor, the relative velocity between equal-sized particles (left curve) and particles with a mass ratio of 100 (right curve) can be calculated. The collisions in the simulation should occupy the parameter space between these two lines. The small deviations are due to the volume filling factor not being exactly 40\% during the simulation.}
  \label{lowd_cont}
\end{figure*}

\subsection{The low density model}
\label{sec:lowd}

In this disk model, the gas density at 1 AU is $2.4 \times 10^{-11}$ g cm$^{-3}$, the turbulence parameter is $\alpha=10^{-4}$, and the critical mass ratio is $r_{\mathrm{m}}=100$. As shown in Fig. \ref{relcont4}, the particles reach the fragmentation velocity (1 m s$^{-1}$) at sizes smaller than a millimeter because the particles in low gas density environment decouple from the gas at these small radii.

Figure \ref{lowd_pics} shows the evolution of the mass distribution (a), the porosity distribution (b), and the collision frequency of the various collision types (c). The x-axis shows the time on a logarithmic scale. The y-axis of Fig. \ref{lowd_pics}a, b shows the mass and enlargement distributions, respectively. The intensity of the color reflects the number density of representative particles, which, as explained in ZsD08, indicates the mass density of the distribution. Thus, in Fig. \ref{lowd_pics}a the intensity levels directly reflect the mass density, while in Fig. \ref{lowd_pics}b the colors indicate the mass-weighted enlargement parameter. The black lines show the average of these quantities over the particle distribution. The y-axis in Fig. \ref{lowd_pics}c represents the nine collision types used in this chapter. Every stripe shows the total collision rate of the collision types at a given time.

Figure \ref{lowd_cont} represents the collision history in the eight collision regimes. The x-axis is the velocity, the y-axis shows the mass of the projectile. A mass-velocity grid is created and we calculate how many collisions occurred within each grid cell during which our `local approach' assumption is correct. The different collision types and their border lines, as well as the areas covered by the laboratory experiments (indicated by grey colors) are plotted. For more details on the experiments, we refer to Chapter \ref{chp:paper2}. 

In the {\cc} panel, we indicate two curves with dotted lines. These curves are evolution tracks. The left curve is obtained by calculating the relative velocity between equal-sized particles with an enlargement parameter of 2.5 (volume-filling factor of 40\%). The right curve represents the relative velocity between particles with a mass ratio of 100. These two curves serve as a guide to our results, as collisions should occur between these two curves in the {\cc} panel. The lower part of the left curve, where the relative velocity decreases with increasing mass, is a sign that relative velocities between equal-sized particles are dominated by Brownian motion. For higher masses, the relative velocity is dominated by turbulence. These curves do not precisely match the contours because we assumed a constant enlargement parameter of 40\% when calculating the evolution tracks, whereas $\Psi$ is a free parameter in the simulation.


\subsubsection{Early evolution}
We discuss the evolution of the distribution functions until the `local approach' assumption becomes invalid ($4\times 10^5$ yr). The long-term evolution of the dust is discussed in Section \ref{subsec:longt}.

We distinguish between two distinct phases. During the first 300 yr, particles grow by the means of the {\Sa} mechanism. The second phase is dominated by {\Ba}; the particles leave the S1 regimes. During this phase, the mass of the particles slowly decreases and the enlargement parameter asymptotically reaches a minimum value of 2.23. As discussed in Chapter \ref{chp:paper2}, keeping the bouncing velocity of a particle constant, the porosity of the aggregate will asymptotically reach a maximum value, $\phi_{\mathrm{max}}$ (see Chapter \ref{chp:paper2}). The relative velocity of a particle is a function of the friction time (Eq. \ref{eq:ts1-4}), which depends on the ratio of the mass to surface area, $m/A$. Since particle growth is halted at this point in the simulation ($m$ remain constant), only a decrease in $A$ caused by compaction can further increase the velocity between particles. The particle radius can decrease until either $\phi_{\mathrm{max}}$ for the given relative velocity is reached or particles reach the maximum compaction possible. The latter limit, random close packing (RCP), corresponds to an enlargement parameter of 1.6 (volume-filling factor of $\sim$60\%).

We find that fragmentation does not play a role during the evolution of these particles indicated by Fig. \ref{lowd_pics}c. As can be seen in Fig. \ref{lowd_cont}, their evolution is halted by bouncing before the particles are able to reach the fragmentation barrier. The two dominant collision types are {\Sa} and {\Ba}.\\

\subsubsection{Termination of growth}
As we can see from Fig. \ref{lowd_cont}, sticking at higher energies than the {\Sa} limit is possible only inside the {\pP} regime. As soon as we no longer have collisions inside this regime or the S1 regimes, the growth is halted. There can be two reasons why this occurs 1.) All particles become compact, i.e., there are simply no collisions in the {\pP} regime. 2.) The width of the particle mass distribution is smaller than the critical mass ratio ($r_m$), such that all collisions take place in the equal-size regimes (e.g., {\pp}, {\pc}).

In the case of the current simulation, the small particles have been `consumed'. Once the heavy particles grow into the {\Ba} area of the {\pp} regime, their growth in the {\pp} regime stops. The heavy particles collect the small ones via collisions in the {\pP} regime and by doing so, the width of the distribution is reduced to a value smaller than $r_m$. Therefore, before particles are able reach the fragmentation barrier, growth is halted. Because of B1, particles become compact and collisions in the {\cc}, {\cp}, and {\pc} regimes appear.

\subsubsection{Long-term evolution}
\label{subsec:longt}
Before discussing the long-term evolution of the distribution functions, we must consider how long our initial assumptions (`local approach' and constant gas density) hold true.

Using Eq. \ref{eq:drift}, we calculate that a particle with Stokes number $10^{-4}$ drifts a distance of 1 AU in roughly $4\times 10^5$ yr. This is the drift timescale beyond which the `local approach' assumption (discussed in Sect. \ref{sec:impl}) is no longer valid and particles become separated from each other on this timescale.

Another process by means of which particles separate is viscous spreading. We assume that the viscous timescale of the disk at 1 AU is given by:
\begin{equation}
t_{\mathrm{vis}}=r^2/ \nu_T,
\end{equation}
where $r$ is the distance from the central star (1 AU), $\nu_T$ is defined in Eq. \ref{eq:nuT4}. The viscous timescale in our model, using $\alpha = 10^{-4}$, is of the order of $10^6$ yr. 

One has to consider the results of the simulation with caution for longer times than the drift or viscous timescales. The equilibrium or final state of the particles is reached when mass decrease during {\Ba} and {\Bb}, and mass increase by {\Sa} and {\Sb} are in equilibrium. In other words, the final state is reached when the evolution of the average mass and the enlargement parameter can only be determined by the stochastic fluctuations of the simulation. We find that the equilibrium state of the particles is hardly reached within these timescales. Upon neglecting these warnings, we find that the equilibrium state of the dust is reached at $t=4 \times 10^5$ yr. The equilibrium is reached between the bouncing collisions, resulting in breakage and {\Sa} (see Fig. \ref{lowd_pics}c). The equilibrium average mass and porosity of the particles are $ \bar{m}_{\mathrm{fin}}=2\times 10^{-8} \mbox{ g}$, and $\bar{\Psi}_{\mathrm{fin}}=2.77$.\\

To be able to compare the distribution functions of different runs, we define some quantities using the mean of the distribution functions indicated with black lines in Figs. \ref{lowd_pics}a and b: $\max (\bar{m})$, the maximum of the mean mass; $\max (\bar{\Psi})$, the maximum of the mean enlargement parameter; $\Psi _{min}$, the minimum mean enlargement parameter when particles can no longer be compacted anymore; $t_{\mathrm{noc}}$, the time when $\Psi _{min}$ is reached, which is when  the time derivative of $\bar{\Psi}$ becomes zero ($d\bar{\Psi}/dt = 0$); and $\max(\bar{St})$, the maximum average Stokes number reached during the simulation. The values of these quantities are listed in Table \ref{tab:res} (model id `Lt1d-4m100'). In this table, Col. 1 describes the model names, `L' representing the low density model, `M' being the MMSN model, `H' being the high density model, the letter `t' and the following number indicates the value of the turbulence parameter, and the letter `m', and the number providing the used critical mass ratio values. Columns 2, 3, and 4 show the gas density, turbulence parameter, and the critical mass ratio respectively. Columns 5, 6, 7, 8, and 9 list the parameters defined to characterize the distribution functions. These are $\max (\bar{m})$ in Col. 5, $\max (\bar{\Psi})$ in Col. 6, $\Psi _{min}$ in Col. 7, $t_{\mathrm{noc}}$ in Col. 8, and finally $\max(\bar{St})$ in Col. 9. 

\subsection{The MMSN model}
\label{sec:mmsn}

In the MMSN model, the gas density at 1 AU at the midplane is $1.4\times 10^{-9}$ g cm$^{-3}$, $\alpha = 10^{-4}$, and the critical mass ratio is 100. As shown in Fig. \ref{relcont4}, the particles become larger than in the low density model, because they are more tightly coupled to the gas and the relative velocities are suppressed. As in the previous section, we first discuss the evolution of the distribution functions for as long as the `local approach' assumption holds true ($6\times 10^5$ yr in this model).

Figure \ref{mmsn_pics} again shows the time evolution of the mass (a), the enlargement parameter (b), and the collision frequency (c). Figure \ref{mmsn_cont} shows the collision history. These figures depict a rather different evolution than that of the previous model.

\subsubsection{Early evolution}
\label{subsec:mmsn_early}
\begin{figure*}
\centering
  \includegraphics[width=0.9\textwidth]{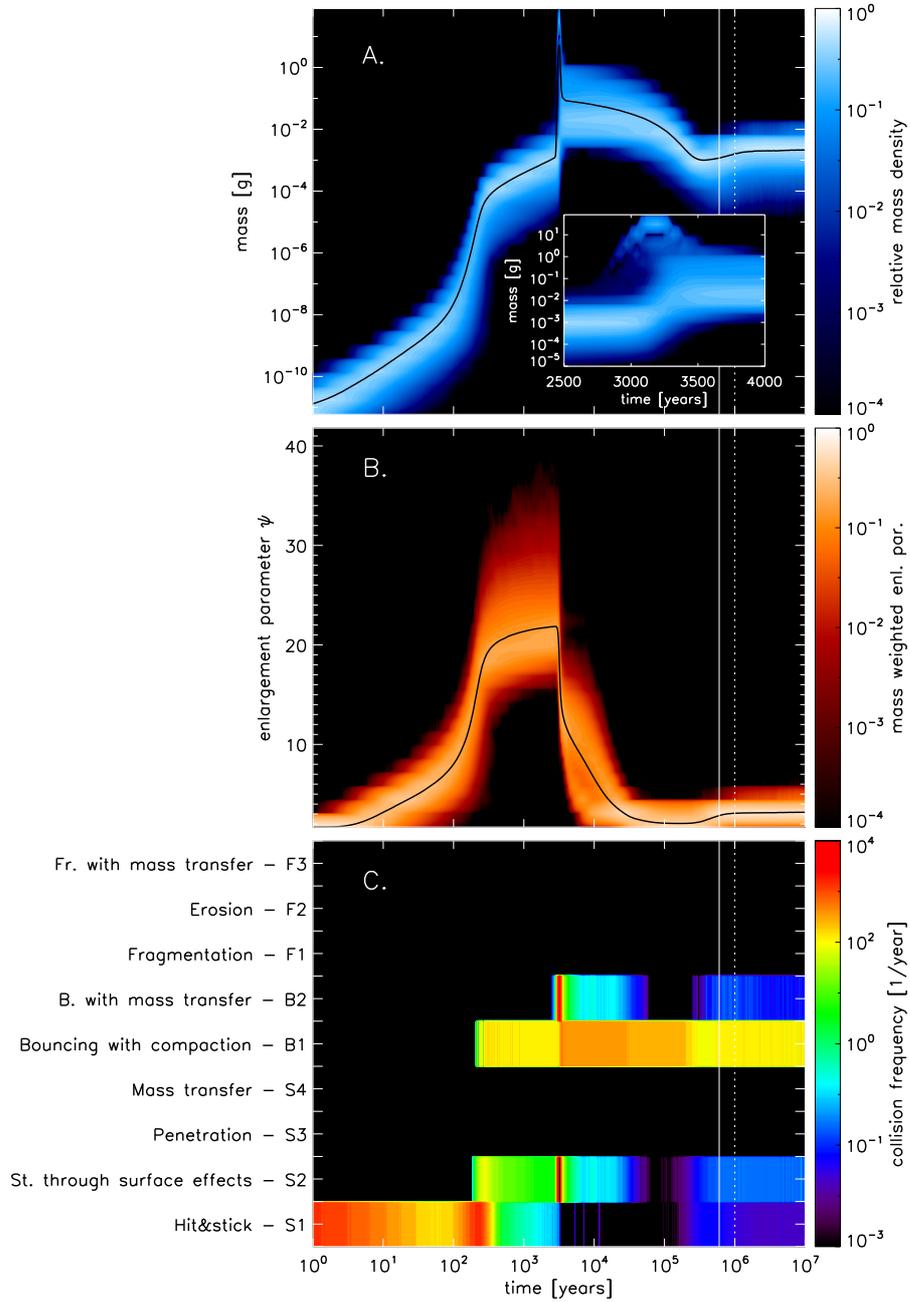}
  \caption{Same as Fig. \ref{lowd_pics} but for the MMSN model. We magnify the spike of the mass distribution at $\sim 2.5\times 10^{3}$ yr in Fig. (a). Four phases can be distinguished here. Initially (first 300 yr) particles grow purely by {\Sa}. After this, the growth slows down because Bouncing with compation (B1) starts and all particles leave the {\Sa} regime. Between $3\times 10^3$ and $10^4$ yr, particles enter the transition regime between {\Sb} and {\Bb} on the {\pP} regime. Some particles reach masses of 1 g, but their masses are lowered rapidly by B2. The last phase is dominated by {\Ba}. The solid/dotted white lines indicate how long our `local approach' assumptions remains valid (discussed in Sect. \ref{sec:impl}).}
  \label{mmsn_pics}
\end{figure*}

\begin{figure*}
\centering
  \includegraphics[width=0.9\textwidth]{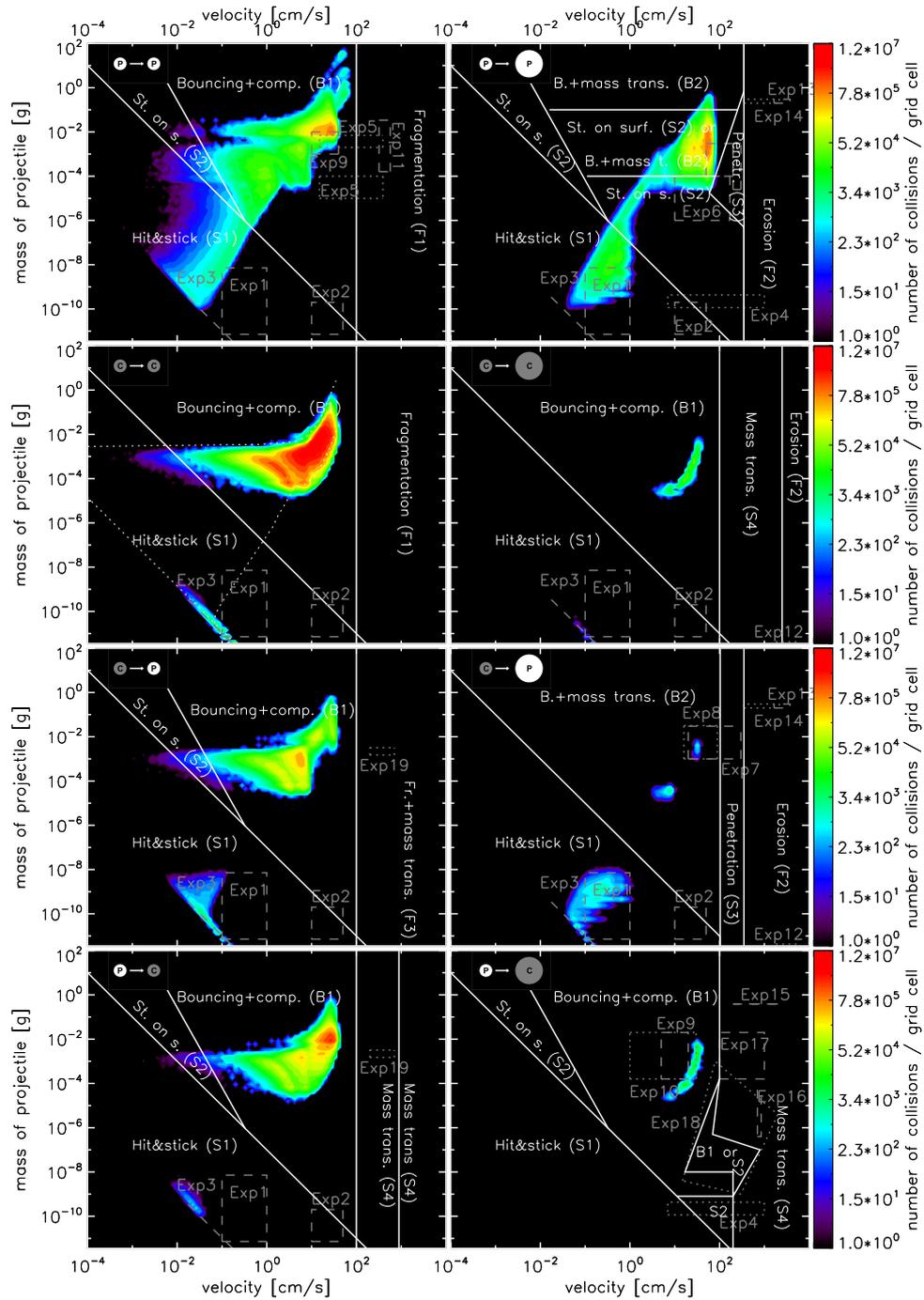}
  \caption{Same as Fig. \ref{lowd_cont} but for the MMSN model. The particles are more tightly coupled to the gas due to the higher gas density. Therefore, they grow to larger sizes than in the low density model.}
  \label{mmsn_cont}
\end{figure*}

During the fractal growth regime, we find that the collision rate of {\Sa} is much higher than in the low density model (Fig. \ref{mmsn_pics}c). This is because of the higher dust densities. We can see from Fig. \ref{mmsn_cont}, for the {\cc} regime, that growth begins with Brownian motion because the relative velocity decreases with increasing particle mass for particle masses less than $10^{-9}$ g. As a result of these low velocity collisions, some particles reach enlargement parameter values of higher than 30 (volume-filling factors less than of 3.3\%). At 200 yr, some particles grow above the border line of {\Sa} and enter the area of {\Sb} in the {\pP} plot, and {\Ba} in the {\pp} plot. Growth caused by both S1 and S2 continues until different-sized particles enter the transition regime in the {\pP} plot. One can see in Fig. \ref{mmsn_pics}a, that some particles reach 1 g in mass. However, when particle collisions enter the transition regime between {\Bb}, and {\Sb} in the {\pP} plot, their masses are equalized because of the mass transfer that occurs during the B2 collisions and the collisions shift to the similar-sized regime (B1). After roughly $10^4$ yr, we find that particles mostly bounce and become compact. The enlargement parameter reaches a minimum value of 1.85 (54\% volume filling factor), the mass distribution function slowly decreases because of the low probability of breakage. Collisions at this point occur mainly in the {\cc} regime. \\

\begin{figure}
\centering
  \includegraphics[width=0.7\textwidth]{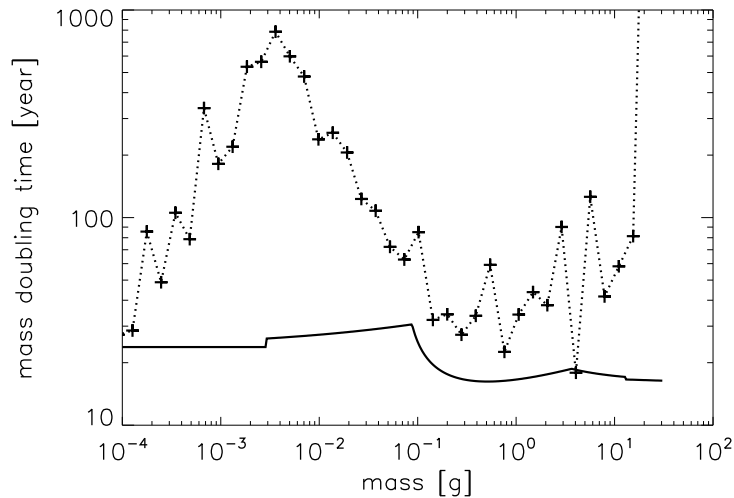}
  \caption{The dotted line and the `+' signs represent the growth timescale of the heaviest particle in the MMSN simulation with $\alpha=10^{-4}$ and $r_m = 100$. As a comparison, we show the minimum growth timescale a particle can have in this simulation (solid line).}
  \label{fig:acc_mmsn}
\end{figure}

A peculiar feature of Fig. \ref{mmsn_pics}a is a peak at $t=2.5\times 10^3$ yr, which is accompanied by a fast decrease in the enlargement parameter in Fig. \ref{mmsn_pics}b and an increased collision rate of {\Sb} and {\Bb} in Fig. \ref{mmsn_pics}c. At this point, the relative velocity produced by turbulence increases. As discussed in Sect. \ref{sec:vrel}, particles leave the `tightly coupled particle' regime and enter the `intermediate particle' regime (see the relative velocity bump in Fig. \ref{relcont4}b). We calculate the growth timescale of the heaviest particle with mass $M$ in the simulation to be:
\begin{equation}
t_{\mathrm{gr}}=\left( \frac{1}{M}\frac{dM}{dt} \right)^{-1},
\end{equation}
which is illustrated with a dotted line in Fig. \ref{fig:acc_mmsn}. As a comparison, we also calculate the minimum growth timescale that a particle can have (solid line), which
\begin{equation}
t_{\mathrm{max}}= \left( \frac{M}{\rho_d \Delta v \sigma_M}\right)^{-1},
\end{equation}
where $\sigma_M$ is the cross-section of the largest particle. Here, we assume that the `swept up' particles have masses of $M/100$, therefore we use the relative velocity curve presented in Fig. \ref{relcont4}b, dotted line. The effects of both the relative velocity bump and the increased growth rate can be seen at 0.1 g. 

The relative velocity `boost' happens shortly after the particles enter the transition regime of S2 and B2 in the {\pP} plot. The heaviest particle, which experiences the velocity transition the earliest, acquires higher relative velocities, leading to a higher rate of collision with the other particles. As the particles are initially in the lower part of the S2-B2 transition regime (with masses of $10^{-3}$ g, see Fig. \ref{mmsn_pics}a), the heaviest particle experiences rapid growth reaching masses of 30 g. The simulated timescale, however, does not reach the minimum growth timescale because of the {\Bb} collisions, which reduce the mass of the heaviest particle. The remainder of the particle population increase in mass because of B2, and the growth rate of the heaviest particle decreases. Eventually, the rapid growth of the heaviest particle is halted, and the growth timescale at $m=30$ g becomes infinity. From this point on, the heaviest particle reduced in mass, and B2 equalizes the masses of the particles.\\

\subsubsection{Long-term evolution}
We calculate the drift and viscous timescales to determine how long our assumptions of `local approach' and constant gas density remain valid. Assuming Stokes number $10^{-4}$ particles, we find that the drift timescale is of the order of $6\times 10^5$ yr, and the viscous timescale is $10^6$ yr. These timescales are indicated by solid and dotted white lines in Fig. \ref{mmsn_pics}.

We find that the final equilibrium is reached at $t=2\times 10^6$ yr, which is longer than the drift and viscous timescales. The equilibrium is reached between the growth mechanisms of {\Sa}, {\Sb} and the destruction mechanisms of bouncing resulting in breakage and {\Bb}. The final average mass and porosity of the particles are $\bar{m}_{\mathrm{fin}}=2\times 10^{-3} \mbox{ g}$, $\bar{\Psi}_{\mathrm{fin}}=3.3$.\\

We conclude that the dust evolution is more complex in the MMSN model than in the low density model because the complex interaction of the velocity field and the collision kernel is apparent in this model. As in the previous model, {\Ba} is the most frequent collision type and {\Sa} determines the initial particle growth, but both {\Sb} and {\Bb} are of importance in this model. The final equilibrium is not reached within the drift and viscous timescales. 

\subsection{The high density model}
\label{sec:desch}
The gas density in this model is $2.7\times 10^{-8}$ g cm$^{-3}$ at the midplane of the disk at a distance of 1 AU from the central star. The values of $\alpha$, $r_m$, and the dust-to-gas ratio are the same as in the previous models.

Figure \ref{relcont4}, dashed line, shows the relative velocity field of fluffy aggregates in this model. As already discussed in Sect. \ref{sec:vrel}, the aggregates reach 1 m s$^{-1}$ relative velocities at similar masses as the MMSN model, because of the Stokes drag. Therefore, we expect that the final aggregate sizes and masses will be similar to the particles produced in the MMSN model.

\begin{figure*}
\centering
  \includegraphics[width=0.9\textwidth]{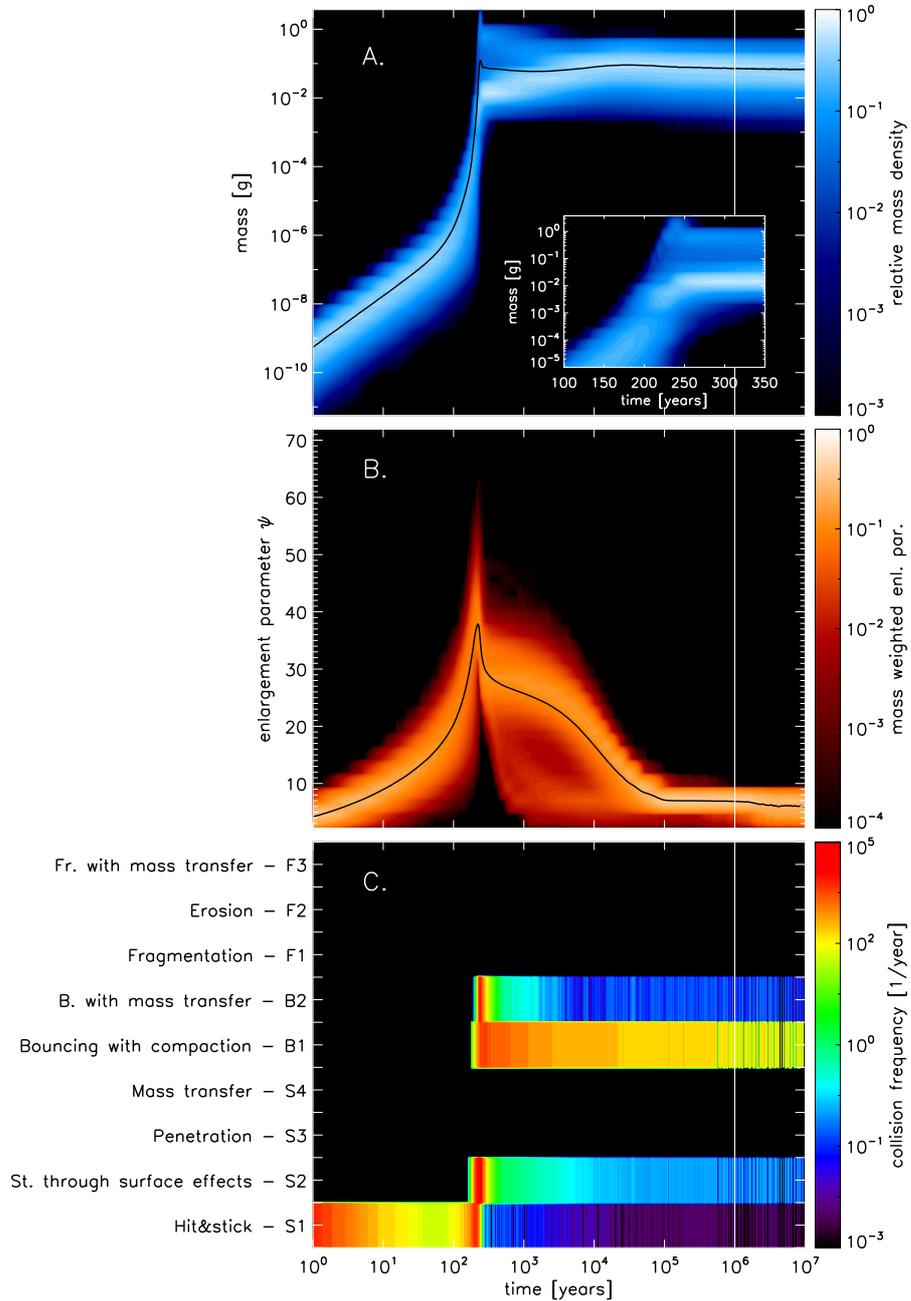}
  \caption{Same as Fig. \ref{lowd_pics} but for the high density model. As in Fig. \ref{mmsn_pics}a, we zoom in on the peak at the mass distribution. The solid white line indicate how long our `local approach' assumptions remains valid at $t=10^6$ yr (discussed in Sect. \ref{sec:impl}).}
  \label{desch_pics}
\end{figure*}

\begin{figure*}
\centering
  \includegraphics[width=0.9\textwidth]{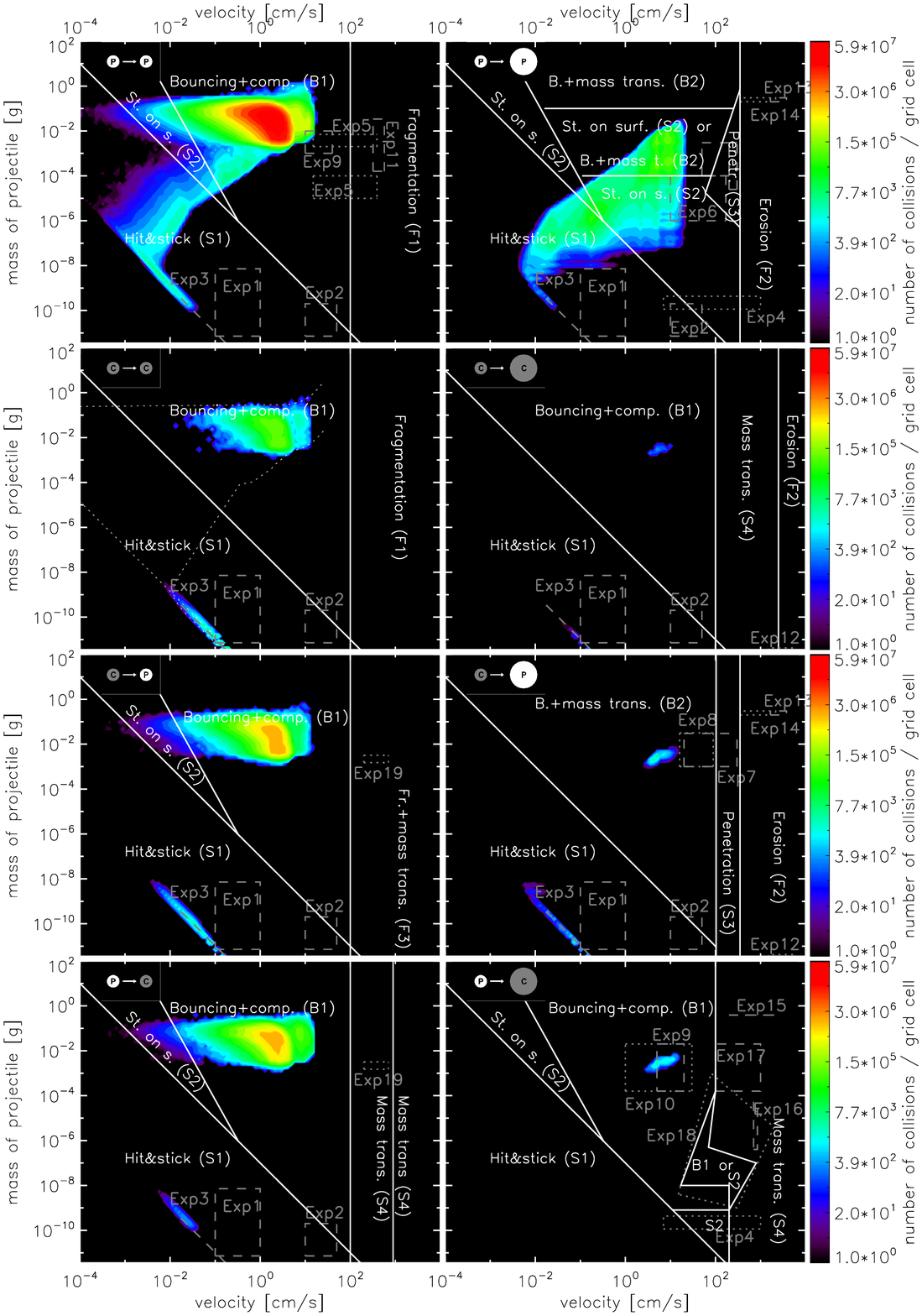}
  \caption{Same as Fig. \ref{lowd_cont} but for the high gas density model.}
  \label{desch_cont}
\end{figure*}

Figure \ref{desch_pics} shows the time evolution of the mass (a), enlargement parameter (b), and the collision frequency (c). Figure \ref{desch_cont} illustrates the collision history. 

\subsubsection{Early evolution}
As seen in Fig. \ref{desch_cont}, Brownian motion is the dominant source of relative velocity, as long as particles have masses of lower than $10^{-8}$ g (that is an order of magnitude higher than in the MMSN model). Therefore, the enlargement parameter of the aggregates is also higher than in the MMSN model. As the {\Sa} collisions are more frequent than in the MMSN model because of the higher dust densities, the particles reach the transition regime between {\Sb} -- {\Bb} earlier, at $t=200$ yr. The peak in the mass distribution is not as pronounced as in the MMSN model. The relative velocity boost occurs at heavier aggregate masses ($10^{-2}$ g, see Fig. \ref{relcont4}b) because of the higher gas density of the model. When the rapid growth of the heaviest particle begins, most of the projectiles are already in the transition regime. Here, the B2 collisions soon reduce the mass of the heaviest particle and narrow the mass distribution. 

In contrast to the MMSN model, the mass of the particles is not reduced because of the low probability of breakage in {\Ba}, but is kept nearly constant in time. This is the result of the increased collision rate of {\Sb}. The S2 collision rate is increased because of the low velocity collisions, which occur when particles are in the tightly coupled regime and have similar stopping times. These S2 collisions occur in the {\pp} regime as seen in Fig. \ref{desch_cont}. These collisions cancel out the effect of breakage in B1. 

The maximum Stokes number reached in this model is $3.6\times 10^{-5}$ (see Table \ref{tab:res}, model id `Ht1d-4m100'), which is lower than in the MMSN model. The growth in this model is halted by the {\Bb} collisions in the transition regime of the {\pP} panel. This shows us that particles cannot reach masses much higher than 1 g independently from the exact value of the gas density, because at this point, particles enter the S2-B2 transition regime and the growth is halted. Increasing the gas density yet further would lead to even lower Stokes numbers. 

\subsubsection{Long-term evolution} The drift and the viscous timescales in the high density model are both $10^6$ yr. As seen in Fig. \ref{desch_pics}a, the particle masses do not change significantly after $t=10^3$ yr. The porosity is reduced by {\Ba} and reaches a final value of 5.41 at $t=3\times 10^6$ yr.

\subsection{Varying the turbulence parameter}
\label{sec:turb4}

\begin{table*}
\caption{Overview and results of all the simulations.}
\label{table:lowd}      
\begin{center}   
\small
\begin{tabular}{l l l l l l l l l }        
\hline\hline                 
Model& $\rho_g$ & $\alpha$ & $r_{\mathrm{m}}$	&$\max (\bar{m})$	&$\max (\bar{\Psi})$	&$\Psi _{\mathrm{min}}$	&$t_{\mathrm{noc}}$ & $\max (\bar{St})$\\    
	& [g cm$^{-3}$]& & &[g]	&	&	&[yr]	&\\
(1)	&(2)&(3)&(4)&(5)&(6)&(7)&(8)&(9)\\
\hline                        
Lt1d-3m100	&$2.4 \times 10^{-11}$	&$10^{-3}$	&100	&$8\times 10^{-8}$		&7.27	&1.77	&$2\times 10^{4}$	&$2.5\times 10^{-4}$	\\
Lt1d-4m100	&$2.4 \times 10^{-11}$	&$10^{-4}$	&100	&$9.7 \times 10^{-8}$	&7.12	&2.23	&$8\times 10^{4}$	&$2.2 \times 10^{-4}$	\\
Lt1d-5m100	&$2.4 \times 10^{-11}$	&$10^{-5}$	&100	&$2.66 \times 10^{-7}$	&7.72	&3.78	&$3 \times 10^{5}$	&$2.1 \times 10^{-4}$\\
\hline  
Mt1d-3m100	&$1.4\times 10^{-9}$	&$10^{-3}$	&100	&8.13				&24.41	&3.88	&$10^{4}$		&$5.1\times 10^{-4}$\\		
Mt1d-4m100	&$1.4\times 10^{-9}$	&$10^{-4}$	&100	&4.18				&21.9	&1.85	&$2\times 10^5$	&$2.8 \times 10^{-4}$\\		
Mt1d-5m100	&$1.4\times 10^{-9}$	&$10^{-5}$	&100	&$7.7\times 10^{-2}$	&30.0	&4.13	&$7\times 10^{5}$	&$2.1\times 10^{-4}$\\
\hline
Ht1d-3m100	&$2.7\times 10^{-8}$	&$10^{-3}$	&100	&3.77				&34.1	&5.61	&$10^5$			&$1.4 \times 10^{-4}$\\		
Ht1d-4m100	&$2.7\times 10^{-8}$	&$10^{-4}$	&100	&0.23				&38.0	&5.41	&$3\times 10^6$	&$3.6\times 10^{-5}$\\		
Ht1d-5m100	&$2.7\times 10^{-8}$	&$10^{-5}$	&100	&0.28				&43.9	&4.94	&$4\times 10^6$	&$7.7\times 10^{-5}$\\
\hline
Lt1d-4m10	&$2.4 \times 10^{-11}$	&$10^{-4}$	&10		&$9.2\times 10^{-4}$	&5.88	&2.28	&$10^5$			&$3.8 \times 10^{-3}$	\\
Lt1d-4m100	&$2.4 \times 10^{-11}$	&$10^{-4}$	&100	&$9.7 \times 10^{-8}$	&7.12	&2.23	&$8\times 10^{4}$	&$2.2 \times 10^{-4}$	\\
Lt1d-4m1000	&$2.4 \times 10^{-11}$	&$10^{-4}$	&1000	&$9.7 \times 10^{-8}$	&7.09	&2.29	&$8\times 10^{4}$	&$2.2 \times 10^{-4}$	\\
\hline
Mt1d-4m10	&$1.4\times 10^{-9}$	&$10^{-4}$	&10		&$2.5\times 10^{-2}$	&19.4	&2.1		&$2\times 10^{5}$	&$2.2 \times 10^{-4}$\\
Mt1d-4m100	&$1.4\times 10^{-9}$	&$10^{-4}$	&100	&4.18				&21.9	&1.85	&$2\times 10^5$	&$2.8 \times 10^{-4}$\\		
Mt1d-4m1000	&$1.4\times 10^{-9}$	&$10^{-4}$	&1000	&$9.5 \times 10^{-3}$	&23.1	&2.9		&$2\times 10^{5}$	&$1.3\times 10^{-4}$\\
\hline  
Ht1d-4m10	&$2.7\times 10^{-8}$	&$10^{-4}$	&10		&0.15				&34.6	&2.46	&$2\times 10^6$	&$4.5\times 10^{-5}$\\		
Ht1d-4m100	&$2.7\times 10^{-8}$	&$10^{-4}$	&100	&0.23				&38.0	&5.41	&$3\times 10^6$	&$3.6\times 10^{-5}$\\		
Ht1d-4m1000	&$2.7\times 10^{-8}$	&$10^{-4}$	&1000	&$8.8 \times 10^{-2}$	&40.0	&7.1		&$10^5$			&$3.5\times 10^{-5}$\\
\hline
\label{tab:res}
\end{tabular}
\end{center}
In this table, Col. 1 describes the model names. `L' stands for the low density model, `M' is the MMSN model, `H' is the high density model, the letter `t' and the following number indicates the value of the turbulence parameter, the letter `m' and the number shows the used critical mass ratio values. Columns 2, 3, and 4 shows the gas density, turbulence parameter, and the critical mass ratio respectively. Columns 5, 6, 7, 8, and 9 list the parameters defined to characterize the distribution functions. These are the average maximum mass in Col. 5, the average maximum enlargement parameter in Col. 6, the minimum enlargement parameter in Col. 7, the end of the compaction phase in Col. 8, and the average maximum Stokes number in Col. 9. 
\end{table*}

\begin{figure*}
\centering
  \includegraphics[width=1\textwidth]{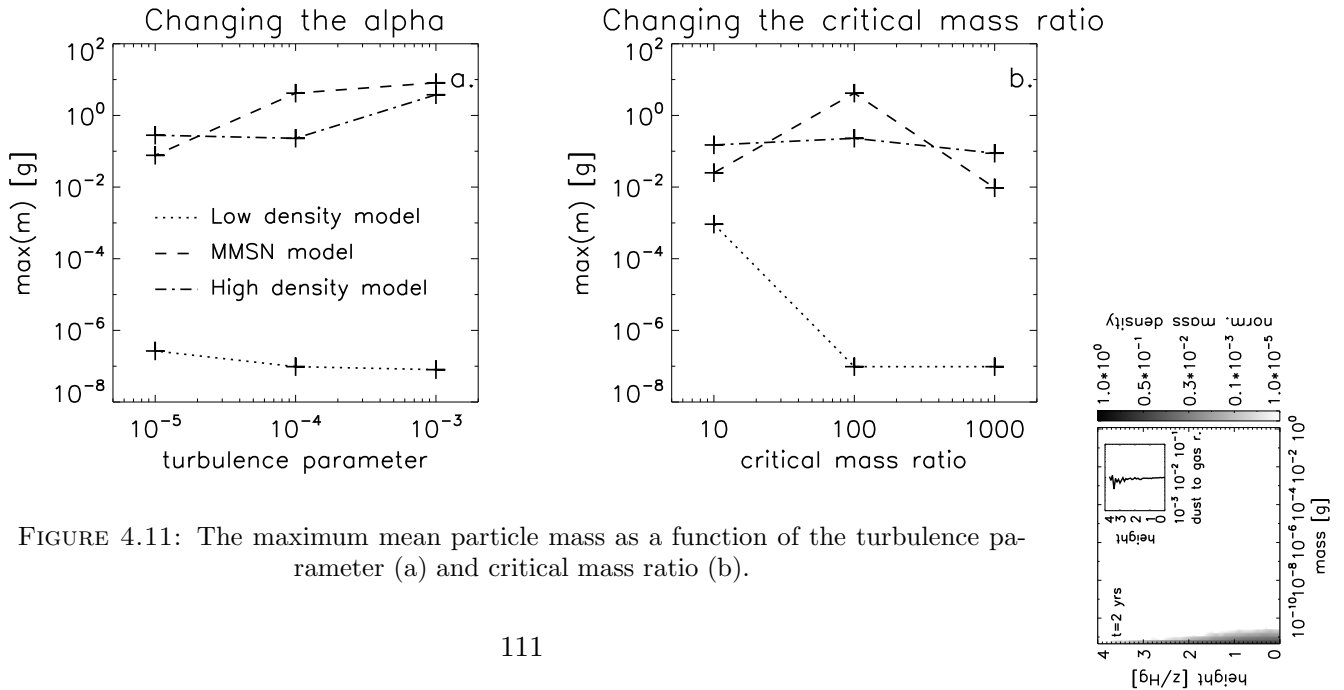}
  \caption{The maximum mean particle mass as a function of the turbulence parameter (a) and critical mass ratio (b).}
  \label{fig:res}
\end{figure*}

To explore the effects of turbulence, we perform two more simulations in each of the disk models. We keep the critical mass ratio fixed (100) and vary only the turbulence parameter ($\alpha$) to obtain values of $10^{-3}$, $10^{-4}$, and $10^{-5}$. The results are shown in Table \ref{tab:res}, for the first nine models, and in Fig. \ref{fig:res}a. 

The work of \cite{Brauer2008a-4} suggests that in situations where fragmentation limits the growth, a lower turbulence strength produces larger aggregates. This, of course, directly reflects the shift in the fragmentation threshold (1 m/s) to larger sizes when $\alpha$ is lower (Fig. \ref{relcont4}). In this study, it is fragmentation that balances the growth, producing a (quasi) steady-state. For the low density models, we do see a decrease in the final particle mass, but it is balanced by bouncing and not fragmentation. In the low density model, particles grow only in the {\Sa} regimes. When particles leave these regimes, the growth stops due to bouncing. The border of the S1 regime corresponds to the collision energy that is lower than $5\times E_{\mathrm{roll}}$, where $E_{\mathrm{roll}}$ is the rolling energy of monomers (see Chapter \ref{chp:paper2}). As the collision energy is $E_{\mathrm{coll}} = 1/2 \mu (\Delta v)^2$, particles in strong turbulence leave the S1 regimes at lower particle masses.

On the other hand, the MMSN and high density models show that the maximum mass of the particles can even increase with $\alpha$. The precise value of the $\max (\bar{m})$ is determined by the intensity of the peak in the mass-density plots (Sect. \ref{subsec:mmsn_early}) and this may vary between simulations. In the `Mt1d-4m100' model, we have argued that the spike is exceptionally pronounced because of the high probability of {\Sb} collisions at the initial part of the rapid growth. However, the main point is that in the MMSN/high-density simulations the maximum particle masses all end up around 1 g, independent of the turbulent strength because of the nature of the S2-B2 transition, which occurs at projectile masses of $10^{-4}$ g in the {\pP} plot. As explained before, collisions in the {\pP} plot are the only way by which particles can grow after the {\Sa} phase is finished. Thus, we require a broad distribution with a high growth rate. However, B2 collisions operates in the opposite way: they transfer mass from the target to the projectile, narrowing the distribution and decreasing the overall probability of the {\pP} process occuring. Thus, once B2 becomes effective, there is a shift from the {\pP} panel to the {\pp} panel. For the MMSN/high-density models this behavior is always present and the important quantities involved (i.e., relative probability of B2 over S2) scale with mass but not with velocity. The result is that the maximum masses that particles reach are $\sim$ 1 g, which is rather insensitive to the strength of the turbulence.

\begin{figure*}
\centering
  \includegraphics[width=1\textwidth]{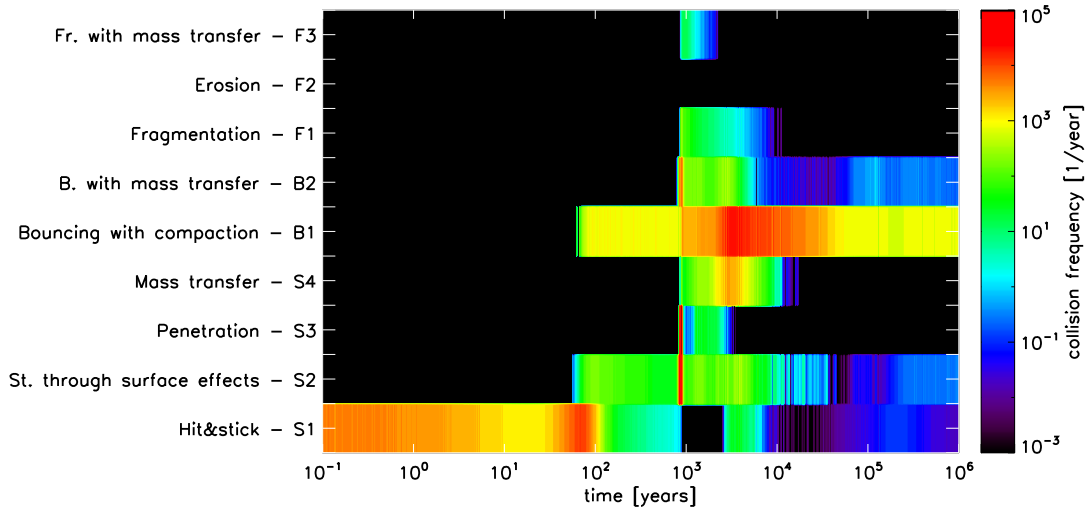}
  \caption{The collision frequencies of the 9 collision types in the MMSN model with $\alpha = 10^{-3}$ and $r_{\mathrm{m}}=100$.}
  \label{fig:mmsn_1d-3}
\end{figure*}

\subsection{Varying the critical mass ratio}
\label{sec:critmass}
We perform simulations in the disk models with $\alpha=10^{-4}$ but a varying critical mass ratio. We explore how the dust distributions change upon using $r_{\mathrm{m}} = 10$, 100, and 1000. Table \ref{tab:res}, lines 10 to 18, shows the parameters that describe the distribution functions, and Fig. \ref{fig:res}b illustrates the maximum particle mass as a function of the critical mass ratio. \\

By examining Table \ref{tab:res}, we see that using $r_{\mathrm{m}} = 10$ in the low density model (`Lt1d-4m10') results in heavier and more compact particles. The low critical mass ratio means that the largest particles in the different-sized regimes can sweep up the projectiles and grow to larger sizes, eventually reaching the fragmentation line, where growth stops. As discussed in Sect. \ref{sec:vrel}, assuming a fragmentation velocity of 1 m s$^{-1}$, the maximum Stokes number of the aggregates is $4.7\times 10^{-3}$, a value almost reached in this model. 

We find that there is no significant difference between the $r_{\mathrm{m}} = 100$ and 1000 simulations in the low density model. The explanation for this can be found by examining the width of the mass distribution in the {\Sa} phase. This initial phase occurs in the same way independently of the critical mass ratio. If the critical mass ratio $r_m$ is equal to or larger than the width of the distribution function, collisions between different-size particles in the {\pP} regime are inhibited. After the S1 phase, the width of the distribution in the low density regime is approximately 100. Therefore, we do not see any difference when the mass threshold is shifted from $r_m$ = 100 to $r_m$ = 1000; in both cases, collisions occur between equal-size particles only, and these are either S1 or (when this stage is over) B1.

For the high density models (MMSN/Desch), we find that the outcome is again similar: growth halts at $\sim$ 0.1 g (within a factor of 10) and no clear dependence on $r_m$ is seen. For the high mass ratios, growth is always in the similar-size regime. Here, it is the gas density that determines the velocity, i.e., whether we have a sticking (S1) or a bouncing (B1) collision. Therefore, if $r_m=1000$, the high density model produces heavier particles than the MMSN model (see Fig. \ref{fig:res}b). For lower $r_m$, it is again the nature of the S2-B2 transition regime that limits the maximum mass.

Thus, the critical mass ratio is an important parameter because it determines the relative likelihood of collisions occurring in the different-size regime, which are in general more conducive to growth. In contrast, for simulations where B2 collisions are important -- which have the effect of narrowing the distribution -- the width of the distribution will correspond to the value of the $r_m$ parameter, although we have also seen that the absolute size/mass is rather insensitive to the $r_m$ parameter. Overall, these arguments indicate that a good knowledge of this parameter is important.

\section{Discussion}
\label{sect:disc}
We have performed simulations of varying turbulence parameter and critical mass ratio values in three disk models with low, intermediate and high gas densities. We have found that {\Sa} and {\Ba} are the most dominant collision types. All simulations show evidence of long-lived, quasi-steady states. Fragmentation is rarely present, but even then, for only a limited time period. The absence of fragmentation is caused by the bouncing collisions.

\subsection{The sensitivity of the results}
\label{subsec:sens}
As presented in Sect. \ref{sec:res4}, the outcome of our simulations is determined by the collision kernel, and the relative velocity field. A significant change in one, or both can alter the evolution of the aggregates. 

We present the results of a test simulation, where the {\Sb} -- {\Bb} transition regime in the {\pP} plot is neglected and replaced by S2 collisions. This alternative transition regime provides a good opportunity to examine the rapid growth presented in Sect. \ref{subsec:mmsn_early}, as the kernel is now simplified. The new kernel also provides us with the possibility to see how much the outcome of our simulations can be altered by changing critical areas of the parameter space. As the transition regime is only constrained by one experiment in a rather small area (see e.g. Fig. \ref{lowd_cont} or Fig. 11 in Chapter \ref{chp:paper2}), this part of the parameter space may require changing in the future. We use the same initial conditions as in the `Mt1d-4m100' model described in Sect. \ref{sec:mmsn}. 

\begin{figure}
\centering
  \includegraphics[width=0.7\textwidth]{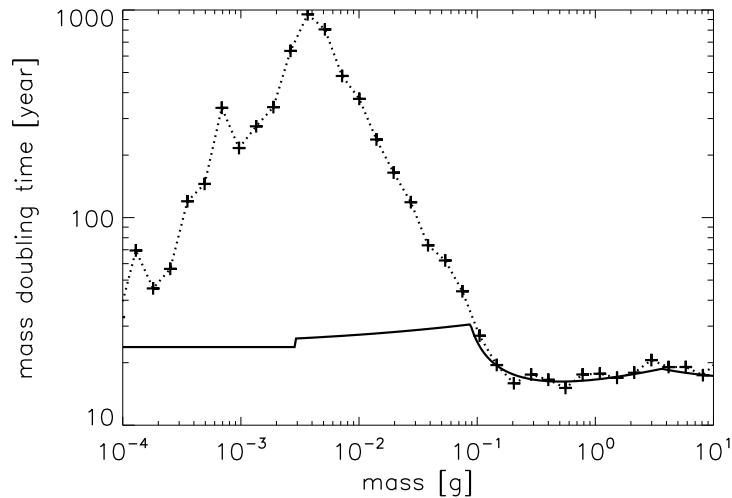}
  \caption{Growth timescale in the test simulation where the S2-B2 transition regime is replaced by S2 collisions only. The dotted line represents the growth timescale of the heaviest particles, the solid line is the minimum growth timescale. In this scenario, the growth timescale reaches its maximum possible value.}
  \label{fig:acc}
\end{figure}

In this case, the heaviest particle experiences increased relative velocities, as soon as it reaches $m=0.1$ g, and the particle undergoes a rapid growth period (as in the original MMSN simulation, Sect. \ref{sec:mmsn}). Figure \ref{fig:acc} indicates the growth timescale of the heaviest particle (dotted line) and the minimum growth timescale possible (solid line). Since none of the B2 collisions can reduce the mass of the heaviest particle, the growth timescale reaches its possible maximum. The heaviest particle increases in mass until the rest of the particle population enters the B2 regimes above 0.1 g in the {\pP} plot. In this simulation, the maximum average mass is 27 g, whereas in the original simulation for the transition regime, the value was 4.18 g.

Together with Chapter \ref{chp:paper2}, this work is the first attempt to calculate dust growth in protoplanetary disks on an empirical, thus more realistic basis. However, a few additional cycles of the feedback loop between the laboratory experiments, the models of the kind described by Chapter \ref{chp:paper2}, and the models described in this chapter have to be conducted before we can obtain a truly reliable model of dust growth in protoplanetary disks. 

\subsection{Retention of small grains}
\cite{Dullemond:2005p78} showed that without a mechanism that reduces the sticking probability of particles in the upper layers of the disk or without a continuous source of small particles, the observed SEDs of TTauri stars should exhibit a very weak infrared excess. The SEDs of TTauri stars have strong IR excess (e.g., \cite{Furlan2005-4}, \cite{Kessler-Silacci2006-4}); therefore, some kind of grain-retention mechanism is needed to explain these SEDs. Previous models of grain growth assumed a continuous cycle of growth and fragmentation, which provides the necessary amount of small particles (see e.g., \cite{Brauer2008a-4}, \cite{Dullemond:2005p78}, \cite{Birnstiel2009-4}). Our simulations, however, have shown that the mass distribution function is narrow. Small, monomer-sized particles are not present and fragmentation is ineffective in providing small particles, which could be transported to disk atmospheres. The question naturally arises: how can small grains be produced in our collision model? \\

One possible solution might be provided by bouncing. \cite{Weidling2009-4} performed bouncing experiments by placing an aggregate onto an oscillating metal plate and measuring the porosity of particles due to collisions with the plate. They observed that approximately 10\% of the projectile mass was eroded during the experiment (see Table 1 of their paper). This mass loss may be caused by the initial collisions, the eroded mass sticking to the baseplate. It is also possible that small pieces of fragments grind off when the aggregates bounce, which cannot be observed in the experiment. These ground-off particles can then diffuse out of the midplane and provide the small particles that we observe to enter to the upper layers of the disk. Future laboratory experiments are needed to quantify the level of ground-off particles created by bouncing collisions.\\

The second possible explanation involves dust growth in the upper layers of the disk. We performed two simulations at four pressure scale-heights in both the low density model and the MMSN model for $\alpha=10^{-4}$. We find that the relative velocity of two monomers in the Brauer model is 2 m s$^{-1}$, thus monomers at these heights do not coagulate, only bounce. The particles in the MMSN model can form aggregates of a maximum of 10 $\mu$m in size. Using a higher value of $\alpha$ (as usually assumed in the upper layers of the disk), we can completely halt even this limited growth. Therefore, bouncing could be the key mechanism reducing the sticking probability of the particles. However, if substantial vertical turbulent mixing takes place, bouncing may not be able to help, because these monomers would then be ``vacuum-cleaned'' away by the larger particles at the interior of the disk. Additional studies of 1D vertical slices of disk models are needed to investigate this scenario. 

\subsection{Implications for planetesimal formation models}
\label{subsec:plf}
It is also evident that coagulation alone is unable to produce planetesimals under the conditions presented in this work. Even if the turbulence parameter is assumed to be zero, relative velocity caused by radial drift prevents particles crossing the so called 'meter-size barrier'. An ideal environment for particle growth is a pressure bump in the dead zone, where both the turbulent and radial relative velocities are reduced. This environment is located around the snow line (\cite{Kretke2007-4}). \cite{Brauer2008b-4} showed that in these pressure bumps relative velocities stayed below a fragmentation threshold of 10 m s$^{-1}$, providing a window through which particles could overcome the m-size barrier, although they assumed perfect sticking (no bouncing) below the fragmentation barrier. Future studies should verify whether planetesimals can be formed with the collision model presented in this study.\\

Another planetesimal-forming mechanism is the gravitational collapse of swarms of boulders (\cite{Johansen:2007p65-4}). This scenario assumes that a large amount of the solid material is presented in dm-sized boulders ($St \ge 0.1$) at the midplane of the disk. These boulders then concentrate in long-lived high pressure regions in the turbulent gas and initial over-densities are amplified further by the streaming instability. This mechanism forms 100 km sized objects on a very short timescale (some orbits). However, our simulations produce particles with $St \approx 10^{-4}$, which is due to {\Ba} and the low (1 m s$^{-1}$) fragmentation velocity of silicates. Using a `stickier' material such as ices or particles with organic mantels may produce larger particles. Molecular dynamical simulations (e.g., \cite{Dominik:1997p89-4}, \cite{Wada2007-4}, \cite{Wada2008-4}) have shown that icy aggregates could have fragmentation velocities of about 10 m s$^{-1}$, although these findings have yet to be confirmed by laboratory experiments. Similarly, it is conceivable that the enhanced sticking capabilities of ices can prevent bouncing, which is so omnipresent for small particles in our simulations, or shifts its proficiency to larger sizes.\\

\cite{Cuzzi2008-4} outlined an alternative concentration mechanism to obtain gravitationally unstable clumps of particles, which can then undergo sedimentation and form a `sandpile' planetesimal. In this model, turbulence causes dense concentrations of aerodynamically size-sorted, chondrule-size particles (\cite{Cuzzi2001-4}) or more precisely, particles of Stokes numbers $St=Re^{-1/2}\approx 10^{-4}$ in our simulations. Since growth in our models is typically halted at these Stokes numbers, this concentration mechanism is an obvious successor to coagulation -- at least where it concerns the conditions adopted in this chapter (1AU, silicates). \\

However, we emphasize that the formation of a gravitationally unstable clump does not imply that planetesimals will form without impediment. An important question is how collisions will affect the collapse. In the \cite{Cuzzi2008-4} scenario, the collapse occurs on a sedimentation timescale and at these high densities, collisions between particles are frequent. Likewise, in the Johansen scenario -- where the collapse occurs on an orbital timescale and involves $St \sim 0.1$ particles -- collisions can be rather violent. Collisional fragmentation or erosion may change the appearance of the collapse, because the small fragments are carried away by the gas. The role of collisions in these situations is certainly an important question, and our new collision model provides a tool to quantitatively address this issue in future studies.

\subsection{Consequences for laboratory experiments} 
\label{subsec:lab}
From Fig. 11, in Chapter \ref{chp:paper2}, one can see  that only a small part of the relevant parameter space has been covered by experiments. Although laboratory experiments cannot be performed for every point of the parameter space, we suggest future ones based on Figs. \ref{lowd_cont}, \ref{mmsn_cont}, and \ref{desch_cont} to understand dust growth in the early stages of planet formation.
\begin{itemize}
\item More experiments in the {\cc} and {\cC} regimes are needed as particles become compactified toward the end of their evolution. Thus, most of the collisions occur in this regime, at velocities between 0.1 and 100 cm s$^{-1}$, and masses between $10^{-7}$ and 10 g.
\item As seen in Figs. \ref{lowd_cont}, \ref{mmsn_cont}, and \ref{desch_cont}, the `hot spots', where most of the collisions occur, are located in equal-sized regimes, at the left side of the fragmentation line. Therefore, it is important to map these areas of the parameter space in detail.
\item We define a sharp border line between the {\Sa} and {\Ba} collisions. If there is a continuous transition between S1 and B1, the growth of particles should not be halted by bouncing at these low particle sizes. Since many collisions occur in the {\pp} and {\cc} regimes, even a small probability of growth could increase the particle sizes.
\item As seen in Fig. \ref{mmsn_pics}b, particles in high gas density environments can have enlargement parameters that are much higher than 6.6 ($\phi = 0.15$). An interesting question is whether the collision types and regimes are also valid for particles with such low volume-filling factors, or whether these particles have different collision behaviors?
\item The {\Sb} -- {\Bb} transition regime greatly affects the outcome of the simulations (see Sect. \ref{subsec:sens}). However, the transition regime is only mapped at the high velocity and low mass regions. Therefore, it is important to constrain more tightly this part of the parameter space.
\item The critical-mass ratio affects both particle masses and porosities. Experiments are needed to constrain its value.
\item The bouncing model, described in Chapter \ref{chp:paper2}, has important implications for the evolution of dust aggregates in protoplanetary disks but is unfortunately still constrained by too few experiments. Additional experiments are needed to refine the model, as {\Ba} is the most frequent collision type in all of the simulations. 
\end{itemize} 

\section{Summary}
\label{sec:sum}
We have performed simulations of dust growth using the Monte Carlo code of ZsD08 and a dust collision model based on laboratory experiments (Chapter \ref{chp:paper2}). We have performed simulations in the midplane of three disk models at low ($2.4 \times 10^{-11}$ g cm$^{-3}$), intermediate ($1.4 \times 10^{-9}$ g cm$^{-3}$), and high ($2.7\times 10^{-8}$ g cm$^{-3}$) gas densities at 1 AU distances from the central star. We have varied the turbulence parameter ($\alpha$) and the critical-mass ratio ($r_{\mathrm{m}}$) to explore their effects on the mass and porosity distribution functions. Our main results are:
\begin{itemize}
\item Upon using $\alpha = 10^{-4}$, the low-density / MMSN / high-density model produces particles with maximum mean mass of $9.7 \times 10^{-8}$ g / 4.18 g / 0.23 g, respectively, the maximum average enlargement parameter of these particles are 7.12 / 21.9 / 38.0, respectively. The maximum average Stokes numbers are $2.2\times 10^{-4}$ / $2.8\times 10^{-4}$ / $3.6 \times 10^{-5}$, respectively. 
\item We find that particle evolution does not follow the previously assumed growth-fragmentation cycles. Although catastrophic fragmentation is present for a short period of time in some models (typically when $\alpha = 10^{-3}$), it has a fringe effect. Particles in most of the simulations do not reach the fragmentation barrier because their growth is halted by bouncing.
\item We see long-lived, quasi-steady states in the distribution function of the aggregates that are caused by bouncing. The final equilibrium state is not reached within the drift or the viscous timescales.
\item We have performed simulations of varying turbulence strength. We find that the system is `non-linear': the maximum mass of particles is not a decreasing function of the turbulence parameter and is not an increasing function of the gas density.
\item We have explored the effects of the critical mass ratio. We find that different critical mass ratios can affect the particle evolution. Low critical mass ratios can produce heavier particles, while high values of $r_m$ can halt the growth earlier.
\item The maximum Stokes number is almost independent of either the gas density or the strength of the turbulence.
\item The maximum mass of the aggregates is limited to $\approx$ 1 g because of the S2-B2 transition regime.
\item The Stokes number $10^{-4}$ particles can be concentrated by aerodynamical size-sorting, thus planetesimals can form from these particles.
\end{itemize}
 



%% file: Chapters/Chapter5.tex

\chapter{Sedimentation driven coagulation within the snow line} 
\label{chp:sedi}
\lhead{Chapter 5. \emph{Sedimentation}} 
\rhead{}

\section{Introduction}
Due to the vertical component of the stellar gravity, dust particles sediment towards the midplane of the disk. Observational evidence of the vertical sedimentation of grains exists for a large number of disks, although such evidence is usually indirect. 

Sub-micron grains are present at the disk surfaces as shown by scattered light images in the optical and near infrared (NIR) wavelengths. Such multi-wavelength scattered light images provide evidence for grain growth (\cite{Watson2007-5} and references therein). \cite{Pinte2007} showed by reproducing multi-band images of the binary system of GGTau that the dust scale height for 10 micron sized particles is roughly half of that for micron sized particles. 

The spectral energy distribution (SED) is also affected by settling. \cite{D'Alessio2006-5} showed that in order to explain the median SEDs of classical TTauri stars, the dust to gas ratio has to be reduced by a factor of 10 at the disk atmosphere compared to the standard value. There are also indications that the settling of grains is correlated with the age of the disk \citep{Sicilia-Aguilar2007}. However, the connection between the exact shape of the 10 micron feature of SEDs and sedimentation is not well understood \citep{Dullemond2008-5}.

\cite{Apai2005} showed evidence for settling in disks around brown dwarf stars and concluded that growth, crystallization and settling of dust happens around low mass stars in a similar manner as around intermediate and solar mass stars suggesting that planet formation is a robust process.

Sedimentation also affects the vertical temperature structure of the disk. The simulations of \cite{Aikawa2006} showed that as the dust particles sediment towards the midplane, the opacity is reduced, therefore the temperature of the gas decreases. As the stellar radiation can now penetrate deeper in the disk, the temperature at intermediate heights increases. The change in the density and temperature structure naturally influences the chemistry of the disk atmospheres \citep{Bergin2007}. 

Dust sedimentation not only affects the upper layers of the disk. The gas tends to rotate on a sub-Keplerian velocity. The dust particles without the gas would rotate on a Keplerian orbit as the dust particles do not exert pressure on each other. But due to the presence of gas, the small dust particles move with the same (sub-Keplerian) speed as the gas, if the dust to gas ratio is much smaller than unity. However, if the dust to gas ratio approaches unity due to settling, the dust influences the motion of the gas. Therefore the gas at the dusty midplane layer rotates faster than the upper (not so dusty) layers and vertical shear is generated. This shear triggers Kelvin-Helmholtz instability which develops into turbulence. This process was first recognized by \cite{Weidenschilling19805} and it is still an actively researched area (e.g. by \cite{Johansen2006}, \cite{Chiang2008}).

In the framework of this thesis, a collaboration started between the lab-community and the modelers to better constrain the dust evolution in protoplanetary disks using a realistic collision model that is based on the laboratory experiments. In Chapter \ref{chp:paper2} (based on \cite{Guttler2010}), we introduced this collision model. In Chapter \ref{chp:paper3} (based on \cite{Zsom2010}), we used this collision model for the first time in the Monte Carlo (MC) method of \cite{Zsom2008} (henceforth ZsD08). The models of Chapter \ref{chp:paper3} were local box models meaning that the dust evolution was only followed at one location of the disk. These models showed that bouncing plays an important role in dust evolution. 

We further develop these models to simulate a 1D vertical column in the disk thus investigating sedimentation driven coagulation. We want to better understand the process of sedimentation and the role of particular physical phenomena like porosity of the aggregates, collision models and turbulence. 

Previous work by \cite{Dullemond:2005p78-5} (henceforth DD05) showed that without a mechanism that reduces the sticking probability of particles in the upper layers of the disk, or without a continuous source of small particles, the observed spectral energy distributions (SED) of TTauri stars should exhibit a very weak IR excess. In contrast, the observed SEDs of TTauri stars have strong IR excess (e.g. \cite{Furlan2005, Kessler-Silacci2006}) therefore some a grain-retention mechanism is needed to explain the SEDs. 

Previous models of grain evolution assumed a continuous cycle of growth and fragmentation, which provides the necessary amount of small particles (e.g. \cite{Brauer2008a}, \cite{Birnstiel2009}). However, Chapter \ref{chp:paper2} and \ref{chp:paper3} showed that particle evolution is halted by bouncing and no cycle of growth and fragmentation is present. We simulate dust evolution driven by Brownian motion, turbulence, and sedimentation in a 1D vertical column of the inner disk. We investigate the time evolution of sedimentation driven coagulation, and search for ways that can keep a sufficient amount of the small dust particles levitated at several pressure scale heights to explain the observed SEDs of young stars. 

The chapter is organized as follows. In Sec. \ref{sec:num} we describe the numerical method used to follow the particle motion and coagulation. We validate the code and increase the complexity of the model step-by-step in Sec. \ref{sec:test}. We show the results in Sec. \ref{sec:res5}, finally we discuss those results in Sec. \ref{sec:disc} and provide a summary in Sec. \ref{sec:concl}.
 
\section{Numerical method}
\label{sec:num}
\subsection{Basic considerations}
\label{subsec:bas}
The local box approach in Chapter \ref{chp:paper3} is based on two assumptions. 1.) The particles are homogeneously mixed inside the box. 2.) Particles do not enter or leave the box, i.e. it is closed. Due to these two assumptions, it was not necessary to follow the exact location of the particles. 

In the models considered here, however, we place such boxes (or grid cells) on top of each other to simulate a 1D column in the disk and follow how particles settle towards the midplane. Inevitably, particles move from box to box during this process. Therefore the assumption that particles cannot enter or leave the boxes has to be relaxed. The first assumption of the method in Chapter \ref{chp:paper3} is kept, we still assume that the particles inside a given box are homogeneously distributed when we consider coagulation (for sedimentation, the individual positions of the particles are used). The second assumption is modified in the following way. 2.) The simulated column is closed, e.g. particles inside the column can move freely vertically.
However neither do \textit{new} particles enter from the ``outside'', nor do particles from inside the column \textit{leave}. As particles move through the boxes, it is necessary to follow the position of the particles (see Sec. \ref{subsec:pos}) as we must find out in which box a particle is located.

The motion of particles imposes a limit on the time-step of the simulation. We do not want the particles to move more than one box in a time-step. A sedimenting particle should have the possibility to interact with all other particles along its way, it should not skip over boxes thus avoiding the particles in it. Therefore, we use an adaptive time-stepping method. The maximum of all particle velocities is obtained ($v_{\mathrm{max}}$), and since we know the height of the boxes ($h_{\mathrm{box}}$), the maximal (safe) time-step can be determined as
\begin{equation}
\Delta t = C \frac{h_{\mathrm{box}}}{v_{\mathrm{max}}},
\end{equation}
where $C$ is the Courant number which we typically set to be $0.1$. 

The code schematically performs the following steps:
\begin{enumerate}
\item First the velocities of the particles are calculated.
\item A safe time-step is determined to avoid particle `jumps'.
\item The position of the particles is updated using their velocities, their previous positions, and the time-step.
\item We determine the box in which each particle resides.
\item We call the coagulation subroutine described in Chapter \ref{chp:paper3} to calculate the evolution of the particles separately in each box for the given time-step.
\end{enumerate}

\subsection{Initial conditions}
\label{subsec:inicond}
We assume that the gas density profile is constant during the simulation. This assumption is valid if the simulated time is less or comparable to the viscous timescale of the gas. The viscous timescale can be calculated as
\begin{equation}
t_{\mathrm{vis}}=r^2/ \nu_T,
\end{equation}
where $r$ is the distance from the central star, $\nu_T$ is the turbulent viscosity. A typical value for $t_{\mathrm{vis}}$ at 1 AU is $10^3$ - $10^7$ yrs. We assume that turbulence is parameterized by the \cite{Shakura1973-5} $\alpha$ parameter
\begin{equation}
\nu_T=\alpha c_s H_g,
\label{eq:nuT}
\end{equation}
where $c_s$ is the isothermal sound speed, and $H_g$ is the pressure scale height of the gas disk. The turbulence parameter $\alpha$ reflects the strength of the turbulence in the disk. Typical values range between $\alpha=10^{-6}$ and $10^{-2}$, where the former corresponds to the turbulent strength in dead zones, the latter describes turbulence in disk atmospheres. 

The vertical structure of the disk is determined by the equilibrium between the vertical component of the gravitational force and the acceleration due to the vertical pressure gradient in the gas. If the disk mass ($M_{\mathrm{disk}}$) is much smaller than the mass of the star ($M_*$), and the vertical thickness of the disk ($H_g$) is a small fraction of the radial distance (both conditions are safely met for the disk parameters described below), then the vertical density can be approximated as
\begin{equation}
\rho_g (r,z) = \frac{\Sigma(r)}{\sqrt{2 \pi} H_g}\exp(-z^2/2H_g^2),
\end{equation}
where $\Sigma(r)$ is the gas surface density at distance $r$, and $z$ is the height above the midplane. In this chapter we choose $M_* = 0.5 M_\odot$, $r=1$ AU, $\Sigma(1$ AU$)=100$ g/cm$^2$ similarly to DD05. The pressure scale height can be calculated as
\begin{equation}
H_g = c_s/\Omega,
\end{equation}
where $\Omega$ is the orbital frequency at 1 AU. The isothermal sound speed is
\begin{equation}
c_s = \sqrt{\frac{k_B T}{\mu m_p}},
\end{equation}
where $k_B$ is the Boltzmann constant, $\mu$ is the molecular weight, which is 2.3 for molecular gas, $m_p$ is the mass of the proton, and $T$ is the temperature of the gas, which is 200 K for the stellar and disk parameters considered above (see DD05). We assume the temperature to be constant as a function of height. This is a reasonable assumption if the temperature of the gas is solely determined by the stellar irradiation. 

We simulate 4 pressure scale heights (0.16 AU above the midplane), use 40 evenly spaced boxes and $10^5$ particles in the simulations unless otherwise stated. The number of boxes and the number of particles are chosen by taking into account two points. 1.) The upper box is initially not empty, it contains at least a few particles. 2.) Simulations, which are performed with the exact same initial set-up for the gas and using the same collision model, but using different initial positions for the dust (e.g. using a different seed for the random number generator), do not differ qualitatively. Differences are expected due to the intrinsic stochasticity of the Monte Carlo implementation. Every particle represents the same portion of the total dust mass, therefore more particles are present in the lowest box and only a few in the upper box. The initial particle positions are determined randomly, therefore the initial gas to dust ratio is noisy. The noise is lower closer to the midplane, and gradually increases with height. However, the mean initial gas to dust ratio is constant, 1:100.

\subsection{Position update}
\label{subsec:pos}
The position of the particles are determined by vertical settling and turbulent diffusion. In principle, Brownian motion also contributes to the change of particle positions, but its effect is negligible compared to the other two effects.

The equation governing the diffusion and settling of the dust in a non-homogenous gas density field is \citep{Dubrulle1995, Fromang:2006p324}
\begin{equation}
\frac{\partial \rho_d}{\partial t} = \frac{\partial}{\partial z} \left[D_d \rho_g \frac{\partial}{\partial z}
\left( \frac{\rho_d}{\rho_g} \right)\right] + \frac{\partial}{\partial z} (\Omega^2 t_s \rho_d z),
\end{equation}
or in a more practical form:
\begin{equation}
\frac{\partial \rho_d}{\partial t} = \frac{\partial^2 D_d \rho_d}{\partial z^2} - \frac{\partial}{\partial z}\left( \rho_d \times D_d \frac{1}{\rho_g}\frac{\partial\rho_g}{\partial z}\right) + \frac{\partial}{\partial z} (\rho_d \times z \Omega^2 t_s)
\label{eq:diff}
\end{equation}
where $\rho_d$ is the dust density, $D_d$ is the diffusion coefficient of the dust (for the calculation of $D_d$, see the next paragraph) and $t_s$ is the stopping time of the particle. The stopping time is the timescale a particle needs to react to the changes of the surrounding gas. 
We define the dimensionless Knudsen number being
\begin{equation}
Kn = \frac{a}{\lambda_{\mathrm{mfp}}},
\end{equation}
where $a$ is the size of the aggregate, and $\lambda_{\mathrm{mfp}}$ is the mean free path of the gas. A particle is in the Epstein regime if $Kn < 1$ (to be more precise, if $a < \frac{9}{4}\lambda_{\mathrm{mfp}}$), where the stopping time is (\cite{Epstein1924-5}):
\begin{equation}
t_{s} = t_{\mathrm{Ep}} = \frac{3 m}{4 v_{\mathrm{th}} \rho_g A},
\label{eq:ts1}
\end{equation}
where $m$ and $A$ are the mass and the aerodynamical cross-section of the particle, and $v_{\mathrm{th}}$ is the thermal velocity. If the Knudsen number is greater than 1 (at high gas densities where the mean free path is low or in the case of large particles), the first Stokes regime applies and the stopping time becomes
\begin{equation}
t_s = t_{\mathrm{St}} = \frac{3 m}{4 v_{\mathrm{th}} \rho_g A} \times \frac{4}{9} \frac{a}{\lambda_{\mathrm{mfp}}}.
\label{eq:ts2}
\end{equation} 

The first term on the right hand side of Eq. \ref{eq:diff} is the well-known diffusion term. Using only this term, particles with $t_s = 0$ (tracers) would be homogeneously distributed as a function of height over several diffusion timescales. The first and the second term together on the right hand side ensures that the tracer particles will be distributed according to the background gas density field. The third term describes the settling of the particles. Equation \ref{eq:diff} is valid if the motion of the dust does not influence the motion of the gas (the back-reaction from the dust to the gas is negligible). This condition is met if the dust to gas ratio is $\ll 1$.

We note that we do not solve for this dust density field directly. We follow the motion of dust \textit{particles}, each of which represents a portion of the total dust mass inside the column, thus we derive the corresponding velocities (or fluxes) for the first, second, and third terms of Eq. \ref{eq:diff} to calculate the position update of the particles.

We calculate $D_d$, the diffusion coefficient of the dust, and define the diffusion velocity, $v_{D1}$. The diffusion coefficient of the gas can be defined as \citep{Dullemond:2004:5p325}
\begin{equation}
D_g =\nu_T =\alpha c_s H_g.
\end{equation}
Based on \cite{Youdin:2007p576}, the diffusion coefficient of the dust can be calculated as
\begin{equation}
D_d = D_g/(1+St^2),
\end{equation}
where $St$ is the Stokes number 
\begin{equation}
St = t_s \Omega.
\end{equation}
The average displacement of a particle in 1D during the time-step of $\Delta t$ then is
\begin{equation}
L = \sqrt{2 D_d \Delta t}.
\end{equation}
The real displacement of the particle ($\Delta z$) is drawn from a Gaussian distribution which has zero mean and a half width of $L$. The ``diffusion velocity'' can then be calculated as
\begin{equation}
v_{D1} = \Delta z/\Delta t.
\end{equation}
This velocity component tries to smear out dust concentrations. It is important to note that the real, physical velocity of the particle during turbulent diffusion changes randomly every time the aggregate interacts with a turbulent eddy. The diffusion velocity defined above is a numerical construct to calculate the time-averaged velocity of the particle during a time-interval $\Delta t$.

The second term on the right side of Eq. \ref{eq:diff} results in a systematic velocity term which pushes particles towards the density maxima of the gas. This velocity can be determined by
\begin{equation}
v_{D2} = D_d\frac{1}{\rho_g}\frac{\partial\rho_g}{\partial z}. 
\end{equation}
Using these two velocity components ($v_{D1}$ and $v_{D2}$), the particles with non-zero stopping times will be distributed according to the gas density profile in a timescale longer than the diffusion timescale.

The fact that particles with $t_s>0$ settle towards the midplane and have a scale height less than $H_g$ is the result of the third term of Eq. \ref{eq:diff}, the settling velocity. The settling velocity of a particle can be determined by
\begin{equation}
v_{\mathrm{set}}=-z \Omega^2 t_s.
\label{eq:set}
\end{equation}

The new position of the particles can then be determined by using these three velocity terms:
\begin{equation}
z = z_{\mathrm{old}} + (v_{D1}+v_{D2}+v_{\mathrm{set}})\Delta t.
\end{equation}

\subsection{Coagulation}
\label{subsec:coag}
The collision model used in this work is similar to the one used in Chapter \ref{chp:paper3}. There are however two differences. The first difference is the additional source of relative velocity due to differential vertical settling (see Eq. \ref{eq:set}):
\begin{equation}
\Delta v_S = | v_{\mathrm{set1}}-v_{\mathrm{set2}} |.
\end{equation}
One could use the individual height of the particles (as calculated in Sec. \ref{subsec:pos}) to obtain $\Delta v_S$, but that would violate the first assumption of the coagulation model (namely that the particles are uniformly mixed inside the box). When particles collide, they have a zero distance, therefore their height must be identical. If the (non-representative) particles are uniformly mixed, two colliding particles with identical stopping times do not have differential settling velocity. If, however, one uses the individual heights of the representative particles, it is implied that the particles are not uniformly mixed in the box. Their height can be different which would result in a non-zero settling relative velocity if $t_{s1} = t_{s2}$. Therefore, we must use an averaged height, the height of the box, to reliably calculate $\Delta v_S$.

The second difference is in the calculation of the aerodynamical cross section which is used to calculate the stopping time. In Chapter \ref{chp:paper3} we used geometrical cross section of the particles \citep{Ormel5:2007p93}
\begin{equation}
A=r_c^2 \pi \Psi^{2/3},
\label{eq:cross1}
\end{equation}
where $r_c$ is the compact radius of the aggregate (assuming that the mass of the particle is contained in a compact sphere of radius $r_c$), and $\Psi$ is the enlargement parameter defined in e.g. \cite{Ormel5:2007p93}. This formula works well for particles with fractal dimension above 2. However, we will also use the porosity model of \cite{Okuzumi2009a}, and their model produces aggregates with fractal dimension below 2. If one calculates the stopping time of such an aggregate in the Epstein regime (Eq. \ref{eq:ts1}) using the formula in Eq. \ref{eq:cross1}, one gets that the stopping time is less than the stopping time of a monomer. This is clearly unphysical. The reason for this low stopping time (large area) is that Eq. \ref{eq:cross1} does not take into account the empty space between the `fractal branches'. To avoid such unphysical results for aggregates with low fractal dimensions, we use the aerodynamical cross section as defined in Eq. 47 in \cite{Okuzumi2009a}.

\section{Results}
\label{sec:res5}
We perform 11 simulations, in which we gradually use more realistic collision models, investigate the effects of different porosity models and turbulence. The IDs and parameters of these simulations are shown in Tab. \ref{table:sedi}. First we compare our model against the results of DD05 (model DD in Tab. \ref{table:sedi}). Then we use the porosity model of \cite{Ormel5:2007p93} and \cite{Okuzumi2009a} to investigate the effects of porosity (models DDa and DDb, respectively). So far we assume that the aggregates stick together at all relative velocities and the turbulence parameter $\alpha$ is zero. In the next step we construct a more realistic collision model with sticking, bouncing and fragmentation (models SB1, SB2). We call this collision model the ``simplified Braunschweig model'' because it uses only three collision types out of 9, which is described in Chapter \ref{chp:paper2} (the complete Braunschweig model). In the next step we turn on turbulence (models SB3-6) to examine the effects of turbulent stirring. Finally we use the complete Braunschweig model with turbulence (models FB1, FB2). 

\begin{table*}
\caption{Overview and results of all the sedimentation simulations.}
\label{table:sedi}      
\begin{center}   
\small
\begin{tabular}{l l l l l l l l}        
\hline\hline                 
ID		&Coll. model	&Por. model	&$\alpha$	&$t_{\mathrm{rain}}$	&$m_{\mathrm{rain}}$	&$\Psi_{\mathrm{rain}}$  &$H_p$  \\ 
		&			&			&		&[yrs]	&[g]			&		&[$H_g$]\\
(1)		&(2)			&(3)			&(4)		&(5)		&(6)			&(7)		&(8)\\
\hline                        
DD		&hit\&stick	&--			&0		&500		&$10^{-2}$	&1 		&0\\
DDa		&hit\&stick	&Ormel		&0		&600		&$10^{0}$		&10 		&0\\
DDb		&hit\&stick	&Okuzumi	&0		&900		&$10^{7}$		&$10^6$ 	&0\\
SB1		&simpl. Br.	&Ormel		&0		&3000	&$10^{-5}$	&50 		&0\\
SB2		&simpl. Br.	&Okuzumi	&0		&6000	&$10^{-2}$	&$10^4$ 	&0\\
SB3		&simpl. Br.	&Ormel		&$10^{-6}$&2000	&$10^{-5}$	&30 		&0.1\\
SB4		&simpl. Br.	&Okuzumi	&$10^{-6}$&4000	&$10^{-2}$	&$10^4$ 	&0.1\\
SB5		&simpl. Br.	&Ormel		&$10^{-4}$&500	&$10^{-6}$	&10 		&0.5\\
SB6		&simpl. Br.	&Okuzumi	&$10^{-4}$&700	&$10^{-4}$	&$10^3$ 	&0.5\\
FB1		&compl. Br.	&Ormel		&$10^{-4}$&500	&$10^{-6}$	&10 		&0.25\\
FB2		&compl. Br.	&Okuzumi	&$10^{-4}$&700	&$10^{-4}$	&$10^3$ 	&0.25\\
\hline
\end{tabular}
\end{center}
Col. 1 is the ID of the simulations, col. 2 describes the used collision model (hit\&stick, simplified Braunschweig model, or complete Braunschweig model), col. 3 indicates the used porosity model (based on \cite{Ormel5:2007p93} or \cite{Okuzumi2009a}), col. 4 describes the value of the turbulence parameter $\alpha$, col. 5 is the time the rain-out particle reaches the midplane of the disk in years, col. 6 and 7 are the mass and the enlargement parameter of the rain-out particles, respectively, and col. 8 is the scale height of the dust expressed in the scale height of the gas. Note that for the FB1 and FB2 simulations, the $m_{\mathrm{rain}}$ values are not the final masses of the particles, but the masses the rain-out particles have when they first reach the midplane. The final masses of the particles are $10^{-2}$ g for both the FB1 and FB2 simulations. 
\end{table*}

\begin{figure*}
\centering
  \includegraphics[width=\textwidth]{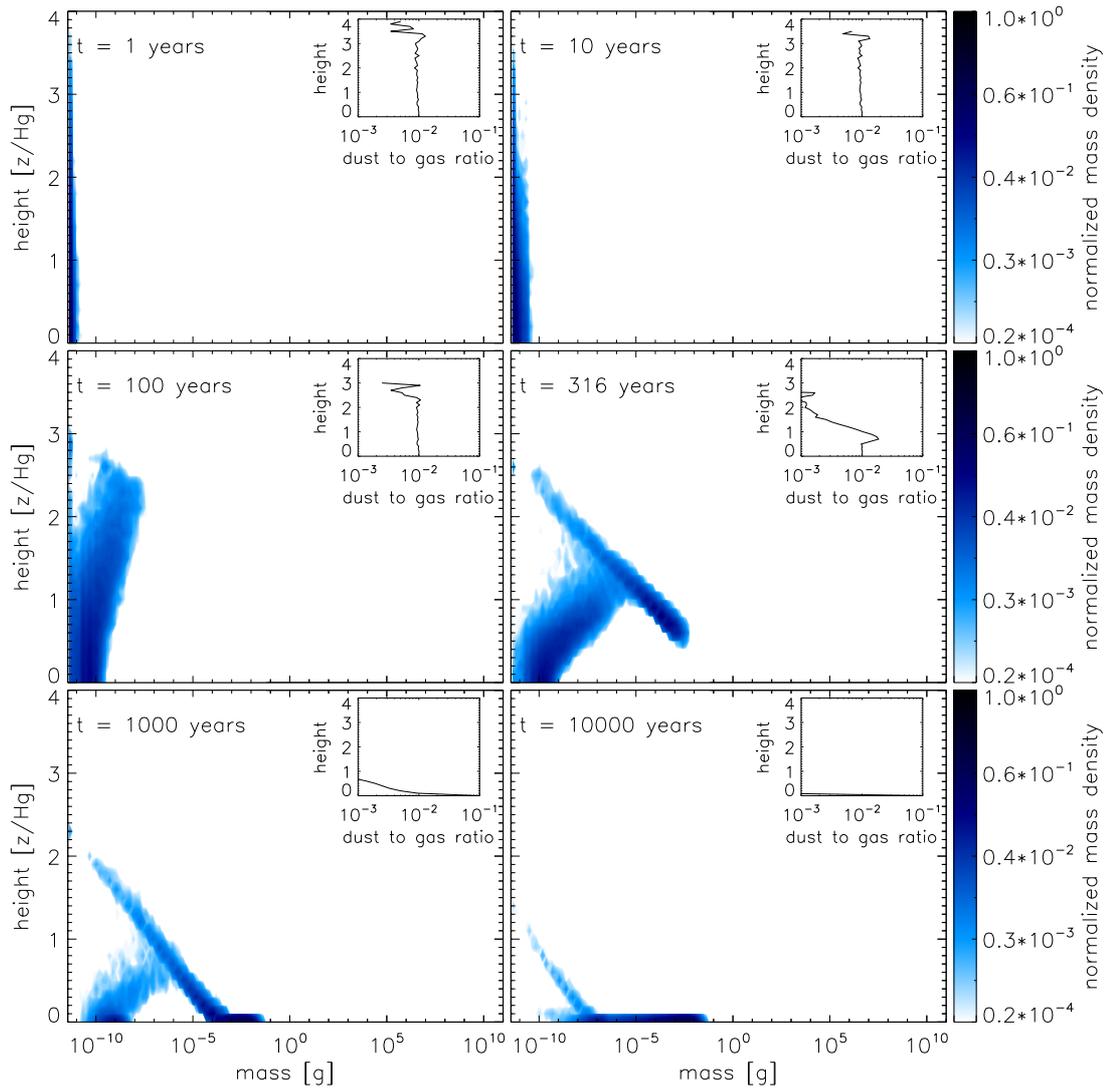}
  \caption{The mass distribution at $t=1$, 10, 100, 316, $10^3$ and $10^4$ yrs for the DD05 model. The x axis is the mass of the aggregates in grams, the y axis is the height above the midplane expressed in units of the pressure scale-height. The contours represent the normalized mass density of the dust. The sub-figures illustrate the dust to gas ratio (x axis) as a function of height above the midplane (y axis)}
  \label{fig:mass_DD05}
\end{figure*}

\subsection{Test: comparison with the DD05 model}
\label{sec:test}
DD05 performed a simulation (S2 in their paper, DD in this Chapter), where the disk model is the same as the one described in Sec. \ref{subsec:inicond}: the particles are compact, upon collision particles stick together at all collision energies, and the only source of relative velocity is Brownian motion and differential settling. They found that the `rain-out' particles reach the midplane in 500 yrs having attained sizes of a millimeter (10$^{-2}$ g in mass). 

We performed the exact same simulation to validate our code. We find that the `rain-out' particles in our simulation reach the midplane also at 500 yrs and have masses of $10^{-2}$ g, therefore we can conclude that our code works properly. We illustrate the mass evolution of particles as a function of their height at six different snapshots ($t=1$ yr, 10 yrs, 100 yrs, 316 yrs, 10$^3$ yrs, 10$^4$ yrs) in Fig. \ref{fig:mass_DD05}. 

Brownian motion is essential in our simulations because growth by Brownian motion kicks in the sedimentation driven coagulation. The reason is that we have initially a \textit{mono-disperse} particle size-distribution (meaning that all monomers have the same size and mass), therefore the aerodynamical properties of the monomers are identical, thus there is no relative velocity due to settling between the monomers at a given height. If growth due to Brownian motion was not initiated (e.g. growth by Brownian motion did not introduce aggregates with different aerodynamical properties than that of the monomers), the monomers would simply sediment to the midplane without any growth. Although DD05 included Brownian motion, this effect would not have been present as that simulation started with a (narrow, but not infinitely narrow) size distribution.

As shown in Fig. \ref{fig:mass_DD05}, growth by Brownian motion is faster at the midplane due to the higher gas and dust densities (at $t=1$ and 10 yrs). Once particles at the upper layer also start growing by Brownian motion, sedimentation driven coagulation starts and particles at the upper layers grow much faster than the aggregates at the midplane (at $t=100$ yrs). The heaviest particles sweep up the smaller particles while they sediment and further increase their settling velocity, resulting in a rain-out at $t=500$ yrs. Once the first rain-out particles reach the midplane, they could only grow by Brownian motion, because at the midplane, the settling velocity of any particle is zero (see Eq. \ref{eq:set}). But the relative velocity due to Brownian motion for such heavy aggregates is low, therefore the particles that have reached the midplane, do not increase in mass significantly during the simulation. 

\subsection{The effects of porosity}
In the previous section we used compact particles. However fluffy particles couple better to the gas. Therefore we perform simulations with two different porosity models to investigate the effects of porosity. We include growth by Brownian motion and settling only, and assume that particles stick together at all collisional energies. 

\begin{figure*}
\centering
  \includegraphics[width=1\textwidth]{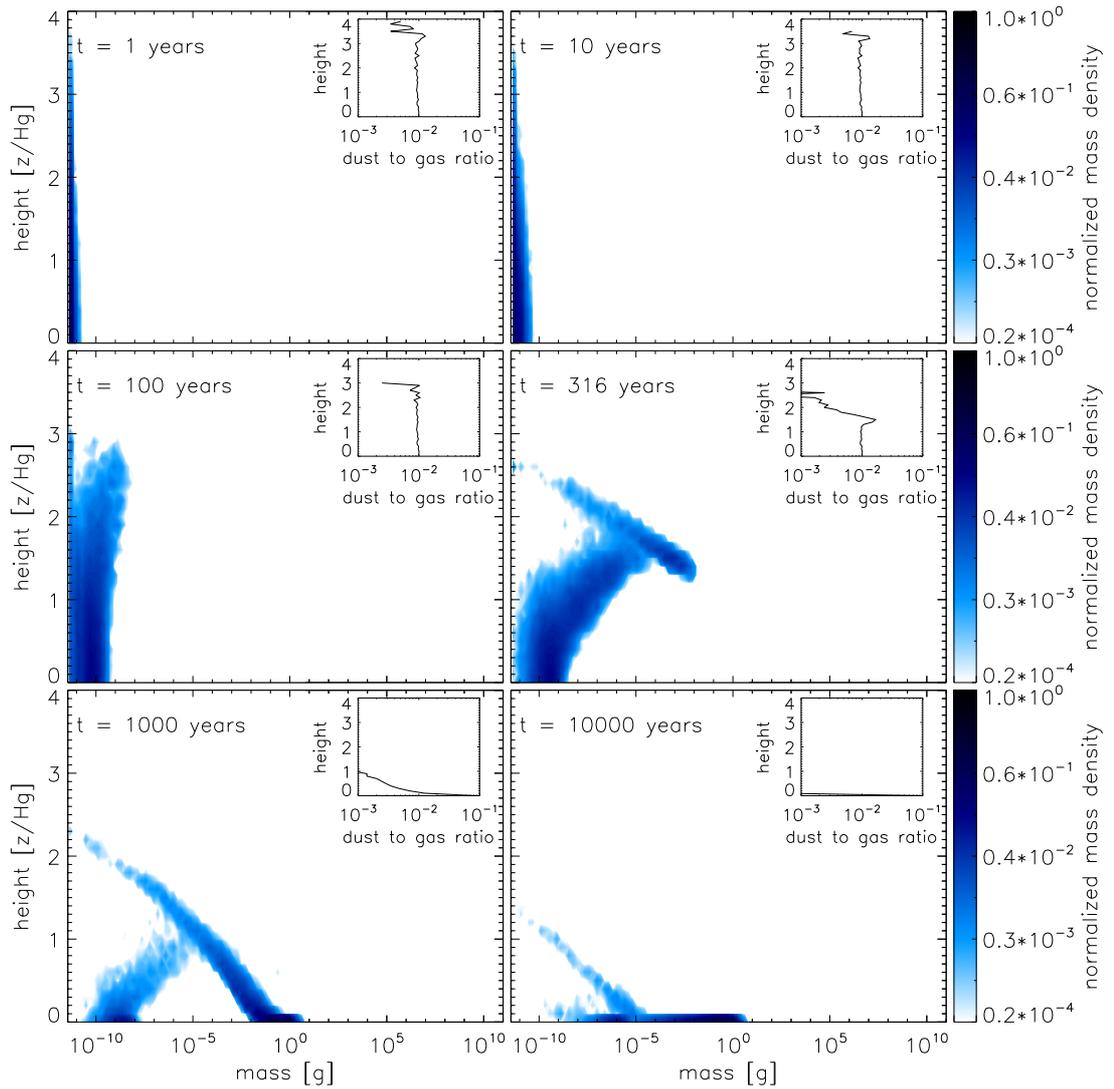}
  \caption{The mass distribution for the DD05 model using the Ormel porosity model. The axes and contours are the same as in Fig. \ref{fig:mass_DD05}.}
  \label{fig:mass_DD05_or}
\end{figure*}

\begin{figure*}
\centering
  \includegraphics[width=1\textwidth]{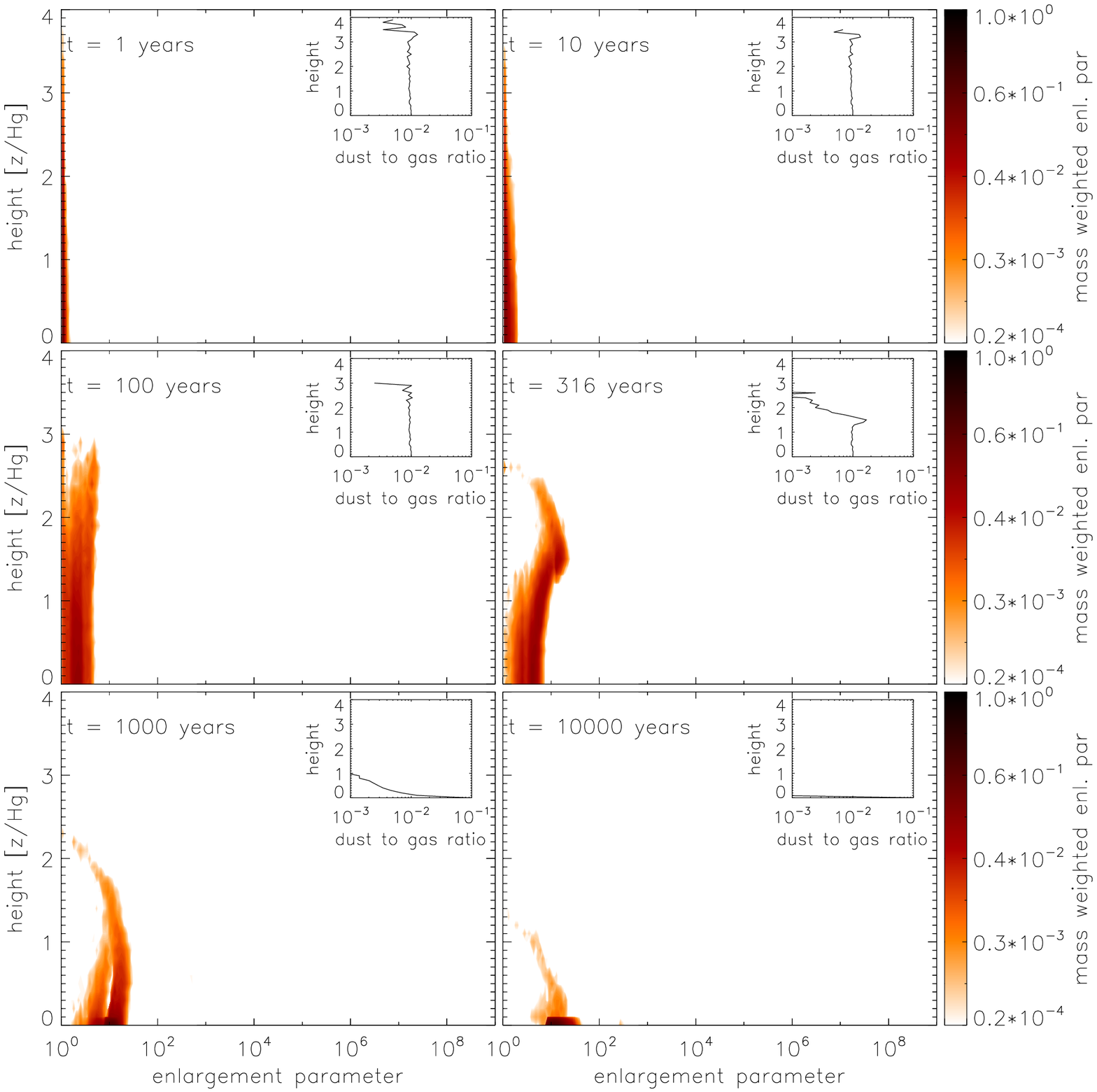}
  \caption{Enlargement parameter distribution for the DD05 model using the Ormel porosity model. The x axis here represents the enlargement parameter of the aggregates, and the contours show the normalized mass-weighted enlargement parameter of the particles.}
  \label{fig:enpar_DD05_or}
\end{figure*}

\paragraph{The Ormel model - DDa}\label{para:ormel} First we use the porosity model of \cite{Ormel5:2007p93}. This porosity model is based on both theoretical and experimental investigations of the microphysics of dust aggregates. It is essentially incorporates PCA-like collisions (particle-cluster aggregation - collisions between particles and clusters) and CCA-like collisions (cluster-cluster aggregation - collisions between clusters of similar size) and interpolates between these two. In this porosity model, compaction can occur if the collision energy is $E_{\mathrm{coll}} > 5 E_{\mathrm{roll}}$, where $E_{\mathrm{roll}}$ is the rolling energy, which is the energy needed to roll two monomers by 90 deg. However, we do not include this regime to stay consistent with the porosity model of \cite{Okuzumi2009a}, as their model also do not treat compaction.

The evolution of the mass can be seen in Fig. \ref{fig:mass_DD05_or}, the evolution of the enlargement parameter is illustrated in Fig. \ref{fig:enpar_DD05_or}. The evolution of these porous dust particles shows a similar behavior to the evolution of compact particles. The rain-out happens somewhat later at $t=600$ yrs, and the particles that reach the midplane have masses of 1 g, two orders of magnitude higher than in the previous model, which agrees with the results of \cite{Ormel5:2007p93}. 

The porosity evolution (Fig. \ref{fig:enpar_DD05_or}) shows some interesting features. The enlargement parameter naturally increases during Brownian motion (at $t=1$, 10 yrs) and also during the initial phases of settling (at $t=100$ yrs). However, as the particles approach the midplane, they become more compact (at $t=10^3$ yrs). The particle population that the rain-out particles can sweep up is getting smaller as they approach the midplane (see Fig. \ref{fig:mass_DD05_or} at $t=10^3$ yrs). Therefore the rain-out particles collide with ever smaller particles and these small particles can `fill up' the holes of the rain-out particles, the collisions are more PCA-like and the enlargement parameter decreases. 

\begin{figure*}
\centering
  \includegraphics[width=1\textwidth]{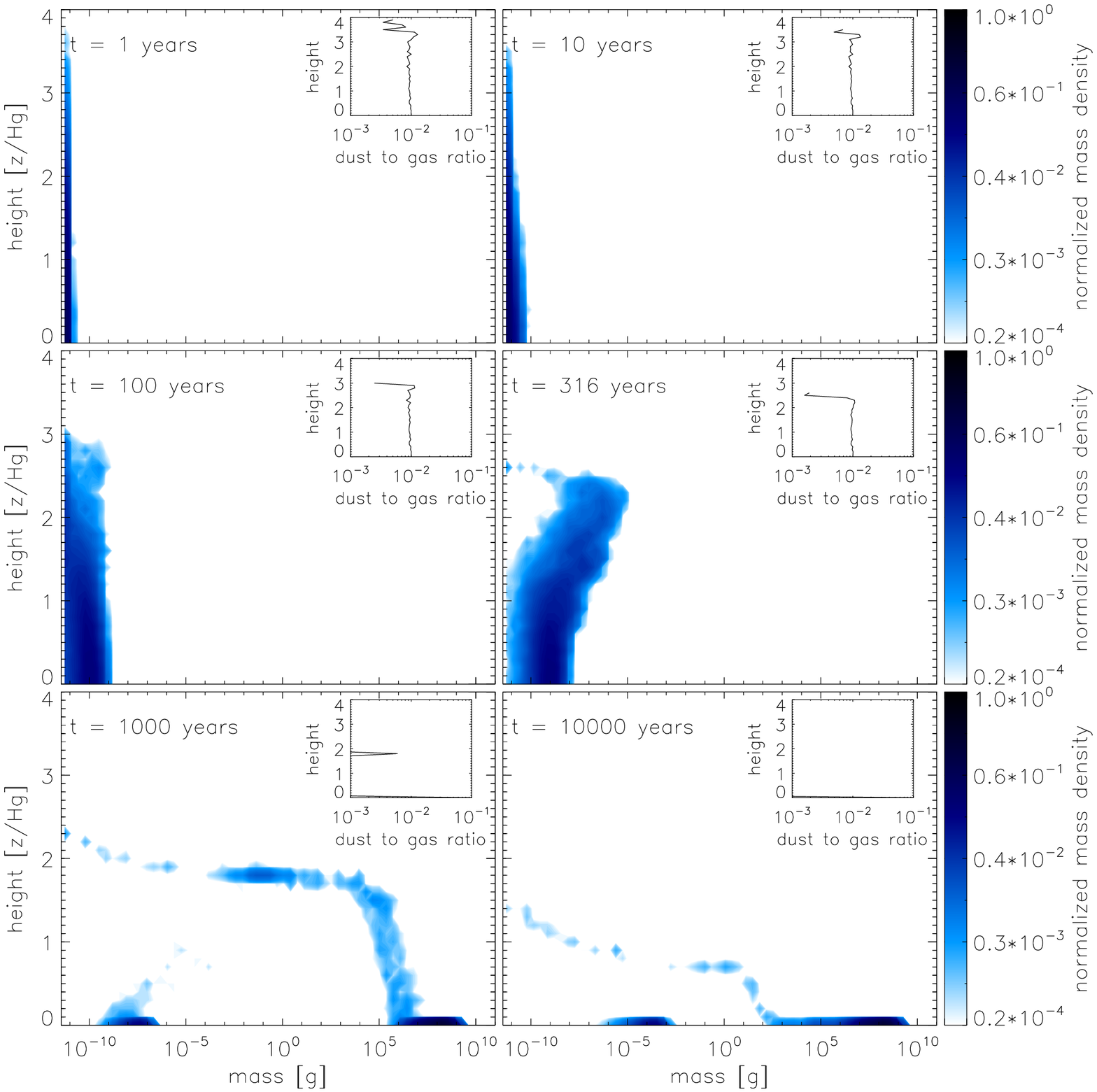}
  \caption{Mass distribution for the DD05 model using the Okuzumi porosity model. The axes and contours are the same as in Fig. \ref{fig:mass_DD05}.}
  \label{fig:mass_DD05_ok}
\end{figure*}

\begin{figure*}
\centering
  \includegraphics[width=1\textwidth]{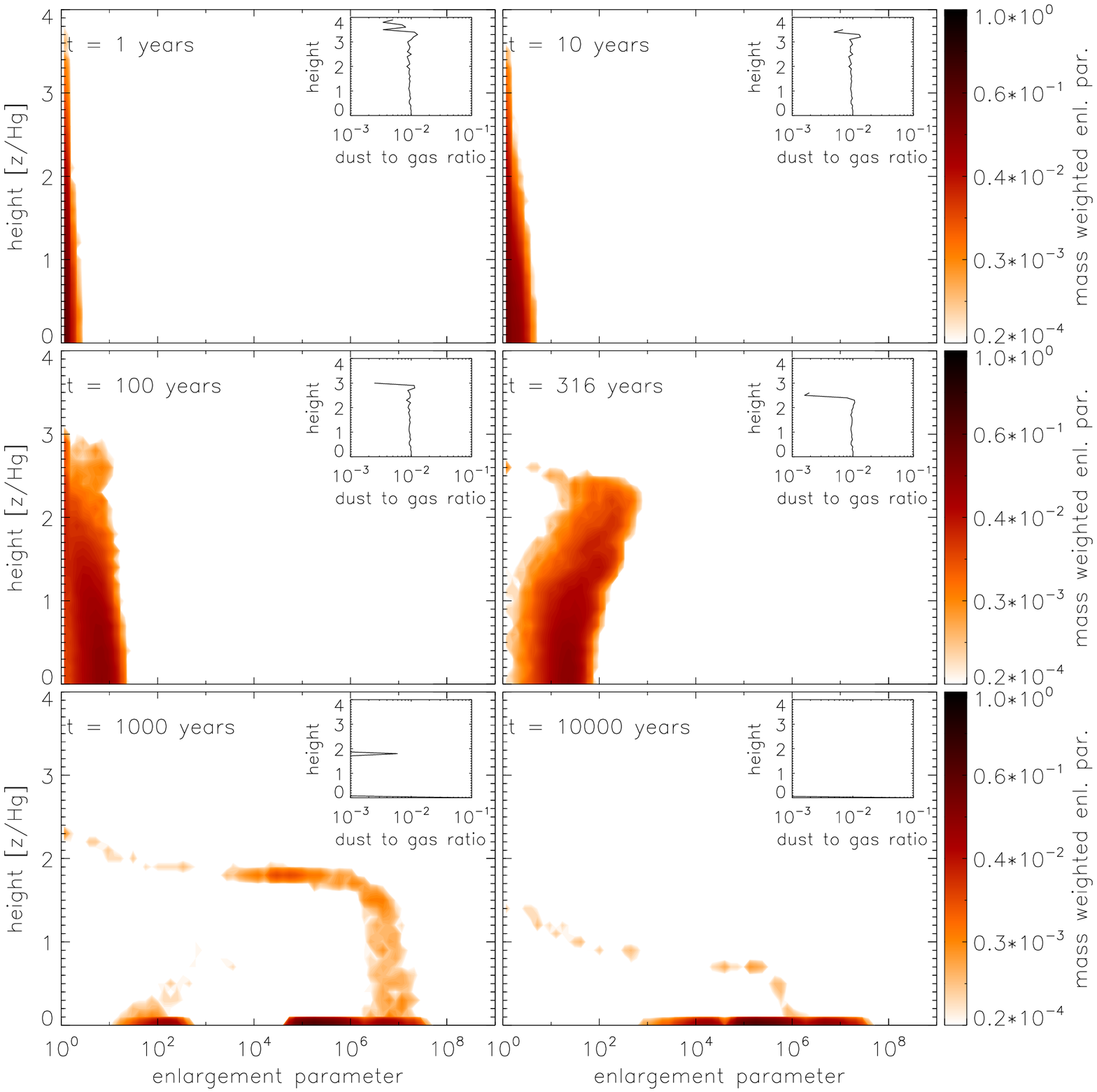}
  \caption{Enlargement parameter distribution for the DD05 model using the Okuzumi porosity model. The axes and contours are the same as in Fig. \ref{fig:enpar_DD05_or}.}
  \label{fig:enpar_DD05_ok}
\end{figure*}

\paragraph{The Okuzumi model - DDb} The second porosity model we use is the one constructed by \cite{Okuzumi2009a}. This collision model defines a third type of aggregation next to the PCA and CCA collisions, that is QCCA (quasi cluster-cluster aggregation - collisions between clusters with a predefined mass ratio). This model is based on numerical models of geometrical sticking (e.g. no restructuring of monomers happens during a collision). 

The mass and the porosity evolution are shown in Figs. \ref{fig:mass_DD05_ok} and \ref{fig:enpar_DD05_ok}. The most striking property of this simulation is the maximum mass and porosity of the particles. We end up with particles of $10^{10}$ g in mass having an enlargement parameter of almost $10^8$ (the compact radius of such an aggregate is some meters, however, the enlarged radius is several kilometers). The stopping time in the Epstein regime (Eq. \ref{eq:ts1}) is proportional to $m/A$. As the Okuzumi-model produces aggregates with a fractal dimension of $\sim 2$ (the mass scales with $a^2$, where $a$ is the particle radius), the stopping time only slightly increases with mass. Therefore particles settle slowly and produce extremely fluffy structures. At one point, however, the particle size becomes larger than the mean free path of the gas, and the aggregate enters the Stokes regime (Eq. \ref{eq:ts2}). As the stopping time is now proportional to $ma/A$, the stopping time more strongly increases with mass. This transition from Epstein to Stokes regimes happens at t=900 yrs for the particles located at 1.7 $H_g$ above the midplane. Once the transition happens for a given particle, it settles to the midplane in a matter of years due to the heavy mass of the aggregate. Therefore $a=\lambda_{\mathrm{mfp}}$ is a natural upper size limit to expect rain-out in this model. Such huge particles are in reality probably very fragile and would break up due to the smallest perturbations in the gas (e.g. turbulent eddies, or the ``fall'' to the midplane after entering the Stokes regime). 

If one uses only the Epstein stopping time (although this is not physical), the particles do not sediment so rapidly to the midplane, their fractal growth goes on unhindered.  

Although porosity can somewhat delay sedimentation (the particles in the DDa and DDb simulations reach the miplane later than in the DD simulation), this delay is limited. A monomer has the lowest available stopping time, as it has the lowest settling velocity. In the disk model used in this chapter, a monomer needs $1.5 \times 10^{3}$ yrs to settle from 4 $H_g$ to 2 $H_g$. No matter what porosity model we use, the sedimentation timescale down to 2 $H_g$ cannot be longer than this.

We must emphasize that any collision model containing exclusively sticking is only valid, if no significant restructuring happens during the collisions (e.g. the collision energy is less than $5 E_{\mathrm{roll}}$, the rolling energy, which is the energy needed to roll two monomers by 90 deg). This condition is clearly not met at all times in our simulations, e.g. the rain-out particles can have collision velocities with the swept up particles as high as several 10 m/s in these simulations. Such collisions would result in catastrophic fragmentation. Therefore the results presented in this section should be considered as toy models.

\subsection{A simplified Braunschweig model}
In this section we construct a simplified version of the collision model described in Chapter \ref{chp:paper2}. We assume sticking, if the collision energy is smaller than $5 E_{\mathrm{roll}}$. Bouncing with compaction is used if the collision energy is greater than $5 E_{\mathrm{roll}}$, but the relative velocity of the two aggregates is less than 1 m/s. Fragmentation occurs if the relative velocity of two aggregates is greater than 1 m/s. The recipe for mass and porosity evolution for bouncing and fragmentation is taken from Chapter \ref{chp:paper2} (our hit \& stick (S1), bouncing with compaction (B1), and fragmentation (F1) collision types). We still assume that the particles grow by Brownian motion and settling only (the effects of turbulence is discussed in the next section), but we use both porosity models discussed in the previous section. 

\begin{figure*}
\centering
  \includegraphics[width=1\textwidth]{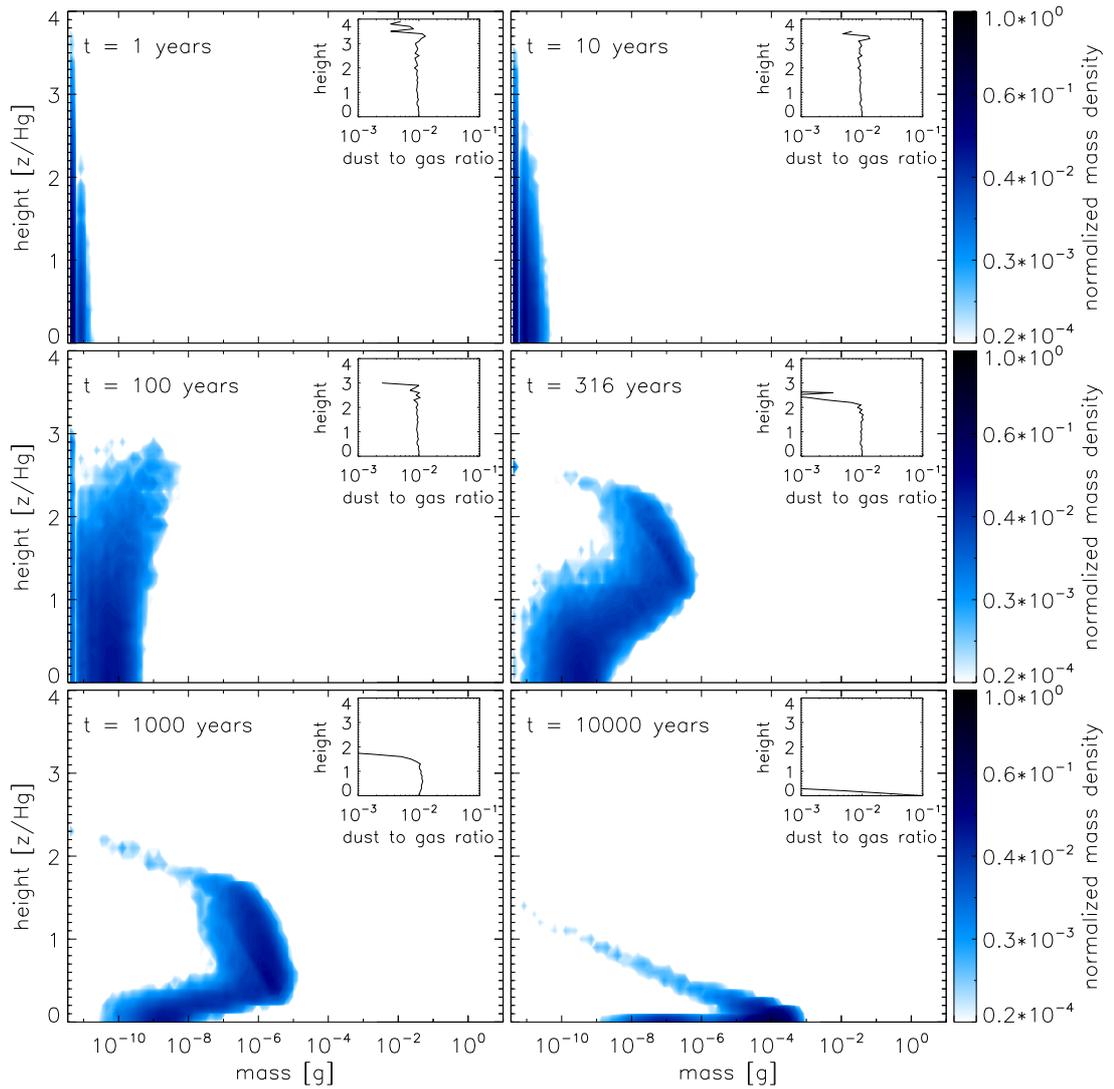}
  \caption{Mass distribution for the simplified Braunschweig model using the Ormel porosity. The axes and contours are similar to Fig. \ref{fig:mass_DD05}. Notice the x axis ranges from $10^{-12}$ g until 10 g in this figure.}
  \label{fig:mass_simpl_not_or}
\end{figure*}

\paragraph{The Ormel model - SB1} The evolution of the mass in case of the Ormel porosity model is shown in Fig. \ref{fig:mass_simpl_not_or}. The mass distributions at $t=1$, 10, 100 yrs are identical to Fig. \ref{fig:mass_DD05_or}. At $t=316$ and $1000$ yrs we see the effects of bouncing at the intermediate energies. The rain-out particles cannot increase their mass, if they suffer bouncing collisions. Therefore the rain-out particles only reach masses of $10^{-5}$ g when they arrive at the midplane. As the rain-out particles are smaller, they settle slower, therefore they reach the midplane only at $t=3000$ yrs. The enlargement parameter is affected by bouncing, but the typical enlargement parameter is between 10 and 100 as in the DD05a model in Fig. \ref{fig:enpar_DD05_or}.

\begin{figure*}
\centering
  \includegraphics[width=1\textwidth]{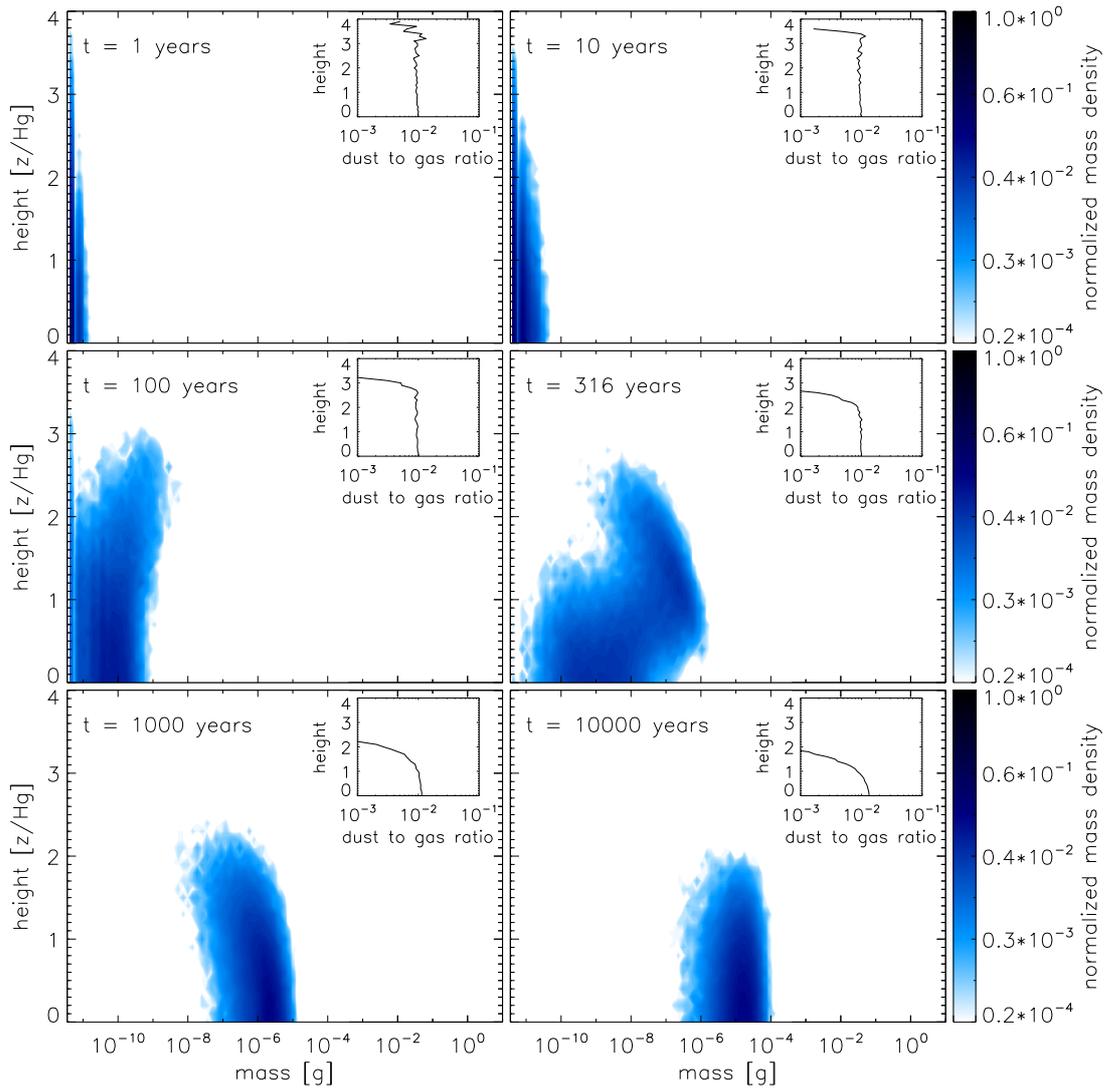}
  \caption{Mass distribution for the simplified Braunschweig model using the Ormel porosity with $\alpha=10^{-4}$. The axes and contours are the same as in Fig. \ref{fig:mass_simpl_not_or}.}
  \label{fig:mass_simpl_1d-4_or}
\end{figure*}

\begin{figure*}
\centering
  \includegraphics[width=1\textwidth]{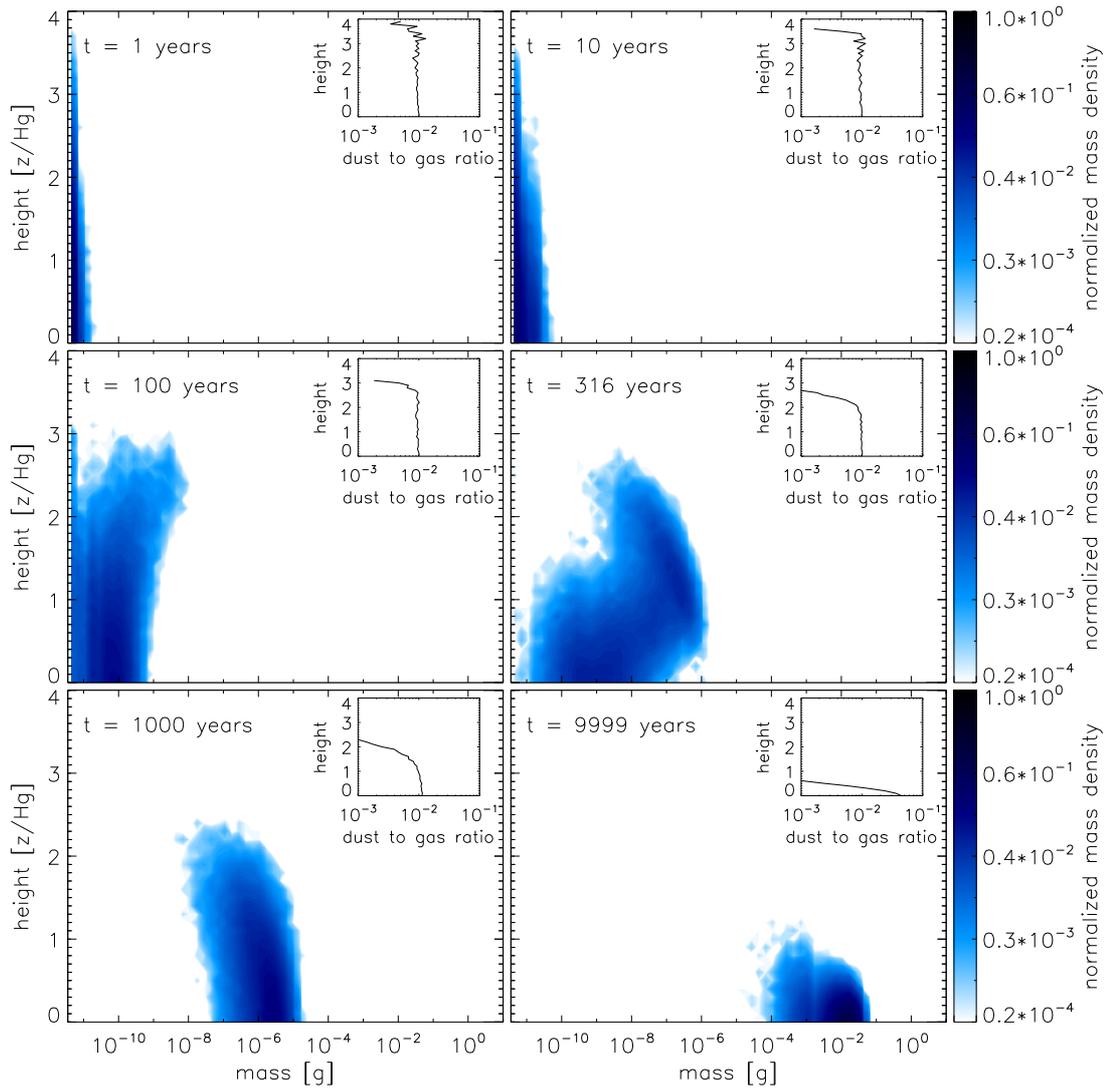}
  \caption{Mass distribution for the complete Braunschweig model using the Ormel porosity with $\alpha=10^{-4}$. The axes and contours are the same as in Fig. \ref{fig:mass_simpl_not_or}. The snapshots of this simulation is presented at the lower right corners of the thesis as a flip-cartoon.}
  \label{fig:mass_br_1d-4_or}
\end{figure*}

\paragraph{The Okuzumi model - SB2} The Okuzumi model produces fluffier particles. As a result the particles are heavier (10$^{-2}$ g) but arrive to the midplane later (at $t=6000$ yrs) than in the previous simulation.  


\subsection{The effects of turbulence}
\label{sec:turb5}

So far all particles sooner or later ended up at the midplane because there was no effect that could counteract settling. In this section we examine the effects of a non-zero turbulence parameter, which can stir particles back up. 

A small turbulence parameter ($\alpha = 10^{-6}$) does not affect significantly the masses of the rain-out particles compared to models SB1-2 of the previous section (see models SB3-4 in Tab. \ref{table:sedi}). As particles do not only settle but also diffuse downward (and upward) due to turbulence, the time the rain-out particles reach the midplane is somewhat shorter than for models SB1 and 2. In all previous simulations an infinitely dense dust layer formed at the midplane of the disk. However, even this low level of turbulence can prevent the formation of this layer and introduce a non-zero (although small) dust scale-height.

The influence of turbulence is more pronounced if $\alpha=10^{-4}$. The mass evolution of the aggregates using the Ormel porosity model is shown in Fig. \ref{fig:mass_simpl_1d-4_or}. The first rain-out particles reach the midplane already at $t=500$ yrs due to downward diffusion, although these particles have lower masses than in model SB1 ($10^{-6}$ g -- therefore, in the absence of turbulence, these particles would reach the midplane later than the particles in SB1). The dust distribution reaches a steady state at $t=10^{4}$ yrs.  

We see that the particle mass is constant as a function of height at $t=10^4$ yrs. As turbulence effectively mixes the particles, and as bouncing prevents further growth or fragmentation (the dust growth is halted), both the masses and porosities of the aggregates are similar at all heights. This has an important consequence for observations. If the turbulence parameter is constant as a function of height (which might not be true -- see \cite{Gammie1996-5} and Sec. \ref{sec:alpha}), we expect that the particles observed at the disk atmosphere have the same properties as the ones located at the disk midplane. If however $\alpha$ is some function of the height (e.g. high at the disk atmosphere and low at the dead zone), we might not be able to constrain the particle properties at the midplane of the disk, unless the disk is optically thin at the given wavelength.  

We also see from these simulations that a higher turbulence value reduces the mass of particles and increases the dust scale height. If turbulence is strong enough ($\alpha \simeq 1$), the dust scale height can be similar to the gas scale height and the disk atmosphere remains dusty at all times. However, such high turbulence value prevents any significant dust growth, which is not a fertile environment for planet formation. 

\begin{figure*}
\centering
  \includegraphics[width=0.9\textwidth]{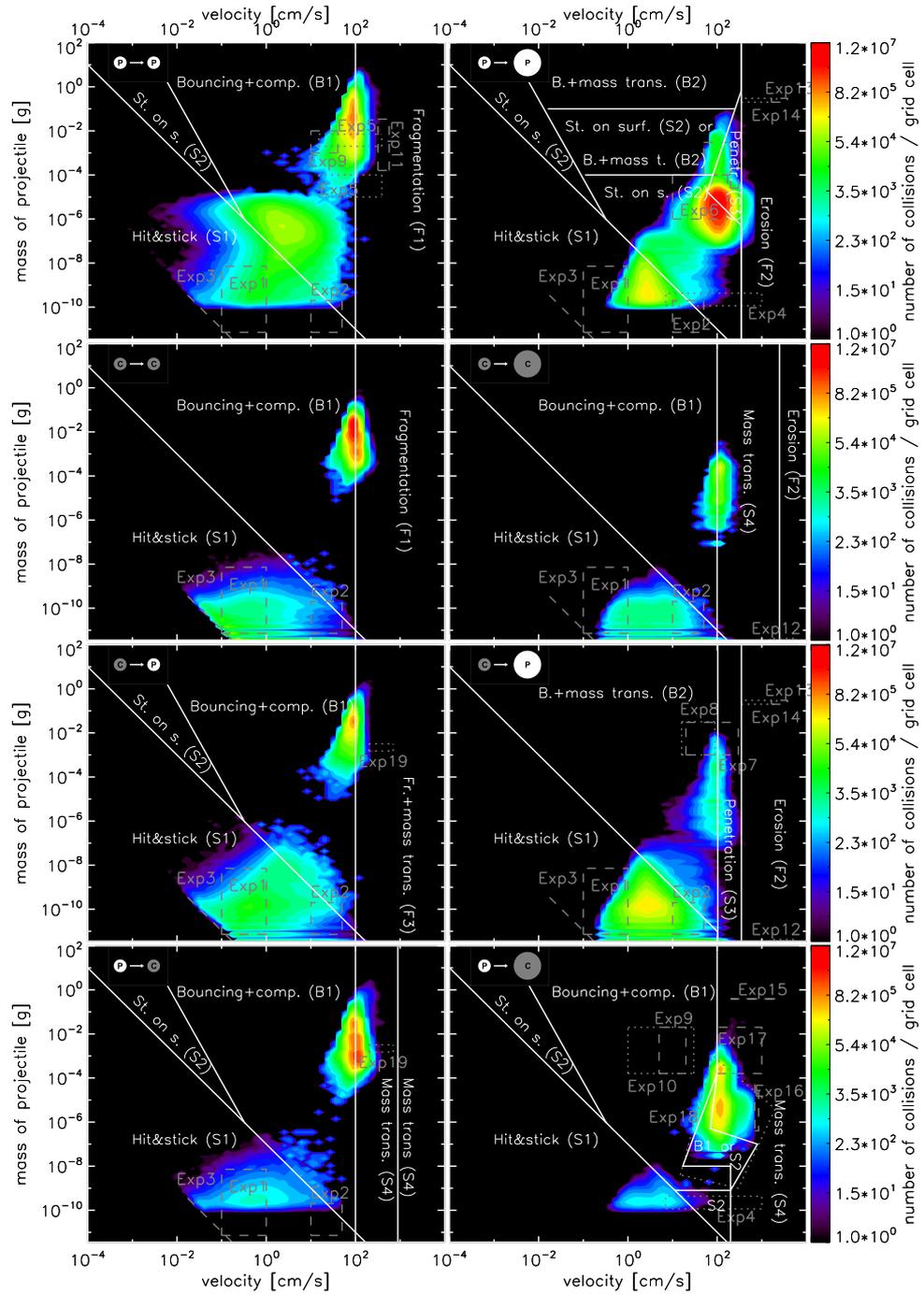}
  \caption{The collision history for the FB1 simulation. The eight regimes of the complete Braunschweig model are shown with the corresponding border lines of the nine different collision regimes (white solid lines). The x axis is the relative velocity of the particles in $cm/s$, the y axis is the mass of the projectile in gram units. The grey boxes indicate the areas that are covered with laboratory experiements (see Chapter \ref{chp:paper2} for more details). The colors indicate how many collisions happened at the given part of the parameter space during the simulation. The red and yellow areas are `hot spots', where most of the collisions take place.}
 \label{fig:coll_hist_br_1d-4}
\end{figure*}

\begin{figure*}
\centering
  \includegraphics[width=1\textwidth]{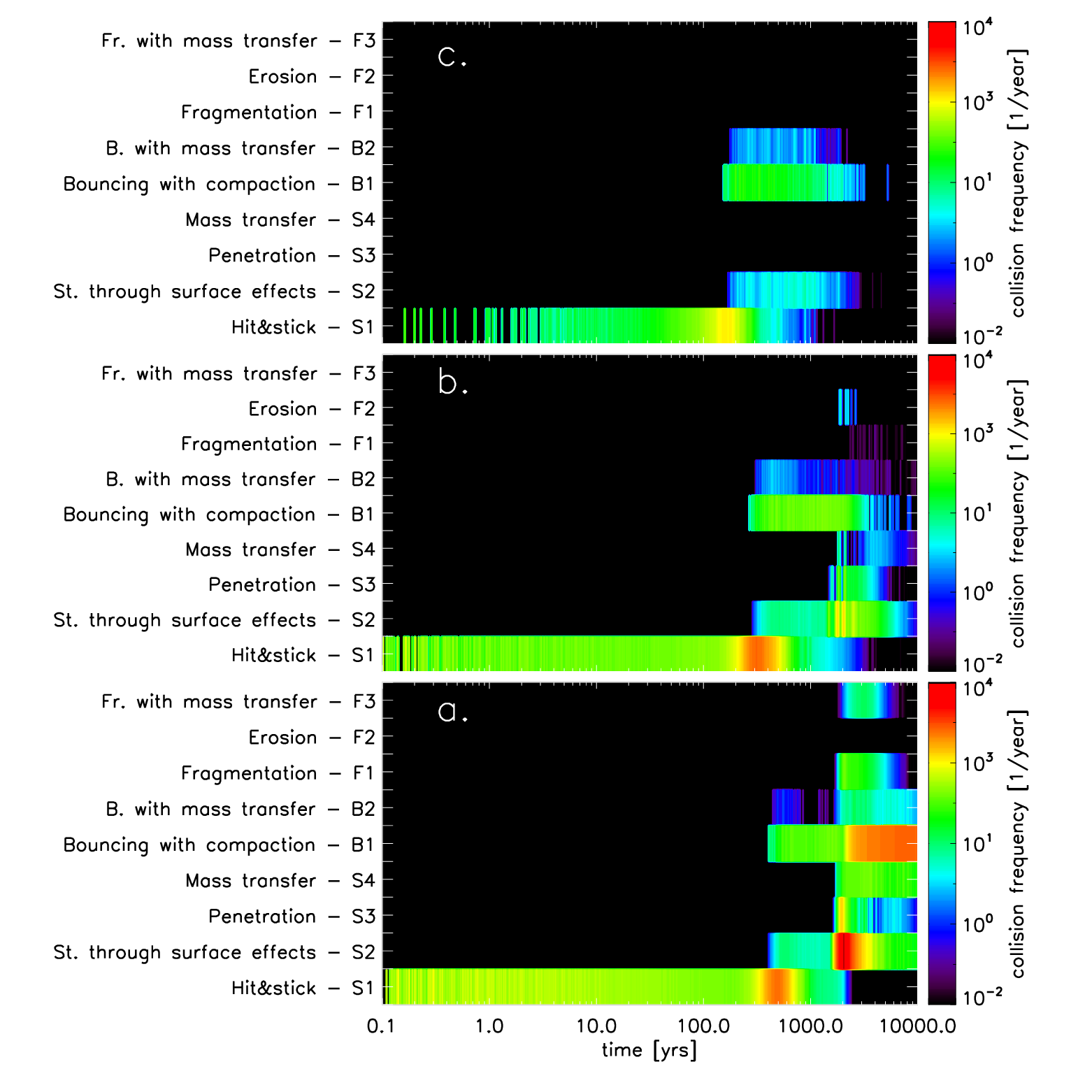}
  \caption{The collision frequency of the nine collision types. The x axis is the time in years, the y axis indicates the collision types. The colors of the stripes indicate the collision frequency (e.g. the number of collisions per year). The collision frequency is shown at the midplane (a.), at 1 pressure scale height above the midplane (b.), and at 2 $H_g$ (c.).}
  \label{fig:coll_hist_br_1d-4_or}
\end{figure*}

\subsection{The complete Braunschweig collision model}
\label{sec:comp_br}
In this Section we use the complete Braunschweig model (see Chapter \ref{chp:paper2} for details), the value of the turbulence parameter is $\alpha=10^{-4}$ and calculations are performed with both the Ormel porosity model (FB1) and the Okuzumi porosity model (FB2).

In the simplified Braunschweig collision model, the growth is halted by bouncing immediately if the particles enter the bouncing regime. However, in the complete Braunschweig model, there are possible ways for growth beyond the hit\&stick border line (that is where $E_{\mathrm{coll}} > 5 E_{\mathrm{roll}}$). The most important area is in the {\pP} regime where a small porous projectile collides with a heavy porous target (see Fig. \ref{fig:colored_regimes}). Due to these ``green'' areas at intermediate collision energies, particles in the FB1 and FB2 simulations grow to higher masses than in the SB5 and SB6 simulations. As a consequence, the scale height of the dust is lower in these simulations, as heavier particles are more difficult to stir up by turbulence. We illustrate the mass distribution at $t=1$, 10, 100, 316, $10^3$, $10^4$ yrs in Fig. \ref{fig:mass_br_1d-4_or}. 

The particle evolution has two phases in these simulations. The first 1000 yrs are identical for the SB5/SB6 and FB1/FB2 simulations, respectively (see also the first five snapshots of Figs. \ref{fig:mass_simpl_1d-4_or} and \ref{fig:mass_br_1d-4_or}). In this phase, particles start sedimenting, and the rain-out particles reach the midplane. The dust evolution in the SB5 and SB6 models halts at this point as only bouncing collisions happen. However, during the second phase of the FB1 and FB2 simulations, a short run-away growth appears as the relative velocities are ``boosted'' at this point (see Sec. \ref{subsec:mmsn_early} for a detailed explanation). Due to the rapid growth, particles reach $10^{-2}$ g in mass for the FB1 and FB2 simulations.

Figure \ref{fig:coll_hist_br_1d-4} illustrates the collision history of the FB1 simulations. If we compare this figure to Figs. \ref{lowd_cont}, \ref{mmsn_cont}, and \ref{desch_cont}, we see that the features are more smeared out in Fig. \ref{fig:coll_hist_br_1d-4} than in the other figures. As we simulate here several boxes at different heights above the midplane, the physical conditions (e.g. gas density and sedimentation velocity) at the midplane and at the upper scale heights of the disk are different, which is responsible for the smeared out features of Fig. \ref{fig:coll_hist_br_1d-4}. 

It is more interesting to investigate the collision frequency of the nine collision types as a function of time (see Fig. \ref{fig:coll_hist_br_1d-4_or}). If one compares this figure with Figs.  \ref{lowd_pics}c, \ref{mmsn_pics}c, and \ref{desch_pics}c, we immediately see that the diversity of occurring collision types is much greater in the FB1 (and also in the FB2) simulation, although the strength of the turbulence is the same in all cases ($\alpha=10^{-4}$). This can be explained by the presence of sedimentation. The particle population in a given box is not fixed as in Chapter \ref{chp:paper3}. During the rain-out process, heavy particles coming from 1-2 $H_g$ `hit' the particle population at the midplane. Due to this process, the relative velocity between the small, midplane particle population and the generally larger rain-out population is increased, thus collision types can occur that require larger collision energies.

\section{Discussion and future work}
\label{sec:disc}

\subsection{Porosity models}
\label{sec:por models}
We examined the effects of two porosity models on settling in this Chapter. Here we critically discuss the strong and weak points of these models.

The porosity model of \cite{Ormel5:2007p93} only treats PCA and CCA collisions (particle-cluster and cluster-cluster aggregation, respectively) and ``constructed'' semi-analytical recipes for intermediate size ratios. \cite{Okuzumi2009a} improved on this by modeling the quasi-CCA collisions (QCCA), where two clusters with a given mass ratio can collide. QCCA growth generally produces fluffier particles. 

However, the porosity model of \cite{Okuzumi2009a} assumes collisions of pure geometrical sticking (no restructuring). This assumption is correct if the collision energy is \textit{always} much lower than the rolling energy, e.g. during particle growth by only Brownian motion. In that case the collision velocity of the aggregates is decreasing as the particles grow, which ensures that the collision energy is always much smaller than the rolling energy. However, if particles grow also by differential settling, turbulence, or differential radial drift, the assumption might not be correct. If for example a collision happens with an energy just slightly smaller than $5 E_{\mathrm{roll}}$, the restructuring for a single collision can be negligible. However, if an aggregate experiences many such collisions, the cumulative effect of the small restructuring might not be negligible anymore, as the restructuring does not behave as a random error, but it is additive in nature.

The porosity model of \cite{Ormel5:2007p93} is more robust because restructuring at energies $\sim E_{\mathrm{roll}}$ can be included (but we do not include it, see Sec. \ref{para:ormel}), however it most probably underestimates the enlargement parameter by not taking into account the QCCA collisions. 

We propose the development of a porosity model that combines the advantages of both models, e.g. it incorporates QCCA collisions, but does not assume geometrical sticking. 

\subsection{Where can $\alpha$ be constant as a function of height?}
\label{sec:alpha}
\cite{Gammie1996-5} proposed the concept of layered accretion disks. If the ionization fraction of the gas is not sufficient to support magneto-rotational instability (MRI - \cite{Balbus1991}), the turbulence parameter drops and a dead-zone forms at the midplane of the disk. The extent of the dead-zone is uncertain, as the ionization processes of the gas are not well-constrained. For typical TTauri disks it can extend between 0.1 - 4 AU \citep{D'Alessio1998-5}. Inside 0.1 AU, the thermal radiation from the star can keep the dust sufficiently ionized for MRI, and outside 4 AU, the gas surface density is typically below 100 g/cm$^2$, therefore cosmic rays can penetrate the disk and keep it MRI active at all heights. 

If a disk can be observationally resolved down to 4 AU or inside 0.1 AU and one can be sure that no dead-zone is present at the resolved location of the disk, the dust properties at the whole column can be constrained even if the disk is not optically thin at the given wavelength (see Sec. \ref{sec:turb5}).

It would be however interesting to investigate how dust evolves in a layered disk model. Small dust particles can very efficiently sweep up charges in the gas. As shown by \cite{Turner2010}, the dead-zone can extend to 2 $H_g$ for 1 micron sized particles, but it shrinks below 0.5 $H_g$ for aggregates that are 100 micron in size. In a simulation like the one presented here, this would mean that as the particles grow, the dead-zone shrinks. When the dead-zone disappears, the whole disk becomes MRI active and the particles settled to the midplane might be fragmented and stirred back up. This could lead to an oscillatory process. 

\subsection{Can bouncing keep the disk atmospheres dusty?}
Another way to keep disk atmospheres dusty might be possible via bouncing. In the previous models of dust evolution, the cycle between growth and fragmentation provided a source for small particles (see e.g. \cite{Brauer2008a, Birnstiel2009}). However, the dominant collision type in our simulations is bouncing. 

\cite{Weidling2009} observed that the mass of the particles in their multiple bouncing experiments were reduced by the end of the experiment and it was unclear why this happened. We propose that small pieces can grind off from the aggregates while they bounce. These grind off bits and pieces can provide a continuous source of small particles. 

It is possible to theoretically constrain the critical amount of grind off particles that is necessary to keep the disk atmospheres dusty using the models presented in this Chapter. In more detailed laboratory experiments, it should be possible to measure whether that mass-loss is really present.

\section{Summary}
\label{sec:concl}
We performed simulations in a 1D vertical column of a protoplanetary disk to better understand the process of sedimentation. We simultaneously solved for the particle motion and growth inside this column. The complexity of the models was gradually increased to examine the effects of different processes. The first simulation used a collision model that only contained sticking, we furthermore assumed that the particles were compact, and the turbulence parameter ($\alpha$) was set to zero. Later on we investigated the effects of different porosity models, more realistic collision models (with sticking, bouncing and fragmentation) and turbulence of different strengths. Below we summarize our results.

\begin{itemize}
\item Porosity helps to produce heavier particles, and it can somewhat delay sedimentation (e.g. increase the sedimentation timescale), although this delay is limited. Without the stirring effect of turbulence, particles inevitably settle down to the midplane.

\item Upon using the porosity model of \cite{Ormel5:2007p93}, the enlargement parameter of the particles is generally between 10 and 100. Upon using the \cite{Okuzumi2009a}, the enlargement parameter is between $10^3$ and $10^4$.

\item Bouncing prevents particles to reach masses greater than $\sim 1$ g (the exact value depends on the disk, porosity, and collision models).

\item As bouncing results in a narrow size distribution and halts particle growth, turbulent mixing equalizes the particle properties. The mass and porosity is constant as a function of height above the midplane, if the turbulence parameter is constant. Unless a dead-zone is present at the midplane, the particle properties observed in the disk atmosphere directly reflect those at the midplane. 

\item A higher value of turbulence decreases the particle masses but it increases the dust scale height. Using the simplified Braunschweig model with $\alpha=10^{-6}$ results in a dust scale height of 0.1 $H_g$ and final particle mass of $10^{-2}$ g, however the dust scale height is $0.5 H_g$, and the final particle mass is $10^{-5}$ g when using $\alpha=10^{-4}$. Therefore, a sufficiently high turbulence value can keep the disk atmosphere dusty but the absence of significant dust growth is not favorable for planet formation.

\item When using the most detailed collision model up to date (the Braunschweig collision model), we obtain particle masses of $10^{-2}$ g (with an average radius of 1 mm, and an average Stokes number of $2 \times 10^{-2}$) and a dust scale height of 0.25 $H_g$ in the considered disk model. 

\end{itemize}


%% file: Chapters/Chapter7.tex

\chapter{Conclusions and Outlook} 
\label{chp:concl}
\lhead{Chapter 6. \emph{Conclusions and Outlook}} 
\rhead{}

In this thesis, numerical Monte Carlo simulations of dust growth were performed in local box (0D) and vertical column (1D) models using a laboratory experiment-based collision model to better constrain the initial stages of planet formation. Previous dust coagulation simulations usually used a simplified collision model with sticking and fragmentation. However, the laboratory experiments performed during the last 15-20 years revealed a variety of different collision types (penetration, erosion, bouncing, etc.). The main idea behind this thesis was to construct a collision model using all the available laboratory data for silicate particles and implement this collision model into a dust evolution code. 

We identified nine different collision types based on 19 experiments and constructed eight collision regimes based on the mass ratio and the porosity of the particles (similar sized versus different sized particles, fluffy versus compact particles). The experiments showed that sticking collisions mostly happen at low collision energies. At intermediate energies, bouncing is dominant; and if the relative velocity of the two colliding aggregates is higher than 1 m/s, fragmentation occurs. Naturally the whole parameter space cannot be covered by experiments, therefore we had to extrapolate at areas where no experiments were performed. Upcoming new experiments have to confirm whether these extrapolations are correct or the collision model needs to be modified.

Calculations using previous collision models (based on sticking and fragmentation only) showed that an equilibrium between fragmentation and sticking is reached. In our collision model however this does not happen since the particles enter the bouncing regime and at that point their growth is halted. This result has strong consequences. 1.) As small particles are not produced in a continuous manner by fragmentation, it is not clear what mechanism can keep the disk atmospheres dusty throughout the lifetime of the disk as it is observed. 2.) As the growth of the particles is halted already at intermediate collision energies, the particles produced in our simulations are generally smaller than in previous simulations of dust growth, which is not favorable for planetesimal formation. We performed simulations of three different collision models, using three different turbulence strengths and we found that the Stokes number of these particles is rather insensitive to the disk parameters. It is always around $St=10^{-4}$. It has to be investigated whether these small particles are large enough for particle clumping mechanisms \citep{Cuzzi2008, Johansen:2007p65-7} to form planetesimals. 

Previous models of the vertical structure of disks suggested that in order to keep the dust atmospheres dusty for the observed lifetime of the disk (some 10$^6$ yrs), some kind of grain retention mechanism is needed. However, as discussed earlier, bouncing halts particle growth therefore small particles are not produced by fragmentation. For this reason we revisited the problem of vertical settling using the new collision model and we were searching for ways to delay particle sedimentation. The effects of different porosity models were investigated and we found that porosity alone cannot sufficiently delay sedimentation. A simplified Braunschweig collision model was constructed using sticking, bouncing and fragmentation only. Such a collision model slows down sedimentation as particles are kept small ($10^{-4}$ - 10$^{-3}$ g) and even an intermediate level of turbulence can stir up and keep these particles levitated. Upon using the complete Braunschweig model, we found that particles can reach higher masses ($10^{-2}$ g) and as a results, the scale height of the dust is smaller.

There are various opportunities for future work, most of which are highlighted in the discussion sections of the various chapters. Here I mention a few of them which I believe are of major importance. 

\textbf{Planetesimal formation by coagulation:} \cite{Brauer2008b-7} showed that a pressure bump stops the radial drift of the particles through the disk and in these areas the dust particles are accumulated, therefore they can form planetesimals by coagulation. This problem should be revisited using the collision model described in this thesis as bouncing might prevent the formation of these planetesimals. 

\textbf{The radial drift and stopping time of aggregates:} The radial drift of aggregates through the gas disk is one of the biggest problems of planet formation as the solid material can be rapidly lost to the star. It is therefore astonishing that the most commonly used formulae to calculate the stopping time of aggregates (the coupling strength of the dust to the gas, which determines the radial drift) were derived in the '70 and at earlier times (Epstein regime -- \cite{Epstein1924}; Stokes regime -- \cite{Whipple1972-7}). The basic assumption that enters these models is that the aggregates are spheres. Since then, not so much work was spent on investigating the problem \citep{Meakin1988, Nakamura1998}. It would be worthwhile to study by numerical simulations how the stopping time is altered for e.g. fractal or fluffy aggregates. Using such non-uniform or non-spherical structures, a significant amount of the collision energy between a gas atom/molecule and the aggregate might be converted into rotation energy instead of kinetic energy. Such calculations might show that the radial drift is reduced for aggregates and new, more realistic formulae for the stopping time could be derived.

\textbf{New hit \& stick porosity model:} We investigated two porosity models in Chapter \ref{chp:sedi}. The porosity model of \cite{Ormel:2007p93-7} probably underestimates the fluffiness of the aggregates by not taking into account collisions between different sized aggregate structures. The model of \cite{Okuzumi2009a-7} treats such collisions, however their model is based on geometrical sticking (no restructuring). Therefore their model might overestimate the aggregate porosity. We therefore propose the development of a new porosity model that takes into account collisions between different sized aggregates, and in the same time does not assume geometrical sticking. As the porosity determines the coupling strength between the dust and the gas, better understanding of the porosity is crucial for coagulation calculations. 

\textbf{Properties of icy and organic monomers:} In this thesis we considered the coagulation of silicates with 1.5 micron diameter, however other materials and sizes should be considered, too. Performing laboratory experiments with such materials is a challenge, but their importance for planet formation is potentially great. Such monomers might have much better sticking properties than silicates, therefore these materials can turn out to be favorable for planet formation. Although some experiments were performed on macroscopic bodies (see Sec. \ref{sec:experiments}), we still do not know the properties of microscopic bodies.